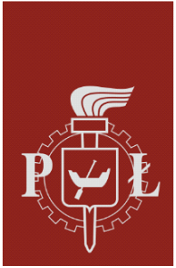
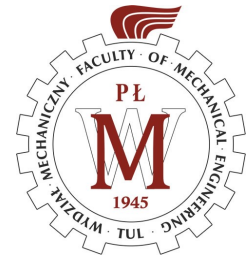

Politechnika Łódzka
Wydział Mechaniczny
Katedra Automatyki,
Biomechaniki i Mechatroniki

mgr inż. Krystian Polczyński

# Rozprawa Doktorska
w dyscyplinie Inżynieria Mechaniczna

## Modelowanie i dynamika układów wahadeł poddanych działaniu niestacjonarnego pola magnetycznego

Promotor pracy:
prof. dr hab. inż. Jan Awrejcewicz

Promotor pomocniczy:
dr inż. Adam Wijata

Łódź 2023 r.

# Spis treści

# Wykaz symboli

| | | |
|---|---|---|
| $A_1, A_2$ | | amplituda drgań wahadła 1 i 2 |
| $C_\mu$ | | stała zależna od przenikalności magnetycznej ośrodka |
| $E_{11}$ | | całkowita energia wahadła 1 odniesiona do jego momentu bezwładności |
| $E_{12}, E_{21}$ | | czynnik energetyczny odniesiony do momentu bezwładności wahadła i informujący o przesunięciu fazowym między nimi |
| $E_{22}$ | — | całkowita energia wahadła 2 odniesiona do jego momentu bezwładności |
| $E$ | | całkowita energia wahadeł 1 i 2 |
| $E_{ij}$ | — | element symetrycznej "macierzy energii" w układzie sprzężonych wahadeł magnetycznych ($i, j = 1, 2$) |
| $E'_u$ | | pochodna po czasie całkowitej energii wahadeł 1 i 2 |
| $F_{exp}$ | | siła w cięgnie podczas eksperymentu wyznaczania momentu od działywania magnetycznego pomiędzy magnesem a cewką elektryczną |
| $F_{mag}$ | | siła wzajemnego oddziaływania pomiędzy odseparowanymi biegunami dipola magnetycznego w modelu Gilberta |
| $G$ | | prawa strona równania ruchu oryginalnego układu pojedynczego wahadła |
| $G_a$ | | prawa strona równania ruchu przybliżonego układu pojedynczego wahadła |
| $I_0$ | | amplituda prostokątnego pulsującego sygnału prądowego cewki |
| $I_{B0}$ | | zmodyfikowana funkcja Bessela pierwszego rodzaju rzędu 0 |
| $I_{B1}$ | — | zmodyfikowana funkcja Bessela pierwszego rodzaju rzędu 1 |
| $J; J_1, J_2$ | | masowy moment bezwładności wahadła względem osi obrotu; masowy moment bezwładności wahadła 1 i 2 względem osi obrotu |

| | |
|---|---|
| $M_{Amag2}$; $M_{Amag4}$; $M_{Amag6}$ | modele wielomianowe momentów magnetycznych o następujących stopniach wielomianu mianownika: 2, 4 i 6 |
| $M_{CM}$ | model coulombowski oporów ruchu z tłumieniem wiskotycznym |
| $M_F$; $M_{F1}$, $M_{F2}$ | — moment oporów ruchu; moment oporów ruchu wahadła 1 i 2 |
| $M_{Fexp}$ | moment oporów ruchu (eksperymentalny) |
| $M_{Fmag}$ | — moment magnetyczny wg modelu Gilberta |
| $M_K$ | moment generowany przez gumowy element podatny |
| $M_{Rmag}$ | model wielomianowego momentu magnetycznego |
| $M_{SE}$ | — model momentu oporów ruchu z efektem Stribecka i tłumieniem wiskotycznym |
| $M_{Sexp}$ | moment skręcający gumowy element podatny (eksperymentalny) |
| $M_c$; $M_{c1}$, $M_{c2}$ | — stała wartość momentu oporów coulombowskich (kinetycznych); stała wartość momentu oporów coulombowskich (kinetycznych) dla wahadła 1 i 2 |
| $M_{exp}$ | moment siły oddziaływania magnetycznego pomiędzy magnesem a cewką elektryczną (eksperyment) |
| $M_{mag}$ | model gaussowski momentu magnetycznego |
| $M_s$ | stała wartość statycznego momentu oporów ruchu |
| $\widehat{M}_{Amag2}$; $\widehat{M}_{Amag4}$; $\widehat{M}_{Amag6}$ | — modele wielomianowe momentów magnetycznych dla zmiennego prądu cewki o następujących stopniach wielomianu mianownika: 2, 4 i 6 |
| $\widehat{M}_{Rmag}$ | model wielomianowego momentu magnetycznego dla zmiennego prądu cewki |
| $\widehat{M}_{mag}$ | model gaussowskiego moment magnetyczny dla zmiennego prądu cewki |
| $P$ | współczynnik podziału energii pomiędzy wahadłami 1 i 2 |
| $P_0$ | — położenie stacjonarne współczynnika $P$ podziału energii |
| $P'_u$ | pochodna po czasie współczynnika podziału energii pomiędzy wahadłami 1 i 2 |
| $Q$ | — "wskaźnik koherencji" zależny od przesunięcia fazowego między wahadłami 1 i 2 |
| $Q_{mag}$ | moment pochodzący od oddziaływania magnetycznego pomiędzy cewką a magnesem |

| | | |
|---|---|---|
| $R_1; R_2$ | | prawa strona równania na pochodną po czasie amplitudy $k$; prawa strona równania na pochodną po czasie przesunięcia fazowego $u$ |
| $R_D$ | | promień polimerowego dysku |
| $T_p$ | | okres drgań swobodnych wahadła |
| $T_t$ | — | okres wymuszenia |
| $V$ | | całkowita energia potencjalna układu |
| $V_{M_{F_{mag}}}$ | | potencjał momentu magnetycznego dla modelu Gilberta |
| $V_{M_{mag}}$ | — | potencjał momentu magnetycznego dla modelu gaussowskiego |
| $V_J$ | | energia potencjalna zachowawczego zlinearyzowanego układu sprzężonych wahadeł odniesiona do masowego momentu bez władności |
| $V_{graw}$ | | energia potencjalna pochodząca od grawitacji dla zlinearyzowanego wahadła |
| $V_{pM}$ | | zlinearyzowana energia potencjalna pola magnetycznego |
| $V_s$ | | suma energii potencjalnej pola grawitacyjnego oraz elementu sprężystego |
| $V_{spr}$ | | energia potencjalna zgromadzona w sprężynie sprzęgającej wahadła |
| $a$ | | parametr modelu gaussowskiego momentu magnetycznego i jego potencjału |
| $a_I$ | | prądowy współczynnik kierunkowy |
| $b$ | | parametr modelu gaussowskiego momentu magnetycznego i jego potencjału |
| $c; c_{w1}, c_{w2}$ | | współczynnik tłumienia wiskotycznego; współczynniki tłumienia wiskotycznego dla wahadła 1 i 2 |
| $c_1$ | | parametr funkcji piłokształtnej $g_u$ |
| $c_2$ | | parametr funkcji piłokształtnej $g_k$ |
| $c_e$ | — | współczynnik tłumienia wiskotycznego gumowego elementu podatnego |
| $c_{et}$ | | współczynnik tłumienia wiskotycznego stalowej sprężyny |
| $d$ | | odległość między środkami odseparowanych biegunów dipola magnetycznego mierzona w położeniu innym niż dolne położenie równowagi wahadła |
| $d_1$ | | parametr funkcji piłokształtnej przesunięcia fazowego $u$ |
| $d_2$ | — | parametr funkcji piłokształtnej amplitudy $k$ |
| $d_g$ | | odległość między środkami odseparowanych biegunów dipola magnetycznego mierzona w dolnym położeniu równowagi wahadła |
| $f$ | | częstotliwość prostokątnego pulsującego sygnału prądowego cewki |
| $f_j$, $f_1$, $f_2$ | | funkcje opisujące tłumienie, sprzężenie mechaniczne i wpływ pola magnetycznego na wahadło 1 i 2 ($j = 1, 2$) |
| $f_N$ | | częstotliwość drgań swobodnych pojedynczego wahadła magnetycznego |
| $g$ | | przyspieszenie grawitacyjne |
| $g_k$ | | funkcja piłokształtna aproksymująca amplitudę $k$ |
| $g_u$ | — | funkcja piłokształtna aproksymująca przesunięcie fazowe $u$ |
| $i; i_1, i_2$ | | sygnał prądu płynącego w cewce elektrycznej; sygnał prądu płynącego w cewce elektrycznej wahadła 1 i 2 |
| $i_0$ | — | parametr modelu prostokątnego pulsującego sygnału prądowego |
| $i_A$ | | amplituda sygnału prądowego |
| $i_p$ | | sygnał prostokątnego pulsującego prądu płynącego w cewce elektrycznej |
| $k$ | | amplituda rozwiązania opisującego ruchu drgający wahadła |
| $k_e$ | | współczynnik sztywności gumowego elementu podatnego |
| $k_{et}$ | | współczynnik sztywności stalowej sprężyny |
| $k_n$ | | liczba naturalna |
| $\langle k \rangle$ | | wartość średnia amplitudy $k$ |
| $m$ | | masa wahadła |
| $m_{odw}$ | | masa odważnika |
| $n$ | | wartość okresowości drgań wahadła w jednym dołku potencjału |
| $p$ | | parametr modelu wielomianowego momentu magnetycznego |
| $p_I$ | | prądowy współczynnik kierunkowy |
| $q$ | | parametr modelu wielomianowego momentu magnetycznego |
| $q_{m1}; q_{m2}$ | | wartości ładunków odseparowanych biegunów dipola magnetycznego |
| $q_t$ | | współczynnik skalujący czas |
| $r$ | | odległość pomiędzy biegunami dipola magnetycznego |
| $s; s_1, s_2$ | | odległość pomiędzy środkiem ciężkości wahadła a jego osią obrotu; odległość pomiędzy środkiem ciężkości wahadła 1 i 2 a osiami obrotu |
| $t$ | | czas |
| $t_0$ | — | parametr modelu prostokątnego pulsującego sygnału prądowego |
| $t_{maxOFF}$ | | czas przez jaki cewka powinna być niezasilana przy maksymalnym wypełnieniu sygnału prądowego, aby spełniony został ruch układu według scenariusza I |
| $t_{maxON}$ | — | czas przez jaki cewka powinna być zasilana przy maksymalnym wypełnieniu sygnału prądowego, aby spełniony został ruch układu według scenariusza I |
| $t_{minOFF}$ | | czas przez jaki cewka powinna być niezasilana przy minimalnym wypełnieniu sygnału prądowego, aby spełniony został ruch układu według scenariusza I |
| $t_{minON}$ | | czas przez jaki cewka powinna być zasilana przy minimalnym wypełnieniu sygnału prądowego, aby spełniony został ruch układu według scenariusza I |
| $u$ | | przesunięcie fazowe rozwiązania opisującego ruch drgający wahadła |
| $\langle u \rangle$ | — | wartość średnia przesunięci fazowego $u$ |
| $v_j, v_1, v_2$ | | prędkość kątowa wahadła 1 i 2, $(j = 1, 2)$ |
| $w$ | — | wypełnienie prostokątnego pulsującego sygnału prądowego cewki |
| $w_{max}$ | | maksymalna wartość wypełnienia sygnału prądowego dająca jednookresowe oscylacje wahadła według scenariusza I |
| $w_{min}$ | — | minimalna wartość wypełnienia sygnału prądowego dająca jednookresowe oscylacje wahadła według scenariusza I |
| $w_{nT}$ | | wartość wypełnienia sygnału prądowego dająca wielookresowe oscylacje wahadła według scenariusza I |
| $\Delta$ | | przesunięcie fazowe między wahadłem 1 i 2 |
| $\Delta_0$ | | położenie stacjonarne przesunięcia fazowego |
| $\Delta'_u$ | | pochodna po czasie przesunięcia fazowego $\Delta$ |
| $\Omega$ | | częstość drgań własnych zlienaryzowanego wahadła |
| $\Omega_{nT}$ | | okres rozwiązania wielookresowego dla wahadła drgającego w jednym dołku potencjału |
| $\Omega_t$ | | częstość drgań rozwiązania opisującego ruch drgający wahadła |
| $\alpha$ | | znormalizowany parametr tłumienia wiskotycznego stalowej sprężyny |
| $\beta$ | | znormalizowany parametr sztywności stalowej sprężyny |
| $\delta$ | | "szybka" faza |
| $\delta'_u$ | — | pochodna po czasie "szybkiej" fazy $\delta$ |
| $\epsilon$ | | mały parametr zaburzenia |
| $\varepsilon_I$ | | parametr regularyzacyjny w modelu prostokątnego pulsującego sygnału prądowego |
| $\varepsilon_c$ | | parametr regularyzacyjny |
| $\zeta_1, \zeta_2$ | | znormalizowane parametry oporów ruchu wahadeł 1 i 2 |
| $\eta$ | — | numeryczny parametr |
| $\theta$ | | faza rozwiązania opisującego ruch drgający wahadła |
| $\kappa$ | | parametr modelu aproksymującego momentu oporów ruchu z efektem Stribecka |
| $\lambda_1, \lambda_2, \lambda_3, \lambda_4$ | | wykładniki Lapunowa |
| $\lambda_{a1}$ | | znormalizowana energia wahadła 1 |
| $\lambda_{a2}$ | | znormalizowana energia wahadła 2 |
| $\mu$ | | parametr modelu aproksymującego momentu oporów ruchu z efektem Stribecka |
| $\nu_s$ | — | prędkość Stribecka |
| $\nu_w$ | | parametr zależny od wypełnienia prostokątnego pulsującego sygnału prądowego |
| $\sigma$ | — | parametr modelu aproksymującego momentu oporów ruchu z efektem Stribecka |
| $\tau$ | | okres prostokątnego pulsującego sygnału prądowego cewki |
| $\tau_{OFF}$ | | czas przez jaki cewka powinna być niezasilana, aby otrzymać rozwiązanie wielookresowe według scenariusza I |
| $\tau_{ON}$ | | czas przez jaki cewka powinna być zasilana, aby otrzymać rozwiązanie wielookresowe według scenariusza I |
| $\tau_t$ | — | bezwymiarowy czas |
| $\tau_w$ | | czas stanu niskiego prostokątnego pulsującego sygnału prądowego cewki |
| $\tau_z$ | — | czas stanu wysokiego prostokątnego pulsującego sygnału prądowego cewki |

| | |
|---|---|
| $\phi$ | przesunięcie fazowe rozwiązania harmonicznego |
| $\varphi;\ \varphi_j,$ $\varphi_1, \varphi_2$ | położenie kątowe wahadła magnetycznego; położenie kątowe wahadła 1 i 2 ($j = 1, 2$) |
| $\varphi_0$ | amplituda wychylenia wahadła podczas oscylacji w jednym dołku potencjału (kąt początkowy) |
| $\varphi_A$ | kąt graniczny występowania strefy aktywnej pola magnetycznego |
| $\varphi_S$ | kąt położenia równowagi wahadła znajdującego się w stanie 2 |
| $\varphi_k$ — | kąt wychylenia wahadła, przy którym następuje włączenie zasilania cewki i zmiana stanu układu z 1 na 2 |
| $\varphi_{lk}$ | kąt wychylenia wahadła, przy którym następuje wyłączenie zasilania cewki i zmiana stanu układu z 2 na 1 |
| $\varphi_{max}$ — | maksymalna wartość wychylenia kątowego wahadła |
| $\dot\varphi;\ \dot\varphi_j,$ $\dot\varphi_1, \dot\varphi_2$ | prędkość kątowa wahadła; prędkość kątowa wahadeł 1 i 2 ($j = 1, 2$) |
| $\varphi';\ \varphi''$ — | prędkość kątowa; przyspieszenie kątowe wahadła przy bezwymiarowym czasie |
| $\chi$ | parametr regularyzacyjny |
| $\omega,\ \omega_1,$ $\omega_2$ — | częstości drgań własnych zachowawczego zlinearyzowanego układu sprzężonych wahadeł |
| $\omega_0$ | początkowa prędkość kątowa wahadła podczas oscylacji w jednym dołku potencjału |
| $\omega_i$ — | częstość kołowa funkcji sinusoidalnej w modelu prostokątnego pulsującego sygnału prądowego |
| $\omega_k$ | prędkość kątowa wahadła, przy której następuje włączenie zasilania cewki i zmiana stanu układu z 1 na 2 |
| $\omega_{lk}$ | prędkość kątowa wahadła, przy którym następuje wyłączenie zasilania cewki i zmiana stanu układu z 2 na 1 |
| $\ell$ | długość ramienia siły oddziaływania magnetycznego między biegunami dipola magnetycznego |

# Streszczenie

Przygotowana rozprawa doktorska poświęcona jest badaniom dynamiki układów złożonych z wahadeł magnetycznych poddanych działaniu niestacjonarnego pola magnetycznego. Przez wahadło magnetyczne rozumie się wahadło fizyczne z magnesem zamocowanym na jego końcu, które umieszczone jest w zewnętrznym polu magnetycznym. Niestacjonarne zewnętrzne pole magnetyczne generowane jest przez cewkę elektryczną umieszczoną pod wahadłem i zasilaną zmiennym w czasie sygnałem prądowym. Przeprowadzone badania dotyczą przede wszystkim drgań pojedynczego wahadła magnetycznego odbywających się w jednym „dołku" potencjału oraz sterowania przepływem energii między dwoma sprzężonymi skrętnie wahadłami magnetycznymi.

W pierwszym rozdziale pracy dokonano przeglądu literatury dotyczącej drgań układów mechanicznych wykorzystujących pole magnetyczne. Przegląd ten został podzielony na dwie części, z której pierwsza dotyczyła układów o jednym stopniu swobody, natomiast druga układów o wielu stopniach swobody. Omawiane prace starano się uporządkować ze względu na chronologię oraz podobieństwo analizowanych problemów badawczych. Badania te najczęściej dotyczyły wahadeł magnetycznych znajdujących się w stacjonarnym polu magnetycznym, a jeśli nie, to niestacjonarne pole magnetyczne generowane było przez sinusoidalny sygnał prądowy płynący w cewce elektrycznej. Przegląd literatury wskazuje na brak badań naukowych dotyczących sterowania przepływem energii między sprzężonymi wahadłami magnetycznymi poddanymi oddziaływaniu pól magnetycznych.

Celem naukowym pracy jest opracowanie nowych modeli matematycznych dla pojedynczego wahadła magnetycznego oraz układu dwóch skrętnie sprzężonych wahadeł magnetycznych, a także zbadanie ich dynamiki nieliniowej pod względem praktycznego zastosowania uzyskanych wyników. W pracy postawiono dwie tezy badawcze. Pierwsza odnosi się do przypadku drgań okresowych pojedynczego wahadła magnetycznego odbywających się w jednym „dołku" potencjału, dla których możliwe jest zmienianie okresowości tych drgań bez wyraźnego naruszenia przebiegu ich trajektorii fazowej. Zmiana okresowości odbywa się poprzez zmianę częstotliwości sygnału prądowego cewki elektrycznej. Okresowość drgań wahadła rozumiana jest jako liczba okresów wymuszenia (sygnału prądowego) przypadająca na jeden okres drgań wahadła. Druga teza wskazuje na możliwość kontrolowania przepływu energii pomiędzy sprzężonymi wahadłami magnetycznymi przy użyciu pola magnetycznego generowanego przez cewki elektryczne znajdujące się pod nimi.

Opracowano modele dynamiczne układów o jednym i dwóch stopniach swobody wykorzystujące różne modele oporów ruchu oraz empiryczne modele od działywania magnetycznego. Badania dynamiki przeprowadzono wykorzystując metody numeryczne oraz analityczno-numeryczne oparte na metodzie uśredniania. Analiza oparta na klasycznych metodach stosowanych w dynamice nieliniowej pozwoliła na wykrycie bogatej dynamiki obu układów z uwzględnieniem drgań chaotycznych i quasiokresowych, co zostało potwierdzone w sposób eksperymentalny na specjalnie zbudowanych stanowiskach badawczych.

Przeprowadzona analiza numeryczna oraz teoretyczna udowodniła pierwszą z postawionych tez badawczych. Badania wykazały, że w przypadku drgań okresowych pojedynczego wahadła magnetycznego odbywających się w jednym „dołku" potencjału, możliwe jest zmienianie okresowości tych drgań bez wyraźnego naruszenia przebiegu ich trajektorii fazowej. Zachowanie to jest możliwe ze względu na szczególny charakter oddziaływania magnetycznego układu i istnienie tzw. strefy aktywnej (strefy faktycznego oddziaływania magnetycznego pary cewka magnes).

W przypadku sprzężonych wahadeł zaproponowano dwie metody sterowania przepływem energii między nimi: otwartą bez sprzężenia zwrotnego i zamkniętą ze sprzężeniem zwrotnym. Podczas drgań w antyfazie energia przemieszcza się z wahadła poddanego odpychającemu polu magnetycznemu do wahadła znajdującego się w przyciągającym polu. Natomiast w przypadku drgań w fazie przepływ energii jest odwrotny. Na podstawie obserwacji oraz analizy numerycznej dynamiki układu wykazano, że podczas sterowania bez sprzężenia zwrotnego koniecznym jest znanie a priori typu drgań wahadeł (tj. w fazie lub w antyfazie) oraz czasu po jakim energia układu zostanie całkowicie rozproszona. W przypadku sterowania ze sprzężeniem zwrotnym koniecznym jest określenie przesunięcia fazowego między wahadłami na podstawie pomiaru ich położeń. Przedstawione badania numeryczne oraz eksperymentalne pokazały, że przy odpowiednim sterowaniu polami magnetycznymi cewek, a z mechanicznego punktu widzenia nieliniową sztywnością poszczególnych wahadeł, możliwe jest zapewnienie kierunkowego transferu energii między sprzężonymi wahadłami. Badania te udowodniły drugą z postawionych tez badawczych.

Badania przeprowadzone dla pojedynczego wahadła magnetycznego mogą stanowić podstawy do nowego sposobu modelowania silników krokowych. Dodatkowo opracowane metody sposobu sterowania przepływem energii w układach połączonych wahadeł magnetycznych, mogą stanowić bazę do przyszłych badań w zakresie tłumienia drgań bądź odzyskiwania energii z drgających struktur składających się z łańcuchów oscylatorów.

# Abstract

The prepared doctoral dissertation focuses on studying dynamics of systems composed of magnetic pendulums subjected to a non-stationary magnetic field. A magnetic pendulum is a physical pendulum with a magnet attached to its end and is placed in an external magnetic field. The non-stationary external magnetic field is generated by an electric coil placed under the pendulum and powered by a time-varying current signal. The presented research mainly concerns the oscillations in one potential well of a single magnetic pendulum, as well as the control of the energy flow between two torsionally coupled magnetic pendulums.

In the theoretical introduction, a review of works on mechanical systems using a magnetic field was carried out. This introduction was divided into two parts, the first one described systems with one degree of freedom, and the second one concerned systems with multiple degrees of freedom. The discussed papers were organized in terms of chronology and similarity of the analyzed research problems. The analyzed studies are mostly aimed at magnetic pendulums placed in a stationary magnetic field, and if not, a non stationary magnetic field was produced by the sinusoidal current signal flowing in an electric coil. In addition, there are no scientific studies focused on the control of the energy flow between coupled magnetic pendulums as a result of generating magnetic fields in their vicinity.

The scientific goal of the work is to develop new mathematical models for a single magnetic pendulum and a system of two torsionally coupled magnetic pendulums, as well as to study their non-linear dynamics in terms of the applicability of the obtained results. Two research theses were formulated in the dissertation. The first one claims that in the case of periodic oscillations of a single magnetic pendulum taking place in one potential well, it is possible to change the periodicity of these oscillations without disturbing their phase trajectory. The change in periodicity is caused by a change in the frequency of the current signal flowing in the electric coil. The periodicity of pendulum oscillations is understood in a classical way, i.e. as the number of excitation periods (current signal) per one period of pendulum oscillation. The second thesis indicates that it is possible to control the energy flow between coupled magnetic pendulums using the magnetic field generated by electric coils located underneath them.

Mathematical models of single and coupled pendulums systems were developed. Various models of motion resistance as well as empirical models of magnetic interaction were tested. Dynamics studies were carried out using numerical and semi analytical methods based on the averaging method. Using classical methods applied in non linear dynamics, a basic dynamic analysis was obtained for both systems, showing their rich dynamics including chaotic or quasi periodic phenomena. These behaviors were confirmed experimentally on specially built experimental rigs.

Numerical and theoretical analysis explained why a significant change in the duty cycle and frequency of the rectangular current signal does not have to affect the phase trajectory of the one well oscillations of a single magnetic pendulum. It was concluded that due to the special nature of the magnetic excitation and the existence of the so-called active zone (the zone of the actual magnetic interaction of the coil magnet pair), it is possible to change the periodicity of pendulum oscillation taking place in one potential well while phase trajectory remains the same.

In the case of coupled pendulums, two methods of controlling the energy flow between them were proposed: open (without feedback) and closed (with feedback). During an antiphase oscillation, energy moves from a pendulum in a repulsive magnetic field towards a pendulum in an attractive field. In contrast, during in-phase oscillations, the flow of energy is reversed. Based on the observations and numerical analysis of the system dynamics, it was found that during control without feedback, it is necessary to know a priori the type of pendulum oscillation (i.e. in-phase or antiphase) and the time after which the energy of the system will be completely dissipated. However, in the case of control with feedback, it is necessary to determine the phase shift between the pendulums via measurements of their positions. Presented numerical and experimental studies shown that with appropriate control of the coil magnetic fields (from the mechanical point of view, the non-linear stiffness of each pendulums is controlled), it is possible to provide directional energy transfer between coupled oscillators.

The studies carried out for a single magnetic pendulum can be the basis for a new method of modelling stepper motors. Additionally, the developed methods of controlling the energy flow in systems of connected magnetic pendulums can be the basis for future research in the field of vibration damping or energy harvesting of structures consisting of oscillator chains.

# Rozdział 1

# Wstęp

W teorii mechaniki rozróżnia się dwa rodzaje wahadeł: wahadło matematyczne (proste) i wahadło fizyczne. Pierwsze z nich definiowane jest jako punkt materialny zawieszony na idealnie wiotkiej, nierozciągliwej i nieważkiej nici (pręcie) mogący wykonywać ruchy w płaszczyźnie pionowej. Drugie rozumiane jest jako ciało mogące wykonywać swobodny ruch obrotowy dookoła poziomej osi nieprzechodzącej przez środek masy tego ciała [1, 2]. Wynika stąd, że każde ciało w rzeczywistości może pełnić rolę wahadła fizycznego, natomiast wahadło matematyczne może stanowić jego wyidealizowany model.

Wahadła towarzyszą ludziom od wieków, chociażby pod postacią huśtawki do zabawy. Jednakże w przeszłości oprócz bycia zabawkami potrafiły one pełnić również bardziej zaawansowane i użytkowe role. W starożytnych Chinach (ok. 132 r.) zaprojektowano mechanizm oparty na wahadle, którego zadaniem było ostrzeganie ludzi o zbliżającym się trzęsieniu ziemi [3, 4]. Mechanizm pełnił więc rolę urządzenia nazywanego dzisiaj sejsmografem i został uznany jako jedno z największych osiągnięcie technologicznych tamtej epoki.

Największą popularność wahadło zyskało jednak jako element wykorzystywany do pomiaru czasu. Jednym z pierwszych badaczy, który zainteresował się teorią ruchu wahadeł był Galileusz. Zaobserwował on i opisał przybliżony izochronizm[1] wahań i na tej podstawie w 1637 r. wykorzystał go do odmierzania czasu. Próby budowy zegara przez Galileusza oraz jego syna odniosły fiasko ze względu na trudności wykonania mechanizmu wyzwalania [5]. Dopiero w 1673 r. holenderski uczony C. Huygens opracował pierwszy projekt mechanizmu przypominającego ten, który znajduje się we współczesnych zegarach wahadłowych [5, 6].

Obserwacje ruchu wahadła przysłużyły się również do potwierdzenia teorii obrotu ziemi dookoła własnej osi. W 1851 r. francuski fizyk Foucault przymocował do kopuły Panteonu w Paryżu wahadło, którego ramię miało długość 67 metrów, a na jego końcu zamocowano mosiężną kulę o masie 28 kilogramów. Trajektorie wykreślone przez poruszającą się kulę dały do zrozumienia, że pionowa płaszczyzna wahań zmienia swoje położenie względem Ziemi na skutek jej obrotu [7].

---

[1]izochronizm [gr.], własność układu drgającego polegająca na niezależność okresu drgań własnych od amplitudy tych drgań.

Obecnie wahadła wykorzystywane są np. jako tłumiki drgań wysokich budynków, które szczególnie podatne są na działanie obciążeń dynamicznych pochodzących od wiatru czy trzęsień ziemi. Największe drapacze chmur takie jak Taipei 101 (509,2 m) czy Burj Al Arab (321 m) posiadają w swoich konstrukcjach ogromne wahadła mające tłumić ich drgania wywołane wyżej wymienionymi czynnikami środowiskowymi [8, 9].

Wahadło stanowi przykład najprostszego oscylatora nieliniowego, a ze względu na swoja prostą budowę jest ono wykorzystywane przez naukowców w badaniach empirycznych jako analogia innych oscylatorów spotykanych w fizyce, np. sprężyn z nieliniowościami [10], drgających atomów [11] czy zjawiska tunelowania Josephsona występującego między nadprzewodnikami lub w układach elektrycznych [5, 12, 13].

Ze względu na szybki rozwój technologii i chęć tworzenia połączeń synergicznych pomiędzy różnymi dziedzinami techniki, coraz więcej układów czysto mechanicznych poddawanych jest działaniu pola elektromagnetycznego. Najbardziej powszechnym przykładem takiego urządzenia mechatronicznego, które zrewolucjonizowało przemysł jest silnik elektryczny [14, 15]. W silniku elektrycznym ruch wirnika spowodowany jest powstaniem siły oddziaływania magnetycznego pomiędzy uzwojeniem wzbudzenia a uzwojeniem twornika. Układy prostych wahadeł mechanicznych również poddano rozwojowi technologicznemu. Od początku pierwszej połowy XX wieku zaczęły pojawiać się pierwsze teoretyczne prace naukowe, których przedmiotem badań były wahadła wykonane z ferromagnetyków bądź wyposażone w magnesy, a następnie umieszczane w polu magnetycznym generowanym przez cewkę elektryczną lub inny magnes. Układy tego rodzaju otrzymały nazwę *wahadeł magnetycznych* [16].

*Przedmiotem badań niniejszej pracy jest układ pojedynczego wahadła magnetycznego oraz układ dwóch torsyjnie sprzężonych wahadeł magnetycznych. Przeprowadzone badania obejmują zarówno symulacje numeryczne jak i eksperymenty. W wyniku badań opracowane zostały modele matematyczne wyżej wymienionych układów z uwzględnieniem użytkowych modeli oddziaływania magnetycznego, a także wykazano istnienie ciekawych zjawisk dynamicznych oraz procesów, które mogą znaleźć zastosowanie w technice.*

## 1.1 Dotychczasowy stan wiedzy o układach wahadeł magnetycznych

Przedstawiona rozprawa poświęcona jest dynamice układu pojedynczego wahadła magnetycznego (układ o jednym stopniu swobody) i układu dwóch sprzężonych wahadeł magnetycznych (układ o dwóch stopniach swobody). Jak już zostało wspomniane wcześniej we Wstępie, wahadło magnetyczne to wahadło fizyczne z zamocowanym na końcu ramienia magnesem lub ferromagnetykiem i umieszczone w zewnętrznym polu magnetycznym pochodzącym od innego magnesu lub cewki elektrycznej. Ze względu na różne stopnie swobody układów analizowanych w pracy, przegląd literatury zawarty w tym podrozdziale został podzielony na układy o jednym i wielu stopniach swobody.

### 1.1.1 Układy o jednym stopniu swobody

Jedną z najstarszych prac jaką udało się znaleźć autorowi dysertacji na temat pojedynczego wahadła magnetycznego jest praca francuskiego inżyniera J. Bethenoda [17]. Przedmiotem jego badań teoretycznych było wahadło, na końcu którego zamocowano ferromagnetyczną kulkę. Pod wahadłem umieszczono cewkę elektryczną, której oś rdzenia pokrywała się z osią wahadła w spoczynku. Cewka zasilana była prądem o wysokiej częstotliwości. Badany układ został rozdzielony na część elektryczną i mechaniczną. Układ elektryczny opisany został poprzez zwyczajne równanie różniczkowe pierwszego rzędu, natomiast równanie ruchu wahadła zamodelowano poprzez zlinearyzowane równanie różniczkowe drugiego rzędu. Czynnikiem sprzęgającym dwa układy była indukcyjność cewki będąca funkcją wychylenia wahadła. Jego badania wykazały, że istnieje pewien próg napięcia oraz częstotliwość prądu cewki, dla których wahadło będzie wykonywało drgania o zadanej amplitudzie. Doprowadziło to do wniosku, że układ ten mógłby służyć jako „transformator" zamieniający drgania elektryczne o wysokiej częstotliwości na drgania mechaniczne o niskiej częstotliwości. Badania te były jednak przeprowadzone z dużą liczbą przybliżeń oraz rozpatrywały tylko kilka wybranych przypadków ruchu w stosunku do przyjętych parametrów układu. W kolejnych latach pojawiły się prace rozszerzające nieznacznie te badania [18, 19]. Autorzy tych prac przybliżyli indukcyjność cewki szeregami potęgowymi, tworząc proste nieparzyste funkcje zależne od położenia wahadła. Ponadto, podali dodatkowe warunki jakie muszą spełniać parametry układu (rezystancja, pojemność, indukcyjność, tłumienie wiskotyczne, częstość prądu, częstość własna wahadła), aby drgania utrzymywały zadaną amplitudę. Inny pogląd na ten problem zaprezentowany został przez Minorsky'iego [20–22], który zbadał symetryczny rozkład indukcyjności cewki względem położenia zerowego wahadła. Intuicyjnie sprowadził on problem do równania Mathieu zakładając, że parametrem zmieniającym się periodycznie w czasie jest długość wahadła. Pozwoliło to na wyznaczenie wystarczającego warunku występowania stacjonarnych drgań wahadła.

Praca Minorsky'iego spotkała się jednak z krytyką ze strony Kesavamurthy i in. [23]. Autorzy zarzucili Minorsky'iemu, że błędnie sformułował równanie prądu cewki poprzez pominięcie składowych o wysokich częstotliwościach oraz, że stosując równanie Mathieu nie mógł poprawnie zbadać zjawiska indukowania się dodatniego tłumienia w przypadku dominującej rezystancji układu elektrycznego. Zaproponowali oni również paraboliczny rozkład indukcyjności cewki, a następnie przeprowadzili analizę teoretyczną i eksperymentalną ruchu wahadła.

W latach 70. i 80. XX wieku pojawiło się kilka prac, które bezpośrednio nie dotyczyły wahadeł magnetycznych, ale były ich analogicznymi odpowiednikami. W pracy [24] skupiono się na opracowaniu ogólnego sposobu analizy drgań parametrycznych urządzeń elektromechanicznych, w których występują siły magnetyczne. Jako przykłady tych urządzeń podano rezonatory VHF używane do przyspieszania cząsteczek oraz obrotowy parametryczny silnik elektryczny. W rezultacie wykazano, że układy te maja tendencję do podtrzymywania swoich drgań. Do podobnych wniosków doszedł autor pracy [25] badając teoretycznie i eksperymentalnie silnik reluktancyjny opisany nieliniowym równaniem różniczkowym drugiego rzędu z okresowo zmiennymi parametrami. Russel i in. [26] zbadali jakie warunki musi spełniać taki silnik, aby mógł poprawnie pracować oraz wyznaczyli jego optymalne częstotliwości pracy. Blakley [27] natomiast skupił się na zachowaniu energii w tego rodzaju układach. Zauważył on, że energia pomiędzy układem elektrycznym a mechanicznym wymieniana jest tylko w wąskim zakresie ich wzajemnego położenia, dlatego transfer energii pomiędzy układami aproksymował przy pomocy impulsu. Pozwoliło to na zapisanie układu nieliniowych nieautonomicznych równań ruchu w postaci liniowego autonomicznego równania drugiego rzędu i wykreślenie przybliżonych trajektorii na płaszczyźnie fazowej.

Występowanie nieokresowego ruch w układzie wahadła magnetycznego zaprezentowane zostało przez Moona i in. [28, 29]. Badane przez nich wahadło magnetyczne składało się z ferromagnetycznej belki wspornikowej umieszczonej pomiędzy dwoma magnesami, której ruch wymuszany był przez harmoniczne przemieszczanie się całego układu w poziomie. Umieszczenie wahadła pomiędzy magnesami generowało symetryczny dwudołkowy potencjał pola magnetycznego. W wyniku przeskakiwania układu pomiędzy dwoma lub trzema stanami równowagi wykazywał on zachowania chaotyczne potwierdzone teoretycznie i eksperymentalnie, a także autorzy wykryli dziwne atraktory chaotyczne wraz z ich fraktalną budową. Zmodyfikowany układ takiego wahadła analizowano w pracy [30]. Autorzy tej pracy opisali ruch układu równaniem Duffinga i wprowadzić zmienną magnetyzację belki modelując ją nieliniowością Preisacha. W wyniku badań numerycznych doszli do wniosku, że histereza w postaci nieliniowości Preisacha w większości przypadków odgrywała rolę dodatkowego czynnika tłumiącego. Radons i in. [31] również analizowali prototypowy układ wahadła magnetycznego charakteryzujący się złożoną naturą histerezową opartą na operatorze Preisacha. Układ ten miał symulować zmiany zachodzące w mikrostrukturze materiałów magnetycznych, stopów z pamięcią kształtu i materiałów porowatych. Wyniki badań numerycznych wykazały, że regularne i chaotyczne zachowania układu wykazują fraktalne zależności od parametrów. Struktura fraktalna występuje również w zależności od warunków początkowych, co według autorów pracy wydaje się być unikatową cechą dla tego rodzaju układów. Kumar i in. [32] uznali, że zwykłe równanie Duffinga nie uwzględnia wszystkich nieliniowych efektów oddziaływania pomiędzy magnesami a wahadłem magnetycznym w analizowanym układzie. Dlatego opracowali analityczny model takiej interakcji, zapewniając możliwość zbadania wpływu nieliniowych efektów szóstego stopnia na dynamikę układu. Przeprowadzili badania numeryczne oraz eksperymentalne w celu wyjaśnienia bifurkacji położeń równowagi. Następnie, informacje pozyskane z analizy bifurkacyjnej zostały wykorzystane do śledzenia zmian konfiguracji stabilności oscylatora z monostabilnego do tristabilnego, z tristabilnego do bistabilnego, itd.

Siahmakoun [33] także zaobserwował dwudołkowy charakter potencjału badanego przez siebie wahadła magnetycznego, które wymuszane było dodatkowo zewnętrzną siłą harmoniczną. Eksperymentalne i numeryczne portrety fazowe wykazały istnienie atraktorów jednodołkowych (odpowiadającym prawemu i lewemu dołkowi potencjału), a także atraktorów międzydołkowych [34, 35]. W wynikach pojawiły się również typowe dla układów nieliniowych skoki amplitudy i histereza rezonansowa. Kwuimy i in. [36] rozważali wpływ asymetrycznego rozkładu potencjału na dynamikę wahadła magnetycznego. Asymetryczny potencjał uzyskano dzięki umieszczeniu wahadła pomiędzy magnesami na równi pochyłej, której powierzchnia poruszała się ruchem harmonicznym. Stosując metodę Mielnikowa wyprowadzili analitycznie warunek na przejście dynamiki układu z okresowej na chaotyczną. Wyniki tych analitycznych przewidywań zostały przetestowane i zweryfikowane poprzez analizę fraktalnych i regularnych kształtów basenów przyciągania. Baseny o regularnym kształcie wskazywały na okresową dynamikę układu, podczas gdy baseny o nieregularnym (fraktalnym) kształcie związane były z dynamiką nieokresową. Kryterium energetyczne „ucieczki" wahadła magnetycznego z asymetrycznych studni potencjału podano w pracy [37]. Procedura zawierała określenie poziomu energii jaką musi osiągnąć układ do przedostania się przez lokalną barierę potencjału. Ze względu na asymetrię potencjału, konieczne były obliczenia kryterium „ucieczki" dla każdej z dwóch różnych studni tego potencjału. Analityczne i eksperymentalne badania dynamiki nieliniowej wahadła magnetycznego umieszczonego między dwoma magnesami i wymuszanego silnikiem DC przedstawiono w pracach [38, 39]. Autorzy analizowali okresowe i chaotyczne ruchy w odniesieniu do różnych wartości parametrów kontrolnych takich jak amplituda i częstotliwość zewnętrznego wymuszenia. Na wykresach bifurkacyjnych widoczne było, że chaos w układzie powstaje poprzez podwajanie się okresu. Wskazano, że układ ten mógłby znaleźć zastosowanie w mieszaniu cząstek materii [40], gdzie występowanie dynamiki chaotycznej pozwoliłoby na równomierne rozproszenie cząstek i uniknięcie kumulowania się w jednym miejscu.

Teoretyczne, numeryczne i eksperymentalne badania nad chaosem tłumionego wahadła magnetycznego przeprowadził Khomeriki [41]. Wykorzystując równanie Mathieu do opisu dynamiki układu wyznaczył granice istnienia rezonansu parametrycznego, a także wykazał w oparciu o obliczenie wartości wykładników Lapunowa, że w jego układzie niestabilność parametryczna zawsze prowadzi do zachowań chaotycznych.

Elementy sterowania ruchem wahadła magnetycznego zostały przedstawione w pracach [42, 43]. Sterowanie polegało na odpowiednim zasilaniu dwóch cewek umieszczonych w dolnym i górnym położeniu równowagi wahadła magnetycznego w zależności od jego położenia kątowego. Eksperymentalne i teoretyczne badania dotyczyły drgań samowzbudnych, wymuszenia parametrycznego oraz drgań w trzech studniach potencjału. Dzięki wprowadzeniu sterowania z dodatnim i ujemnym sprzężeniem zwrotnym zbadane zostały drgania samowzbudne, gdzie straty energii kompensowane były energią pochodzącą od pola magnetycznego. Do analizy teoretycznej tego zjawiska użyto równania Van der Pola. Zjawisko wymuszenia parametrycznego wytłumaczono opierając się na równaniu Mathieu ze zmienną sztywnością układu. Przy odpychającym oddziaływaniu pomiędzy cewką a magnesem, dla dużych wartości prądu cewki obserwowano drgania w dwóch studniach potencjału, natomiast dla małych wartości prądu drgania odbywały się w trzech studniach potencjału.

Tran i in. [44] wykryli dwa atraktory chaotyczne w układzie wymuszanego wahadła magnetycznego oraz poprzez porównanie danych eksperymentalnych z numerycznymi zademonstrowali jakościowe i ilościowe zdolności przewidy-wania struktury atraktora chaotycznego poza zakresami ruchu opracowanego modelu układu. Wykazali też zależność wartości parametrów układu od częstotliwości jego wymuszenia oraz wykazali, że zjawisko to występuje w przypadku, gdy model układu zawiera tarcie suche.

Wahadło magnetyczne dzięki występowaniu zachowań chaotycznych spełnia obecnie rolę taniego, popularnego i łatwego w budowie układu do badań naukowych i edukacyjnych. W pracach [45–49] przedstawione zostały szczegóły dotyczące budowy takich stanowisk oraz programów do analizy symulacyjnej i doświadczalnej. Wahadła z oddziaływaniem magnetycznym znalazły również zastosowanie jako aktuatory [50] mające na celu wykonywanie ruchu wahadłowego z różnymi ograniczeniami i wymogami określonymi przez operatora. Tsubono i in. [51] zaprojektowali potrójne wahadło magnetyczne jako system wibroizolacyjny dla interferometru laserowego. W tym przypadku autorzy skupili się na tłumieniu drgań ramion wahadeł siłami magnetycznymi, a nie na ich wymuszaniu. Opierając się na regule Lenza i zjawisku indukcji prądów wirowych w metalu poruszającym się w sąsiedztwie magnesu opracowali formułę tłumienia magnetycznego. Eksperymentalne i obliczeniowe badania udowodniły dobre własności tłumiące układu, zapobiegające niepożądanemu zjawisku sprzęgania się pionowego i poziomego ruchu luster interferometru.

Stanowiska wahadeł magnetycznych używane są również do weryfikacji spełnienia prawa indukcji elektromagnetycznej. Jang i in. [52] wyprowadzili analitycznie dwie funkcje aproksymujące indukcję magnetyczną dla dowolnego punktu z otoczenia metalowej obręczy oraz solenoidu, w których płynie prąd. Następnie opierając się na tych funkcjach, wyznaczyli siły magnetyczne działające na magnes przemieszczający się nad obręczą lub solenoidem. Obliczyli też moment pary sił generowany pomiędzy magnesem a cewką znajdujących się w położeniu niewspółosiowym. Do sprawdzenie poprawności swoich obliczeń wykorzystali model wahadła magnetycznego, gdzie doświadczalnie [53] wykonali pomiary napięcia indukowanego w cewce przy różnej intensywności pola magnetycznego, początkowej prędkości wahadła oraz liczby zwojów cewki.

W pracy [54] dokonano analizy stabilności wahadeł znajdujących się w polu magnetycznym. Badane wahadło zamodelowano poprzez jedną lub dwie zamknięte pętle elektryczne umieszczone w zmiennym polu magnetycznym. Teoretyczne badania oparte były na asymptotycznym rozwiązaniu równań Lagrange'a-Maxwella, opisujących dynamikę układów. W zależności od parametrów i warunków początkowych układu, ruch wahadła z jedną pętlą elektryczną dążył do jednego z położeń równowagi lub do cykli granicznych usytuowanych blisko tych położeń. Badania układu z dwoma pętlami elektrycznymi pokazały, że istnieje możliwość obracania się wahadła. Jeśli dolne położenie równowagi jest niestabilne to możliwe jest wystąpienie dwóch stabilnych ruchów obrotowych (w różnych kierunkach). Jeśli zaś jest ono stabilne, to wahadło może się nie obracać lub może wykazywać cztery przypadki ruchu obrotowego, gdzie dwa z nich są stabilne, a pozostałe dwa niestabilne.

Stabilnością wahadła magnetycznego zajmowali się również naukowcy w pracy [55, 56]. Badany przez nich układ składał się z wahadła z zamocowanym na końcu magnesem i harmonicznie przybliżającą się (oddalającą się) od dołu metalową płytą. Poruszająca się płyta jest w stanie przekazać energię do wahadła poprzez pole magnetyczne prądów wirowych indukujących się w jej wnętrzu. Problem stabilności układu rozważany był jak dla klasycznego równania Mathieu z współczynnikiem tłumienia. Wykorzystując analityczną metodę bilansu harmonicznych określone zostały warunki wystąpienia rezonansu parametrycznego w zależności od amplitudy drgań płyty oraz jej odległości od magnesu wahadła. Badania wykazały, że niestabilność układu jest możliwa tylko wtedy, gdy amplituda drgań i odległość między płytą a magnesem spełniają pewne wymagania związane z jednoczesnym wzbudzaniem i tłumieniem wahadła przez poruszającą się płytę.

W pracy [57] analizowano dynamikę przejściową niewymuszanego i tłumionego przestrzennego wahadła magnetycznego. Na wahadło działały siły przyciągania magnetycznego pochodzące od trzech magnesów położonych pod nim i rozmieszczonych w wierzchołkach trójkąta równobocznego. Wykazano, że ze względu na brak długotrwałego ruchu, zachowanie analizowanego układu jest zupełnie inne niż to, które obserwowano wcześniej dla układów z wymuszeniem. Pracę podsumowują następujące wnioski: i) obliczony wymiar granic basenu przyciągania może być niecałkowity, a wykładniki Lapunowa w skończonym czasie mogą być dodatnie we wszystkich skalach; ii) granice basenu mają fraktalne kowymiary 1; iii) prawdopodobieństwo utrzymania się trajektorii z dala od atraktorów spada superwykładniczo w czasie.

Analiza dynamiki globalnej przestrzennego wahadła magnetycznego została przeprowadzona przez Qin i in. [58]. Dotyczyła ona przede wszystkim wrażliwości układu na warunki początkowe, oszacowania liczby położeń równowagi w zależności od różnych odległości magnesów oraz analizy stabilności. Ewolucja fraktalnych basenów przyciągania została przeprowadzona numerycznie i potwierdzona eksperymentalnie. Pokazano, że położenie magnesu wahadła w czasie jego ruchu wprost wpływa na topologię basenu przyciągania. Wynika to z faktu, że zakres przyciągania atraktora, który jest najbliżej zostaje znacząco zwiększony, natomiast wpływ strefy innych atraktorów jest pomniejszany.

Pewną modyfikację wahadła magnetycznego typu Bethenoda zaproponował Duboshinski [21, 59]. Postanowił on zmienić orientację cewki znajdującej się pod wahadłem w taki sposób, że jej oś była prostopadła do osi wahadła w spoczynku. W układzie tym zaobserwowano zjawisko „kwantyzacji amplitudy" [60, 61]. Polegało ono na tym, że dla danej częstotliwości prądu i położenia początkowego, wahadło wykazywało się kilkoma stabilnymi drganiami o różnej amplitudzie, przy czym położenie początkowe nie mogło być mniejsze od tego, które odpowiadało drganiom o najmniejszej amplitudzie [62]. Za stabilizację tych drgań odpowiedzialne jest występowanie w układzie tłumienia, którego straty energii są równoważone przez energię pola magnetycznego. Energia tego pola musi jednak znajdować się w ściśle określonym przedziale. Zbyt mała wartość energii pola magnetycznego dostarczona do układu nie będzie w stanie zrównoważyć strat energii tłumienia i drgania zgasną. Natomiast zbyt duża wartość energii spowoduje chaotyczne „przeskakiwanie" pomiędzy różnymi stabilnymi drganiami wahadła. Układ zazwyczaj wybiera najbliższą stabilną amplitudę drgań dla danych warunków początkowych. Według Luo i in. [63] występowanie wielu „dyskretnych" („skwantowanych") rozwiązań okresowych w tym układzie jest wynikiem powstawania rezonansu subharmonicznego. Częstotliwość rezonansu subharmonicznego związana jest z symetrią siły wymuszającą. Nieparzyste rezonanse subharmoniczne występują, gdy symetryczna funkcja siły wymuszającej jest parzysta i odwrotnie. W pracy [64, 65] wahadło to zakwalifikowano do klasy tzw. samoadaptujących układów wzbudzanych impulsowo (ang. *kick-excited self adaptive dynamical systems*). Wynika to z faktu, że magnetyczna siła wymuszająca układ działa tylko w pewnym, ograniczonym zakresie położeń wahadła względem cewki. Przeprowadzone badania analityczne i numeryczne dotyczyły punktów osobliwych oraz basenów przyciągania. Analizowane były również wielokrotne bifurkacje zbiorów atraktorów, a także ich ewolucje i fraktalne postacie.

Drgania, które charakteryzują się skwantowanymi amplitudami nazywane są argumentacyjnymi (ang. *argumental oscillations*). Badania nad nimi prowadził Cintra i in. [66, 67] wykorzystując kilka różnych wahadeł Duboshinskiego. Nazwa wynika z faktu, że wymuszenie oscylatora jest zależne od „argumentu", którym jest jego położeniem w przestrzeni. Traktując układ jak oscylator Duffinga, opracowali analityczne wyrażenie na kryterium jego stabilności. Wykorzystując metodę uśredniania, autorzy wyznaczyli rozwiązanie analityczne równania ruchu dla układu bez tłumienia, a w przypadku układu z tłumieniem podali tylko rozwiązanie przybliżone tego równania. Eksperymentalne, numeryczne i analityczne wyniki swoich badań przedstawili w postaci biegunowych wykresów Van der Pola, gdzie wolno zmieniająca się amplituda drgań stanowiła promień, a faza drgań była kątem obrotu.

Ciekawa analogia do wahadła magnetycznego przedstawiona została w pracach [68, 69]. Autorzy tych prac zaobserwowali eksperymentalnie drgania tzw. ściany domenowej (ang. *domain wall*) i porównali je do drgań wahadła poddanego zmiennej grawitacji. Ściana domenowa powstaje w materiale, gdy w jego strukturze spotkają się dwa przeciwnie namagnetyzowane regiony. Badana ściana miała szerokość 50 nm i została wytworzona w zakrzywionym drucie niklowo-żelazowym, przez który przepuszczano bardzo mały prąd. Ruch ściany jest równoważny drganiom wahadła, gdzie masa ściany domenowej odpowiada masie wahadła. Natomiast oddziaływanie magnetostatyczne zachodzące pomiędzy polem magnetycznym wytwarzanym przez płynący w drucie prąd a polem magnetycznym ściany domenowej odpowiada grawitacji. Analizowane były drgania ściany domenowej podczas przepuszczania przez drut zmiennego prądu o różnych częstotliwościach. Autorzy wspomnianych prac na podstawie badań doświadczalnych wykazali, że drgania rezonansowe ściany domenowej są dostatecznie duże, a zarazem pobierają bardzo mało prądu, co daje możliwości do opracowania i stworzenie nowych układów elektronicznych o niskim poborze mocy stanowiących konkurencję dla technologii CMOS.

Kolejną analogię do wahadła magnetycznego opisano w pracy [70]. Autorzy wykazali, że wahadło magnetyczne może stanowić klasyczny przykład zdegenerowanej teorii zaburzeń w mechanice kwantowej. Degeneracją (zwyrodnieniem) nazywa się sytuację, gdzie jednej wartości energii układu odpowiadają różne stany kwantowe. Zwykłe wahadło przestrzenne ma cylindryczną symetrię wokół cięgna zawieszenia oraz dwie postacie drgań normalnych, tzn. wahadło może oscylować w dowolnym kierunku z tą samą częstotliwością. Naukowcy postanowili „złamać" tę symetrię poprzez dołączenie magnesu do ramienia wahadła i umieszczenie go w niejednorodnym polu magnetycznym innego magnesu. In-terakcja magnetyczna przesunęła położenie równowagi układu w wyniku czego częstotliwości drgań w dwóch prostopadłych kierunkach, nie były już takie same. Eksperymentalne i analityczne badania pokazały, że drgania wahadła mierzone wzdłuż danego kierunku charakteryzują się dudnieniami wynikający z transferu energii na drgania w kierunku prostopadłym. Wykazano również, że okres dudnień w stosunku do okresu drgań wahadła jest proporcjonalny do piątej potęgi odległości między środkami magnesów.

Warto w tym miejscu wspomnieć, że w dotychczasowej literaturze istnieje też szereg prac, których obiektem badań nie jest bezpośrednio wahadło magnetyczne, ale układ liniowego oscylatora poddanego działaniu sił magnetycznych. Ze względu na charakter równania ruchu wspomnianego oscylatora uznać go można za układ opisujący ruch uproszczonego (zlinearyzowanego) wahadła magnetycznego. Z tego powodu warto również prześledzić dotychczasowe prace z zakresu tego rodzaju układów dynamicznych.

Darula i in. [71] dokonali szerokiej analizy zjawiska rezonansu w układzie elektromagnesu przyciągającego ferromagnetyczny element. Stosując analityczną metodę wielu skal określili stany ustalone oscylatora wymuszanego harmonicznie z częstością bliską rezonansowej. Na wykresach amplitudowo-częstościowych widoczny był wpływ nieliniowego oddziaływania magnetycznego, który „przechylał" krzywą rezonansową (ang. *softening effect*). Dodatkowo wykazano analitycznie w jaki sposób parametry obwodu elektrycznego mogą zmieniać bądź tłumić drgania elementu ferromagnetycznego. Rozszerzone badania nad tłumieniem elektromagnetycznym w układzie nieliniowego oscylatora ze sztywnością typu Duffinga i wymuszanego zewnętrzną siłą zaprezentowano w pracy [72]. W rozważanym układzie, do drgającej masy przymocowano na stałe cewkę elektryczną, która w czasie ruchu „przechodziła" przez nieruchomy magnes. Numerycznie przeanalizowano wpływ na dynamikę układu takich parametrów jak: rezystancja, indukcyjność i pojemność cewki. Okazało się, że to rezystancja cewki ma największy wpływ na tłumienie elektromagnetyczne, a wartość pojemności znacząco wpływa na zachowanie dynamiczne układu, co zostało wykazane przy pomocy wykresów bifurkacyjnych. Ruch heterokliniczny i kryteria wystąpienia chaosu Mielnikowa przedstawiono w pracy [73]. Ponadto zaobserwowano, że układ wykazywał ruch okresowy i chaotyczny odpowiednio poniżej i powyżej progu Mielnikowa.

Eksperymentalne i analityczne badania dynamiki oscylatora w postaci metalowej masy zawieszonej na sprężynie i poruszającej się wewnątrz cewki elektrycznej zostały zaprezentowane przez Ho i in. [74]. Korzystając z metody wielu skal, autorzy otrzymali przybliżone rozwiązanie wskazujące na bliską relację między częstotliwością drgań metalowej masy a częstotliwością napięcia przyłożonego do cewki w przypadku innym niż wzbudzenie rezonansowe. Rozwiązanie analityczne ułatwiło identyfikację krytycznych wartości parametrów układu, które miały wpływ na jego odpowiedź. W pracy Bednarka i in. [75] przedstawiony został układ do aktywnego tłumienia drgań oscylatora łożyskowanego aerostatycznie i pod danego działaniu sprężyny elektromagnetycznej składającej się z pary cewka-magnes. Na podstawie doświadczenia wyznaczono formuły matematyczne opisujące sztywność sprężyny elektromagnetycznej w funkcji prądu cewki. Dzięki specjalnie opracowanemu sterowaniu prądem cewki możliwe było natychmiasto-we wytłumienie drgań oscylatora, co potwierdzono w badaniach symulacyjnych i eksperymentalnych.

Witkowski i in. [76] opracowali modele matematyczne opisujące ruch drgający wózka z oddziaływaniami magnetycznymi pozwalające wykonywać względnie szybkie symulacje numeryczne ruchu układu. Wózek oprócz przymocowanych do niego sprężyn mechanicznych wyposażono w magnes, który był odpychany przez magnes utwierdzony do podstawy stanowiska. Odpychające działanie magnesów wprowadzało do równania ruchu sztywność typu Duffinga, dla której autorzy opracowali pięć różnych modeli matematycznych. Jakościowa i ilościowa analiza bifurkacyjna dowiodła, że użycie modelu oddziaływania magnetycznego w formie nieparzystej funkcji wymiernej prowadzi do najlepszej zgodności wyników pomiędzy symulacją a eksperymentem.

Układ magnetycznego oscylatora z uderzeniami został zaprezentowany w pracach [77–79]. Oscylator zbudowany był z masy zamocowanej do dwóch belek pełniących rolę analogiczną do sprężyn. Z jednej strony masy zamocowano magnes, który podczas oddziaływania z nieruchomą cewką wymuszał ruch układu. Natomiast druga strona masy mogła uderzać w nieruchomą przeszkodę. W pracy zaproponowano analityczną metodę obliczania dużych ugięć belki, przy czym zagadnienie brzegowe pozwoliło na wyznaczenie wyrażenia na siłę sprężystości. Wykorzystując prawo Biota-Savarta i metodę analizy nieskończenie małych elementów, autorzy opracowali wzór na oddziaływanie magnetyczne pomiędzy magnesem a cewką o określonej liczbie zwojów. Pracę uzupełniają wykresy rezonansowe i bifurkacyjne obejmujące stabilne oraz niestabilne rozwiązania układu, a także przeprowadzono analizy różnych typów bifurkacji.

### 1.1.2 Układy o wielu stopniach swobody

Liczba prac naukowych odnosząca się do badań nad dynamiką układów o wielu stopniach swobody opartych na wahadłach magnetycznych jest znacznie mniejsza niż miało to miejsce w przypadku układów o jednym stopniu swobody. W tym paragrafie omówione zostaną prace, których wyniki są najbardziej zbliżone do tematu niniejszej rozprawy doktorskiej.

Fradkov i in. [80] analizowali problemem wzbudzenia i synchronizacji drgań w układzie mechatronicznym złożonym z dwóch torsyjnie sprzężonych wahadeł magnetycznych. Skomplikowany układ składał się z dwóch obracających się obręczy, które sprzężono skrętnie elastycznym elementem. Na dole obręczy zamontowano magnesy oddziałujące z cewką elektryczną umieszczoną pod nimi. Dodatkowo, wewnątrz tych obręczy zamocowano obracające się dźwignie, na końcach których przyczepiono magnesy stałe odpychające się z magnesami obręczy. Do zbudowanego stanowiska eksperymentalnego opracowane zostały algorytmy do estymacji wektora stanu układu, a także identyfikacji parametrów oraz określenia poziomu energii układu. Stosując metodę szybkiego spadku opracowano sterowanie pozwalające na osiągnięcie przez wahadła zadanego poziomu energetycznego z wymogiem drgań w przeciwfazie. Dokonano również analizy ukierunkowanej na wyznaczenie wartości współczynnika wzbudzenia, który jest wersją asymptotycznego wzmocnienia mierzącą właściwości rezonansowe układu.

W pracy [81] autorzy skupili się na polepszeniu zdolności odzyskiwania energii z ruchu dwóch skrętnie sprzężonych wahadeł poddanych zewnętrznemu wymuszeniu. Na końcu wałów wahadeł zamocowane były kołowe magnesy z dwoma przeciwstawnymi biegunami, które podczas obracania się generowały napięcie wewnątrz rozmieszczonych wokół nich cewek. Naukowcy wykazali, że układ ten jest odpowiedni do odzyskiwania energii z drgań o niskich częstotliwościach (< 5 Hz). Stosując analityczną metodę bilansu harmonicznych wykryto rozwiązania o małej amplitudzie drgań oraz współistniejące rozwiązania o znacznie większej amplitudzie, które były niewykrywalne w obliczeniach numerycznych. Zaprezentowane wyniki symulacyjne pokazały, że zwiększenie odzyskiwania energii dla szerokiego spektrum częstotliwości drgań wahadeł może zostać zapewnione poprzez wybór odpowiednich warunków początkowych.

Surganova i in. [82] badali wpływ parametrów takich jak masa, moment bezwładności czy sztywność sprzężenia na nieliniowe postacie drgań układu dwóch sprzężonych wahadeł magnetycznych. Korzystając z metody małego parametru oraz metod numerycznych przeprowadzili szereg badań w odniesieniu do różnych warunków początkowych układu. Wykazali, że drgania w fazie występują dla wszystkich wychyleń początkowych wahadeł zarówno przy silnym jak i słabym oddziaływaniu magnetycznym. Dodatkowo, przy wychyleniach początkowych mniejszych niż 4 5° wykryto drgania układu, których trajektorie fazowe odpowiadały rozwiązaniom quasiokresowym.

Russell i in. [83] analizowali zachowanie się łańcucha osiemnastu wahadeł magnetycznych rozmieszczonych wzdłuż okręgu. Czynnikiem sprzęgającym wahadła w układzie były siły magnetyczne. Przedstawione wyniki numeryczne i eksperymentalne dowiodły, że w układzie powstają poruszające się nieliniowe fale (znane w literaturze pod angielską nazwą *moving breathers*), których energia koncentruje się w sposób zlokalizowany i wykazuje charakter oscylacyjny. Fale te mają strukturę obwiedni solitonów i czasami są nazywane solitonami obwiedniowymi. Badania dotyczyły również stabilności tych fal w odniesieniu do małych przypadkowych zakłóceń układu. Według autorów pracy, otrzymane wyniki są ważne z punktu widzenia transferu energii zachodzącego podczas zderzeń pomiędzy atomami (lub jonami) poruszającymi się ze stosunkowo dużymi prędkościami a atomami ciała stałego.

Dynamikę chaotyczną i okresową podwójnego wahadła magnetycznego analizował Wojna i in. [84]. Na końcu dolnego wahadła zamocowano silny magnes, który odpychał się z drugim magnesem zamocowanym w podstawie stanowiska. Wymuszenie realizowane było przez specjalnie skonstruowany silnik elektryczny. Autorzy przedstawili równania ruchu układu uwzględniając złożony model oporów ruchu generowany przez łożyska toczne, a także moment magnetyczny w funkcji różnych położeń kątowych wahadeł. Eksperymentalne i numeryczne wykresy bifurkacyjne, trajektorie fazowe i przekroje Poincarégo wykazały dużą zgodność opracowanego modelu z badaniami doświadczalnymi. Chaos w układzie powstawał przez podwajanie się okresu, co zaobserwowano na wykresach bifurkacyjnych. W pracy [85] skorzystano z występowania w tym układzie zjawiska chaosu w celu efektywniejszego odzyskiwania energii. Magnes znajdujący się w podstawie stanowiska zastąpiono szeregiem cewek elektrycznych rozmieszczonych równomiernie wzdłuż łuku dolnej trajektorii wahadła. Dzięki temu auto-rzy odzyskiwali energie mechaniczną drgań podłoża w postaci energii elektrycznej generowanej w cewkach. Przeprowadzono analizę numeryczną do określenia optymalnych wartości parametrów układu (masy i długości wahadeł oraz masy magnesu) względem odzyskiwanej ilości energii. Badania numeryczne i eksperymentalne wykazały, że nagły wzrost odzyskiwania energii następuje wówczas, gdy ruch dolnego wahadła staje się chaotyczny, a także gdy zwiększona zostaje liczba cewek elektrycznych.

Zhang i in. [86] zaproponowali autoparametryczny pochłaniacz drgań o dwóch stopniach swobody oparty na wahadle magnetycznym. Badany model fizyczny składał się z drgającej pionowo masy poddanej harmonicznemu wymuszeniu i wewnątrz której znajdowało się wahadło magnetyczne. Wahadło umieszczone było pomiędzy dwoma przyciągającymi go magnesami. Numerycznie oraz analitycznie zbadano odpowiedzi układu na siłę wymuszającą. Korzystając z metody małego parametru wyznaczony został optymalny zakres tłumienia drgań w odniesieniu do parametrów układu. Wyniki badań pokazały, że zaprezentowany układ efektywniej tłumi drgania niż układ wyposażony w zwykłe wahadło ze względu na szerszy zakres częstotliwości pracy.

Zaawansowany układ pochłaniacza drgań i jednocześnie urządzenia odzyskującego energię przedstawiono w pracach [87–89]. Układ składał się z drgającej na sprężynach masy, do której przymocowany był silnik elektryczny z wahałem zawieszonym na jego wale. Wzdłuż ramienia wahadła zamocowano cewkę elektryczną, wewnątrz której poruszał się magnes. Układ w czasie ruchu tłumił drgania masy odzyskując przy tym energię na dwa sposoby: (i) przy pomocy samej cewki (układ o 4 stopniach swobody) lub (ii) przy pomocy silnika i cewki (układ o 5 stopniach swobody). Przeprowadzona analiza dynamiczna wykazała, że tzw. „języki" niestabilności parametrycznej występują dla szerokiego pasma częstotliwości wymuszenia i są spowodowane przez bifurkację typu Neimarka-Sackera. Obszary tych niestabilności mogą być kontrolowane przez odpowiednie zmienianie rezystancji w obwodzie silnika. Zaproponowano trzy wskaźniki opisujące skuteczność odzyskiwania energii oraz jeden wskaźnik tłumienia drgań. Analizując wszystkie wskaźniki znaleziono obszar zgodności, w którym występuje jednoczesne tłumienie drgań i odzyskiwanie energii. Badania numeryczne i eksperymentalne wykazały, że silnik elektryczny efektywniej odzyskuje energię niż zespół cewka-magnes zamocowany na wahadle. Z drugiej strony jednak znacząco obniża on zdolność układu do rozpraszania energii drgań masy. Natomiast zespół cewka-magnes pomimo, że jest mniej skuteczny w odzyskiwaniu energii praktycznie nie wpływa na tłumienie drgań masy.

Autorzy pracy [90] zbadali dynamikę zabawki nazywanej Levitronem, w której znaleźli analogię do wahadła magnetycznego. Levitron jest to zabawka o sześciu stopniach swobody, składająca się z wirującego „bąka lewitującego w polu magnetycznym. Na podstawie analizy stabilności ruchu tej zabawki w kierunku pionowym stwierdzono, że równanie ruchu odpowiadające temu kierunkowi jest analogią do wahadła magnetycznego i jego drgań wokół położenia równowagi. Na podstawie zlinearyzowanych równań dynamicznych wyznaczone zostały zakresy stabilności w zależności od prędkości obrotowej „bąka" i jego odległości od podstawy zabawki.

Układy o wielu stopniach swobody złożone z liniowych oscylatorów i podda-ne działaniu sił magnetycznych, podobnie jak w przypadku układów o jednym stopniu swobody, można uznać za uproszczone analogie układów sprzężonych wahadeł magnetycznych poruszających się w zakresie małych kątów. Dlatego podobnie jak wcześniej, opisane zostaną wyniki uzyskane w pracach, których przedmiotem badań są tego rodzaju obiekty.

W pracy [91] przedstawiony i zamodelowany został układ dwóch szeregowo połączonych oscylatorów. Oprócz liniowych sprężyn łączących oscylatory, na jeden z nich działała silnie nieliniowa sprężyna o ujemnej sztywności zbudowana z odpychających się magnesów. Siła oddziaływania magnetycznego została opisana w postaci wielomianu trzeciego stopnia. Badania teoretyczne drgań swobodnych i wymuszonych przeprowadzone zostały w oparciu o analizę równania Duffinga. Badania eksperymentalne wykonano dla różnych nieliniowości i sztywności sprężyny magnetycznej, wynikających ze zmiany odległości między magnesami. Obliczono, że istnieje pewna wartość sztywności sprężyny magnetycznej, powyżej której występuje niestabilność układu. Podczas badań z wymuszeniem układu zaobserwowano, że dla drgań przedrezonansowych sprężyna magnetyczna wykazuje tylko liniowe zachowanie. Ponadto wartość częstotliwości rezonansowej układu z nieliniową sprężyną magnetyczną zmniejszyła się w odniesieniu do przypadku, gdyby zastąpić ja klasyczną sprężyną liniową. Zastosowanie sprężyny magnetycznej spowodowało zmniejszenie się własności wibroizolacyjnych układu. Pewną modyfikacje tego układu zaproponowali autorzy pracy [92]. Modyfikacja polegała na dodaniu kolejnej sprężyny magnetycznej do oscylatora, który wcześniej był pod działaniem tylko sprężyny liniowej. W nowym układzie, jedynie sprężyna sprzęgająca dwa oscylatory była liniowa. Badania numeryczne i eksperymentalne skoncentrowane były głównie na dynamice nieliniowej oscylatorów. Zademonstrowano bifurkację będące wynikiem podwajania się okresu drgań, chaotyczne atraktory, zachowania histerezowe, skoki amplitudy i drgania quasiokresowe. Autorzy w procesie walidacji modelu, uzyskali jego wysoką zgodność z danymi eksperymentalnymi. Ponadto nieciągłe funkcje występujące w modelu matematycznym zostały przybliżone funkcjami wymiernymi, co pozwoliło na opracowanie rozwiązań analitycznych badanego układu przy użyciu metody bilansu harmonicznych. W efekcie, znaleziono niestabilne okresowe rozwiązania dynamiczne pominięte w symulacjach numerycznych.

Zhou i in. [93] zaproponowali inną konfigurację montażu sprężyn na specjalnie skonstruowanym stanowisku, niż ta zaprezentowana w pracach [91, 92]. Rozważono przypadek, gdy oscylatory były sprzęgnięte nieliniową sprężyną magnetyczną, a zwykłe sprężyny linowe pełniły role łączników z nieruchomą podstawą stanowiska. Przeanalizowali dynamikę układu ze szczególnym uwzględ nieniem transferu energii z oscylatora wymuszanego do oscylatora pasywnego, w przypadku rezonansu wewnętrznego o stosunku częstości 1:3. Eksperyment i obliczenia analityczne wykazały występowanie zjawiska nasycenia amplitudy drgań oraz jej przeskoków pomiędzy dwoma stanami ustalonymi. Główną koncepcją badań było zastosowanie zjawiska nasycenia, występującego w rezonansie wewnętrznym, do tłumienia rezonansu głównego oscylatora wymuszanego siłą zewnętrzną.

Układ dwudziestu ośmiu sprzężonych ze sobą wahadeł magnetycznych zaprezentowano w pracy [94]. Autorzy przedstawili pomysł na budowę nowego ro-dzaju mechanicznej anteny, pracującej na ultra niskich częstotliwościach (od 300 Hz do 3 kHz). Antena ta składa się z dwudziestu ośmiu walcowych magnesów neodymowych, namagnesowanych w kierunku diametralnym (tzn. magnesowanie odbywa się wzdłuż średnicy) i mogących obracać się wzdłuż osi symetrii. Magnesy umieszczone są blisko siebie wewnątrz prostokątnej cewki elektrycznej, tak że ich własne pola magnetyczne oddziałują na siebie tworząc sprzężenie. Przy założeniu, że magnesy zachowują się jak wahadła magnetyczne, autorzy analizowali ich drgania oraz wydajność przy generowaniu fal elektromagnetycznych z zakresu ultra niskich częstotliwości. W rezultacie okazało się, że wydajność tego rodzaju anteny jest o 7 dB większa niż zwykłej anteny elektrycznej.

## 1.2 Geneza i uzasadnienie tematu pracy

Badania dynamiki układów złożonych z wahadeł magnetycznych prowadzone są przez naukowców od co najmniej pierwszej połowy XX wieku, aż do dnia dzisiejszego. Biorąc pod uwagę interdyscyplinarny charakter podjętej tematyki badań i wywodzącej się z połączenia dziedzin takich jak mechanika i elektrodynamika, przyciąga ona zainteresowanie badaczy z obu tych obszarów nauki. Wynika to z faktu poszukiwania synergizmu i płynących z niego korzyści dla rozwoju technologii oraz poprawy poziomu życia. Pomimo tego zainteresowania, w dostępnej literaturze odnoszącej się do dynamiki układów złożonych z wahadeł magnetycznych wciąż istnieją pewne braki, które przyczyniły się do powstania tej rozprawy. W niniejszym podrozdziale zostaną przedstawione kierunki możliwego dalszego rozwoju badań nad tymi układami, pozwalające poszerzyć obecny stan wiedzy.

Biorąc pod uwagę fakt, że wprawdzie konstrukcyjnie układy wahadeł magnetycznych nie są skomplikowane to jednak wykazują one silnie nieliniową dynamikę. Dzięki temu wykorzystuje się je do badań podstawowych nad typowymi zjawiskami nieliniowymi takimi jak bifurkacje [38, 44, 95], skoki amplitud [33, 39], cykle graniczne [65], zachowania chaotyczne [29, 45, 47] czy ruchy quasiokresowe [96], ale również do badania zjawiska rezonansu parametrycznego [36, 41, 55, 56], czy badań nad multistabilnością [37, 97]. W dotychczasowych badaniach, aby wyjaśnić źródła powstawania tych zjawisk, naukowcy często opierali się na znanych modelach matematycznych takich jak równanie Duffinga [28, 30, 32], równanie Mathieu [20, 21, 41] czy równanie Van der Pola [42, 43]. Jednakże, ze względu na złożoność charakteru zjawiska oddziaływania magnetycznego, modele te nie zawsze się sprawdzają, a ponadto nie uwzględniają innych zjawisk, jak np. tarcie suche występujące w rzeczywistych układach mechanicznych. Wydaje się więc, że dokonanie analizy dynamiki przy wykorzystaniu złożonych modeli matematycznych może doprowadzić do wykrycia wcześniej nieobserwowanych zachowań. Dlatego w niniejszej pracy wyniki badań podstawowych zostały opracowane przy wykorzystaniu modeli matematycznych, w których starano się jak najdokładniej odwzorować naturę oddziaływania magnetycznego oraz oporów ruchu, wykorzystując przy tym badania eksperymentalne.

Dostępne w literaturze badania najczęściej prowadzone były dla wahadeł magnetycznych umieszczonych w stacjonarnym polu magnetycznym [38, 57, 58]. Pole to pochodziło np. od stałych magnesów lub cewek elektrycznych zasilanych prądem stałym. Natomiast wymuszenie tych układów realizowano w sposób mechaniczny, np. kinematycznie [91, 93] (poprzez układ elementów sprężystych), parametrycznie (poprzez ruch punktu zawieszenia [55, 86, 98]) bądź poprzez zewnętrzny moment generowany przez silnik [44, 84]. Reszta dostępnych artykułów poświęcona jest dynamice układów wymuszanych niestacjonarnym polem magnetycznym, pochodzącym od cewki zasilanej zmiennym w czasie prądem elektrycznym. W znacznej większości tych prac analizie poddano układy, w których cewka zasilana była sinusoidalnym sygnałem prądowym [54, 62, 63, 66]. Inne sygnały takie jak np. prostokątny nie były brane pod uwagę. Brak wyników takich badań w dotychczasowej literaturze skłonił autora pracy do zainteresowania się tą problematyką.

Kolejnym słabym punktem obecnego stanu wiedzy jest mała ilość materiałów naukowych dotyczących zagadnień dynamiki układów wahadeł magnetycznych o wielu stopniach swobody. Odnosząc się do prac zawartych w przeglądzie literatury można zauważyć znaczącą dysproporcję pomiędzy liczbą prac dotyczących układów o jednym stopniu swobody a liczbą prac dotyczących układów o wielu stopniach swobody. Dodatkowo tylko w dwóch pracach [80, 81] obiektem badań był układ złożony z wahadeł magnetycznych, których sprzężenie odbywało się przy pomocy skrętnego elementu sprężystego łączącego ich wały. W zamierzeniu autora, wspomniane luki w tym obszarze tematycznym zostały wypełnione badaniami przedstawionymi w tej rozprawie doktorskiej.

## 1.3 Cel naukowy, teza i zakres pracy

Celem naukowym pracy jest opracowanie nowych modeli matematycznych dla pojedynczego wahadła magnetycznego oraz układu dwóch skrętnie sprzężonych wahadeł magnetycznych, a także zbadanie ich dynamiki nieliniowej pod względem praktycznego zastosowania uzyskanych wyników. Opracowane modele wykorzystywać będą empiryczne funkcje opisujące oddziaływanie magnetyczne między komponentami układu oraz uwzględniać będą opory ruchu występujące w stanowiskach doświadczalnych. Dzięki temu stanowić będą alternatywę dla prostych modeli proponowanych dotychczas w literaturze.

Ze względu na fakt, że w prezentowanej rozprawie doktorskiej przedmiotem badań są dwa układy składające się z wahadeł magnetycznych, dla każdego z nich zostanie przyjęta inna teza badawcza. Pierwszy układ złożony jest z pojedynczego wahadła, na końcu którego zamocowany jest magnes. Pod wahadłem umieszczona jest cewka elektryczna zasilana prostokątnym sygnałem prądowym o regulowanej częstotliwości i tzw. współczynniku wypełnienia. Oddziaływanie magnetyczne pomiędzy magnesem a cewką powoduje ich wzajemne odpychanie się. Drugi układ składa się z dwóch wahadeł, na końcach których zamocowane są magnesy, a ponadto wały tych wahadeł połączone są ze sobą elementem sprężystym. Pod każdym z wahadeł znajduje się cewka elektryczna zasilana sygnałem prądowym o zadanym kształcie. Sygnały prądowe płynące w tych cewkach, mogą generować pola magnetyczne przyciągające jak i odpychające magnesy wahadeł.

Na podstawie wstępnych badaniach eksperymentalnych i symulacyjnych poczynionych dla układu pojedynczego wahadła magnetycznego, można postawić

następująca tezę:

> *W przypadku drgań okresowych pojedynczego wahadła magnetycznego odbywających się w jednym dołku potencjału, możliwe jest zmienianie okresowości tych drgań bez wyraźnego naruszenia przebiegu ich trajektorii fazowej poprzez zmianę częstotliwości prostokątnego sygnału prądowego cewki elektrycznej.*

Okresowość drgań wahadła rozumiana jest jako liczba okresów wymuszenia (sygnału prądowego) przypadająca na jeden okres drgań wahadła.

W przypadku układu złożonego z dwóch słabo sprzężonych wahadeł poddanych działaniu pól magnetycznych, można postawić następującą tezę badawczą:

> *Możliwym jest kontrolowanie przepływu energii pomiędzy sprzężonymi wahadłami magnetycznymi przy użyciu pól magnetycznych generowanych przez cewki elektryczne znajdujące się pod nimi.*

Zakres prac potrzebnych do realizacji celu badawczego i uzasadnienia przyjętych tez jest następujący:

(i) gruntowny przegląd literatury w zakresie prowadzonych badań;

(ii) opracowanie modeli matematycznych badanych układów i identyfikacja ich parametrów;

(iii) modyfikacja stanowiska badawczego (powstałego jeszcze w ramach pracy inżynierskiej autora [99]) w zakresie układu zasilania cewek elektrycznych;

(iv) symulacyjne badania dynamiki nieliniowej pojedynczego wahadła magnetycznego dążące do udowodnienia tezy o niezmiennym przebiegu trajektorii fazowej drgań pomimo znaczących zmian parametrów sygnału prądowego cewki;

(v) symulacyjne badania dynamiki nieliniowej dwóch sprzężonych wahadeł magnetycznych dążące do udowodnienia tezy o sterowaniu przepływem energii miedzy wahadłami poprzez odpowiednie generowanie pól magnetycznych cewek;

(vi) walidacja opracowanych modeli matematycznych na bazie przeprowadzonych badań eksperymentalnych.

## 1.4 Wkład wyników pracy w dyscyplinę naukową

Wydawać by się mogło, że dynamika wahadeł, które towarzyszą nam od wieków została przez te wszystkie lata skrupulatnie zbadana przez naukowców i niczym nowym nas nie zaskoczy. Jednak, w przedziale lat 2010 2021 baza *Scopus* odnotowała, aż 17 066 artykułów, dla których słowem kluczowym było „wahadło" (ang.

*pendulum*). Liczba ta pokazuje, że pomimo powszechności wahadeł wciąż są one obiektem badań podstawowych. Badania te skupiają się głównie na nieliniowych aspektach dynamiki układów o jednym i wielu stopniach swobody, złożonych ze sprzężonych wahadeł i poddanych różnym rodzajom wymuszeń [100–103]. Mniejszą popularnością wśród naukowców cieszy się „wahadło magnetyczne" (ang. *magnetic pendulum*), które w bazie *Scopus* ma 51 prac, a aż 33 opublikowane zostały w latach 2010-2022. Trzeba przyznać, że w dzisiejszym świecie nauki i przy obecnych możliwościach prowadzenia badań naukowych, liczba 51 prac nie jest zadowalająca.

Okazuje się, że już samo bogactwo dynamiki nieliniowej prezentowanej przez wahadła magnetyczne może stanowić powód do rozpoczęcia badań naukowych [55, 58, 98, 104, 105], a także być źródłem danych wejściowych używanych podczas np. testowania złożonych metod topologicznej analizy danych [106]. W naukach inżynieryjno-technicznych wahadła magnetyczne dobrze odnajdują się w prężnie rozwijającej się gałęzi nauki, jaką jest odzyskiwanie energii. Energia zgromadzona w ruchu wahadła będąca wynikiem drgań podłoża, może być przekształcana poprzez indukcję elektromagnetyczną lub piezoelektryki na energię elektryczną gromadzoną w akumulatorach i gotową do ponownego użycia [107 111]. Ponadto układy wahadeł magnetycznych ze względu na możliwość łatwej zmiany parametrów oddziaływania magnetycznego wykorzystywane są jako tłumiki niepożądanych drgań mechanicznych np. w systemach gromadzących energię opartych na kołach zamachowych [112]. Badania dynamiki wahadeł magnetycznych mogą również stanowić źródło inspiracji dla nowych modeli matematycznych i symulacyjnych sprzęgieł elektromagnetycznych [113, 114] czy silników elektrycznych [115, 116]. Pakiety sprzężonych wahadeł magnetycznych wykorzystywane są także w radiotechnice do budowy nowoczesnych anten pracujących na ultra niskich częstotliwościach [117 119].

## 1.5 Struktura pracy

Przedstawiona praca składa się z czterech rozdziałów podzielonych tematycznie.

W rozdziale pierwszym zawarto informacje o dotychczasowym stanie wiedzy na temat układów wahadeł magnetycznych charakteryzujących się jednym oraz wieloma stopniami swobody. Podano genezę i uzasadnienie podjęcia tematu pracy. Określono cele naukowe, sformułowano tezy pracy, a także jej zakres. W części końcowej rozdziału przedstawiono jaki wkład mogą mieć wyniki pracy w dyscyplinę naukową inżynieria mechaniczna.

W rozdziale drugim przedstawiono badania symulacyjne i eksperymentalne dla pojedynczego wahadła magnetycznego. Omówiona została konstrukcja i działanie stanowiska badawczego. W następnej kolejności opracowano model matematyczny układu. Przedstawiono stosowane w badaniach modele oporów ruchu, model prostokątnego sygnału prądowego oraz empiryczne modele momentu siły oddziaływania magnetycznego. Opisano też sposoby identyfikacji parametrów układu. Dokonano numerycznych i analityczno-numerycznych badań dynamiki nieliniowej, a następnie skupiono się na przypadku drgań wahadła w jednym dołku potencjału.

W rozdziale trzecim przedstawiono badania symulacyjne i eksperymentalne układu dwóch sprzężonych torsyjnie wahadeł magnetycznych. Opisano konstrukcję stanowiska badawczego opartego na stanowisku pojedynczego wahadła. Opracowano model matematyczny układu w oparciu o wcześniejsze modele używane w układzie o jednym stopniu swobody. Przeprowadzono również analizę bifurkacyjną dynamiki układu oraz sterowania przepływem energii między wahadłami przy użyciu pola magnetycznego.

W rozdziale czwartym podsumowano uzyskane wyniki pracy oraz sformułowano wnioski wynikające z przeprowadzonych badań teoretycznych, numerycznych, analityczno-numerycznych i doświadczalnych. Wyróżniono innowacyjne elementy pracy oraz podano dalsze kierunki możliwych badań.

# Rozdział 2

# Układ pojedynczego wahadła magnetycznego

Rozdział poświęcony jest opisowi stanowiska badawczego pojedynczego wahadła magnetycznego, jego modelowaniu matematycznemu oraz wynikom badań udowadniających pierwszą z postawionych tez badawczych.

## 2.1 Stanowisko badawcze

Podrozdział ten poświęcony jest opisowi budowy stanowiska badawczego pojedynczego wahadła magnetycznego, na którym wykonywane były badania eksperymentalne.

Zdjęcie całego stanowiska badawczego, na którym odbywały się eksperymenty zarówno dla pojedynczego wahadła magnetycznego jak i układu dwóch sprzężonych wahadeł magnetycznych zostało pokazane na rys. 2.1. Jego pod stawowymi elementami są: komputer z oprogramowanie LabVIEW, zasilacz laboratoryjny, generator sygnału, karta pomiarowa i przede wszystkim rekonfigurowalny układ wahadeł magnetycznych. Układ ten w podstawowej wersji daje możliwość badania ruchu pojedynczego wahadła magnetycznego, a po stosunkowo prostej rekonfiguracji można na nim badać układ składający się z dwóch skrętnie sprzężonych wahadeł magnetycznych. Rekonfiguracja polega na sprzęgnięciu wałów wahadeł poprzez umieszczenie elementu podatnego np. sprężyny w specjalnych uchwytach znajdujących się na ich końcach. Ponieważ rozdział poświęcony jest układowi pojedynczego wahadła magnetycznego, podczas badań stanowisko skonfigurowane było w sposób przedstawiony na rys. 2.2. Wahadło oznaczone zostało numerem (1). Ramię wahadła zbudowane jest z elementu wykonanego z materiału kompozytowego (tekstolit), zaś uchwyt którym przymocowane jest ono do mosiężnego wału (3) wykonany został z aluminium. Na końcu ramienia wahadła zamocowany został magnes neodymowy (2) o średnicy 22 mm i wysokości 10 mm (niewidoczny na zdjęciu). Wał wahadła podparty jest przy pomocy dwóch standardowych łożysk kulkowych, zamkniętych w oprawie przymocowanej do aluminiowej ramy stanowiska (11). W specjalnym uchwycie pod wahadłem umieszczono cewkę elektryczną (4) o następujących parametrach:

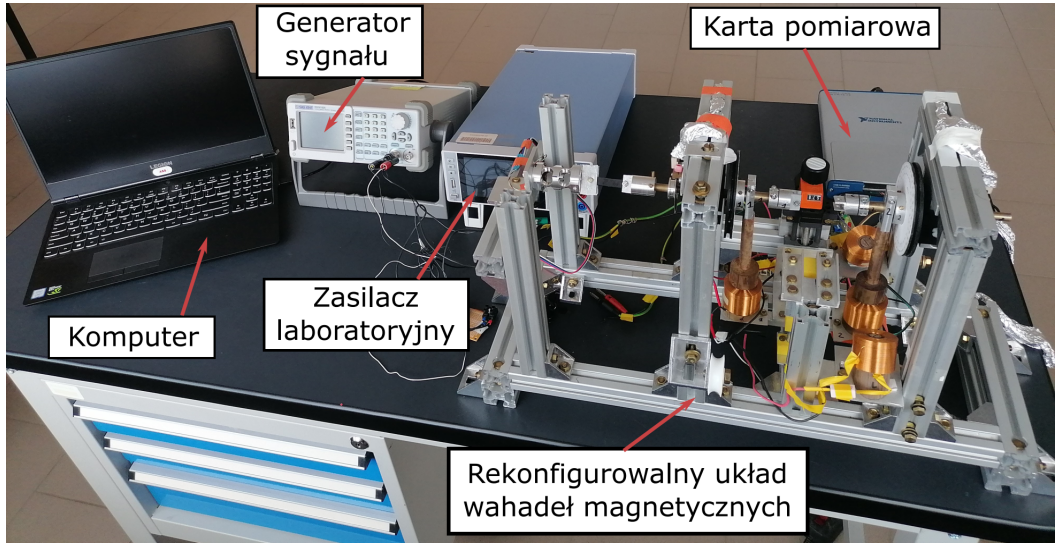

**Rys. 2.1.** Rekonfigurowalne stanowisko eksperymentalne pojedynczego wahadła magnetycznego.

indukcyjność – 22 mH, rezystancja drutu – 10.6 Ω, średnica drutu – 0.5 mm, średnica zewnętrzna cewki – 40 mm, średnica otworu cewki – 17 mm i wysokość cewki – 31 mm. Uchwyt ten przymocowany jest do liniowego prowadnika (5), przy pomocy którego w łatwy sposób można regulować odległość pomiędzy cewką a magnesem w kierunku pionowym. Podczas prowadzonych badań, odległość pomiędzy czołem cewki a powierzchnią magnesu ustawiona była na 1.6

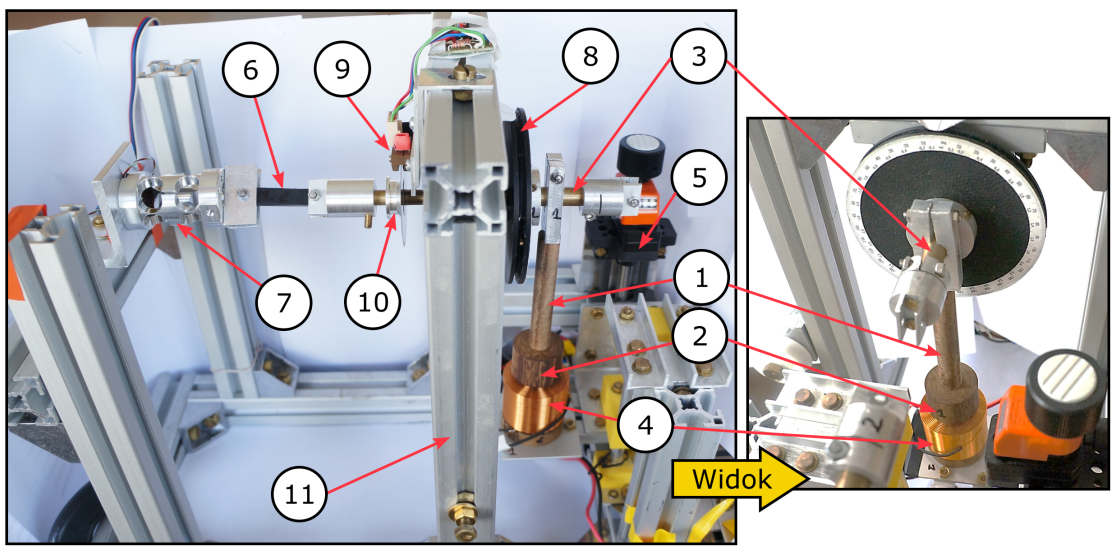

**Rys. 2.2.** Stanowisko eksperymentalne pojedynczego wahadła magnetycznego, gdzie: 1 – wahadło, 2 – magnes neodymowy (niewidoczny na zdjęciu), 3 – mosiężny wał, 4 – cewka elektryczna, 5 – prowadnik liniowy, 6 – gumowy element podatny, 7 – czujnik momentu skręcającego, 8 – polimerowy dysk, 9 - inkrementalny czujnik optyczny, 10 - tarcza kodowa, 11 - aluminiowa rama.

mm. Na jednym z końców wału zamocowano element podatny (6), który sprzęga go z tensometrycznym czujnikiem momentu skręcającego (7). Element ten wykonany jest z gumy o wymiarach 40.3×8.5×6.5 mm i został dodany do układu w celu zwiększenia momentu przywracającego wahadło do położenia równowagi. Czujnik momentu FUTEK TFF325 FSH04055 wykorzystywany jest do badania charakterystyki dynamicznej elementu podatnego. Polimerowy dysk (8) o średnicy 79.6 mm przymocowany jest na stałe do wału i wykorzystywany będzie do pomiarów eksperymentalnych oddziaływania magnetycznego pomiędzy cewką a magnesem. Położenie kątowe wahadła mierzone jest przy pomocy inkrementalnego czujnika optycznego HEDS-9040#J00 oznaczonego numerem (9) oraz tarczy kodowej (10) posiadającej 1000 kres na obwodzie. W rezultacie rozdzielczość sensora mierzącego położenie kątowe wahadła wynosi 0.36 °. Sygnały pochodzące z czujnika położenia wahadła zbierane są przez kartę pomiarową NI USB 6341 i przetwarzane przez program napisany w środowisku LabVIEW.

Podczas badań cewka elektryczna zasilana była prostokątnym pulsującym sygnałem prądowym o zadanej amplitudzie $I_0$, częstotliwości $f$ i tzw. współczynniku wypełnienia $w$ (nazywanym dalej po prostu „wypełnieniem"), tak jak pokazano to na rys. 2.3. Częstotliwość definiowana jest jako odwrotność okresu sygnału, $f = 1/\tau$, natomiast wypełnienie jest stosunkiem czasu załączenia cewki i okresu sygnału, wyrażonym w procentach zgodnie ze wzorem $w = (\tau_z/\tau) \cdot 100\%$. Pro-

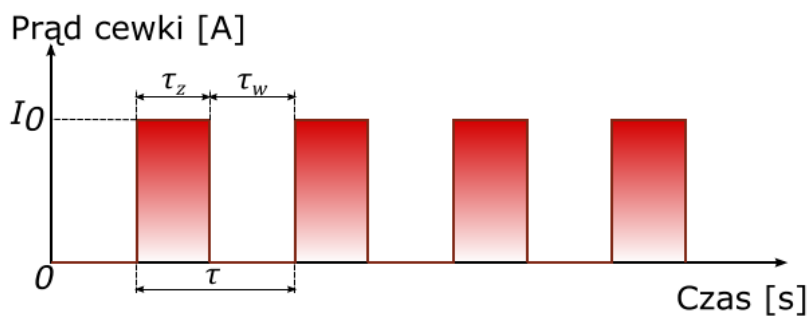

**Rys. 2.3.** Prostokątny pulsujący sygnał prądowy płynący w obwodzie cewki, gdzie: $I_0$ — amplituda sygnału prądowego, $\tau_z$ — czas w jakim prąd płynie przez cewkę elektryczną, $\tau_w$ — czas w jakim prąd nie płynie przez cewkę elektryczną i $\tau$ — okres prostokątnego sygnału prądowego. Częstotliwość sygnału definiowana jest jako $f = 1/\tau$, a wypełnienie $w = \frac{\tau_z}{\tau} \cdot 100\%$.

stokątny kształt sygnału prądowego powstaje poprzez cykliczne zwieranie i rozwieranie obwodu cewki elektrycznej. Układ realizujący zwieranie i rozwieranie obwodu został pokazany na rys. 2.4. Źródło zasilania cewki stanowi laboratoryjny zasilacz prądu stałego KA3005D KORAD pracujący jako źródło prądowe, tzn. wymusza on w obwodzie cewki przepływ prądu o zadanej wartości natężenia niezależnie od zmian rezystancji obwodu. Cykliczne zwieranie i rozwieranie obwodu cewki odbywa się przy użyciu elektronicznego układ kluczującego opartego na tranzystorze IRF540n. Układ ten zwiera i rozwiera obwód zgodnie z prostokątnym sygnałem napięciowym pochodzącym z generatora sygnału SIGLENT SDG1025 i dostarczanym na bramkę (G) tranzystora IRF. W skrócie, prostokątny pulsujący sygnał napięciowy o zadanej częstotliwości i wypełnieniu wysyłany przez

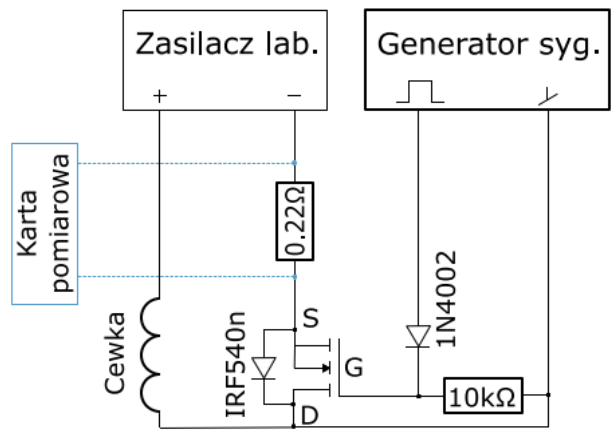

**Rys. 2.4.** Schemat ideowy układu wytwarzającego prostokątny pulsujący sygnał prądowy w obwodzie cewki elektrycznej znajdującej się pod wahadłem.

generator sygnału jest zamieniany dzięki układowi elektronicznemu i zasilaczowi na prostokątny pulsujący sygnał prądowy płynący przez cewkę elektryczną. Mierzony przez kartę pomiarową spadek napięcia na rezystorze 0.22 Ω służy do obserwacji (w sposób pośredni) przebiegu sygnału prądowego w obwodzie cewki elektrycznej przez zastosowanie zależności wynikającej z prawa Ohma. Płynący w obwodzie cewki prąd wytwarza wokół niej pole magnetyczne, które oddziałuje z polem magnetycznym magnesu wahadła. W zależności od kierunku przepływu prądu w cewce jej pole magnetyczne może odpychać bądź przyciągać magnes. Podczas prowadzonych badań założono, że dodatnie natężenie prądu powoduje wzajemne odpychanie się pól magnetycznych magnesu i cewki, natomiast ujemne natężenie prądu powoduje ich wzajemne przyciąganie się. Ze względu na pulsujący charakter prostokątnego sygnału prądowego stosowanego podczas badań, prąd płynący w cewce może być albo dodatni, albo ujemny, a co za tym idzie, cewka i magnes mogą się tylko odpychać, albo tylko przyciągać. Zaprezentowane w tym rozdziale badania uwzględniają tylko taką polaryzację prądu cewki, że powoduje ona odpychanie magnesu wahadła. W zależności od wartości natężenia prądu cewki oddziaływanie to może być silniejsze bądź słabsze. Elementy konstrukcyjne stanowiska wykonane zostały z materiałów niemagnetycznych (stopy aluminium, brąz, austenityczna stal nierdzewna, tworzywa sztuczne), aby ograniczyć wpływ niepożądanych oddziaływań magnetycznych mogących pojawić się pomiędzy magnesem bądź polem magnetycznym cewki elektrycznej a resztą elementów stanowiska.

Prototyp stanowiska badawczego powstał w 2017 roku w ramach inżynierskiej pracy dyplomowej autora [99] pod kierunkiem dr. inż. Grzegorza Wasilewskiego w Katedrze Automatyki, Biomechaniki i Mechatroniki. Zawarty jest w niej dokładny opis budowy poszczególnych elementów stanowiska, połączeń elektrycznych i prototypowych programów napisanych w środowisku LabVIEW.

## 2.2 Modelowanie matematyczne

### 2.2.1 Równanie ruchu

W celu napisania dynamicznego równania ruchu badanego układu posłużono się modelem fizycznym zaprezentowanym na rys. 2.5. Wahadło poddane jest sile grawitacji $mg$ przyłożonej w środku ciężkości oddalonym o $s$ od osi obrotu, momentowi oporów ruchu $M_F$, momentowi $M_K$ pochodzącemu od elementu podatnego oraz momentowi $Q_{mag}$ wynikającemu z oddziaływania magnetycznego pomiędzy magnesem a cewką elektryczną zasilaną prądem $i(t)$. Wykorzystując prawa Newtona oraz siły i momenty uwzględnione w modelu fizycznym, równanie dynamiczne układu ma postać

$$J\ddot{\varphi} + mgs\sin\varphi + M_F(\dot{\varphi}) + M_K(\varphi, \dot{\varphi}) = Q_{mag}(\varphi, i(t)), \qquad (2.1)$$

gdzie $J$ jest masowym momentem bezwładności wahadła względem osi obrotu, natomiast $\ddot{\varphi}$ i $\dot{\varphi}$ to przyspieszenie i prędkość kątowa wahadła.

Badania symulacyjne i eksperymentalne prowadzone na rzecz tej pracy rozciągały się na przestrzeni ponad czterech lat. Tak długi czas powodował, że stanowisko doświadczalne, w pewnym stopniu, zmieniało się na skutek zużycia oraz prowadzonych modernizacji. Zmiany te miały wpływ na badania symulacyjne, ponieważ za każdym razem należało dostosowywać równanie ruchu oraz parametry układu w odniesieniu do obecnego stanu stanowiska, tak aby jak najlepiej odwzorowywały jego zachowanie. Najczęściej zmiany te dotyczyły czynników wpływających na opory ruchu w łożyskach, a w mniejszym stopniu parametrów elementu podatnego czy oddziaływania magnetycznego. Dlatego też poszczególne wyniki badań zawarte w tej pracy, będą opatrzone konkretnym równaniem ruchu wraz z parametrami, dla których zostały otrzymane. Pomimo różnych modeli dynamicznych stosowanych podczas badań, wyniki otrzymane na ich podstawie są jakościowo i ilościowo podobne, dzięki czemu można je ze sobą zestawiać i porównywać.

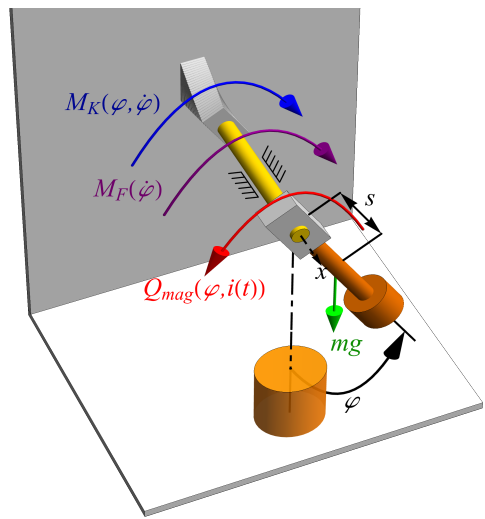

**Rys. 2.5.** Model fizyczny układu pojedynczego wahadła magnetycznego.

### 2.2.2 Modele oporów ruchu

Badany układ posiada dwa źródła oporów ruchu, które opisane są w równaniu (2.1) przez moment $M_F(\dot{\varphi})$. Pierwszym źródłem jest typowe tłumienie wiskotyczne pochodzące głównie od powietrza, natomiast drugim są łożyska, w których zamocowany jest wał wahadła. Występowanie samego tłumienia wiskotycznego, które jest proporcjonalne do prędkości wahadła powodowałoby eksponencjalny charakter zmniejszania się kolejnych amplitud wychyleń wahadła. Jednakże przeprowadzone eksperymenty pokazały, że amplitudy kolejnych wychyleń zmniejszają się przede wszystkim w sposób liniowy, a nie eksponencjalny (rys. 2.6), co wskazuje na pojawianie się w układzie stałych momentów oporów ruchu, gdy tylko prędkość wahadła nie jest równa zeru. Zaproponowano więc dwa modele

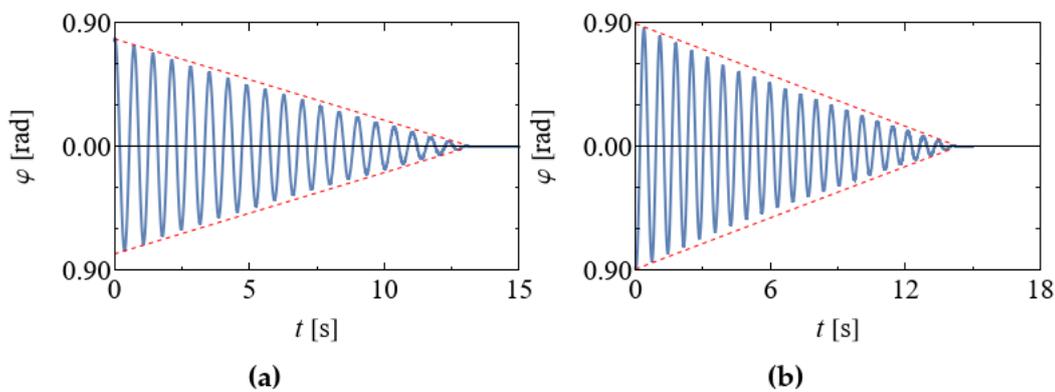

**Rys. 2.6.** Przykładowe eksperymentalne drgania swobodne wahadła, gdzie przerywane czerwone linie odwzorowują liniowy charakter gaśnięcia kolejnych amplitud.

oporów ruchu. Pierwszy model oparty jest na prostym modelu tarcia Coulomba [135] połączonym z tłumieniem wiskotycznym i wyrażony równaniem

$$M_{CM}(\dot{\varphi}) = M_c \operatorname{sgn}(\dot{\varphi}) + c\dot{\varphi} \approx M_c \operatorname{tgh}(\varepsilon_c \dot{\varphi}) + c\dot{\varphi}, \qquad (2.2)$$

gdzie $M_c$ jest stałą wartością momentu oporów Coulomba (kinetycznych), $c$ jest współczynnikiem tłumienia wiskotycznego, a $\varepsilon_c$ jest paramterem regularyzacyjnym. Tangens hiperboliczny stosuje się jako aproksymacje funkcji signum, głównie ze względu na obliczenia numeryczne, dla których występowanie nieciągłych funkcji w modelu dynamicznym może stać się problematyczne. Drugi model oporów ruchu oprócz tłumienia wiskotycznego uwzględnia tzw. efekt Sribecka, czyli efekt przejścia z oporów statycznych do kinetycznych w sposób ciągły i zależny od prędkości wahadła. Model ten opisany jest następującym równaniem

$$M_{SE}(\dot{\varphi}) = \left[ M_c + (M_s - M_c) \exp\left( -\frac{\dot{\varphi}^2}{v_s^2} \right) \right] \operatorname{tgh}(\chi \dot{\varphi}) + c\dot{\varphi}, \qquad (2.3)$$

gdzie analogicznie jak w (2.2) parametr $M_c$ jest stałą wartością momentu oporów Coulomba, $M_s$ jest momentem statycznym oporów ruchu, $v_s$ jest prędkością Stribecka definiującą jak szybko opór statyczny przejdzie w kinetyczny, $\chi$ jest parametrem regularyzacyjnym, a $c$ jest współczynnikiem tłumienia wiskotycznego. Aproksymacją modelu (2.3) przydatną np. do badań analitycznych może być następującą funkcja wymierna

$$M_{SEa}(\dot{\varphi}) = \frac{\mu}{\kappa}\dot{\varphi}\left(1 - \sigma\frac{\dot{\varphi}^2}{\dot{\varphi}^2 + \kappa^2}\right), \qquad (2.4)$$

gdzie $\mu$, $\kappa$ i $\sigma$ są stałymi parametrami.

Rysunek 2.7 przedstawia dopasowanie modeli (2.2), (2.3) i (2.4) do danych eksperymentalnych $M_{Fexp}$. Dane eksperymentalnego momentu oporów ruchu uzyskano na podstawie prędkości i przyspieszeń zarejestrowanych podczas ruchu swobodnego i przetworzonych przez równanie

$$M_{Fexp}(\dot{\varphi}) = -J\ddot{\varphi} - mgs\sin(\varphi). \qquad (2.5)$$

Analizując dane eksperymentalne łatwo zauważyć występowanie szczytu (ang.

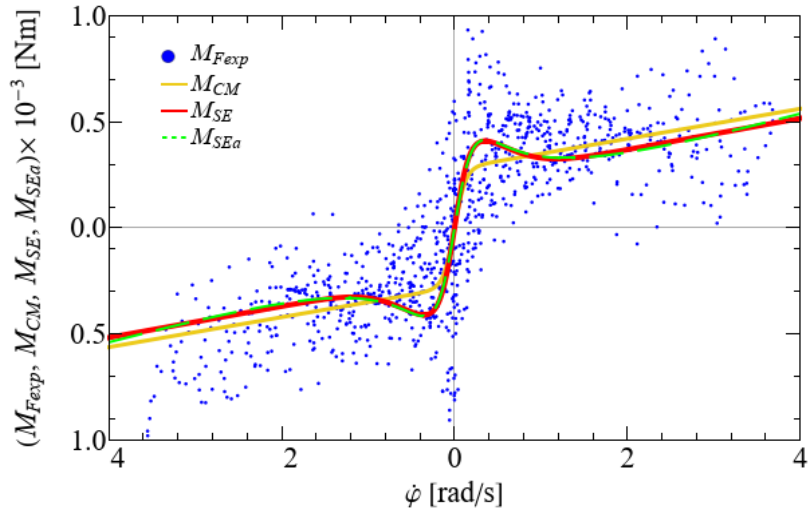

**Rys. 2.7.** Zestawienie modeli $M_{CM}$ oraz $M_{SE}$ opisujących opory ruchu układu z danymi eksperymentalnymi $M_{Fexp}$.

*peak*) momentu statycznego w obrębie małych prędkości, a następnie jego spadek i przejście w moment kinetyczny, co dobrze odwzorowują modele (2.3) i (2.4). Ze względu na małą wartość szczytową momentu statycznego względem wartości momentu kinetycznego, model (2.2) również może być z powodzeniem wykorzystywany podczas modelowania matematycznego badanego układu.

### 2.2.3 Model gumowego elementu podatnego

W badanym układzie element podatny pełni rolę sprężyny o danej sztywności oraz tłumieniu. Podczas ruchu wahadła podlega on odkształceniu i działa na wał momentem oznaczonym w równaniu (2.1) symbolem $M_K$. Założono, że element ten charakteryzuje się stałą sztywnością $k_e$ oraz stałym współczynnikiem tłumienia wiskotycznego $c_e$, zaś sam model momentu jest następujący

$$M_K(\dot{\varphi}, \varphi) = c_e\dot{\varphi} + k_e\varphi. \qquad (2.6)$$

Ponieważ guma może wykazywać się nieliniową sztywnością oraz tłumieniem [136, 137], postanowiono sprawdzić czy zaproponowany model (2.6) wystarczająco dobrze odwzorowuje eksperyment. W tym celu, element podatny poddano badaniom dynamicznego skręcania, tak jak to pokazano na rys. 2.8. Z jednej stro-

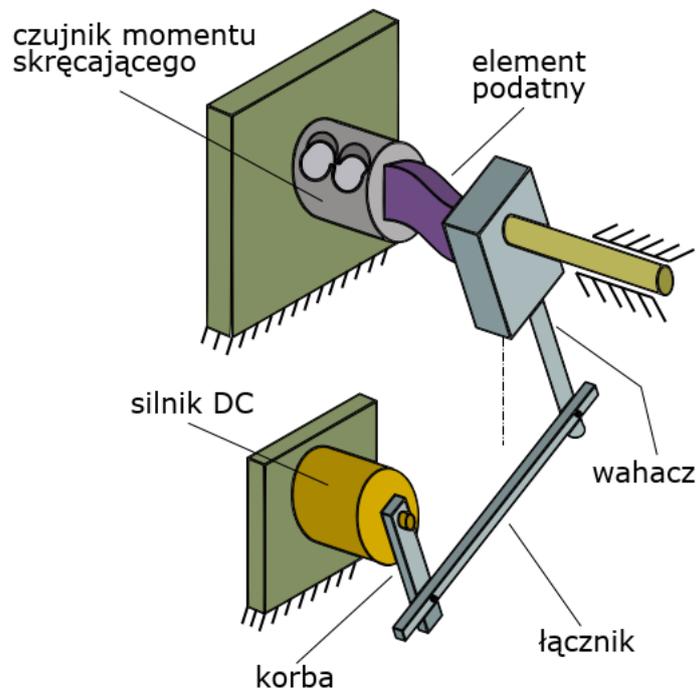

**Rys. 2.8.** Schemat stanowiska eksperymentalnego służącego do wyznaczenia charakteru tłumienia elementu podatnego.

ny element przymocowany jest do tensometrycznego czujnika momentu skręcającego (element (7) na rys. 2.1), zaś z drugiej strony do wahacza mechanizmu korbowo wahaczowego. Mechanizm ten napędzany jest silnikiem prądu stałego i zapewnia skrętne wymuszenie kinematyczne elementu. Podczas badania dokonywano pomiaru momentu $M_{Sexp}$ skręcającego element w zależności od okresowo zmieniającego się w czasie kąta skręcenia. Wyniki eksperymentalne momentu skręcającego oraz odpowiadającej mu symulacji modelu (2.6) pokazano na rys. 2.9. Analizując wyniki symulacji i eksperymentu można stwierdzić, że gumowy element charakteryzuje się pewnymi nieliniowościami przez co symulacja nieznacznie odbiega od eksperymentu, ale nie są to na tyle duże rozbieżności, żeby koniecznym było stosowanie bardziej skomplikowanego modelu momentu $M_K$.

### 2.2.4 Empiryczne modele momentu siły oddziaływania magnetycznego

Istnieje wiele prac [120–133], w których autorzy w sposób teoretyczny opierając się na prawach elektromagnetycznych (Biota Savarta, Gaussa, Coulomba, Ampera, Lorentza, Maxwella) opracowywali modele matematyczne sił i momentów sił, jakie występują podczas oddziaływań magnetycznych pomiędzy parami:

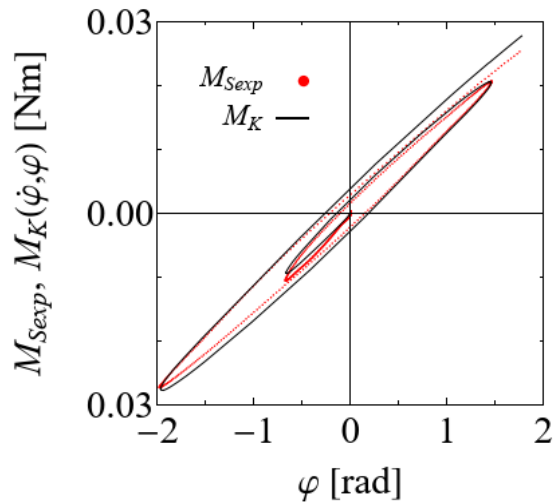

**Rys. 2.9.** Porównanie wyników symulacji modelu matematycznego (2.6) gumowego elementu podatnego z danymi eksperymentalnymi.

magnes-magnes, magnes-cewka i cewka-cewka. Modele te wykorzystują obszerne i skomplikowane formuły matematyczne, które z punktu wiedzenia badań dynamiki układów mechanicznych z oddziaływaniami magnetycznymi komplikują ich modele symulacyjne wydłużając czas obliczeń. Z tego względu podczas badań dynamiki takich układów, modele te zazwyczaj aproksymuje się prostszymi funkcjami o postaci wielomianowej [36, 71, 79, 91]. W tym paragrafie przedstawiono badania eksperymentalne oraz modelowanie matematyczne momentu siły działającego na wahadło, który jest wynikiem oddziaływania magnetycznego zachodzącego pomiędzy magnesem umieszczonym na jego końcu a cewką elektryczną zasilaną dodatnim prądem stałym. Wspomniany moment siły będzie nazywany „momentem magnetycznym". Opracowane modele momentu magnetycznego wykorzystywane będą do dalszych badań dynamiki układu.

**Eksperyment**

Badania rozpoczęto od wyznaczenia w sposób eksperymentalny charakterystyki momentu magnetycznego $Q_{mag}$ jaki działa na wahadło, gdy magnes i cewka elektryczna nawzajem się odpychają. Cewka zasilana jest dodatnim prądem o stałym natężeniu. Schemat układu pomiarowego pokazano na rys. 2.10, a moment magnetyczny oznaczono jako $M_{exp}$. Dokonując tensometrycznego pomiaru siły $F_{exp}$ w cięgnie i kąta wychylenia wahadła $\varphi$, moment magnetyczny przy statycznym pomiarze może zostać wyznaczony z równania równowagi względem punktu zawieszenia. Równanie to można zapisać w następujący sposób

$$M_{exp}(\varphi) = mgs\sin\varphi + F_{exp}R_D, \qquad (2.7)$$

gdzie $m$ jest masą wahadła, $g$ przyspieszeniem ziemskim, $s$ odległością pomiędzy środkiem ciężkości wahadła a osią obrotu, a $R_D$ promieniem dysku. Wykonane w ten sposób pomiary zostaną porównane z opracowanymi w dalszej części rozdziału modelami matematycznymi. Należy wziąć pod uwagę, że pomiary momentu magnetycznego są obarczone niewielkim błędem wynikającym z nieuwzględnienia w równaniu (2.7) momentu oporu powstającego w łożyskach

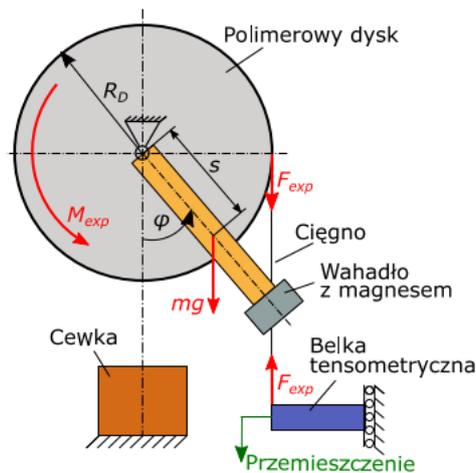

**Rys. 2.10.** Eksperymentalna metoda pomiaru momentu sił oddziaływania magnetycznego.

podczas przemieszczania się wahadła. Wpływ momentu oporu łożysk widać szczególnie dla zerowego kąta wychylenia wahadła, gdzie wg teorii wartość momentu magnetycznego powinna wynosić zero.

**Uproszczony model momentu magnetycznego wg modelu Gilberta**
Pierwsze próby modelowania oddziaływania magnetycznego opierały się na tzw. modelu Gilberta [33, 129, 132, 134]. Model ten jest analogią do coulombowskiej siły powstającej miedzy dwoma ładunkami elektrycznymi lub siły wzajemnego przyciągania się ciał wynikającej z powszechnego prawa ciążenia. Model ten zakłada, że siłę wzajemnego oddziaływania pomiędzy hipotetycznie odseparowanymi biegunami dipola magnetycznego można wyrazić następującym wzorem

$$F_{mag} = C_\mu \frac{q_{m1} q_{m2}}{r^2}, \qquad (2.8)$$

gdzie $C_\mu$ jest stałą zależną od przenikalnością magnetycznej ośrodka, $q_{m1}$ i $q_{m2}$ są wartościami ładunków magnetycznych odseparowanych biegunów dipola magnetycznego, a $r$ jest odległością od środków tych biegunów. Ponadto model ten zakłada, że bieguny magnetyczne mają pomijalnie małe wymiary geometryczne i w przybliżeniu są punktami. Traktując magnes umieszczony na końcu wahadła oraz cewkę elektryczną (zasilaną prądem o stałym natężeniu) jako odseparowane bieguny dipola magnetycznego, tak jak pokazano to na rys. 2.11, siła $F_{mag}$ będzie generowała moment siły względem punktu zawieszenia wahadła, opisany równaniem

$$M_{F_{mag}}(\varphi) = \frac{|\varphi|}{\varphi} \cdot F_{mag} \cdot \ell \cdot \cos\left(\text{arc tg}\frac{d}{\ell \sin \varphi} \quad \varphi\right), \qquad (2.9)$$

gdzie $d_g$ jest odległością pomiędzy środkami biegunów dipola magnetycznego, gdy wahadło jest w spoczynku, $\ell$ jest odległością pomiędzy środkiem magnesu a osią obrotu wahadła, odległość $d = d_g + \ell(1 \quad \cos \varphi)$, natomiast odległość między środkami biegunów dipola magnetycznego w czasie ruchu wahadła wynosi $r = \sqrt{d^2 + (\ell \sin \varphi)^2}$. Pomimo swojej teoretycznej prostoty, model ten nie mógł zostać

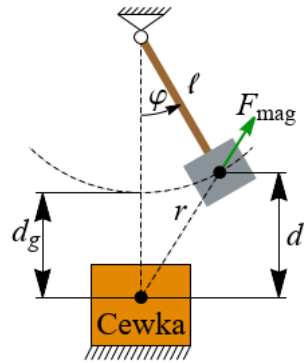

**Rys. 2.11.** Uproszczony schemat oddziaływania magnetycznego dla modelu Gilberta.

zastosowany podczas badań symulacyjnych ze względu na znaczne odbieganie od danych eksperymentalnych, co jest widoczne na rys. 2.12. Dodatkowo, w modelu występuje nieciągłość dla $\varphi = 0$ rad, która mogłaby być problematyczna podczas obliczeń symulacyjnych.

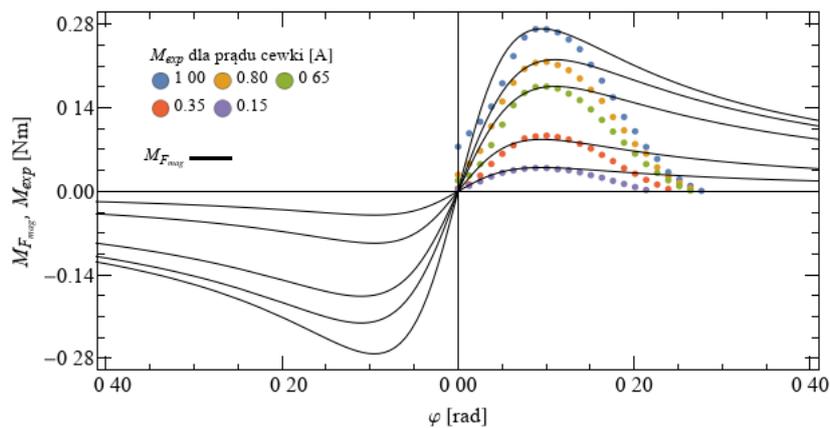

**Rys. 2.12.** Porównanie modelu momentu magnetycznego $M_{F_{mag}}$ z danymi eksperymentalnymi $M_{exp}$ wyznaczonymi dla różnych wartości stałego natężenia prądu cewki.

**Model gaussowski momentu magnetycznego**

Pomimo, że uproszczony model momentu magnetycznego (2.9) nie daje satysfakcjonującego dopasowania do danych eksperymentalnych, posłużył on jako baza do opracowania lepszego modelu nazwanego gaussowskim. Dokonując numerycznej analizy matematycznej momentu $M_{F_{mag}}$, obliczono jego potencjał dla kilku stałych prądów cewki zgodnie ze wzorem [2]

$$V_{M_{Fmag}}(\varphi) = \int M_{F_{mag}}(\varphi)\,d\varphi, \qquad (2.10)$$

a ich przebiegi pokazano na rys. 2.13a. Przebiegi te mają kształty zbliżone do krzywej Gaussa, opisanej równaniem

$$V_{M_{mag}}(\varphi) = a \cdot \exp\left(\frac{\varphi^2}{b}\right), \qquad (2.11)$$

gdzie $a$ i $b$ są parametrami tej funkcji odpowiadającymi za jej kształt.

Z tego też względu, postanowiono w oparciu o funkcję (2.11) opracować tzw. gaussowski model momentu magnetycznego, wyrażony następującym wzorem

$$M_{mag}(\varphi) = \frac{dV_{M_{mag}}(\varphi)}{d\varphi} = \frac{2a}{b} \exp\left(\frac{\varphi^2}{b}\right) \varphi. \tag{2.12}$$

Dopasowanie gaussowskiego modelu momentu magnetycznego do eksperymentu jest znacznie lepsze niż w przypadku modelu Gilberta, co ukazuje rys. 2.14, szczególnie widać to dla kątów wychylenia większych od 0.1 rad. Ponadto, potencjał modelu gaussowskiego jest znacznie węższy w dziedzinie kąta wychylenia niż w modelu Gilberta (rys. 2.13b). Parametry $a$ i $b$ zależą nie tylko od natężenie

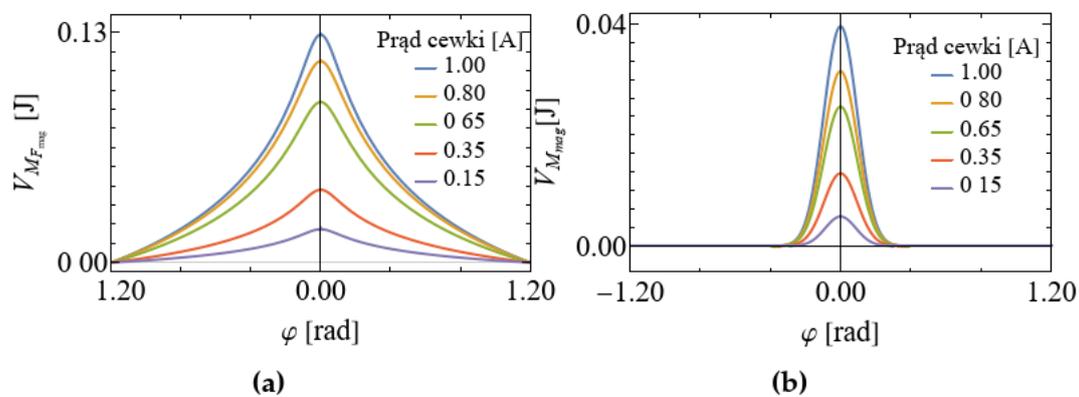

**Rys. 2.13.** Potencjały momentów magnetycznych w przypadku (a) modelu Gilberta $V_{M_{F_{mag}}}$ oraz (b) modelu gaussowskiego $V_{M_{mag}}$, wyznaczone dla różnych wartości stałego natężenia prądu cewki.

prądu cewki, ale także od innych czynników takich jak: parametry geometryczne cewki, odległość magnesu od osi obrotu wahadła, siła pola magnetycznego magnesu, czy odległość pomiędzy magnesem a cewką w czasie spoczynku wahadła. W przypadku badanego układu, tylko pierwszy czynnik jakim jest natężenie prądu cewki $i(t)$ może ulegać zmianie, reszta parametrów jest stała w czasie. Z tego względu, w celu uogólnienia modelu (2.12) warto wyznaczyć zależności tych parametrów od wartości natężenia prądu cewki. Wykres przedstawiający zmiany parametrów $a$ i $b$ w zależności od różnych stałych natężeń prądu cewki, został pokazany na rys. 2.15. Jasno daje on do zrozumienia, że parametr $a$ rośnie w sposób liniowy wraz ze wzrostem natężenia prądu, natomiast parametr $b$ ma stałą wartość, co można zapisać w następujący sposób

$$a(i) = a_I \cdot i(t), \qquad b(i) = b = const, \tag{2.13}$$

gdzie $a_I$ jest prądowym współczynnikiem kierunkowym. Biorąc to pod uwagę, uogólnione równanie (2.12) momentu magnetycznego dla przypadku dowolnego (niekoniecznie stałego) natężenia prądu cewki wyraża się następującym równaniem

$$\widehat{M}_{mag}(\varphi, i(t)) = \frac{2a_I}{b} \exp\left(\frac{\varphi^2}{b}\right) \varphi \cdot i(t). \tag{2.14}$$

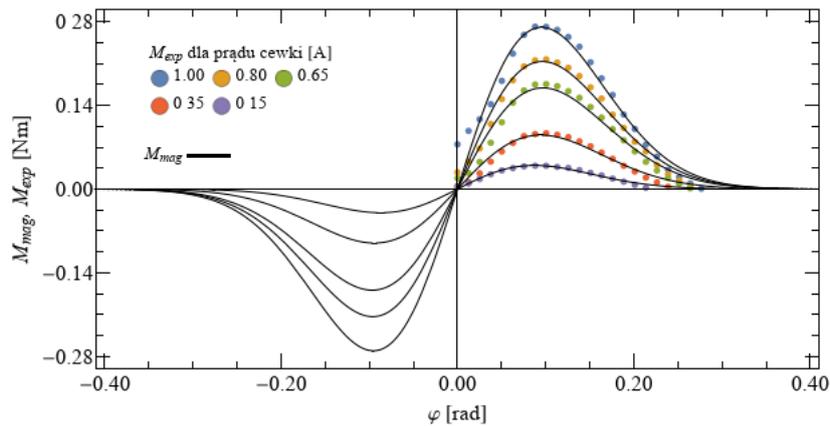

**Rys. 2.14.** Porównanie modelu gaussowskiego momentu magnetycznego $M_{mag}$ z danymi eksperymentalnymi $M_{exp}$ wyznaczonymi dla różnych wartości stałego natężenia prądu cewki.

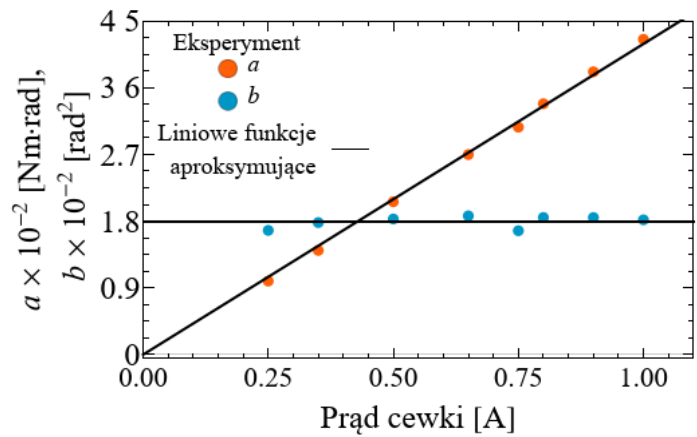

**Rys. 2.15.** Zależność parametrów $a$ i $b$ gaussowskiego momentu magnetycznego $M_{mag}$ od wartości natężenia prądu cewki.

**Modele wielomianowe momentu magnetycznego**

Pierwszym tzw. modelem wielomianowym momentu magnetycznego jest funkcja wymierna w postaci

$$M_{Rmag}(\varphi) = p \frac{\varphi/q}{1 + (\varphi/q)^4}, \qquad (2.15)$$

gdzie $p$ i $q$ są parametrami, które tak jak parametry modelu (2.12) zależą przede wszystkim od natężenia prądu cewki. Dopasowanie modelu $M_{Rmag}$ do danych eksperymentalnych i zależność jego parametrów $p$, $q$ od różnych stałych natężeń prądu cewki zostały pokazane na rys. 2.16. Jak widać, charakter zależność tych parametrów jest taki sam jak w modelu (2.12), tzn. parametr $p$ jest liniową funkcją natężenia prądu $i(t)$, podczas gdy parametr $q$ jest stały, dlatego

$$p(i) = p_I \cdot i(t), \qquad q(i) = q = const, \qquad (2.16)$$

gdzie $p_I$ jest prądowym współczynnikiem kierunkowym. Podstawiając (2.16) do (2.15), otrzymujemy formułę na wielomianowy moment magnetyczny $M_{Rmag}$

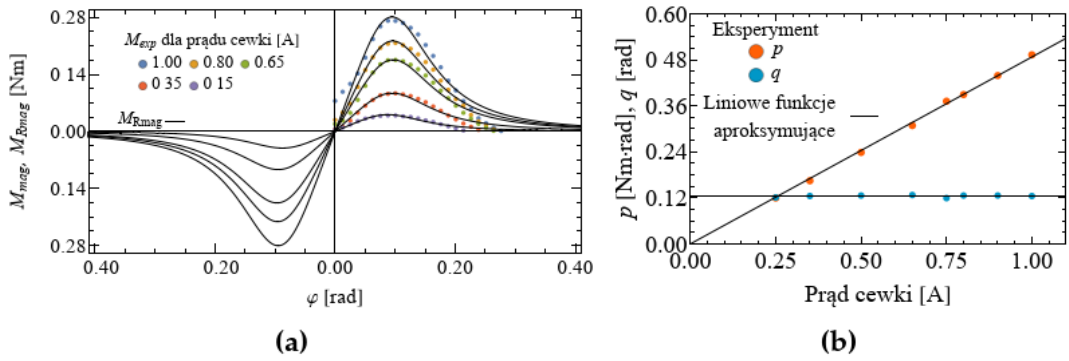

**Rys. 2.16.** Porównanie modelu momentu magnetycznego $M_{Rmag}$ z danymi eksperymentalnymi $M_{exp}$ (a) oraz zależność jego parametrów $p$ i $q$ od wartości stałego natężenia prądu cewki (b).

w zależności od dowolnego sygnału prądowego cewki $i(t)$ w postaci

$$\widehat{M}_{Rmag}(\varphi, i(t)) = p_I \frac{\varphi/q}{1 + (\varphi/q)^4} \cdot i(t). \tag{2.17}$$

Drugi model wielomianowy momentu magnetycznego stanowi funkcja wymierna mająca w mianowniku wielomian stopnia drugiego i jest zapisana jako

$$M_{Amag2}(\varphi) = \frac{2a\varphi}{b + \varphi^2}, \tag{2.18}$$

gdzie parametry $a$ i $b$ są tymi samymi parametrami, co w modelu (2.12). Rys. 2.17 przedstawia dopasowanie tego modelu do danych eksperymentalnych $M_{exp}$ oraz porównanie z modelem $M_{mag}$ dla przypadku jednego natężenia prądu cewki wynoszącego 1 A. Można zauważyć, że dla małych kątów wychylenia wahadła funkcja ta dobrze dopasowuje się do danych eksperymentalnych (z zakresu 0.05-0.1 rad) jak i do modelu gaussowskiego, jednak dla większych kątów zaczyna od nich znacząco odbiegać. Poprawę dokładności dopasowania modelu można osiągnąć poprzez zwiększenie stopnia wielomianu mianownika, mianowicie dla wielomianu czwartego stopnia otrzymamy następujące równanie

$$M_{Amag4}(\varphi) = \frac{2a\varphi}{b + \varphi^2 + \frac{1}{2b}\varphi^4}, \tag{2.19}$$

natomiast dla wielomianu szóstego stopnia

$$M_{Amag6}(\varphi) = \frac{2a\varphi}{b + \varphi^2 + \frac{1}{2b}\varphi^4 + \frac{1}{6b^2}\varphi^6}. \tag{2.20}$$

Wzrost dokładności aproksymacji w zależności od stopnia wielomianu mianownika można zaobserwować na rys. 2.17. Podobnie jak dla modeli (2.14) i (2.17), aby stosować formuły (2.18), (2.19) i (2.20) w przypadku, gdy prąd $i(t)$ płynący przez cewkę ma dowolny przebieg, należy dokonać podstawienia zależności (2.13), w efekcie czego otrzymamy

$$\widehat{M}_{Amag2}(\varphi, i(t)) = \frac{2a_I \varphi}{b + \varphi^2} \cdot i(t), \tag{2.21}$$

$$\widehat{M}_{Amag4}(\varphi, i(t)) = \frac{2a_I \varphi}{b + \varphi^2 + \frac{1}{2b}\varphi^4} \cdot i(t), \qquad (2.22)$$

$$M_{Amag6}(\varphi, i(t)) = \frac{2a_I \varphi}{b + \varphi^2 + \frac{1}{2b}\varphi^4 + \frac{1}{6b^2}\varphi^6} \cdot i(t). \qquad (2.23)$$

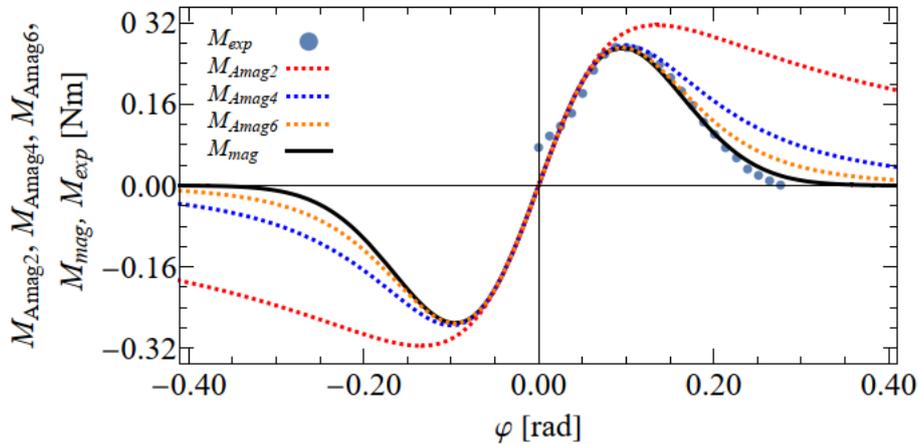

**Rys. 2.17.** Dopasowanie wielomianowych modeli $M_{Amag2}$, $M_{Amag4}$, $M_{Amag6}$ do danych eksperymentalnych oraz porównanie z gaussowsowskim momentem magnetycznym dla stałego prądu cewki równego 1 A.

### 2.2.5 Model prostokątnego pulsującego sygnału prądowego płynącego w cewce elektrycznej

Prąd $i(t)$ płynący w cewce elektrycznej badanego układu ma formę prostokątnego pulsującego sygnału, tak jak to pokazano na rys. 2.3. Opracowany model matematyczny tego sygnału bazuje na funkcji sinus i przedstawiony jest następującą formułą

$$i_p(t) = \frac{1}{2}I_0 \left\{1 + \operatorname{tgh}\left[\varepsilon_i \left(\sin(\omega_i(t + t_0)) + i_0\right)\right]\right\}, \qquad (2.24)$$

gdzie $\varepsilon_i$ jest parametrem regularyzacyjnym, $\omega_i$ jest częstością kołową sinusoidy, $i_0$ pozwala na zmianę wypełnienia prostokątnego sygnału, $t_0$ odpowiada za ustawienie początkowego przesunięcia fazowego sinusoidalnej fali, a $I_0$ jest amplitudą prostokątnego pulsującego sygnału prądowego. Idea konstrukcji tego sygnału została zobrazowana na rys. 2.18. W celu wyrażenia formuły (2.24) w odniesieniu do częstotliwości $f$ i wypełnienia $w$ [%] prostokątnego sygnału prądowego cewki, należy dokonać następujących podstawień

$$\omega_i = 2\pi f, \; i_0 = \sin\left(\frac{1}{2}\left(\pi - \frac{2\pi w}{100}\right)\right), \; t_0 = \frac{1}{2\omega_i}\left(\pi - \frac{2\pi w}{100}\right). \qquad (2.25)$$

Początkowe przesunięcie fazowe sinusoidy $\omega_i t_0$ jest obliczane w taki sposób, żeby spowodować wystąpienie narastającego zbocza prostokątnego sygnału w chwili czasowej $t = 0$ s.

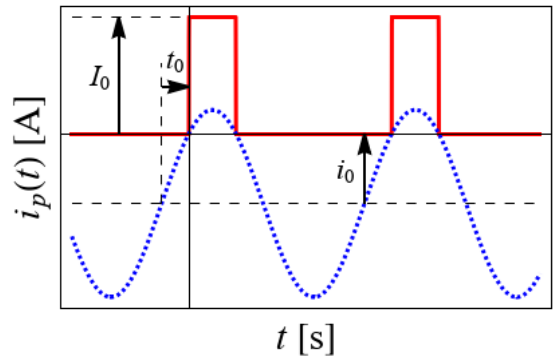

**Rys. 2.18.** Idea modelowania prostokątnego pulsującego sygnału prądowego $i_p(t)$ płynącego w cewce elektrycznej.

### 2.2.6 Identyfikacja parametrów

Identyfikowanie wartość parametrów modeli matematycznych przedstawionych w podrozdziale 2.2 odbywało się wieloetapowo. W pierwszym etapie zajęto się tymi parametrami, których wartości można oszacować bezpośrednio w oparciu o proste eksperymenty. Do parametrów tych należą: iloczyn $mgs$, masowy moment bezwładności wahadła $J$ oraz sztywność gumowego elementu podatnego $k_e$. Wartość iloczynu $mgs$ można oszacować na podstawie eksperymentu przedstawionego na rys. 2.19a. Cięgno przymocowane do polimerowego dysku obciążane jest odważnikami o znanej masie $m_{odw}$. Siła ciężkości $m_{odw}g$ tego odważnika przyłożona w odległości promienia $R_D$ dysku daje moment siły odchylający wahadło. Mierząc kąt wychylenia wahadła oraz znając wartość siły ciężkości odważnika i promień dysku, wartość iloczynu można oszacować ze wzoru $mgs = (m_{odw}gR_D)/(\sin\varphi)$. Następnie, znając $mgs$ można oszacować wartość masowego momentu bezwładności $J$ wahadła w oparciu o jego okres drgań swobodnych $T_p$, korzystając z tzw. „metody wahadła", gdzie $J = (T_p^2 mgs)/(4\pi^2)$. Ostatni parametr jaki można w prosty sposób wyznaczyć z eksperymentu to wartość współczynnika sztywności $k_e$ gumowego elementu podatnego. Można to zrobić w podobny sposób jak w przypadku iloczynu $mgs$. Poprzez statyczne skręcanie elementu znanym momentem i pomiar kąta skręcenia, można wyznaczyć charakterystykę sztywności i oszacować wartość $k_e$, która jest równa współczynnikowi kierunkowemu prostej aproksymującej dane eksperymentalne (rys. 2.19b).

Drugi etap identyfikacji dotyczył parametrów, których wartości nie da się oszacować na podstawie prostych wzorów podpartych eksperymentami. Do takich parametrów należą te zawarte w modelach oporów ruchu (2.2) i (2.3), współczynnik tłumienia wiskotycznego elementu podatnego w modelu (2.6), a także współczynniki w modelach momentów magnetycznych (2.12) i (2.15). W przypadku tych parametrów, do wyznaczenia ich wartości posługiwano się metodami dopasowującymi funkcję (model matematyczny) do danych eksperymentalnych, tak jak zostało to przedstawione na rys. 2.7 i 2.14. Proces dopasowywania modelu matematycznego do danych eksperymentalnych przeprowadzano w programie *Wolfram Mathematica* i przy użyciu jego wbudowanych funkcji LinearModelFit[...] oraz NonlinearModelFit[...], stosowanych odpowiednio do liniowych i nieliniowych modeli aproksymujących.

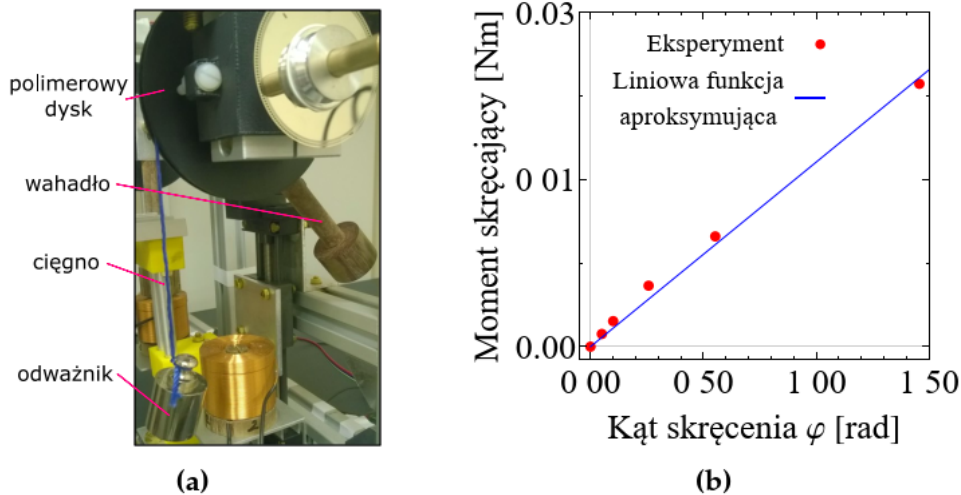

**Rys. 2.19.** Eksperymentalny sposób wyznaczania wartości iloczynu $mgs$ (a) oraz wyniki eksperymentu statycznego skręcania elementu podatnego potrzebne do oszacowania współczynnika $k_e$ sztywności (b).

Ostatni etap identyfikacji polegał na skorygowaniu wyznaczonych na wcześniejszych etapach wartości parametrów w oparciu o eksperymentalne przebiegi czasowe ruchu swobodnego i wymuszonego układu. Dokonywano tego poprzez dopasowanie rozwiązania równania dynamicznego ruchu układu (2.1) do wspomnianych eksperymentalnych przebiegów czasowych. Na tym etapie, proces dopasowania prowadzono tylko przy użyciu funkcji NonlinearModelFit[...] i aby uzyskać jak najlepszą zgodność dopasowania, jako wartości początkowe poszukiwanych parametrów ustawiano wartości oszacowane w pierwszym i drugim etapie. Dopasowanie rozwiązania równania ruchu do eksperymentu ruchu swobodnego, miało na celu ostateczne wyznaczenie wartości parametrów $J$ i $mgs$ oraz wartości parametrów występujących w modelach oporów ruchu i elementu podatnego. Natomiast dopasowywanie rozwiązania do eksperymentu (okresowego) ruchu wymuszonego, miało na celu skorygowanie wartości parametrów występujących w modelu momentu magnetycznego. Na rys. 2.20 pokazano dopasowanie rozwiązań dwóch różnych równań dynamicznych ruchu do danych doświadczalnych ruchu swobodnego. Symulacje zostały przeprowadzone dla równań ruchu (2.1) z zaimplementowanym modelem elementu podatnego (2.6) oraz dwoma modelami oporów ruchu Stribecka (2.3) i Coulomba (2.2), wyniki pokazano odpowiednio na rys. 2.20a,c i rys. 2.20b,d. Przypadek dopasowania dla okresowego ruchu wymuszonego został pokazany na rys. 2.21, gdzie przez cewkę płynął prostokątny sygnał prądowy zamodelowany równaniem (2.24). W tym przypadku, równanie ruchu (2.1) miało zaimplementowany model oporów z efektem Stribecka (2.3), model (2.6) elementu podatnego oraz gaussowski model oddziaływania magnetycznego (2.14).

Z przedstawionych na rys. 2.20 i 2.21 przykładowych porównań pomiędzy danymi eksperymentalnymi a symulacjami można wnioskować, że opracowane modele matematyczne bardzo dobrze odzwierciedlają zachowanie rzeczywistego układu wahadła magnetycznego zarówno podczas drgań swobodnych jak i wymuszonych.

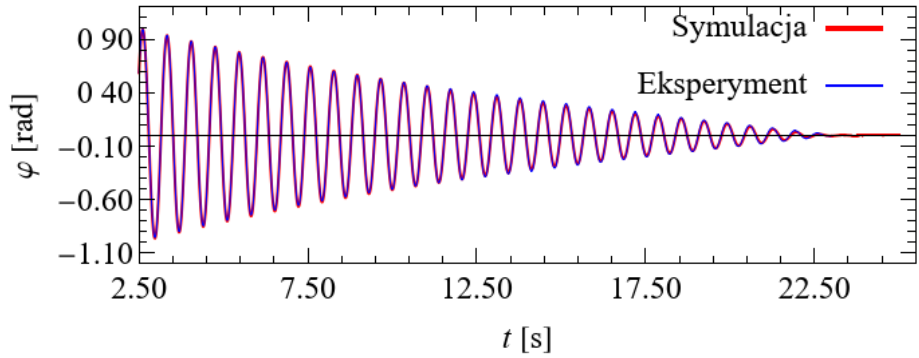

(a)

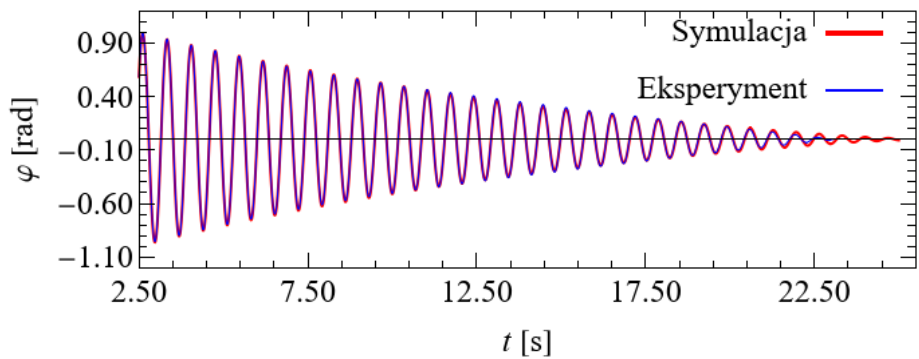

(b)

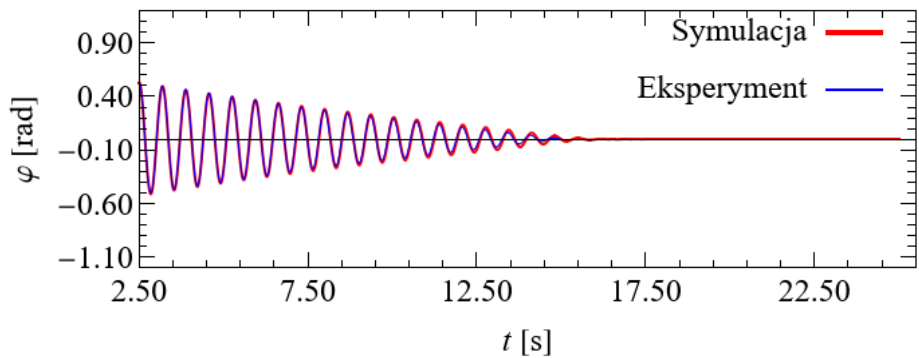

(c)

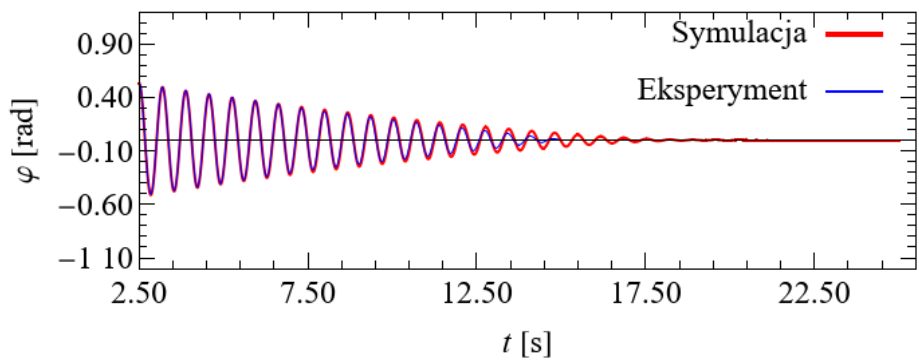

(d)

**Rys. 2.20.** Przykładowe dopasowanie symulacji drgań swobodnych do danych eksperymentalnych. Symulacje przeprowadzone dla równania dynamicznego (2.1) z oporami ruchu Stribecka (2.3) (a)-(c) i Coulomba (2.2) (b)-(d).

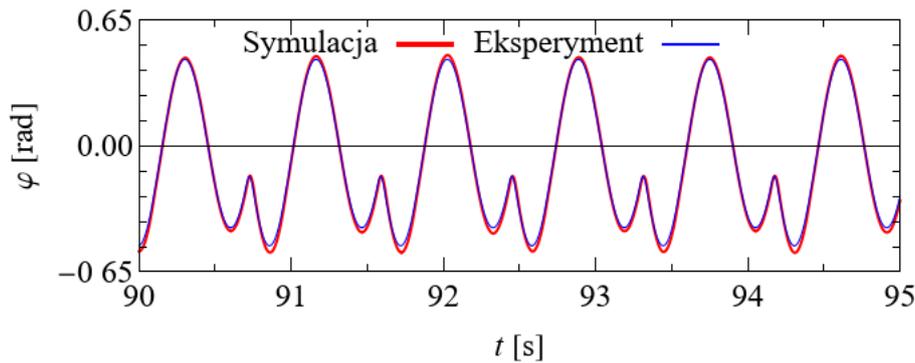

**Rys. 2.21.** Przykładowe dopasowanie symulacji drgań wymuszonych do danych eksperymentalnych. Symulacje przeprowadzone dla równania dynamicznego (2.1) z oporami ruchu Stribecka (2.3), modelem momentu magnetycznego (2.14) i następujących parametrów prostokątnego sygnału prądowego (2.24): $I_0 = 1$ A, $f = 1.16$ Hz, $w = 30\%$.

## 2.3 Dynamika nieliniowa wahadła magnetycznego

Podrozdział zawiera podstawowe badania dynamiki nieliniowej pojedynczego wahadła magnetycznego. Przedstawione zostaną typowe zachowania nieliniowe takie jak chaos, multistabilność czy skoki amplitudy. Badania będą prowadzone przy użyciu metod numerycznych i metody analityczno numerycznej. Metoda analityczno numeryczna oparta jest na analitycznej metodzie uśredniania (ang. *averaging method*) i posłuży do badania okresowych rozwiązań dynamiki układu. Obliczenia numeryczne i symboliczne prowadzone w czasie badań wykonane zostały przy użyciu programu *Wolfram Mathematica*.

Początkowy model matematyczny układu jaki został przyjęty podczas analizy, był następujący

$$J\ddot{\varphi} = mgs\sin\varphi \quad M_{SE}(\dot{\varphi}) \quad M_K(\varphi,\dot{\varphi}) + \widehat{M}_{mag}(\varphi, i_p(t)). \quad (2.26)$$

Model ten w dużej mierze odzwierciedla rzeczywisty układ pokazany na rys. 2.2. Jednak fakt, że zawiera on silnie nieliniowe funkcje $M_{SE}$ oraz $\widehat{M}_{mag}$ powoduje, że prowadzenie badań opartych na metodach analitycznych staje się skomplikowane.

### 2.3.1 Metoda uśredniania dla słabo nieliniowego równania ruchu i wykresy rezonansowe

Rozważmy przypadek małych drgań wahadła, tj. kiedy kąty wychylenia $\varphi$ są małe i możemy założyć, że $\sin(\varphi) \approx \varphi$. Wtedy początkowy model (2.26) można zapisać w zmodyfikowanej formie

$$J\ddot{\varphi} + (mgs + k_e)\varphi = M_{SE}(\dot{\varphi}) \quad c_e\dot{\varphi} + \widehat{M}_{mag}(\varphi, i_p(t)), \quad (2.27)$$

**Tabela 2.1.** Parametry układu pojedynczego wahadła magnetycznego przyjęte podczas badań analityczno-numerycznych i numerycznych.

| | | | |
|---|---|---|---|
| $J$ | $6.786 \cdot 10^{-4}\,\mathrm{kgm^2}$ | $v_s$ | $0.835\,\frac{\mathrm{rad}}{\mathrm{s}}$ |
| $mgs$ | $5.707 \cdot 10^{-2}\,\mathrm{Nm}$ | $\chi$ | $5.759\,\frac{\mathrm{s}}{\mathrm{rad}}$ |
| $k_e$ | $2.264 \cdot 10^{-2}\,\frac{\mathrm{Nm}}{\mathrm{rad}}$ | $c$ | $6.735 \cdot 10^{-5}\,\frac{\mathrm{Nm\,s}}{\mathrm{rad}}$ |
| $c_e$ | $1.282 \cdot 10^{-4}\,\frac{\mathrm{Nm\,s}}{\mathrm{rad}}$ | $a_I$ | $3.615 \cdot 10^{-2}\,\frac{\mathrm{Nm\,rad}}{\mathrm{A}}$ |
| $M_s$ | $3.874 \cdot 10^{-4}\,\mathrm{Nm}$ | $b$ | $1.818 \cdot 10^{-2}\,\mathrm{rad^2}$ |
| $M_c$ | $2.371 \cdot 10^{-4}\,\mathrm{Nm}$ | $\mu$ | $6.886 \cdot 10^{-4}\,\mathrm{Nm}$ |
| $\sigma$ | $9.367 \cdot 10^{-1}\,[\ ]$ | $\kappa$ | $3.125 \cdot 10^{-1}\,\mathrm{rad}$ |
| $\epsilon_i$ | $200\,[-]$ | | |

gdzie za $M_K$ podstawiono formułę (2.6). Wprowadźmy teraz następujące podstawienie zmiennej czasowej $t = \tau_t/q_t$, gdzie $q_t$ jest współczynnikiem skalującym równym $q_t = \sqrt{(mgs + k_e)/J}$, a $\tau_t$ bezwymiarowym czasem. W rzeczywistości, parametr $q_t$ odpowiada częstości drgań swobodnych układu, natomiast ich częstotliwość wyrazić można jako $f_N = q_t/2\pi$ [Hz]. Zatem, zmienne położenia, prędkości i przyspieszenia kątowego wahadła pozostają w następujących zależnościach:

$$\begin{aligned}
\varphi(t) &= \varphi(\tau_t), \\
\dot{\varphi}(t) &= \frac{d\varphi(t)}{dt} = \frac{d\varphi(t)}{d\tau_t/q_t} = q_t\frac{d\varphi(t)}{d\tau_t} = q_t\frac{d\varphi(\tau_t)}{d\tau_t} = q_t\varphi', \\
\ddot{\varphi}(t) &= \frac{d^2\varphi(t)}{dt^2} = \frac{d^2\varphi(t)}{d(\tau_t/q_t)^2} = q_t^2\frac{d^2\varphi(t)}{d\tau_t^2} = q_t^2\frac{d^2\varphi(\tau_t)}{d\tau_t^2} = q_t^2\varphi''.
\end{aligned} \qquad (2.28)$$

Podstawiając zależności (2.28) do równania (2.27) otrzymamy *oryginalny układ*

$$\varphi'' + \varphi = \frac{M_{SE}(q_t\varphi')}{mgs + k_e} - \frac{c_e}{Jq_t}\varphi' + \frac{\widehat{M}_{mag}(\varphi, i_p(\tau_t/q_t))}{mgs + k_e} \equiv G(\varphi, \varphi', \tau_t). \qquad (2.29)$$

Ponadto korzystając z aproksymacji funkcji momentu oporów wyrażonej równaniem (2.4) oraz aproksymacji momentu magnetycznego (2.23) otrzymamy *układ przybliżony*

$$\varphi'' + \varphi = \frac{M_{SEa}(q_t\varphi')}{mgs + k_e} - \frac{c_e}{Jq_t}\varphi' + \frac{\widehat{M}_{Amag6}(\varphi, i_p(\tau_t/q_t))}{mgs + k_e} \equiv G_a(\varphi, \varphi', \tau_t). \qquad (2.30)$$

W zależności od parametrów układu, przybliżona analiza dynamiczna może być zastosowana kiedy prawa strona równań, tj. $G(\varphi, \varphi', \tau_t)$ i $G_a(\varphi, \varphi', \tau_t)$, osiąga małe wartości. Parametry układu jakie przyjęto podczas prezentowanych badań zawarte są w Tabeli 2.1. Biorąc pod uwagę wartości parametrów układu można obliczyć, że $q_t = 10.838$ 1/s, a częstotliwość drgań swobodnych $f_N = 1.725$ Hz.

Do opracowania metody analityczno-numerycznej opartej na metodzie uśredniania, zastosowana zostanie zamiana zmiennych prowadząca do układu ekwiwalentnego zapisanego w postaci normalnej. Zgodnie z tradycyjną metodą uśredniania [138], rozwiązanie odpowiadające jednorodnemu (homogenicznemu) równaniu (2.29) można zapisać w następującej postaci

$$\varphi = k \sin(\Omega_t \tau_t + u), \tag{2.31}$$

gdzie $k$ jest amplitudą, $u$ przesunięciem fazowym, natomiast częstość $\Omega_t = \pi f / q_t$ jest dwukrotnie mniejsza niż częstość wymuszająca $2\pi f / q_t$. Zakładając, że wielkości $k$ i $u$ są funkcjami czasu, równanie oryginalnego układu (2.29) można zapisać w postaci dwóch równań różniczkowych zwyczajnych pierwszego rzędu (zobacz Załącznik A)

$$\begin{aligned} k' &= \frac{\cos\theta}{\Omega_t} \left[ k\sin\theta(\Omega_t^2 \quad 1) + G(k\sin\theta, k\Omega_t\cos\theta, \tau_t) \right] \equiv R_1(k, u), \\ u' &= \frac{\sin\theta}{k\Omega_t} \left[ k\sin\theta(\Omega_t^2 \quad 1) + G(k\sin\theta, k\Omega_t\cos\theta, \tau_t) \right] \equiv R_2(k, u), \\ \theta &= \Omega_t \tau_t + u, \end{aligned} \tag{2.32}$$

gdzie $R_1$ i $R_2$ wyrażają prawe strony równań.

Badania rozpoczęto od przypadku słabego wymuszenia wahadła magnetycznego, tzn. gdy składnik $\widehat{M}_{mag}(\varphi, i_p(t))$ zawarty w modelu początkowym (2.26) osiągał małe wartości. Aby tak się stało, amplituda sygnału prądowego została przyjęta na poziomie $I_0 = 0.04$ A. Ponadto ustalono, że jego częstotliwość będzie równa $f = 2.1$ Hz, a wypełnienie $w = 27\%$. Dla tak przyjętych wartości parametrów sygnału prądowego, obliczono numerycznie przebiegi $k$ i $u$ ekwiwalentnego układu (2.32) i pokazano je na rys. 2.22, przy pomocy niebieskich i czerwonych linii ciągłych. Przebiegi charakteryzują się nieliniowymi profilami, w szczególności dotyczy to przebiegu przesunięcia fazowego $u$. Tradycyjna metoda uśredniania polega na aproksymacji przebiegów $k$ i $u$ poprzez stałe wartości, które są równe średniej wartości danego przebiegu. W prezentowanych badaniach oprócz tradycyjnego podejścia, wartości amplitudy i przesunięcia fazowego zostaną aproksymowane przy użyciu funkcji piłokształtnych [139, 140]. Wybór funkcji piłokształtnej jako funkcji aproksymującej, pozwala w ogólnym przypadku na możliwość stosowania analitycznego całkowania prawych stron $R_1$, $R_2$ równań (2.32) oraz operowanie niewielką liczbą nieznanych parametrów. Warto podkreślić, że dla badanego układu mamy cztery warianty wyboru funkcji aproksymującej przebiegi $k$ i $u$, a mianowicie:

(i) amplituda $k$ i faza $u$ będą aproksymowane przez ich stałe wartości średnie $\langle k \rangle$ i $\langle u \rangle$;

(ii) amplituda $k$ będzie aproksymowana przez jej stałą średnią wartość $\langle k \rangle$, natomiast faza $u$ przez funkcję piłokształtną;

(iii) amplituda $k$ będzie aproksymowana przez funkcję piłokształtną, a faza $u$ przez jej stałą średnią wartość $\langle u \rangle$;

(iv) amplituda $k$ i faza $u$ będą aproksymowane przez funkcje piłokształtne.

W celu rozpoczęcia procedury uśredniania, nieliniowe funkcje $M_{SE}$ oraz $M_{mag}$ wchodzące w skład oryginalnego równania (2.29) zastąpiono przez ich aproksymacje $\widehat{M}_{SEa}$ i $\widehat{M}_{Amag6}$, co jest jednoznaczne z wykorzystaniem równania układu przybliżonego (2.30) do dalszych obliczeń.

**Wariant (i)**. Zarówno amplituda $k$ jak i faza $u$ są aproksymowane przez jej średnie wartości $\langle k \rangle$ i $\langle u \rangle$. Wartości te można otrzymać z przyrównania wyniku uśredniania prawych stron ekwiwalentnego układu (2.32) do zera, zgodnie z równaniami

$$\int_0^{T_t} R_1(\langle k \rangle, \langle u \rangle) \, d\tau_t = 0, \quad \int_0^{T_t} R_2(\langle k \rangle, \langle u \rangle) \, d\tau_t = 0, \tag{2.33}$$

gdzie $T_t = q_t/f$ jest okresem wymuszenia.

Ze względu na to, że w prowadzonej analizie składniki $R_1$ i $R_2$ opisane są w oparciu o funkcję $G_a$ równania (2.30), to ich postacie analityczne nie są możliwe do uzyskania. Dlatego też chcąc obliczyć wartości $\langle k \rangle$ i $\langle u \rangle$ korzystając z zależności (2.33) w programie *Mathematica*, wykorzystano metody całkowania numerycznego oraz funkcję numerycznego znajdowania pierwiastków FindRoot[...]. Wartości jakie otrzymano po obliczeniach są następujące: $\langle k \rangle = 0.09537$ rad i $\langle u \rangle = 7.31994$ rad. Przerywane linie na rysunkach 2.22a,b przedstawiają obliczone wartości średnie przebiegów $k$ i $u$.

**Wariant (ii)**. Amplituda $k$ jest aproksymowana przez jej średnią wartość $\langle k \rangle = 0.09537$ rad, podczas gdy przesunięcie fazowe $u$ jest opisane przez funkcję piłokształtną, która jest funkcją o postaci odcinkowej (ang. *piecewise function*). Biorąc pod uwagę ciągłość i okresowość przebiegu fazy $u$, odcinkowa funkcja piłokształtna wyrażona została jako

$$g_u(\tau_t) = \begin{cases} c_1 \tau_t + d_1 + c_1 \frac{\nu_w T_t}{\nu_w - 1}, & \text{dla } 0 \leq \tau_t < \nu_w T_t, \\ c_1 \frac{\nu_w \tau_t}{\nu_w - 1} + d_1, & \text{dla } \nu_w T_t \leq \tau_t \leq T_t, \end{cases} \tag{2.34}$$

gdzie $\nu_w = w/100\%$, a $c_1$ i $d_1$ są stałymi parametrami. Łatwo zauważyć, że spełniona jest zależność $g_u(0) = g_u(T_t)$ na krańcach dziedziny oraz zależność ciągłości funkcji w punkcie łączenia $\nu_w T_t$, tj. wartość $g_u$ obliczona „idąc" od lewej strony tego punktu jest taka sama jak wartość $g_u$ obliczona „idąc" od prawej strony tego punktu, $g_u(\nu_w T_t-) = g_u(\nu_w T_t+)$.

Do obliczenia wartości parametrów $c_1$ i $d_1$ należy narzucić trzy warunki. Pierwszy warunek to zaprezentowane wcześniej równania (2.33), z których wynikają wartości średnie $\langle k \rangle$ i $\langle u \rangle$. Drugi warunek pochodzi z przyrównania wartości średniej $\langle u \rangle$ do wyrażenia matematycznego na wartość średnią funkcji $g_u$, co można zapisać jako

$$\langle u \rangle = \langle g_u \rangle, \tag{2.35}$$

natomiast samo wyrażenie na wartość średnią funkcji odcinkowej $g_u$ ma postać

$$\langle g_u \rangle = \frac{1}{T_t} \int_0^{T_t} g_u(\tau_t) d\tau_t = d_1 + c_1 \frac{\nu_w(1 + \nu_w)}{2(\nu_w\ 1)} T_t. \tag{2.36}$$

Z równań (2.35) i (2.36) otrzymamy równanie z dwoma niewiadomymi $c_1$ i $d_1$. Trzeci warunek wynika z podstawienia funkcji $g_u$, zdefiniowanej równaniem (2.34),

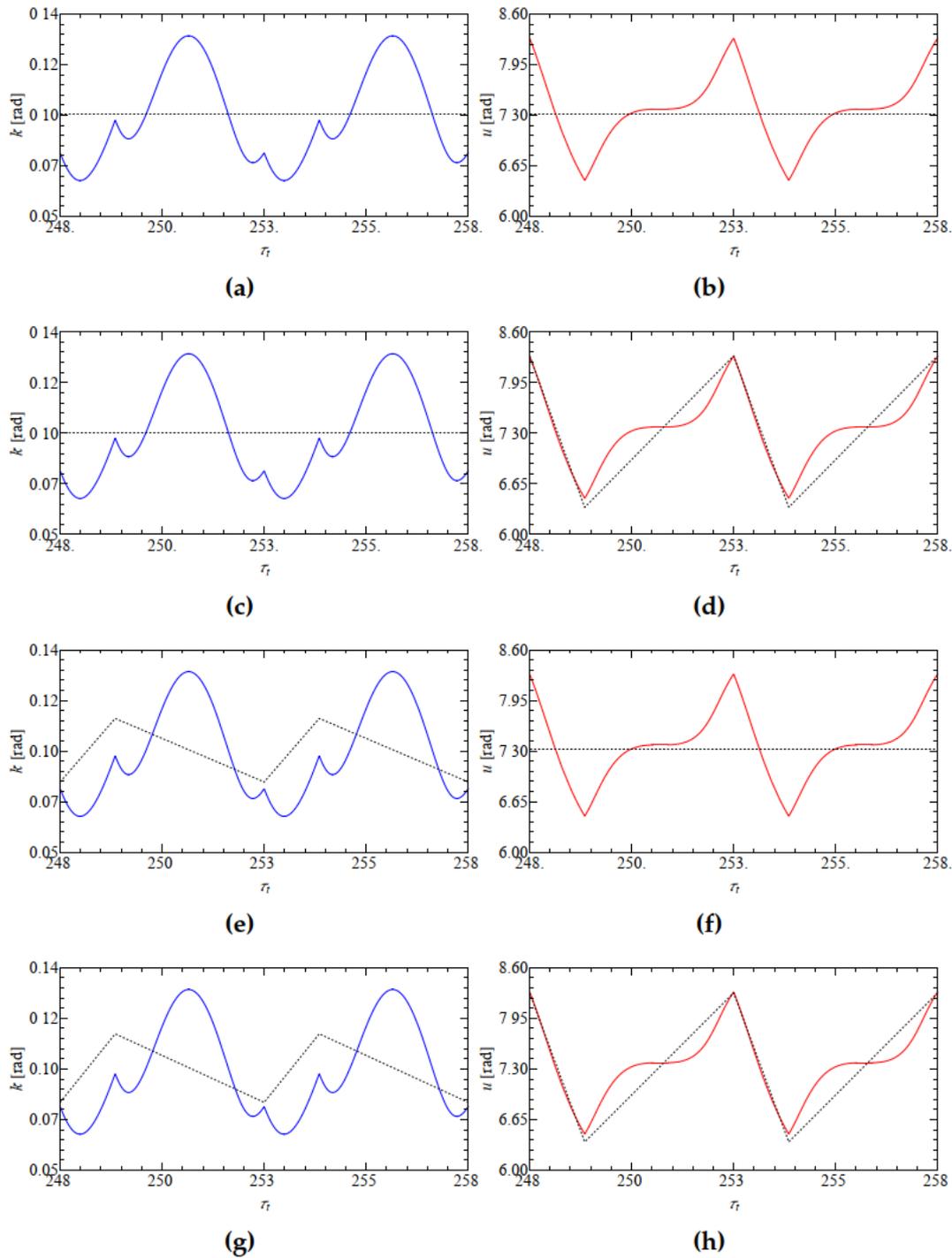

**Rys. 2.22.** Przebiegi amplitudy $k$ i fazy $u$ (odpowiednio niebieskie i czerwone linie) będące numerycznym rozwiązaniem ekwiwalentnego układu (2.32) oraz ich aproksymacje (czarne przerywane linie) dla różnych przypadków: (a)-(b) wariant (i), (c)-(d) wariant (ii), (e)-(f) wariant (iii), (g) (h) wariant (iv).

do drugiego równania układu (2.32) z założeniem, że funkcja ta wyzeruje wynik dla wszystkich $\tau_t$. Prowadzi to do następującej zależności

$$\int_0^{\nu_w T_t} \{c_1 \quad R_2(\langle k \rangle, g_u)\} \tau_t d\tau_t + \int_{\nu_w T_t}^{T_t} \left\{\frac{\nu_w c_1}{\nu_w \quad 1} \quad R_2(\langle k \rangle, g_u)\right\} \frac{\nu_w \tau_t}{\nu_w \quad 1} d\tau_t = 0. \quad (2.37)$$

Obliczone na podstawie układu równań (2.36) i (2.37) wartości parametrów $c_1$ i $d_1$ są następujące: $c_1 = \quad 1.39569$ rad i $d_1 = 5.62819$ rad. Odpowiadające im przybliżenia zostały pokazane na rys. 2.22c,d.

**Wariant (iii)**. Teraz przyjmijmy, że przesunięcie fazowe $u$ jest aproksymowane przez średnią wartość $\langle u \rangle = 7.31994$ rad obliczoną na podstawie równań (2.33), podczas gdy amplituda $k$ jest opisana przez funkcję piłokształtną

$$g_k(\tau_t) = \begin{cases} c_2 \tau_t + d_2 + c_2 \frac{\nu_w T_t}{\nu_w - 1}, & \text{dla } 0 \leq \tau_t < \nu_w T_t, \\ c_2 \frac{\nu_w \tau_t}{\nu_w - 1} + d_2, & \text{dla } \nu_w T_t \leq \tau_t \leq T_t. \end{cases} \quad (2.38)$$

Do obliczenia wartości parametrów $c_2$ i $d_2$ posłużą nam równania podobne do (2.35) i (2.37), których wyrażenia są następujące

$$\langle k \rangle = 0.09537 \text{ rad} = d_2 + c_2 \frac{\nu_w(1+\nu_w)}{2(\nu_w \quad 1)} T_t,$$

$$\int_0^{\nu_w T_t} \{c_2 \quad R_1(g_k, \langle u \rangle)\} \tau_t d\tau_t + \int_{\nu_w T_t}^{T_t} \left\{\frac{\nu_w c_2}{\nu_w \quad 1} \quad R_1(g_k, \langle u \rangle)\right\} \frac{\nu_w \tau_t}{\nu_w \quad 1} d\tau_t = 0. \quad (2.39)$$

Po obliczeniach otrzymujemy, że $c_2 = 0.02028$ rad i $d_2 = 0.11996$ rad, a odpowiadające im przybliżenia zostały pokazane na rys. 2.22e,f.

**Wariant (iv)**. Ostatni wariant zakłada, że zarówno $u$ i $k$ są aproksymowane przez funkcje piłokształtne zdefiniowane przez (2.34) i (2.38). Wtedy niewiadome $c_{1,2}$ i $d_{1,2}$ muszą spełniać układ poniższych równań

$$\int_0^{T_t} R_1(\langle g_k \rangle, \langle g_u \rangle) d\tau_t = 0, \quad \int_0^{T_t} R_2(\langle g_k \rangle, \langle g_u \rangle) d\tau_t = 0,$$

$$\int_0^{\nu_w T_t} \{c_1 \quad R_2(g_k, g_u)\} \tau_t d\tau_t + \int_{\nu_w T_t}^{T_t} \left\{\frac{\nu_w c_1}{\nu_w \quad 1} \quad R_2(g_k, g_u)\right\} \frac{\nu_w \tau_t}{\nu_w \quad 1} d\tau_t = 0, \quad (2.40)$$

$$\int_0^{\nu_w T_t} \{c_2 \quad R_1(g_k, g_u)\} \tau_t d\tau_t + \int_{\nu_w T_t}^{T_t} \left\{\frac{\nu_w c_2}{\nu_w \quad 1} \quad R_1(g_k, g_u)\right\} \frac{\nu_w \tau_t}{\nu_w \quad 1} d\tau_t = 0,$$

gdzie wyrażenie $\langle g_u \rangle = d_1 + c_1 \frac{\nu_w(1+\nu_w)}{2(\nu_w \quad 1)} T_t$, natomiast $\langle g_k \rangle = d_2 + c_2 \frac{\nu_w(1+\nu_w)}{2(\nu_w \quad 1)} T_t$. W rezultacie obliczone wartości są następujące: $c_1 = \quad 1.37382$ rad, $d_1 = 5.65469$ rad, $c_2 = 0.02184$ rad, $d_2 = 0.12185$ rad, a przybliżone przebiegi amplitudy i fazy zostały pokazane na rys. 2.22g,h.

Rysunek 2.23 przedstawia porównanie przebiegów ruchu wahadła, otrzymanych na podstawie równania (2.31) z zaimplementowanymi aproksymacjami amplitudy $k$ i przesunięcia fazowego $u$ (według czterech analizowanych powyżej wariantów) oraz rozwiązania numerycznego otrzymanego dla przybliżonego układu (2.30). Spośród czterech zaproponowanych wariantów aproksymacji dla przebiegów amplitudy i przesunięcia fazowego przyjęto, że to wariant (iv) powinien w najlepszy sposób odwzorowywać wynik rozwiązania numerycznego ze

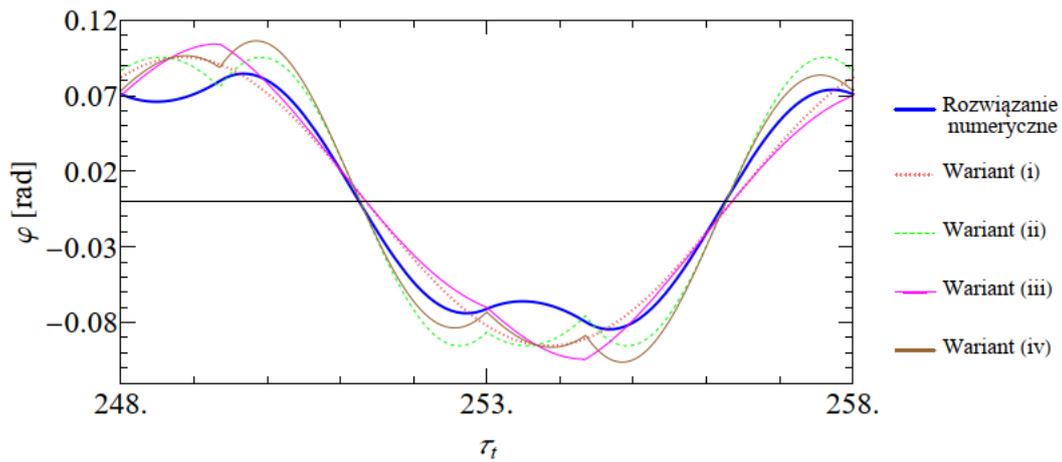

**Rys. 2.23.** Porównanie przebiegów równania (2.31) dla różnych wariantów aproksymacji amplitudy $k$ i przesunięcia fazowego $u$ oraz rozwiązania numerycznego układu (2.30).

względu na wykorzystywanie funkcji piłokształtnej do przybliżenia obu z tych nieliniowych przebiegów.

Kolejnym krokiem było wyznaczenie krzywych amplitudowo-częstotliwościowych badanego układu. Zostały one opracowane dla ruchów ustalonych z pominięciem ruchu przejściowego. Obliczenia wykonano dla słabego momentu wymuszającego w przedziale częstotliwości $f \in (1.5; 3.5)$ Hz (2.24a) i silnego momentu wymuszającego w przedziale częstotliwości $f \in (1; 3.5)$ Hz (2.24b). Następnie, obliczone numerycznie krzywe zestawiono z danymi eksperymentalnymi. Na początku rozważono przypadek słabego wymuszenia układu, tzn. gdy $I_0 = 0.04$ A oraz $w = 27\%$. W tym przypadku rozwiązania modelu początkowego (2.26) i oryginalnego układu (2.29), otrzymane po ich numerycznym scałkowaniu są nie do odróżnienia. Dlatego postanowiono rozważyć tylko krzywą amplitudowo-częstotliwościową oryginalnego modelu (2.29) zawierającego niewielomianowe modele oporów ruchu i momentu magnetycznego, a jej przebieg zaznaczono niebieskimi punktami na rys. 2.24a. W ten sam sposób obliczone zostały krzywe dla układu przybliżonego (2.30) wykorzystującego wielomianowe modele momentu magnetycznego drugiego (2.18), czwartego (2.19) i szóstego stopnia (2.20). Krzywe odpowiadające tym układom zaznaczono fioletowymi, brązowymi i czerwonymi punktami. Oczywistym jest, że krzywa amplitudowo-częstotliwościowa obliczona dla układu przybliżonego z wielomianowym momentem magnetycznym szóstego stopnia, daje najlepsze odwzorowanie w stosunku do krzywej układu oryginalnego (2.29). Punkty w postaci czarnych krzyżyków znajdujące się na wykresie odzwierciedlają wartości maksymalnych wychyleń kątowych wahadła i obliczone zostały analityczno-numeryczną metodą wykorzystującą wariantu (iv) do aproksymacji amplitudy i przesunięcia fazowego rozwiązania (2.31).

Zauważyć można, że obliczone krzywe amplitudowo-częstotliwościowe posiadają pionową asymptotę, zaznaczoną na rysunku czarną przerywaną linią. Asymptota ta występuje dla częstotliwości $2f_N = 3.449$ Hz, która jest zbieżna z podwojoną częstotliwością drgań swobodnych wahadła i powoduje przerwanie ciągłości tych krzywych. Na rysunku zaznaczono również zielonymi i pomarań-

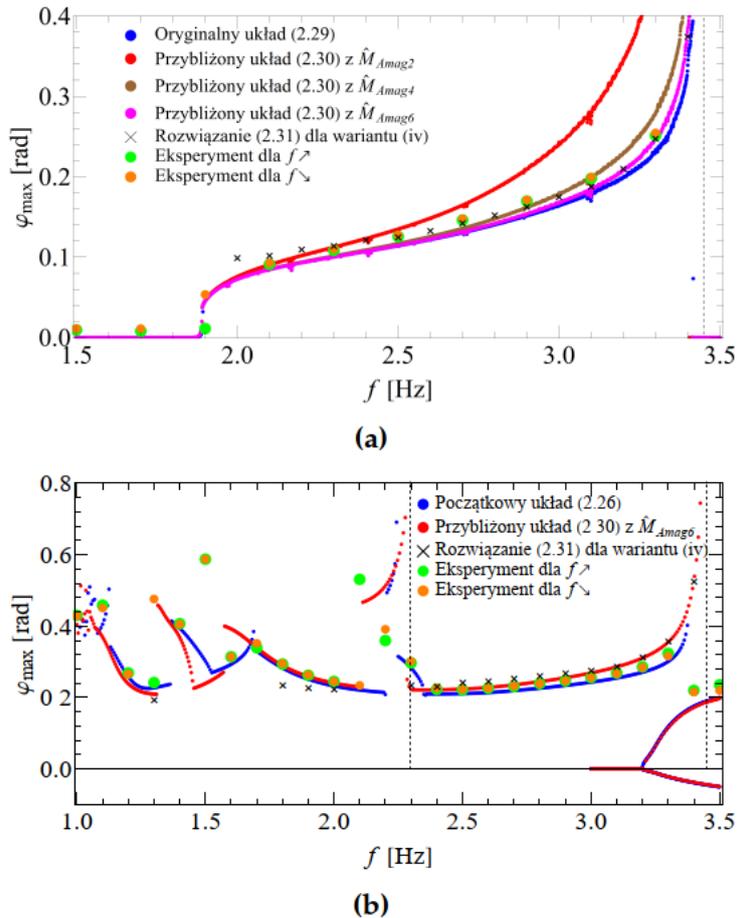

**Rys. 2.24.** Krzywe amplitudowo-częstotliwościowe dla układu poddanego (a) słabemu oraz (b) silnemu momentowi magnetycznemu.

czowymi punktami dane eksperymentalne, otrzymane dla rosnącej i malejącej częstotliwości sygnału prądowego. Dane eksperymentalne pokrywają się z symulacyjnymi krzywymi, choć najmniejsza zbieżność widoczna jest dla przypadku związanego z najmniejszym stopniem mianownika wielomianu momentu magnetycznego.

Podczas badań prowadzonych dla przypadku silnego wymuszenia układu ($I_0 = 0.2$ A, $w = 27\%$), struktura krzywych amlitudowo-częstotliwościowych staje się bardziej złożona. Niebieskie punkty na rysunku 2.24b odpowiadają krzywej amlitudowo-częstotliwościowej otrzymanej dla modelu początkowego (2.26), natomiast czerwone punkty odwzorowują krzywą obliczoną dla układu przybliżonego (2.30). Podczas obliczania krzywych dla układu przybliżonego, posłużono się równaniem (2.30) zawierającym tylko najwyższy szósty stopień wielomianu w modelu momentu magnetycznego (2.20). Punkty w postaci czarnych krzyżyków odpowiadają maksymalnym wychyleniom kątowym wahadła i obliczono je tak jak wcześniej metodą analityczno-numeryczną. Podobnie jak dla przypadku słabego wymuszenia układu, w sąsiedztwie częstotliwości $4f_N/3 = 2.299$ Hz i $2f_N$ (pionowe przerywane linie) można zaobserwować wzrost amplitudy drgań. Dla analizowanych parametrów wymuszenia, układ wykazuje się multistabilnością w zależności od zadanych warunków początkowych, co jest widoczne w zakresie częstotliwości $f \in (3.2; 3.5)$ Hz, gdzie pojawiają się trzy współistniejące rozwiązania. Ponadto układ charakteryzuje się nagłymi skokami amplitud. Różnice pomiędzy obliczonymi krzywymi a eksperymentem mogą mieć kilka przyczyn. Przybliżony układ ze względu na swoje uproszczenia, przede wszystkim linearyzację funkcji sinusoidalnej, znacznie bardziej odbiega od eksperymentu szczególnie dla dużych kątów wychylenia wahadła. Różnice mogą być również spowodowane zbyt krótkimi czasami przejściowymi przyjętymi podczas eksperymentu lub wynikiem niedokładności wykonania stanowiska, np. luzów w łożyskach czy lekko uginającej się podstawy cewki elektrycznej, co nie było brane pod uwagę podczas modelowania matematycznego.

Podsumowując, warto zauważyć, że wyniki otrzymane przy pomocy opracowanej metody analityczno-numerycznej są jakościowo bardzo zbliżone do wyników otrzymanych przy pomocy metod numerycznych oraz eksperymentu. W porównaniu z tradycyjną metodą uśredniania, która ogranicza się do obliczenia uśrednionej i stałej w czasie wartości przebiegu amplitudy i fazy rozwiązania okresowego, opracowana procedura pozwala na aproksymację tych przebiegów funkcją piłokształtną. Wpływa to na dokładniejsze opisanie przebiegu dynamicznego wspomnianej amplitudy i przesunięcia fazowego dla modeli z pulsującym wymuszeniem, kiedy to w rozwiązaniu okresowym pojawiają się „podskoki" amplitudy (rys. 2.23). Przedstawione w tym paragrafie badania zostały opublikowane w pracy [141]. Dodatkowo w artykule przeanalizowano przypadek zastosowania zaprezentowanej metody analityczno-numerycznej dla silnie nieliniowego układu, w którym nie przeprowadzono linearyzacji funkcji sinus.

Szersza analiza odpowiedzi układu na zmiany parametru kontrolnego, za który przyjęto częstotliwość $f$ sygnału prądowego $i_p(t)$, została pokazana na eksperymentalnych i symulacyjnych wykresach bifurkacyjnych (rys. 2.25 i 2.26). Wykresy symulacyjne obliczone zostały dla układu startowego (2.26) i parametrów podanych w tabeli 2.2.

Badania przeprowadzone były dla silnego wymuszenia, gdzie amplituda sygnału prądowego wynosiła $I_0 = 1$ A, a wypełnienie $w = 30\%$. Podczas eksperymentu częstotliwość sygnału prądowego narastała (lub malała) w sposób liniowy w czasie, przez co sygnał prądowy w tym przypadku można uznać za tzw. sygnał świergotowy (ang. *sweep chrip signal*). Wspomniane narastanie i zmniejszanie się częstotliwości w czasie, odbywało się wolno ($25 \cdot 10^{-4}$ Hz/s), tak aby zapewnić układowi względnie najdłuższy czas na osiągnięcie ruchu ustalonego. Przykładowe przebiegi tego rodzaju sygnału prądowego pokazane zostały na rys. 2.25a i 2.26a. Punkty tworzące wykresy bifurkacyjne widoczne na rys. 2.25b i 2.26b rejestrowane były co okres sygnału prądowego i w chwili czasowej odpowiadającej początkowi jego narastającej krawędzi.

**Tabela 2.2.** Parametry układu pojedynczego wahadła magnetycznego przyjęte podczas analizy bifurkacyjnej.

| | | | |
|---|---|---|---|
| $J$ | $6.786 \cdot 10^{-4}$ kgm$^2$ | $v_s$ | $0.733 \frac{\text{rad}}{\text{s}}$ |
| $mgs$ | $5.800 \cdot 10^{-2}$ Nm | $\chi$ | $5.759 \frac{\text{s}}{\text{rad}}$ |
| $k_e$ | $1.742 \cdot 10^{-2} \frac{\text{Nm}}{\text{rad}}$ | $c$ | $7.369 \cdot 10^{-5} \frac{\text{Nm s}}{\text{rad}}$ |
| $c_e$ | $1.282 \cdot 10^{-4} \frac{\text{Nm s}}{\text{rad}}$ | $a_I$ | $3.615 \cdot 10^{-2} \frac{\text{Nm rad}}{\text{A}}$ |
| $M_s$ | $4.436 \cdot 10^{-4}$ Nm | $b$ | $1.818 \cdot 10^{-2}$ rad$^2$ |
| $M_c$ | $2.223 \cdot 10^{-4}$ Nm | $\epsilon_i$ | $200\,[-]$ |

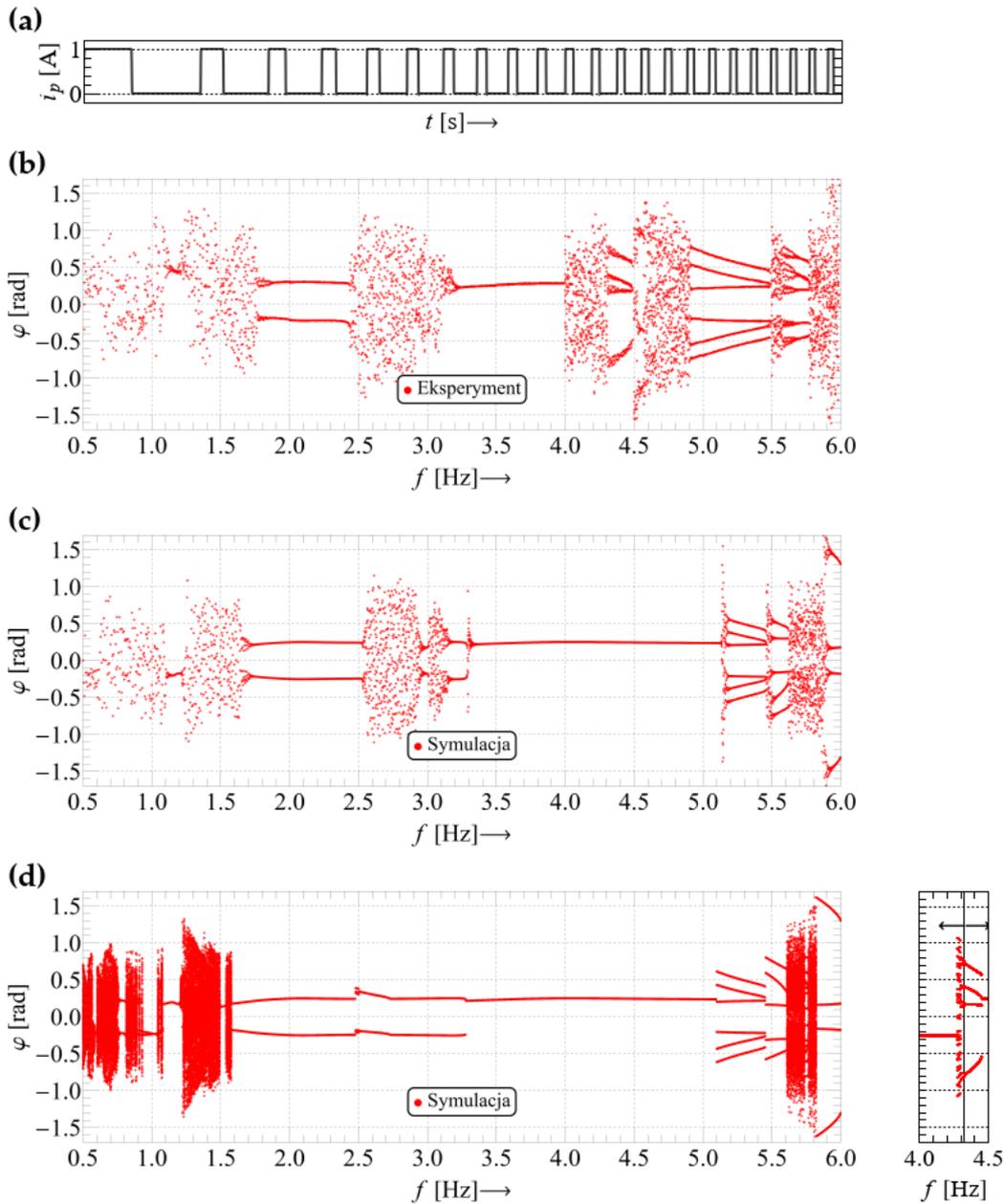

**Rys. 2.25.** Wykresy bifurkacyjne dla rosnącej częstotliwości $f$ sygnału prądowego: (a) przykład sygnału prądowego używanego podczas eksperymentu o rosnącej liniowo częstotliwości, (b) eksperyment, (c) symulacja na podstawie eksperymentalnego sygnału prądowego, (d) „klasyczna" symulacja. Parametry sygnału prądowego: $I_0 = 1$ A, $w = 30\%$.

Bifurkacyjne wykresy symulacyjne przedstawione na rys. 2.25c i 2.26c odwzorowują warunki eksperymentu, tzn. wykonano je dla liniowo narastającej lub malejącej częstotliwość sygnału prądowego. Natomiast wykresy 2.25d i 2.26d obliczone zostały w „klasyczny" sposób, który polegał na całkowaniu numerycznym

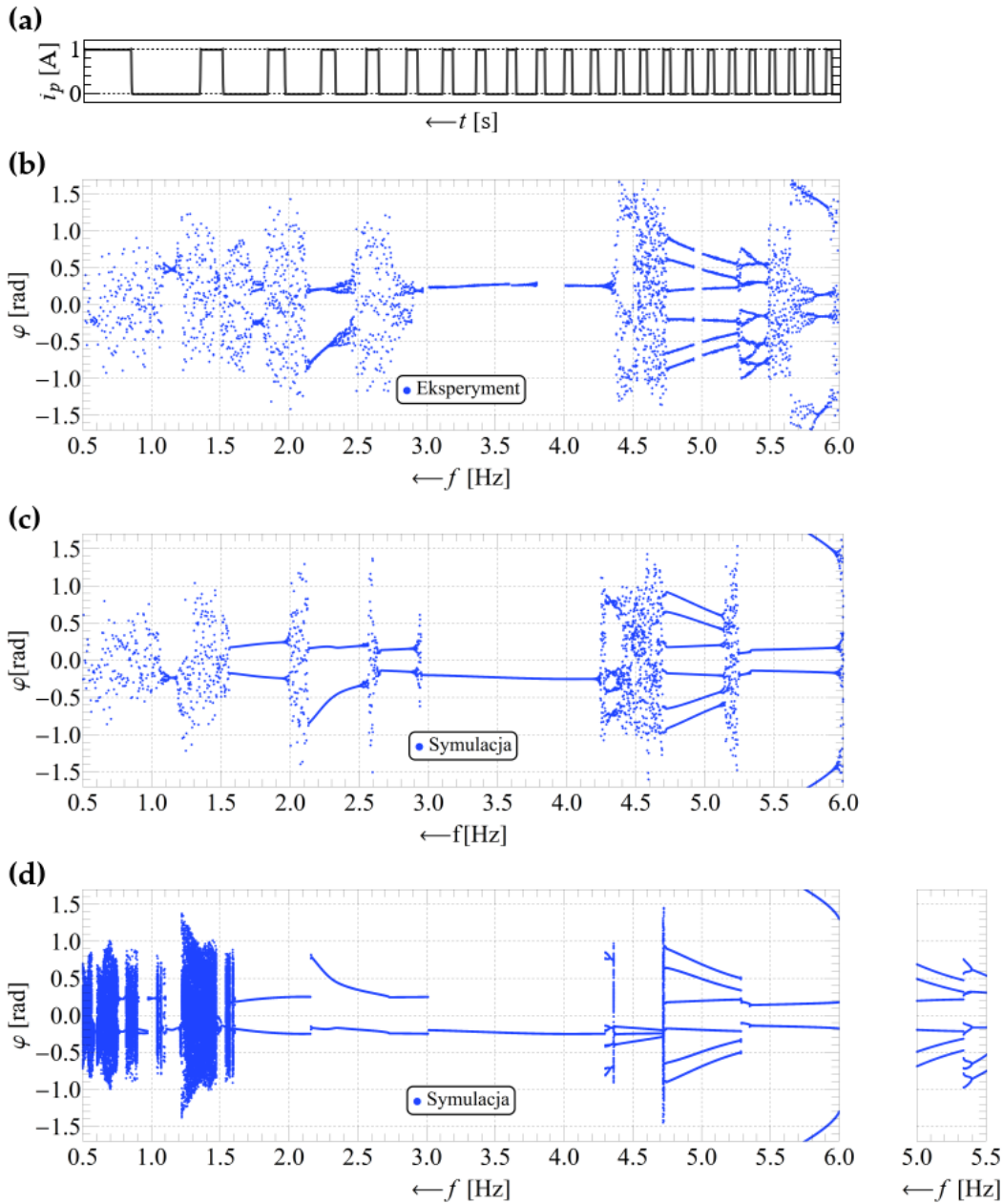

**Rys. 2.26.** Wykresy bifurkacyjne dla malejącej częstotliwości $f$ sygnału prądowego: (a) przykład sygnału prądowego używanego podczas eksperymentu o malejącej liniowo częstotliwości, (b) eksperyment, (c) symulacja na podstawie eksperymentalnego sygnału prądowego, (d) „klasyczna" symulacja. Parametry sygnału prądowego: $I_0 = 1$ A, $w = 30\%$.

modelu matematycznego układu dla danej stałej wartości częstotliwość i w czasie odpowiadającym 1200 okresom sygnału wymuszającego. Pierwsze 1000 okresów uznawano za czas przejściowy i pomijano, a punkty z ostatnich 200 okresów prezentowano na wykresie. Ponadto podczas symulacji prowadzonej dla ustalonej częstotliwości jako warunki początkowe przyjmowano wartość położenia i prędkości kątowej odpowiadające ostatniemu punktowi obliczonemu dla poprzedzającej częstotliwości. Biorąc pod uwagę wyniki analizy bifurkacyjnej zauważalnym jest, że ruch układu może być zarówno chaotyczny jak i regularny. Ruch regularny (rozumiany jako okresowy) może przyjmować różne okresowości: 1-okresowość, 2-okresowość, 4-okresowość, 6-okresowość, a także 10-okresowość. przypomnijmy, że pojęcie okresowości drgań rozumiane jest jako liczba okresów wymuszenia (sygnału prądowego) przypadająca na jeden okres rozwiązania. Obliczone numerycznie wykresy bifurkacyjne odwzorowują prawie wszystkie zachowania zarejestrowane podczas eksperymentu. Zaobserwować można również występowanie współistniejących atraktorów np. dla częstotliwości $f$ = 2.25 Hz na obu wykresach z rys. 2.25b i 2.26b pojawiają się rozwiązania 2-okresowe, ale są to dwa różne rozwiązania. Rodzaj rozwiązania (atraktora) zależy od warunków początkowych. Na wykresach symulacyjnych mających odzwierciedlać eksperyment (rys. 2.25c i 2.26c), istnienie niektórych okien przypominających rozwiązanie chaotyczne może być spowodowane zbyt długim czasem trwania ruchu przejściowego, który nie został pominięty. Problem ten nie występuje na wykresach bifurkacyjnych wykonanych w sposób „klasyczny" (rys. 2.25d i 2.26d), dla przykładu 2-okresowe rozwiązanie widoczne dla $f$ = 2.85 Hz na rys. 2.26d zostało zweryfikowane eksperymentalnie dla stałej częstotliwości i pokazane na portrecie fazowym z przekrojem Poincarégo na rys. 2.27f.

Pawie wszystkie odpowiedzi układu zarejestrowane eksperymentalnie zostały znalezione podczas symulacji numerycznych. Brakujące odpowiedzi zdołano ujawnić dopiero po zmianie warunków początkowych; do takich odpowiedzi należą te pokazane na rys. 2.25d w zakresie częstotliwości 4 ÷ 4.5 Hz (skrajny prawy wykres) oraz 2.26d w zakresie częstotliwości 5 ÷ 5.5 Hz (skrajny prawy wykres). Przykładowe symulacyjne i eksperymentalne wykresy fazowe z przekrojami Poincarégo przedstawione zostały na rys. 2.27. Na różnice pomiędzy symulacyjnymi a eksperymentalnymi trajektoriami fazowymi wpływa w pewnym stopniu sposób przetwarzania danych eksperymentalnych, ponieważ w czasie eksperymentu rejestrowano tylko położenie kątowe, natomiast prędkość otrzymywana była poprzez jego numeryczne różniczkowanie, co generowało pewne zakłócenia, które następnie należało odfiltrować. Wpływ tego błędu można zaobserwować np. na rys. 2.27b, gdzie punkt Poincarégo powinien znajdować się dokładnie w miejscu nagłego spadku prędkości (tak jak na symulacji) spowodowanego pojawieniem się bariery w postaci pola magnetycznego. W przypadku rys.2.27j,n, na różnice pomiędzy symulacją a eksperymentem największy wpływ miał fakt, że dla badanych częstotliwości układ posiada różne współistniejące rozwiązania. Dlatego w przypadku pojawienia się najmniejszych niedokładności (zaburzeń) w układzie rzeczywistym wynikających np. z luzów w łożyskach, rozwiązania te były bliskie „przeskakiwania" między sobą, co objawiało się brakiem idealnego pokrycia trajektorii fazowej w kolejnych okresach drgań.

Zaprezentowana analiza bifurkacyjna potwierdziła znakomitą różnorodność

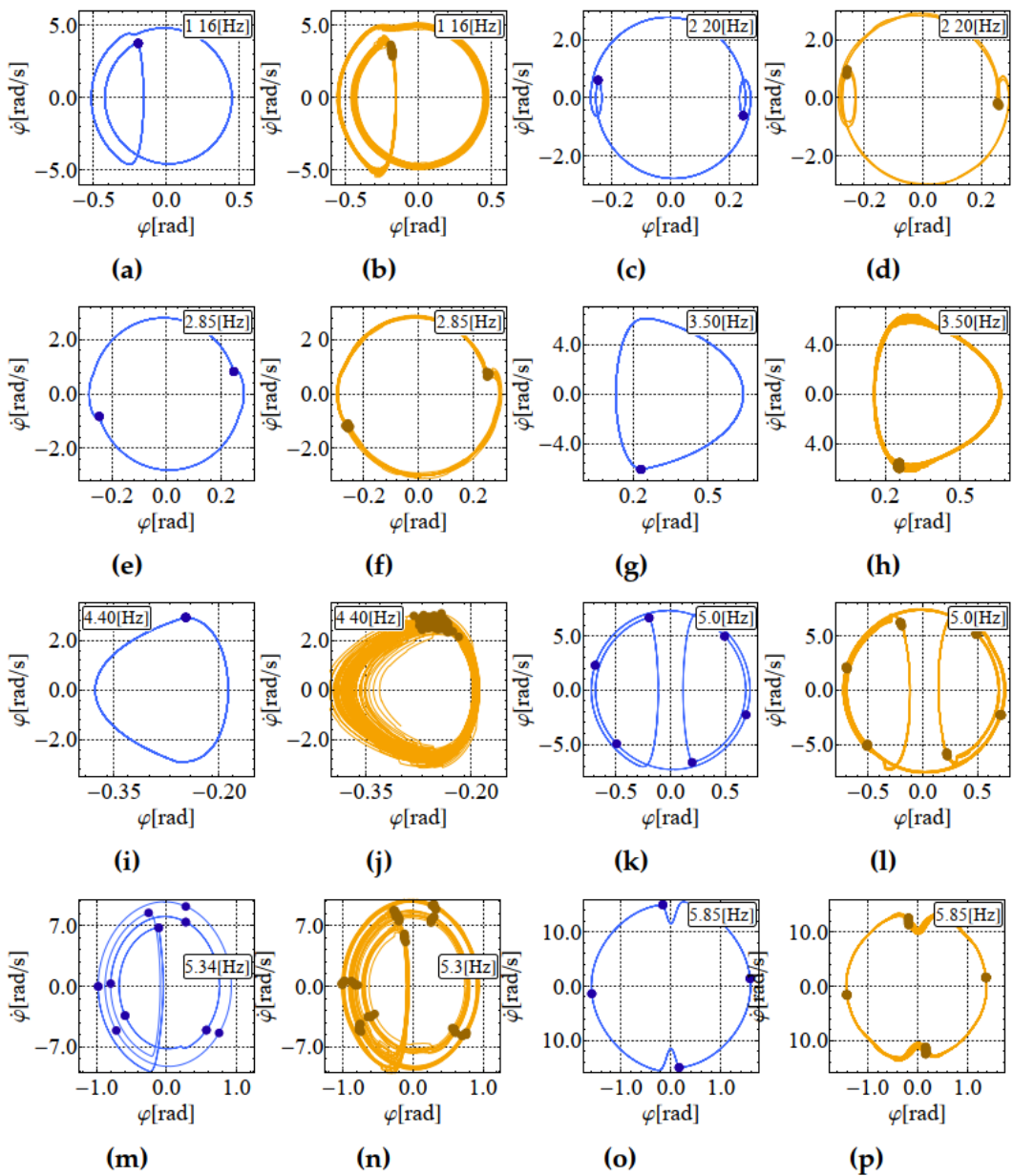

**Rys. 2.27.** Portrety fazowe wraz z przekrojami Poincarégo obliczone numerycznie (a, c, e, g, i, k, m, o) oraz zarejestrowane eksperymentalnie (b, d, f, h, j, l, n, p) dla różnych częstotliwości $f$ sygnału prądowego. Parametry sygnału prądowego: $I_0 = 1$ A, $w = 30\%$.

zachowań układu pojedynczego wahadła magnetycznego, która była już wcześniej odnotowywana przez innych badaczy i przedstawiona we Wstępie (Rozdział 1).

## 2.3.2 Badanie zjawiska drgań w jednym „dołku" potencjału

W tym paragrafie, dokładnej analizie dynamicznej poddane zostaną szczególne drgania pojedynczego wahadła magnetycznego. Mianowicie analizowane będą

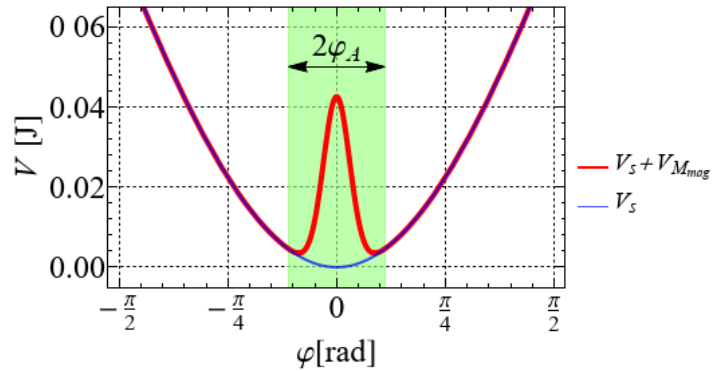

**Rys. 2.28.** Jednodołkowa (niebieska linia) i dwudołkowa (czerwona linia) energia potencjalna układu. Obszar wyróżniony zielonym kolorem to strefa aktywna pola magnetycznego, generowanego przez cewkę elektryczną zasilaną prądem o amplitudzie $I_0 = 1$ A.

drgania układu odbywające się tylko w jednym „dołku" potencjału [35], czyli takie jak na rys. 2.27g,i. Drgania te nazywane będą też jednostronnymi, ponieważ drgające wahadło w czasie ich wykonywania nie przechodzi przez dolne położenie równowagi, a osiągane przez niego wychylenia kątowe są tego samego znaku.

Analizę warto rozpocząć od wprowadzenia nowych spostrzeżeń dotyczących modelowania matematycznego układu. Charakter badanego wymuszenia magnetycznego jest dość nietypowy, ponieważ nie zależy tylko od czasu, ale także od położenia wahadła. Gdy przez cewkę elektryczną płynie prąd, powoduje on powstanie dookoła niej bariery magnetycznej, od której magnes wahadła może się odbić lub ją „przełamać". Bariera ta może być rozumiana jako energia potencjalna, która „pojawia" się i „znika" w układzie zgodnie z pulsującym sygnałem prądowym. W przypadku, gdy przez cewkę nie płynie prąd, układ zachowuje się jak zwykłe wahadło swobodne, którego energia potencjalna zobrazowana jest niebieską linią na rys. 2.28. Energia ta jest sumą energii pola grawitacyjnego oraz energii kumulowanej w elemencie sprężystym i zapisać ją można jako

$$V_s(\varphi) = mgs(1 - \cos\varphi) + \frac{1}{2}k_e\varphi^2. \tag{2.41}$$

W układzie bez pola magnetycznego występuje jeden dołek potencjału. Natomiast, gdy cewka zasilana jest prądem o przyjętym natężeniu $I_0 = 1$ A, wartość energii potencjalnej układu jest sumą energii (2.41) oraz energii potencjalnej pola magnetycznego (2.11). Suma tych energii przedstawiona jest czerwoną linią na rysunku 2.28. Obecność pola magnetycznego powoduje powstanie w układzie dwudołkowego potencjału. Energie potencjalne obliczone dla układu swobodnego i układu poddanego działaniu pola magnetycznego różnią się od siebie, jednak znaczące różnice między nimi występują tylko w bardzo wąskim zakresie kątów wychylenia wahadła opisanych jako $\pm\varphi_A$. Strefa wyznaczona przez te kąty została określona mianem „*strefy aktywnej*" i zwizualizowana jest kolorem zielonym na rys. 2.28. Aby sprecyzować wielkość tej strefy przyjęto, że jej graniczne wartości muszą spełniać następującą zależność

$$\frac{V_s(\varphi_A)}{V_s(\varphi_A) + V_{M_{mag}}(\varphi_A)} = 0.999. \tag{2.42}$$

Na podstawie tej zależności określono wartość $\varphi_A = \pm 0.348$ rad.

Przejście z układu o potencjale jednodołkowym do dwudołkowego (i na odwrót) dyktowane jest przez sygnał prądowy, który w badanym przypadku jest prostokątny i pulsujący. Ze względu na skokowy charakter tego sygnału uznać można, że przejście pomiędzy potencjałami układu następuje prawie natychmiast. Fakt ten umożliwia wprowadzenie pewnego uproszczenia w opisie dynamiki układu. Mianowicie można przyjąć, że układ posiada dwa stany:

(1) kiedy cewka nie jest zasilana prądem, wtedy mamy do czynienia z układem jednodołkowym;

(2) kiedy cewka jest zasilana prądem o natężeniu $I_0$, wtedy mamy do czynienia z układem dwudołkowym.

Model matematyczny takiego układu można zapisać jako układ dwóch równań różniczkowych, z których każde odpowiada innemu stanowi. Pierwsze równanie modelujące stan 1 układu zapisać można jako

$$J\ddot{\varphi} + mgs \sin\varphi + M_{SE}(\dot{\varphi}) + M_K(\varphi, \dot{\varphi}) = 0, \tag{2.43}$$

natomiast drugie równanie odpowiadające stanowi 2 wyraża się

$$J\ddot{\varphi} + mgs \sin\varphi + M_{SE}(\dot{\varphi}) + M_K(\varphi, \dot{\varphi}) \quad \widehat{M}_{mag}(\varphi, I_0) = 0. \tag{2.44}$$

Oba stany w odniesieniu do ich energii potencjalnej zostały zobrazowane na rys. 2.29a,b. Z dynamicznego punktu widzenia, dolne położenie równowagi wahadła będące punktem osobliwym dla każdego z dwóch analizowanych stanów ma inny charakter. Gdy przez cewkę elektryczną nie płynie prąd (stan 1), punkt ten jest ogniskiem stabilnym, co pokazano na rys. 2.29c. W sytuacji, gdy cewka elektryczna jest zasilana i wahadło jest odpychane, punkt osobliwy staje się siodłem tak jak to przedstawiono na 2.29d. Przełączanie pomiędzy równaniami (2.43) i (2.44) następuje w odpowiednich chwilach czasowych zależnych od okresu $\tau = 1/f$ i wypełnienia $w$ sygnału prądowego, co zobrazowano na rys. 2.30a i zapisać można w następujący sposób:

$$\begin{aligned} &\text{Stan 1} \to \text{Stan 2} \quad \text{dla} \quad t = k_n\tau, \, k_n \in \mathbb{N}; \\ &\text{Stan 2} \to \text{Stan 1} \quad \text{dla} \quad t = \tau\frac{w}{100\%} + k_n\tau, \, k_n \in \mathbb{N}. \end{aligned} \tag{2.45}$$

Jak już wspomniano, układ będąc w stanie 1 posiada jedno stabilne położenie równowagi $(\varphi, \dot{\varphi}) = (0,0)$, podczas gdy przechodząc do stanu 2 położenie to zamienia się w niestabilne siodło. W sąsiedztwie tego siodła powstają dwa stabilne położenia równowagi symetryczne względem kąta $\varphi = 0$ rad. Z tego względu dynamikę układu można rozpatrywać w kategorii periodycznie zmieniającego się charakteru dolnego położenia równowagi, co zobrazowano na wykresie biegunowym na rys. 2.30b. Rys. 2.30c przedstawia baseny przyciągania obliczone dla dwóch stabilnych położeń równowagi układu znajdującego się w stanie 2, czyli opisanego równaniem (2.44). Trajektorie znajdujące się w zielonym obszarze, dążą do zielonego punktu równowagi, a te znajdujące się w fioletowym obszarze do fioletowego punktu równowagi.

W przeciwieństwie do początkowego „ciągłego" modelu matematycznego (2.26), model matematyczny opisany równaniami (2.43)-(2.45) można nazwać „dyskretnym", ponieważ zawiera on dwa różne i oddzielne stany w jakich może znajdować się układ. Każdy z tych stanów opisany jest autonomicznym równaniem różniczkowym, ale cały układ jest nieautonomiczny, ponieważ przełączanie się pomiędzy tymi stanami zależy od zmiennej niezależnej jaką jest czas.

W czasie, gdy cewka jest zasilana i tworzy barierę w postaci energii potencjalnej pola magnetycznego, można potraktować ją jak przeszkodę dla wahadła. Jest to sytuacja analogiczna do tej spotykanej w układach z uderzeniami. Na rys. 2.31 przedstawiono trzy różne zachowania wahadła magnetycznego, znajdującego się w strefie aktywnej:

(i) kiedy cewka jest niezasilana, wahadło swobodnie przechodzi przez strefę aktywną, ponieważ nie napotyka bariery magnetycznej. Zachowanie to przedstawia niebieski fragment trajektorii na rys. 2.31a,d);

(ii) kiedy cewka jest zasilana, ale wahadło ma niewystarczającą ilość energii kinetycznej, żeby przełamać barierę energii potencjalnej pola magnetycznego i odbija się od niej. Zachowanie to przedstawiono na rys. 2.31b,e).

(iii) kiedy cewka jest zasilana i wahadło ma wystarczającą ilość energii kinetycznej, żeby przełamać barierę energii potencjalnej pola magnetycznego.

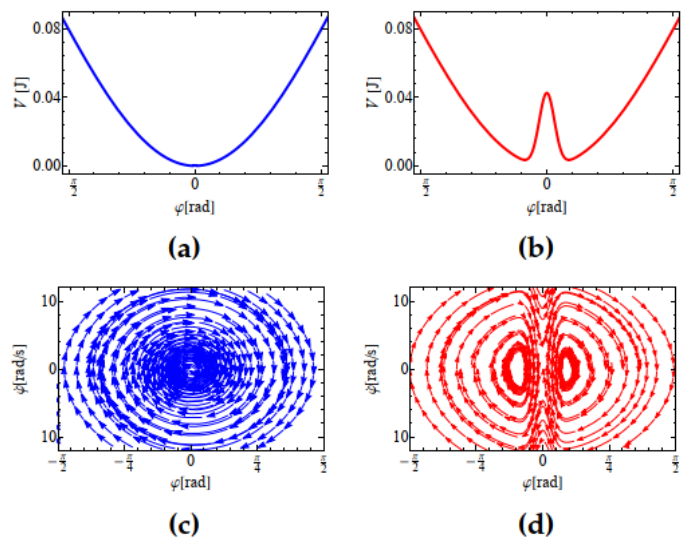

**Rys. 2.29.** Energia potencjalna układu wahadła magnetycznego w przypadku (a) stanu 1 oraz (b) stanu 2. Rodzaj punktu osobliwego $(\varphi, \dot{\varphi}) = (0, 0)$ (dolnego położenia równowagi wahadła) dla stanu 1 — ognisko stabilne (c) oraz stanu 2 — siodło (d).

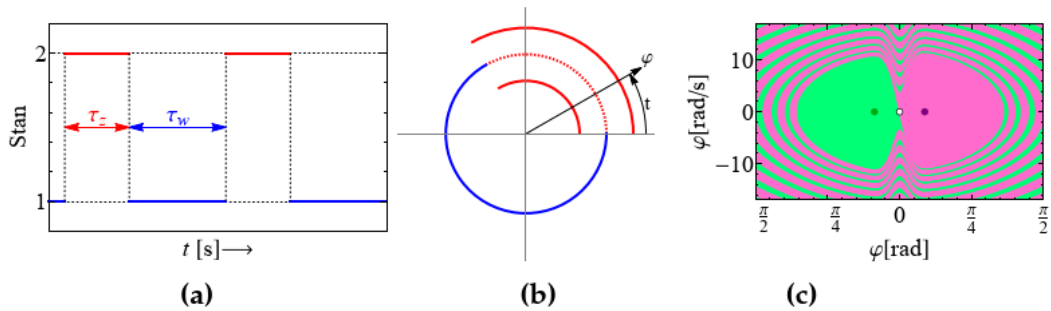

**Rys. 2.30.** (a) Sygnał przełączający układ pomiędzy stanami 1 i 2: $\tau_z$ – cewka jest zasilana, $\tau_w$ cewka nie jest zasilana. (b) Ciągłe linie odpowiadają stabilnym rozwiązaniom, a przerywane niestabilnym; kolor niebieski odpowiada stanowi 1, a kolor czerwony odpowiada stanowi 2. (c) Baseny przyciągania obliczone dla stanu 2, gdzie fioletowy i zielony punkt jest stabilnym położeniem równowagi, a biały punkt jest niestabilnym położeniem równowagi.

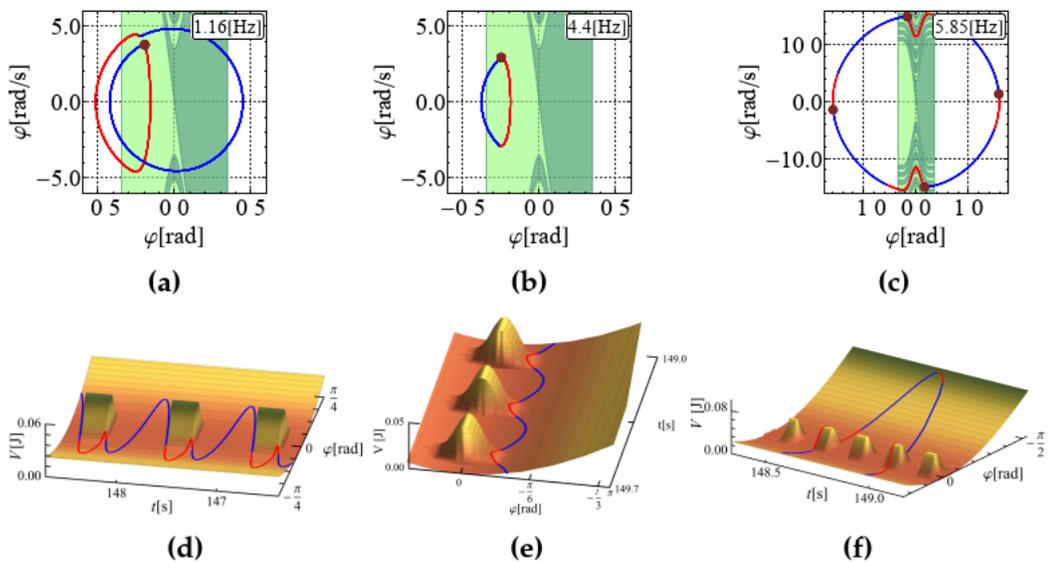

**Rys. 2.31.** Wykresy fazowe (a-c) oraz wykresy energii potencjalnej (d-f) obliczone dla trzech rożnych rozwiązań okresowych przy stałym wypełnieniu $w = 30\%$ sygnału prądowego.

Zachowanie to przedstawiono rys. 2.31c,f);

Kolorem niebieskim zaznaczono chwile czasowe, w których cewka elektryczna była niezasilana, natomiast kolorem czerwonym zaznaczono chwile, w których cewka była zasilana. Obszar działania bariery magnetycznej, który jest tożsamy ze strefą aktywną zaznaczono na wykresach fazowych przy pomocy fragmentu basenów przyciągania, obliczonych dla stabilnych położeń wahadła znajdującego się w stanie 2 (rys. 2.30c).

Przyjmując, że wartość energii potencjalnej pola magnetycznego poza strefą aktywną spada do zera, dostarczanie energii do układu może odbywać się tylko, gdy wahadło będzie znajdowało się w strefie aktywnej i cewka zostanie zasilana.

Wracając do badań nad drganiami układu w jednym dołku potencjału, tzn.

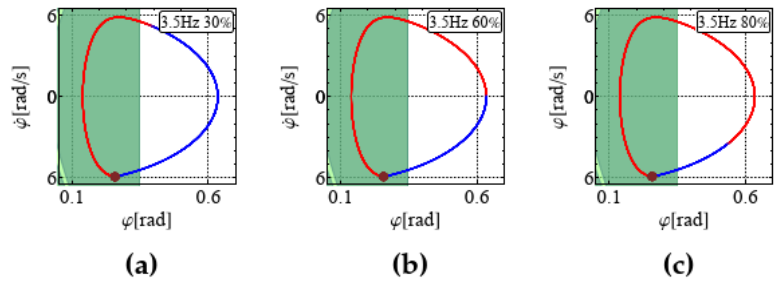

**Rys. 2.32.** Symulacyjne wykresy fazowe jednookresowych drgań wahadła w jednym dołku potencjału wyznaczone dla stałej częstotliwości $f = 3.5$ Hz, ale różnych wartości wypełnienia $w \to \{30\%, 60\%, 80\%\}$.

takich dla których wahadło odbija się od bariery potencjału magnetycznego i nie przełamuje jej, mają one swój początek w obserwacji eksperymentalnej. Podczas eksperymentu zauważono, że dla danej stałej wartości częstotliwości prądu $f$ generującej drgania w jednym dołku potencjału, zmiana wypełnienia $w$ sygnału prądowego nie zawsze miała zauważalny wpływ na ruchu wahadła, tak jak to pokazano na rys. 2.32. Obserwacje te narzuciły postawienie sobie następującego pytania: *„jakie są graniczne wartości wypełnienia „w" przy stałej częstotliwości „f" sygnału prądowego, dla których trajektoria fazowa drgań wahadła w jednym dołku potencjału nie zmieni się?"*

Odpowiedź na to pytanie uzyskana zostanie na przykładzie analizy ruchu jednookresowego, występującego w układzie dla przedziału częstotliwości $f \in (2.85, 3.62)$ Hz (rys. 2.26 i 2.25).

Rozważania zaczniemy od podstawowego rozwiązania pokazanego na rys. 2.33, który nazwiemy *scenariuszem I*. Cykl ruchu rozpoczyna się poza strefą aktywną w punkcie początkowym o współrzędnych $(\varphi_0, \omega_0 = 0)$, czyli gdy układ znajduje się w stanie 1 (cewka jest niezasilana – niebieskie strzałki i linie). Kiedy wahadło po czasie $\tau_w$ znajdzie się w strefie aktywnej, układ tylko raz zmieni swój stan z 1 na 2 (cewka jest zasilana – czerwone strzałki i linie). Punkt, w którym następuje ta zmiana ma współrzędne $(\varphi_k, \omega_k)$ i zostanie nazwany *punktem wymuszenia*, ponieważ powoduje on natychmiastowe podniesienie energii układu. Następnie wahadło odbija się od bariery energii potencjalnej i po czasie $\tau_z$ wraca do punktu początkowego $(\varphi_0, 0)$, gdzie następuje powrót układu ze stanu 2 do 1. Przyjmując punkt $(\varphi_0, \omega_0)$ za punkt zwrotny, jego prędkość musi wynosić $\omega_0 = 0$ rad/s. Wartości współrzędnych, jakie może przyjmować punkt wymuszenia na płaszczyźnie fazowej, mają następujące ograniczenia:

**Ograniczenie 1** – wymuszenie (zmiana stanu z 1 na 2) musi nastąpić w strefie aktywnej, $\varphi_k \leq \varphi_A$. Wynika to z założonego scenariusza I.

**Ograniczenie 2** – przyjmując, że układ przełącza się ze stanu 2 do stanu 1 poza strefa aktywną, minimalna wartość kąta $\varphi_0$ może wynosić $\varphi_A$.

**Ograniczenie 3** – układ po wymuszeniu nie może uciec z dołka potencjału, w którym drga. Z dynamicznego punktu widzenia oznacza to, że graniczne wartości prędkości i położenia punktu wymuszenia muszą należeć do trajektorii będącej granicą basenu przyciągania układu znajdującego się w stanie 2.

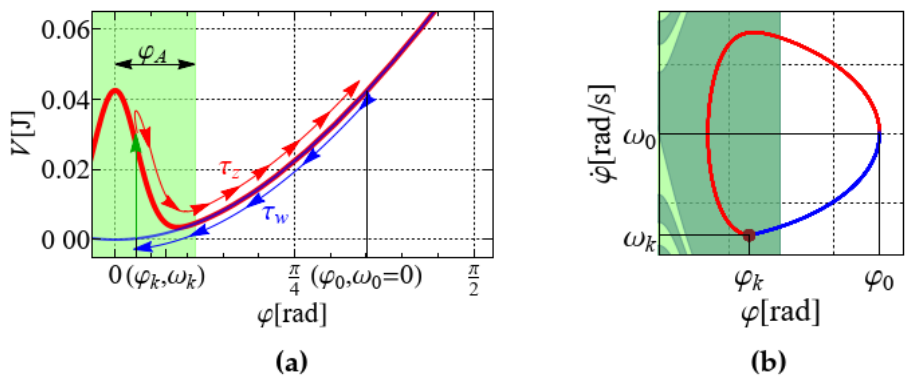

**Rys. 2.33.** Wykres zmian energii potencjalnej (a) i wykres fazowy (b) analizowanego ruchu jednookresowego według scenariusza I.

Wymienione powyżej ograniczenia definiują zamknięty obszar na płaszczyźnie fazowej (jasnoniebieski obszar na rys. 2.34a), wewnątrz którego należy poszukiwać współrzędnych $(\varphi_k, \omega_k)$ spełniających założony scenariusz I.

Poszukiwane współrzędne zostały obliczone numerycznie poprzez wyznaczenie dla obu stanów układu zbiorów punktów $(\varphi, \dot{\varphi}, \varphi_0)$. Obliczenia numeryczne zbioru $(\varphi, \dot{\varphi}, \varphi_0)$ dla układu znajdującego się w stanie 1, polegały na całkowaniu wstecz w czasie równania (2.43) z warunków początkowych wybranych z jasnoniebieskiego obszaru na rys. 2.34a. Punkty $(\varphi, \dot{\varphi})$ osiągane przez układ były rejestrowane, aż do chwili pojawienia się punktu $(\varphi = \varphi_0, \dot{\varphi} = 0)$. Analogiczne obliczenia wykonano dla układu znajdującego się w stanie 2 z tą różnicą, że całkowano w przód w czasie równanie (2.44) z warunków początkowych wybieranych z jasnoniebieskiego obszaru. Obliczenia te również trwały, aż do chwili osiągnięcia przez układ punktu $(\varphi = \varphi_0, \dot{\varphi} = 0)$, a wszystkie osiągnięte do tego czasu przez układ punkty $(\varphi, \dot{\varphi})$ były rejestrowane. Tak powstałe zbiory utworzyły powierzchnie przedstawione na rys. 2.34b, których część wspólna (zielona linia na przecięciu się powierzchni) jest zbiorem wartości punktów wymuszenia $(\varphi_k, \omega_k)$ dla ruchu jednookresowego według scenariusza I. Ponadto dla każdego punktu wymuszenia obliczono czasy $\tau_z$ i $\tau_w$, które zaznaczone są na rys. 2.33a. Znając czasy trwania obu stanów układu w ciągu całego cyklu ruchu wahadła, w łatwy sposób można obliczyć parametry wypełnienia $w$ i częstotliwości $f$ sygnału prądowego, korzystając z następujących zależności:

$$f = \frac{1}{\tau_z + \tau_w}, \qquad w = \frac{\tau_z}{\tau_z + \tau_w}. \tag{2.46}$$

Obliczone wartości tych parametrów pokazane zostały na rys. 2.35a poprzez niebieską krzywą. Jak już wcześniej wspomniano, wpływ pola magnetycznego na wahadło magnetyczne znajdujące się poza strefą aktywną jest pomijany. Wynika z tego, że układ może przełączać się pomiędzy stanami poza strefą aktywną i nie będzie miało to wpływu na okresową trajektorię fazową. Z tego względu, wartość wypełnienia $w$ może zostać zredukowana poprzez przełączenie się układu ze stanu 2 do stanu 1 dokładnie w momencie opuszczania przez wahadło strefy aktywnej (rys. 2.36a), wynika to ze skracania się czasu $\tau_z$ i wydłużania $\tau_w$. Analogicznie można dokonać zwiększenia wartości wypełnienia na skutek przełączenia się układu ze stanu 2 do stanu 1 tuż przed wejściem wahadła do strefy aktywnej,

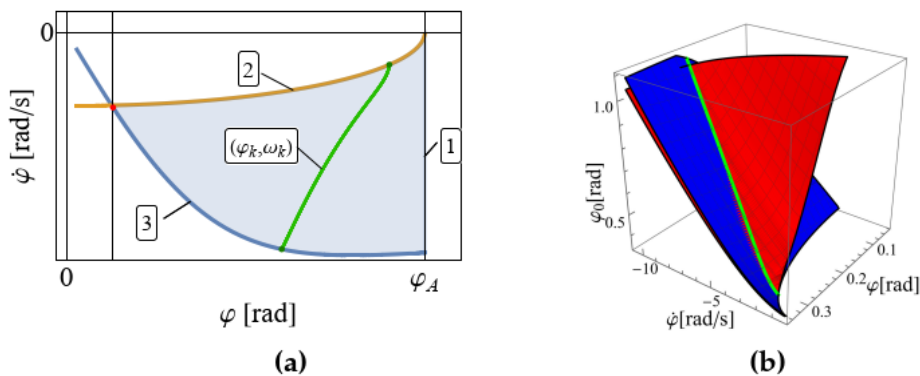

**Rys. 2.34.** Fragment płaszczyzny fazowej (a), gdzie linie (1 3) odpowiadają ograniczeniom opisanym w tekście powyżej. Zielona linia odpowiada współrzędnym punktów wymuszenia $(\varphi_k, \omega_k)$ według scenariusza I. Linia ta jest wynikiem przecięcia się dwóch powierzchni (b) będących zbiorem punktów $(\varphi, \dot{\varphi}, \varphi_0)$ obliczonych dla stanu 1 (niebieska powierzchnia) i 2 (czerwona powierzchnia).

wtedy wydłużeniu ulega czas $\tau_z$ i skróceniu $\tau_w$ (rys. 2.36b). Obliczone zredukowane i zwiększone wartości wypełnienia $w$ zostały zaznaczone na rys. 2.35a, odpowiednio czerwoną i zieloną krzywą. W rezultacie, krzywe te tworzą zaznaczony kolorem zielonym obszar definiujący w jakich granicach mogą zmieniać się wartości wypełnienia $w$ dla danej częstotliwości $f$, dla których ruch układu odbywa się w jednym dołku potencjału, jest jednookresowy, a przebieg jego trajektorii fazowej jest taki sam.

Patrząc na granicę zielonego obszaru z rys. 2.35a od lewej strony, pojawia się ona nagle przy częstotliwości ok. 2.85 Hz, podczas gdy od prawej strony ma ona „gładki" charakter przy $f$ równej ok. 4.6 Hz. Powodem jej nagłego pojawienia się jest fakt, że drgania wahadła dla częstotliwości bliskiej 2.85 Hz są bliskie przełamania granicy basenu przyciągania dla stanu 2 i ucieczki wahadła ze studni potencjału. Wahadło posiada wtedy duże wartości amplitudy drgań $\varphi_0$, tak jak to pokazano na rys. 2.35b, dzięki czemu energia zgromadzona w jego ruchu osiąga wartość bliską energii potencjalnej bariery magnetycznej. Gładki charakter prawej strony jest wynikiem drgań o trajektorii fazowej dalekiej od granicy basenu przyciągania. Punkt zwrotny $(\varphi_0, 0)$ trajektorii fazowej znajduje się wtedy poza strefą aktywną, ale jest bliski minimalnej amplitudzie ruchu założonej dla scenariusza I, tj. $\varphi_A$ (rys. 2.35b). Rozwiązania z amplitudami drgań mniejszymi od $\varphi_A$ będą rozważane w dalszej części paragrafu. Rys. 2.35c przedstawia zmiany wartości współrzędnych $(\varphi_k, \omega_k)$ punktu wymuszenia, obliczone dla ruchu według scenariusza I w zależności od częstotliwości $f$ sygnału prądowego.

W kolejnym kroku, dokonano walidacji eksperymentalnej i weryfikacji numerycznej trajektorii fazowych otrzymanych dla wybranych konfiguracji parametrów $f$ i $w$, których wartości wybrano z obszaru oznaczonego kolorem zielonym na rys. 2.35a. Rys. 2.37a przedstawia symulacyjne trajektorie fazowe, obliczone z równania (2.26) dla wybranych konfiguracji $f$ i $w$, które potwierdzają zgodność współrzędnych $(\varphi_k, \omega_k)$ punktów wymuszenia z współrzędnymi wyznaczonymi na podstawie przecięcia się dwóch powierzchni z rys. 2.34a,b. W celu łatwiejszej

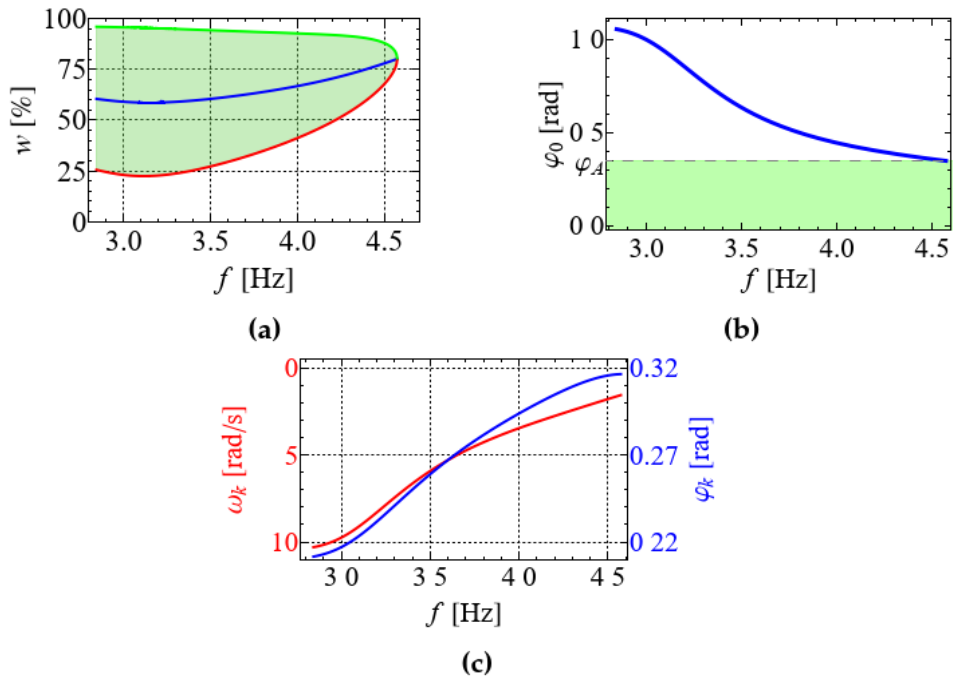

**Rys. 2.35.** (a) Zależność pomiędzy parametrami $f$ i $w$ prostokątnego pulsującego sygnału prądowego dla jednookresowego ruchu układu według scenariusza I (niebieska krzywa) i jego modyfikacji w odniesieniu do wypełnienia (czerwona i zielona krzywa). Zmiana amplitudy $\varphi_0$ drgań (a) i współrzędnych $(\varphi_k, \omega_k)$ punktu wymuszenia (c) w zależności od częstotliwości $f$ dla ruchu według scenariusza I.

weryfikacji tej zgodności, fragment płaszczyzny fazowej z rys. 2.34a nałożono na wykres z rys. 2.37a. Badania eksperymentalne przeprowadzono dla trzech konfiguracji parametrów $\{f, w\} = \{3.5, 80\}, \{4.0, 80\}, \{4.46, 71\}$. Porównanie wybranych pomiarów wraz z symulacjami numerycznymi pokazano na rys 2.37b.

Podczas eksperymentu dla częstotliwości $f$ bliskich 3.063 Hz bardzo trudno było utrzymać przez dłuższy czas drgania założone przez scenariusz I. Wahadło podczas drgań z dużą prędkością uderzało w barierę magnetyczną, co powodo-

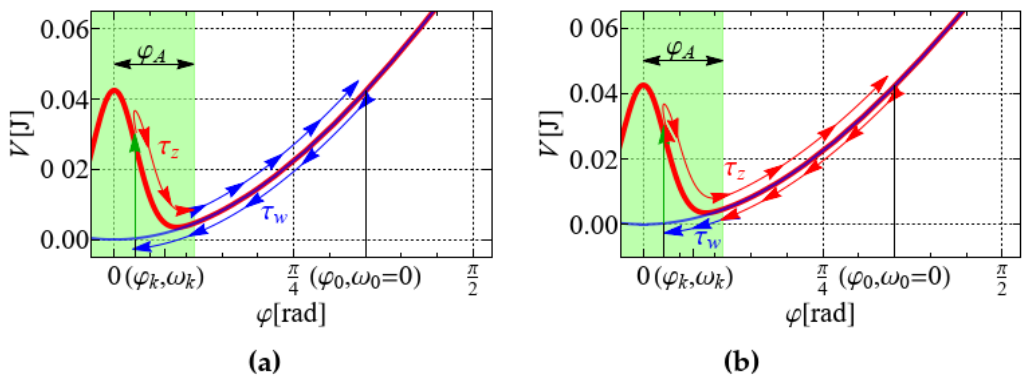

**Rys. 2.36.** Wykres energii potencjalnej analizowanego ruchu jednookresowego, gdy wartość wypełnienia $w$ jest zredukowana (a) i zwiększona (b).

wało uginanie się podstawy cewki oraz znaczne podnoszenie się wału wahadła w łożyskach na skutek obecności luzów. Zmiany te miały na tyle duży wpływ na dynamikę układu, że po chwili następowała ucieczka wahadła z jamy potencjału, w której miało ono drgać.

Analiza dynamiki jednookresowego ruchu układu według scenariusza I, potwierdza słuszność założenia braku wpływu przełączania się układu pomiędzy stanem 1 i 2 poza strefą aktywną na jego ruch. Założenie to prowadzi do wniosku, że możliwa jest zmiana wartości okresowości badanych drgań, bez wpływu na ich trajektorię fazową poprzez przełączanie stanów układu poza strefą aktywną. W celu znalezienia zakresów wartości parametrów $f$ i $w$ dla wspomnianych rozwiązań wielookresowych, wykorzystane zostaną ekstremalne wartości częstotliwości i wypełnienia obliczone dla scenariusza I. Rys. 2.38 przedstawia wykresy czasowe trzech prostokątnych sygnałów prądowych. Amplitudy tych sygnałów

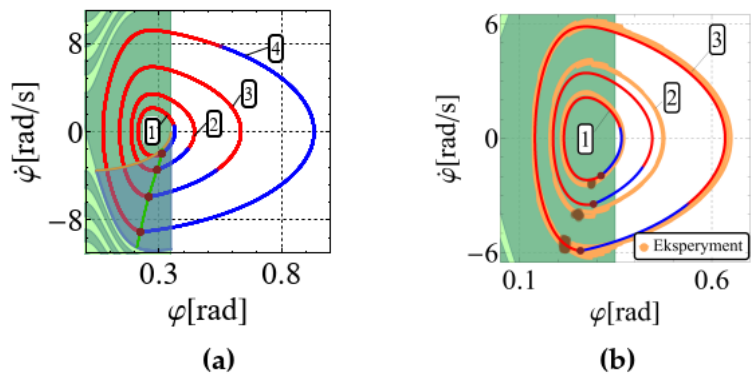

**Rys. 2.37.** Weryfikacja symulacyjna (a) współrzędnych punktów wymuszenia otrzymanych na podstawie równania (2.26) z współrzędnymi otrzymanymi na podstawie przecięcia się powierzchni z rys. 2.34b. Walidacja eksperymentalna (b) symulacyjnych trajektorii fazowych. Badania przeprowadzono dla następujących parametrów $\{f[\text{Hz}], w[\%]\}$: (1) → $\{4.46, 71\}$, (2) → $\{4.0, 80\}$, (3) → $\{3.5, 80\}$, (4) → $\{3.1, 30\}$.

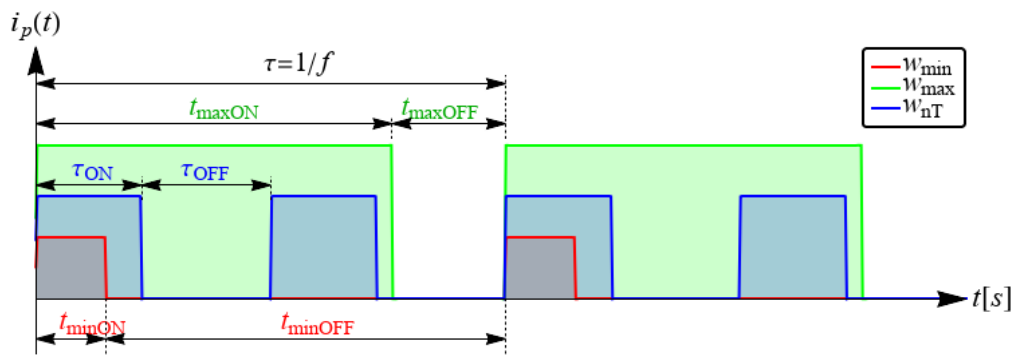

**Rys. 2.38.** Wykresy czasowe sygnałów prądowych odnoszące się do ekstremalnych wartości wypełnień obliczonych według scenariusza I (rys. 2.35a). Czerwony sygnał – minimalna wartość wypełnienia, zielony sygnał – maksymalna wartość wypełnienia, niebieski sygnał – przykładowe rozwiązanie wielookresowe.

zostały sztucznie zmienione, w celu lepszej wizualizacji ich częstotliwości i wypełnień. Czerwony przebieg odpowiada sygnałowi prądowemu z minimalnym wypełnieniem $w_{min}$ (czerwona linia na rys. 2.35a). Natomiast zielony przebieg odwzorowuje sygnał prądowy o maksymalnym wypełnieniu $w_{max}$ (zielona linia na rys. 2.35a). Niebieski przebieg odzwierciedla sygnał prądowy z wypełnieniem $w_{nT}$, który prowadzić ma do wielookresowego rozwiązania o takiej samej trajektorii fazowej jak dla ruchu jednookresowego. Aby tak się stało, wartość wypełnienia $w_{nT}$ rozwiązania wielookresowego musi spełniać następujące warunki:

- okres $\tau$ rozwiązania jednookresowego jest wielokrotnością okresu $\Omega_{nT}$ rozwiązania wielookresowego, dlatego

$$\tau = n \cdot \Omega_{nT} \quad \text{dla} \quad n \in \mathbb{N}^+ \setminus \{1\}. \tag{2.47}$$

- zależność pomiędzy czasami sygnałów o wypełnieniach $w_{min}$, $w_{max}$ i $w_{nT}$ są następujące:

$$\begin{aligned} t_{minON} &\leq \tau_{ON}, \\ t_{maxOFF} &\leq \tau_{OFF}. \end{aligned} \tag{2.48}$$

Teraz można zapisać równania, które pozwolą na przejście z wartości czasów na wartości wypełnień:

$$\begin{aligned} \tau_{ON} &= w_{nT} \cdot \Omega_{nT}, \\ \tau_{OFF} &= \Omega_{nT} - \tau_{ON} = \Omega_{nT}(1 - w_{nT}), \\ t_{minON} &= w_{min} \cdot \tau, \\ t_{maxON} &= w_{max} \cdot \tau, \\ t_{maxOFF} &= \tau - t_{maxOFF} = \tau(1 - w_{max}). \end{aligned} \tag{2.49}$$

Korzystając z równań (2.47) (2.49), otrzymujemy następujące zależności jakim podlega wypełnienie $w_{nT}$ rozwiązania wielookresowego w stosunku do wypełnień granicznych $w_{min}$ i $w_{max}$ rozwiązania jednookresowego:

$$\begin{aligned} w_{nT} &\geq n \cdot w_{min}, \\ w_{nT} &\leq 1 - n(1 - w_{max}). \end{aligned} \tag{2.50}$$

Obliczenia numeryczne przeprowadzone dla nierówności (2.50) wykazały, że dla badanego układu są one spełnione tylko dla $n = 2$ i $n = 3$, co odpowiada ruchom dwu- i trój-okresowym w jednym dołku potencjału.

Zakresy parametrów $f$ i $w$ dla dwu- i trój-okresowych rozwiązań pokazane zostały na rys. 2.39a. W celu weryfikacji obliczeń z obszarów dwu- i trój-okresowości wybrano po dwa punkty odpowiadające różnym $f$ i $w$ (kołowe znaczniki na rys. 2.39a), a następnie przeprowadzono dla nich symulacje, których wyniki zestawiono z eksperymentem, co pokazano rys. 2.39b,c. Warto podkreślić, że w strefie aktywnej znajduje się tylko jeden punkt Poincarégo, natomiast kolejne punkty zwiększające okresowość znajdują się poza nią. Wnioskować więc można, że w tego typu układzie ze zlokalizowanym wymuszeniem magnetycznym, można w sposób „sztuczny" zmieniać okresowość rozwiązania bez wpływu na przebieg trajektorii fazowej tego rozwiązania.

Rys. 2.40 przedstawia analizowane zjawisko na symulacyjnych wykresach fazowych. Wykonano je dla stałego wypełnienia $w = 75\%$, ale różnych częstotliwości $f \to \{1f, 2f, 3f\}$ odpowiadających jedno , dwu  i trój okresowym rozwiązaniom opisanym jako 1T, 2T, 3T. Wartości tych częstotliwości zawierają się w odpowiednich regionach z rys. 2.39a. Przebiegi trajektorii fazowych dla wszystkich trzech rozwiązań są do siebie bardzo zbliżone, co przedstawiono na rys. 2.40d.

Z wykresu pokazanego na rys. 2.35a można odczytać, że dla wartości wypełnienia $w = 30\%$, jednookresowy ruch wahadła w jednym dołku potencjału jest możliwy tylko dla częstotliwości $f \in (2.85, 3.62)$ Hz. Jednakże, na wykresach bifurkacyjnych pokazanych na rys. 2.25 i 2.26 można dostrzec, że ruch ten jest obecny, aż do wartości $f = 4.8$ Hz. Wynika to z faktu, że ruch wahadła prezentowany przez układ poza tym zakresem posiada inne scenariusze niż ten, który został zaprezentowany jako scenariuszu I.

Scenariusze dla tych rozwiązań pokazane są na rys. 2.41. Oba scenariusze zakładają wystąpienie w strefie aktywnej dwóch przełączeń między stanami układu, a nie jednego jak miało to miejsce w scenariuszu I. *Scenariusz II* zakłada, że cewka elektryczna jest niezasilana przez przez większość czasu trwania cyklu (rys. 2.41a), natomiast *scenariusz III* ma odwrotne założenie (rys. 2.41b). Punkt, w którym układ przełącza się ze stanu 2 do stanu 1 posiada współrzędne $(\varphi_{lk}, \omega_{lk})$. Do określenia wartości współrzędnych punktów $(\varphi_k, \omega_k)$ i odpowiadających im

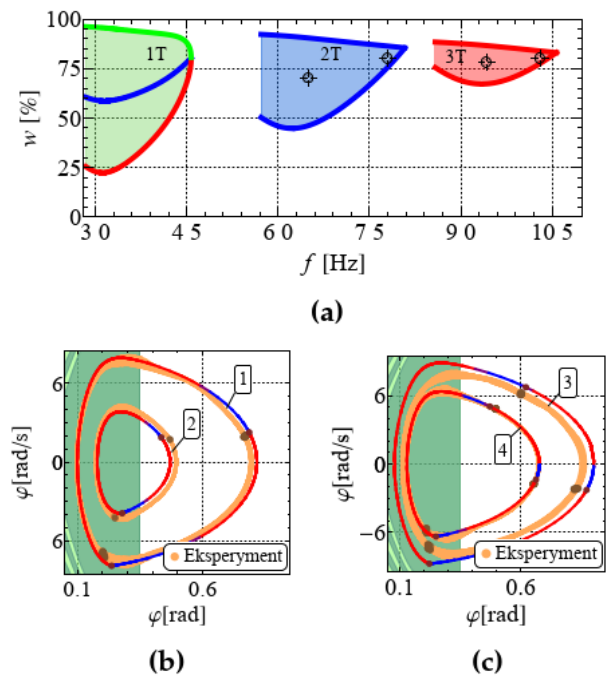

**Rys. 2.39.** (a) Zależność pomiędzy parametrami $f$ i $w$ prostokątnego pulsującego sygnału prądowego dla jedno, dwu  i trój okresowego ruchu układu obliczone na podstawie scenariusza I. Eksperymentalne i symulacyjne wykresy fazowe z przekrojami Poincarégo ruchu dwuokresowego (b) i trój okresowego (c), obliczone dla $\{f[Hz], w[\%]\}$: (1) $\to \{6.5, 70\}$, (2) $\to \{7.8, 80\}$, (3) $\to \{9.42, 78\}$, (4) $\to \{10.31, 80\}$   wartości te odpowiadają kołowym znacznikom z podrysunku (a).

punktów ($\varphi_{lk}, \omega_{lk}$) spełniających założone scenariusze II i III wykorzystano metody numeryczne. Procedura polegała na numerycznym całkowaniu równań (2.43) i (2.44). Warunkiem początkowym dla obu równań były zadane współrzędne punktu wymuszenia ($\varphi_k, \omega_k$), przy czym całkowanie równania (2.43) odbywało się wstecz w czasie, a równania (2.44) w przód. Obliczane w ten sposób trajektorie fazowe po pewnym czasie symulacji przecinały się, a współrzędne punktów przecięcia odpowiadały poszukiwanym dla danego scenariusza punktom ($\varphi_{lk}, \omega_{lk}$). Obliczenia przeprowadzone były dla warunków początkowych ($\varphi_k, \omega_k$), których wartości wybierane były z obszaru płaszczyzny fazowej ograniczonego krzywymi: $\dot{\varphi} = 0$, $\varphi = \varphi_A$ oraz krzywą będącą granicą basenu przyciągania oznaczoną jako (3) na rys. 2.34a.

W rezultacie, na podstawie powyższej analizy numerycznej otrzymano obszary wartości ($\varphi_k, \omega_k$), dla których istnieją i nie istnieją rozwiązania przewidziane scenariuszami II i III. Obszary te pokazano na rys. 2.42. Punkty wymuszenia, dla których rozwiązania istnieją tzn. trajektorie obliczone dla równań (2.43)-(2.44) przecinają się, zaznaczono jasnoniebieskim obszarem, natomiast te, dla których rozwiązania według scenariusza II i III nie istnieją zaznaczono obszarem czerwonym. Ciągłą zielona krzywa widoczna na wykresie i leżąca na granicy oddzielającej jasnoniebieski i czerwony obszar, odpowiada wartościom ($\varphi_k, \omega_k$) spełniającym założenia wystąpienia ruchu według scenariusza I. Wynika to z faktu,

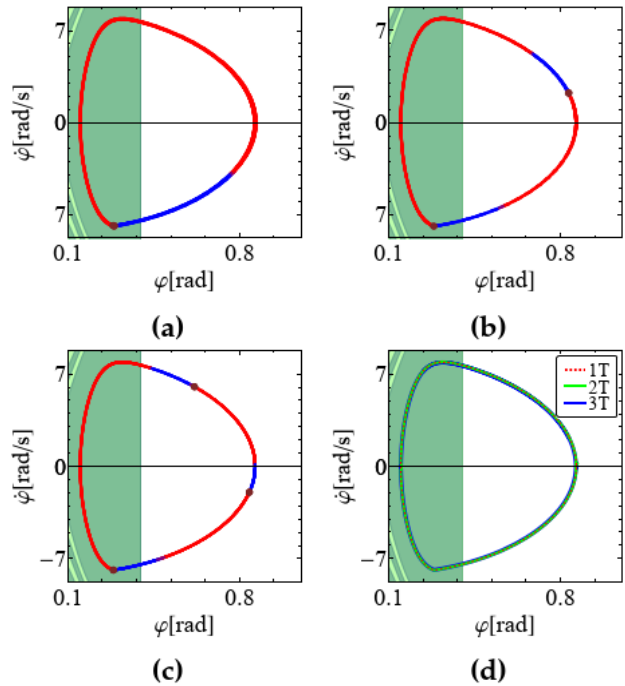

**Rys. 2.40.** Symulacyjne wykresy fazowe wykonane dla stałego wypełnienia $w = 75\ \%$ sygnału prądowego i różnych częstotliwości $f$. Trajektorię fazową rozwiązania jednookresowego (a) otrzymano dla $f = 3.25$ Hz, dwuokresowego (b) dla $f = 2 \cdot 3.25 = 6.5$ Hz i trójokresowego (c) dla $f = 3 \cdot 3.25 = 9.75$ Hz. Wszystkie trzy trajektorie fazowe (a c) nałożono na siebie i pokazano na (d).

że dla ruchu według scenariusza I, energia rozpraszana w układzie podczas każdego okresu drgań jest idealnie równoważona przez energię dodaną do układu podczas wymuszenia, czyli w chwili przejścia ze stanu 1 do 2. Jeśli punkt wymuszenia wystąpi później w czasie niż ten wyznaczony według scenariusza I, to układ otrzyma więcej energii niż będzie w stanie rozproszyć. Dlatego, aby utrzymać założony ruch nadwyżka tej energii musi zostać odebrana z układu poprzez zamianę stanów z 2 na 1 w punkcie $(\varphi_{lk}, \omega_{lk})$. Oczywiście w przypadku punktów $(\varphi_k, \omega_k)$ wybranych z czerwonego obszaru, energia dostarczana do układu i odbierana z niego nie jest taka sama, przez co rozwiązanie okresowe nie istnieje (trajektorie na płaszczyźnie fazowej obliczone dla równań (2.43)-(2.44) nie przecinają się). Przykładowa zależność pomiędzy punktami $(\varphi_k, \omega_k)$ i $(\varphi_{lk}, \omega_{lk})$ została

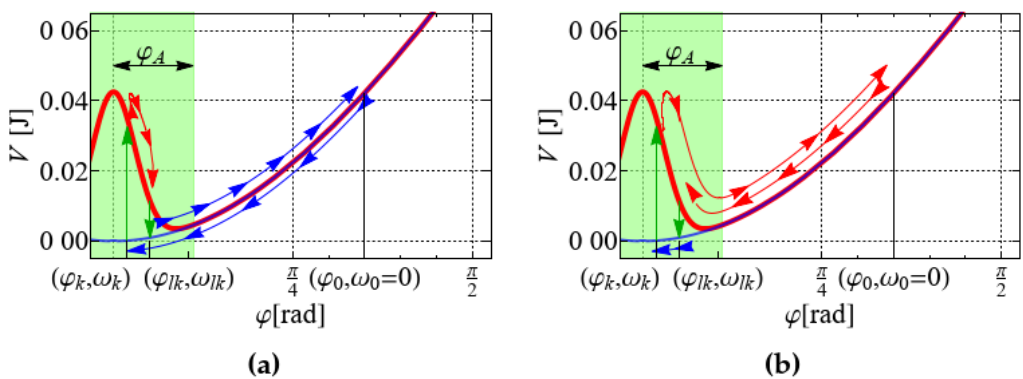

Rys. 2.41. Wykresy zmian energii potencjalnej dla ruchu według (a) scenariusza II i (b) scenariusza III.

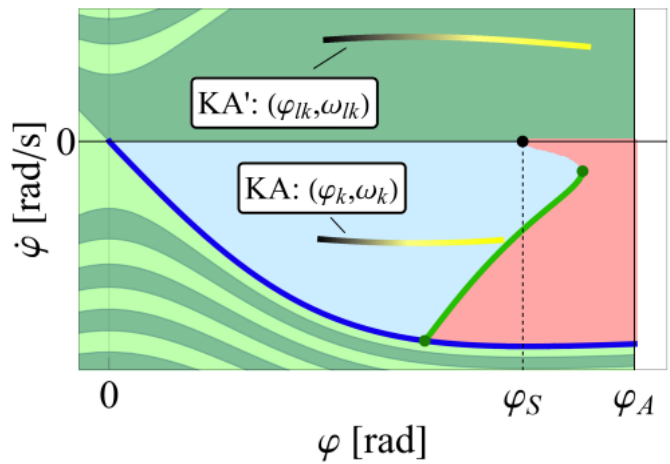

Rys. 2.42. Fragment płaszczyzny fazowej, gdzie jasnoniebieski obszar oznacza punkty wymuszenia $(\varphi_k, \omega_k)$, dla których istnieją rozwiązania według scenariuszy II i III, natomiast czerwony obszar to punkty wymuszenia, dla których takie rozwiązania nie istnieją. Zielona krzywa to zbiór punktów wymuszenia odpowiadających rozwiązaniom według scenariusza I. Krzywe gradientowe KA i KA′ obrazują przykładowe zależności pomiędzy punktami $(\varphi_k, \omega_k)$, a odpowiadającymi im punktami $(\varphi_{lk}, \omega_{lk})$ w przypadku ruchu według scenariuszy II i III.

zwizualizowana na rys. 2.42, przy pomocy krzywych oznaczonych jako KA i KA'. Krzywa KA jest zbiorem wartości punktów ($\varphi_k, \omega_k$), podczas gdy krzywa KA' jest zbiorem wartości punktów ($\varphi_{lk}, \omega_{lk}$). Ton koloru krzywych przyporządkowuje konkretny punkt ($\varphi_k, \omega_k$) do konkretnego punktu ($\varphi_{lk}, \omega_{lk}$). Dodatkowo, punkt $\varphi_S = 0.274$ rad znajdujący się na granicy obszarów i w miejscu, gdzie styka się ona z linią $\dot\varphi = 0$ rad/s jest punktem równowagi dla wahadła znajdującego się w stanie 2.

Rysunek 2.43 przedstawia przykładowe przecięcia się trajektorii fazowych na płaszczyźnie fazowej obliczone dla scenariuszy II i III. Czerwone linie odpowiadają układowi znajdującemu się w stanie 2 (równanie (2.44)), a niebieskie odpowiadają układowi znajdującemu się w stanie 1 (równanie (2.43)). Na wykresach zaznaczone zostały punkty ($\varphi_k, \omega_k$) i ($\varphi_{lk}, \omega_{lk}$). Trajektorie fazowe pokazane na rys. 2.43a,b odzwierciedlają drgania układu według scenariusza II, natomiast te pokazane na rys. 2.43c,d według scenariusza III. Na wszystkich czterech wykresach fazowych, trajektorie wyznaczone dla układu w stanie 2 (czerwone linie) są takie same, podczas gdy trajektorie wyznaczone dla układu w stanie 1 (niebieskie linie) są różne. Wykresy na rys. 2.43a,c pokazują dwa różne przypadki ruchów okresowych otrzymanych dla tych samych punktów ($\varphi_k, \omega_k$), ale rożnych punktów ($\varphi_{lk}, \omega_{lk}$), to samo dotyczy rys. 2.43b,d. Zauważyć można, że zgodnie z wcześniejszym założeniem, trajektorie obliczone dla układu w stanie 1 i 2 poza strefa aktywną są niemal identyczne, a wewnątrz niej znacząco się od siebie różnią. Uwagę mogą przyciągnąć drgania pokazane na rys. 2.43a,c, których trajektorie fazowe posiadają „ostre" przejścia pomiędzy fragmentami odpowiadającymi różnym stanom układu. Z punktu widzenia analizy dynamicznej opierającej się na przełączaniu pomiędzy równaniami (2.43)-(2.44); takie rozwiązania są możliwe, aczkolwiek nie zawsze są one możliwe do otrzymania na podstawie początkowego „ciągłego" modelu matematycznego (2.26). Przyczyna ta tkwi prawdopodobnie w pominięciu przez model „dyskretny" tj. opisany równaniami (2.43)-(2.44), ciągłości zjawiska fizycznego jakim jest narastanie i opadanie sygnału prądowego w obwodzie cewki elektrycznej. W przeciwieństwie do modelu dyskretnego, w którym występują tylko dwa stany zmieniające się w sposób nagły, w modelu ciągłym ze względu na stopniową zmianę sygnału prądowego można wyróżnić stany pośrednie, co zostało zobrazowane na rys. 2.44a. Zauważyć można, że dla chwili przełączenia się stanów z 1 na 2 w modelu dyskretnym (krzywa 2), sygnał prądowy dla modelu ciągłego (krzywa 1) osiąga dopiero połowę swojej amplitudy, co zostało zaznaczone czerwonym punktem na krzywej 1. Punkt ten traktowany jest w modelu ciągłym jako punkt wymuszenia. Fakt ten powoduje, że podczas używania modelu dyskretnego, część informacji o wpływie narastania i opadania sygnału prądowego na dynamikę układu jest pomijana. Na rys. 2.44b,c,d uwypuklono wpływ stopniowej zmiany sygnału prądowego na trajektorie fazowe układu dla analizowanych wcześniej scenariuszy ruchu. Można zauważyć, że trajektorie odpowiadające modelowi ciągłemu (krzywe 1) są „gładsze" w sąsiedztwie punktu wymuszenia niż w przypadku modelu dyskretnego (krzywe 2). Dlatego też, rozwiązanie obliczone z modelu dyskretnego i przedstawione na rys. 2.43c jest niemożliwe do otrzymania (w takiej formie) z modelu ciągłego układu. Ponad to na rys. 2.44c widać wyraźnie, że zmiana stanu układu z 1 na 2 dla modelu dyskretnego powoduje nagłą zmianę prędkości wahadła i następuje dokładnie

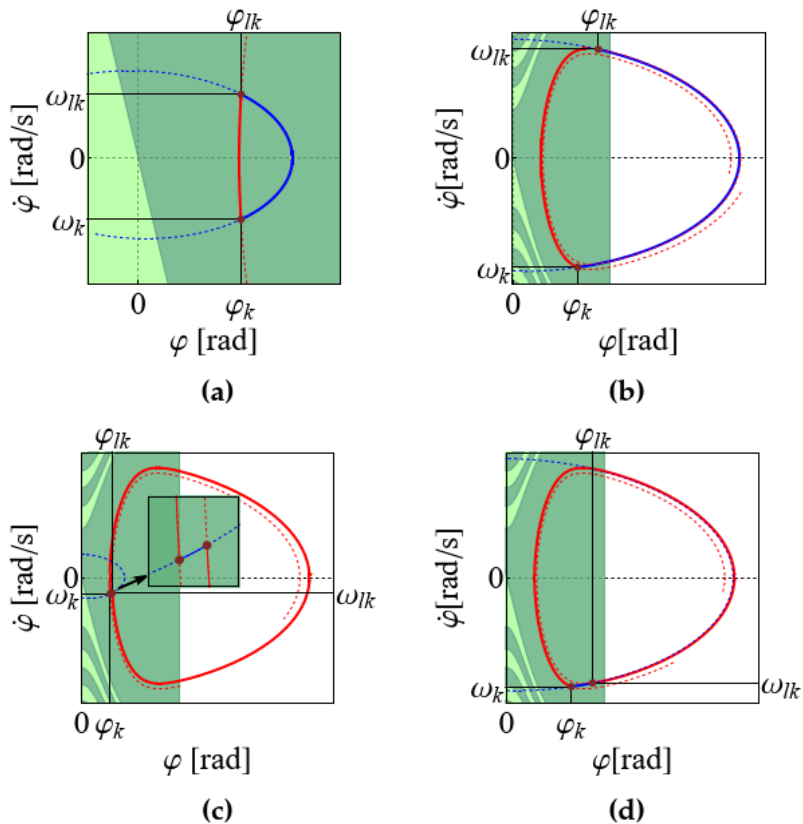

**Rys. 2.43.** Wykresy fazowe ruchu okresowego układu według scenariusza II (a,b) i scenariusza III (c,d) obliczone na podstawie przecięcia się trajektorii otrzymanych z równań (2.43)-(2.44).

w punkcie wymuszenia, podczas gdy dla modelu ciągłego zmiana prędkości następuje wcześniej niż przyjęty poziom zmiany stanu układu.

Podsumowując powyższe badania, przeanalizowany został specyficzny rodzaj jednookresowych drgań wahadła magnetycznego w jednym dołku potencjału. Okresowość drgań wahadła rozumiana jest jako liczba okresów wymuszenia (sygnału prądowego) przypadająca na jeden okres drgań wahadła. Wprowadzono założenie istnienia tzw. „strefy aktywnej" pola magnetycznego generowanego przez cewkę, tzn. przyjęto, że wymuszenie wahadła może odbywać się tylko w ściśle określonym zakresie kątów wychylenia. Po opuszczeniu przez wahadło tej strefy, pole magnetyczne cewki nie wpływa na jego ruch. Założenie to pozwoliło na przekształcenie ciągłego modelu matematycznego układu, będącego w postaci jednego nieautonomicznego równania różniczkowego, na model dyskretny będący w postaci układu dwóch przełączających się między sobą autonomicznych równań różniczkowych. Przedstawiono trzy różne scenariusze analizowanego ruchu jednookresowego oraz określono warunki ich istnienia w postaci warunków początkowych (położenie i prędkość kątowa wahadła) dla równań różniczkowych. Następnie przy pomocy tych warunków, dla pierwszego ze scenariuszy wyznaczono parametry sygnału prądowego takie jak wypełnienie i częstotliwość, dla których ruch ten występuje, a także zbadano jego dwu i trój okresowe modyfikacje. Przeprowadzona analiza numeryczna oraz teoretyczna wyjaśniła, dlacze-

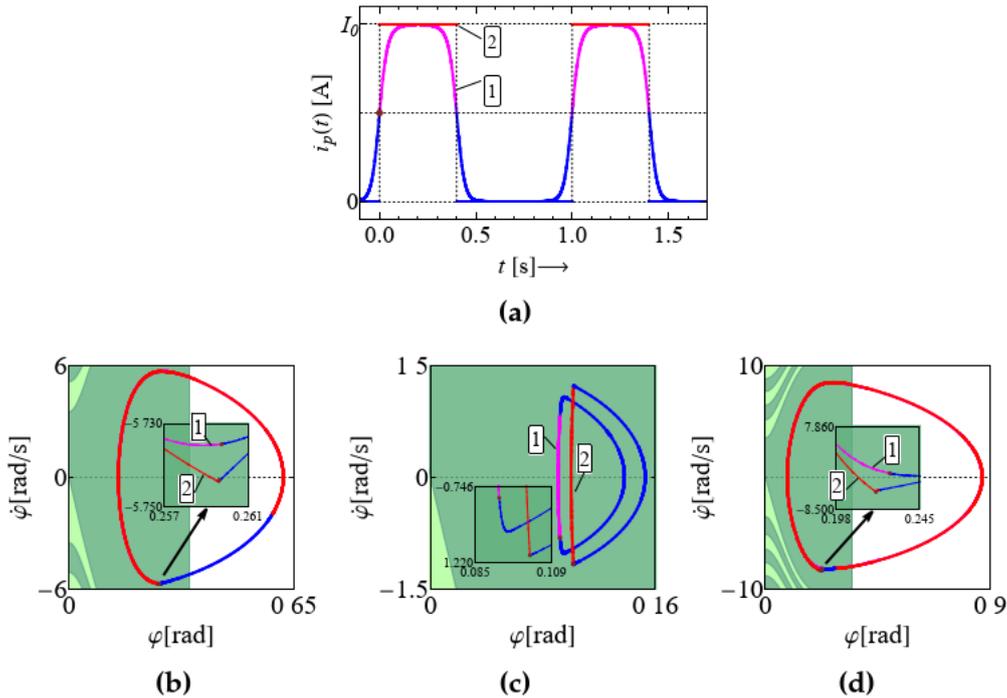

**Rys. 2.44.** Różnice pomiędzy ciągłym a dyskretnym modelem sygnału prądowego (a) oraz pomiędzy trajektoriami fazowymi obliczonymi dla (b) scenariusza I, (c) scenariusza II i (d) scenariusza III. Sygnały prądowe oraz trajektorie fazowe oznaczone jako (1) obliczono z modelu ciągłego (2.26), a oznaczone jako (2) obliczono z modelu dyskretnego (2.43)-(2.44).

go znaczna zmiana parametru wypełnienia i częstotliwości sygnału prądowego nie musi wpływać na zmianę przebiegu trajektorii fazowej badanego ruchu. Jako wniosek z tych badań uznać można, że ze względu na szczególny charakter oddziaływania magnetycznego układu i istnienie tzw. strefy aktywnej, w przypadku drgań okresowych wahadła magnetycznego odbywających się w jednym dołku potencjału, możliwe jest „sztuczne" zmienianie okresowości tych drgań bez wyraźnego naruszenia przebiegu ich trajektorii fazowej. Wniosek ten udowadnia pierwszą z postawionych w pracy tez badawczych. Wyniki badań zaprezentowane w tym podrozdziale zostały opublikowane w pracy [142].

# Rozdział 3

# Układ dwóch sprzężonych wahadeł

Rozdział poświęcony jest badaniom dwóch układów składających się z dwóch sprzężonych torsyjnie wahadeł. Pierwszy układ zbudowany jest z wahadła magnetycznego połączonego z drugim wahadłem, które jest wahadłem fizycznym. Drugi układ składa się z dwóch słabo sprzężonych wahadeł magnetycznych. Przedstawione zostaną badania numeryczne i eksperymentalne dynamiki tych układów, a ponadto dla układu z dwoma wahadłami magnetycznymi opracowane zostanie sterowanie przepływem energii między nimi przy wykorzystaniu pola magnetycznego. Wyniki zawarte w tym rozdziale udowadniają drugą z postawionych w pracy tez badawczych.

## 3.1 Dynamika nieliniowa dwóch sprzężonych wahadeł

Podrozdział zawiera opis stanowiska badawczego, modelowanie matematyczne oraz badania dynamiki układu składającego się z dwóch sprzężonych torsyjnie wahadeł, z których tylko jedno jest wahadłem magnetycznym.

### 3.1.1 Stanowisko badawcze

Zdjęcie stanowiska badawczego pokazane zostało na rys. 3.1. Stanowisko to bazuje na stanowisku pojedynczego wahadła magnetycznego (rys. 2.1). Z tego względu opis budowy stanowiska zostanie ograniczony tylko do najważniejszych elementów potrzebnych do zrozumienia jego działania. Wahadła oznaczone zostały numerami (1 2). Wahadło (2) wyposażone zostało w magnes neodymowy (3) o średnicy 22 mm i wysokości 10 mm, podczas gdy wahadło (1) wyposażone zostało w element mosiężny (4) o takich samych wymiarach i masie jak magnes wahadła (2). Pod wahadłami umieszczone zostały dwie takie same cewki elektryczne (5). Ze względu na fakt, że wahadło (1) nie posiada magnesu, tylko element niemagnetyczny, cewka znajdująca się pod nim nie będzie zasilana sygnałem prądowym

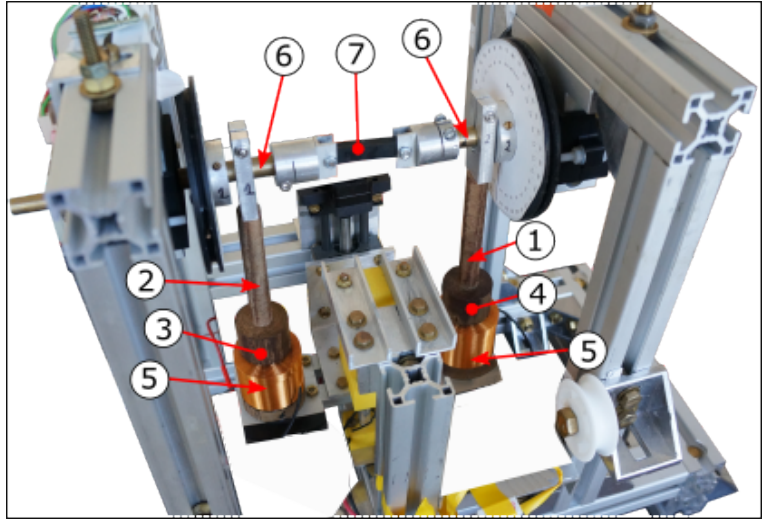

**Rys. 3.1.** Stanowisko eksperymentalne dwóch sprzężonych wahadeł, z których jedno jest magnetyczne; gdzie: 1,2 – wahadła, 3 – magnes neodymowy, 4 – mosiężny element niemagnetyczny, 5 – cewka elektryczna, 6 – wał i 7 - gumowy element podatny.

podczas prowadzonych badań. Na końcach mosiężnych wałów (6) wahadeł zamontowano specjalne uchwyty i połączono je ze sobą przy pomocy gumowego elementu podatnego (7). Jest to ten sam element, który został użyty w stanowisku pojedynczego wahadła magnetycznego.

Podczas badań, cewka wahadła 2 zasilana była takim samym prostokątnym pulsującym sygnałem prądowym jak w przypadku pojedynczego wahadła magnetycznego (rys. 2.3).

### 3.1.2 Modelowanie matematyczne

Model fizyczny analizowanego układu zaprezentowano na rys. 3.2. Przyjmując, że wahadła pod względem masy i geometrii są takie same uznano, że ich siły ciężkości wynoszą $mg$ i leżą w odległościach $s_{1,2} = s$ od osi obrotu. Wstępne badania eksperymentalne pokazały, że pomimo działań mających na celu utrzymanie symetrycznych własności mechanicznych pomiędzy wahadłami, momenty oporów ruchu $M_{F1,2}$ nie były takie same pod względem wielkości tłumienia drgań. Gumowy element podatny sprzęga wały momentem oznaczonym jako $M_K$. Jak już wspomniano w opisie stanowiska eksperymentalnego, tylko wahadło 2 poddane jest działaniu momentu pola magnetycznego $Q_{mag}$, generowanego przez cewkę elektryczną zasilaną pulsującym prostokątnym sygnałem prądowym $i_p$.

Na podstawie przyjętego modelu fizycznego oraz wykorzystując prawa Newtona zapisać można dynamiczne równania ruchu układu, w następującej postaci

$$J_1\ddot{\varphi}_1 + mgs\sin\varphi_1 + M_{F1}(\dot{\varphi}_1) - M_K(\varphi_1, \varphi_2, \dot{\varphi}_1, \dot{\varphi}_2) = 0,$$
$$J_2\ddot{\varphi}_2 + mgs\sin\varphi_2 + M_{F2}(\dot{\varphi}_2) + M_K(\varphi_1, \varphi_2, \dot{\varphi}_1, \dot{\varphi}_2) = Q_{mag}(\varphi_2, i_p(t)), \quad (3.1)$$

gdzie $J_{1,2}$ są masowymi momentami bezwładności wahadeł względem osi obrotu,

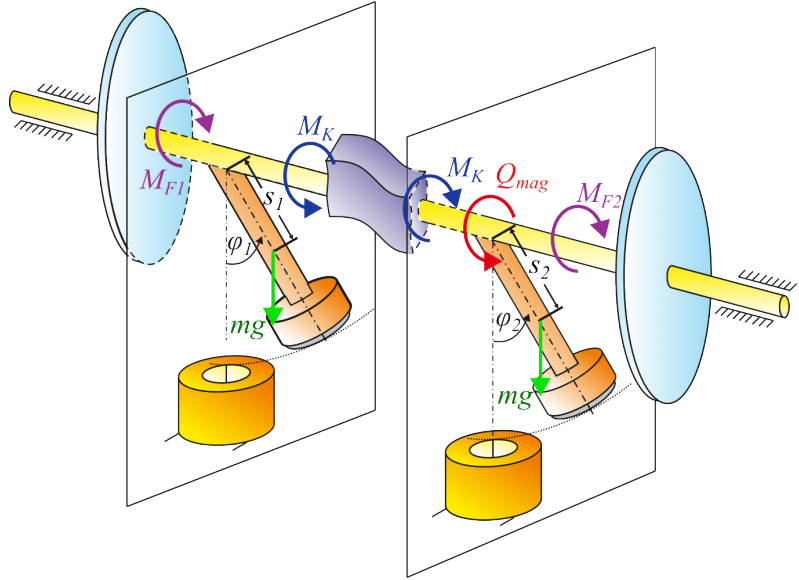

**Rys. 3.2.** Model fizyczny dwóch sprzężonych wahadeł, z których wahadło 1 jest wahadłem fizycznym, a wahadło 2 magnetycznym.

natomiast $\ddot{\varphi}_{1,2}$ i $\dot{\varphi}_{1,2}$ to przyspieszenia i prędkości kątowe wahadeł. Ze względu na zwiększenie się liczby stopni swobody układu, podczas badań numerycznych postanowiono wykorzystać mniej skomplikowany model oporów ruchu $M_{F1,2}$ niż ten wykorzystywany w układzie pojedynczego wahadła magnetycznego. Mianowicie zaimplementowano prosty model Coulomba $M_{CM}$ wraz z tłumieniem wiskotycznym, który opisany jest równaniem (2.2). Ponadto zaimplementowano model (2.6) jako moment $M_K$ elementu gumowego, gaussowski model (2.14) momentu magnetycznego jako moment $Q_{mag}$ i prostokątny model sygnału prądowego $i_p(t)$ wyrażony równaniem (2.24). Po wstawieniu wymienionych modeli matematycznych do równań ruchu (3.1), otrzymamy następujący model matematyczny dwóch sprzężonych torsyjnie wahadeł

$$J_1\ddot{\varphi}_1 + mgs\sin\varphi_1 + M_{c1}\operatorname{tgh}(\varepsilon_c\dot{\varphi}_1) + c_{w1}\dot{\varphi}_1 + c_e(\dot{\varphi}_1 - \dot{\varphi}_2) + k_e(\varphi_1 - \varphi_2) = 0,$$
$$J_2\ddot{\varphi}_2 + mgs\sin\varphi_2 + M_{c2}\operatorname{tgh}(\varepsilon_c\dot{\varphi}_2) + c_{w2}\dot{\varphi}_2 + c_e(\dot{\varphi}_2 - \dot{\varphi}_1) + k_e(\varphi_2 - \varphi_1) =$$
$$= \widehat{M}_{mag}(\varphi_2, i_p(t)). \quad (3.2)$$

Wyrażenia $c_{w1}$, $c_{w2}$ pochodzą z modelu Coulomba i wyrażają tłumienia wiskotyczne poszczególnych wahadeł.

Wartości parametrów użyte podczas obliczeń numerycznych zostały umieszczone w tabeli 3.1. Rys. 3.3 przedstawia porównanie wyników symulacji modelu (3.2) układu i eksperymentu dla przypadku ruchu swobodnego oraz wymuszonego, gdy parametry sygnału prądowego były następujące: $I_0 = 1$ A, $f = 2$ Hz, $w = 25\%$. Zauważyć można, że pomimo zastosowania uproszczonego modelu oporów ruchu w równaniach dynamicznych układu, zbieżność wyników symulacyjnych i eksperymentalnych jest wysoka.

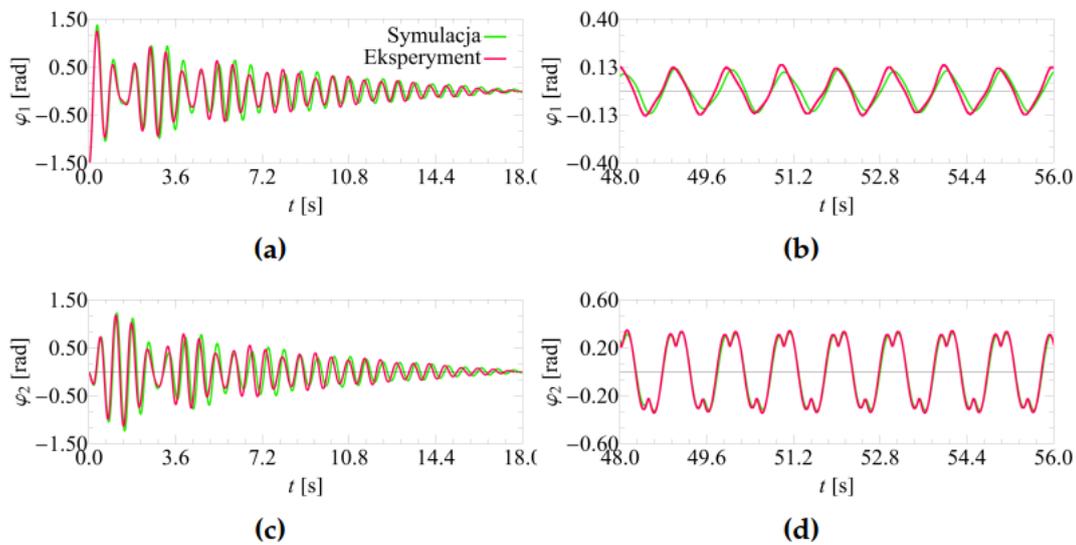

**Rys. 3.3.** Eksperymentalne i symulacyjne wykresy czasowe ruchu swobodnego (a,c) i wymuszonego (b,d) dla układu dwóch sprzężonych wahadeł.

### 3.1.3 Analiza bifurkacyjna

Analiza bifurkacyjna została przygotowana dla częstotliwości $f$ jako parametru kontrolnego, a pozostałe parametry sygnału prądowego były następujące: $I_0 = 1$ A, $w = 30\%$. Rys. 3.4 przedstawia eksperymentalne i symulacyjne diagramy bifurkacyjne opracowane dla rosnącego i malejącego parametru kontrolnego. Symulacje oraz eksperymenty wykonano dla zakresu częstotliwości $f \in (1.5, 6)$ Hz. Podczas eksperymentu, parametr kontrolny zmieniał swoją wartość w sposób liniowy z prędkością narastania (bądź opadania) równą 0.12 Hz/min. Pomiędzy wynikami symulacyjnymi a eksperymentalnymi widoczne jest podobieństwo zarówno jakościowe jak i ilościowe. Niewielkie różnice, takie jak przesunięcia występowania okien okresowych czy chaotycznych w dziedzinie częstotliwości $f$, mogą wynikać z zastosowanych uproszczeń modelu matematycznego i faktu, że numeryczna analiza bifurkacyjna została wykonana w sposób „klasyczny", tzn. wartość $f$ nie narastała (opadała) liniowo w czasie symulacji tak, jak ma to miejsce w eks-perymencie tylko przyjmowała stałe wartości. Dlatego rozwiązania jedno-, dwu- i sześcio-okresowe występujące na symulacyjnych wykresach bifurkacyjnych, ale nie występujące na wykresach eksperymentalnych zostały potwierdzone dopiero po ustawieniu stałych wartości częstotliwości $f$ podczas eksperymentu. Przykładowe rozwiązania okresowe dla stałych częstotliwości sygnału prądowego pokazano na rys. 3.5.

**Tabela 3.1.** Parametry układu dwóch torsyjnie sprzężnych wahadeł, z których jedno jest wahadłem magnetycznym.

| | | | |
|---|---|---|---|
| $J_1$ | $6.802 \cdot 10^{-4}$ kgm² | $M_{c1}$ | $2.745 \cdot 10^{-4}$ Nm |
| $J_2$ | $6.710 \cdot 10^{-4}$ kgm² | $M_{c2}$ | $2.782 \cdot 10^{-4}$ Nm |
| $mgs$ | $5.780 \cdot 10^{-2}$ Nm | $c_{w1}$ | $3.334 \cdot 10^{-5}$ $\frac{\text{Nms}}{\text{rad}}$ |
| $k_e$ | $1.451 \cdot 10^{-2}$ $\frac{\text{Nm}}{\text{rad}}$ | $c_{w2}$ | $7.437 \cdot 10^{-5}$ $\frac{\text{Nms}}{\text{rad}}$ |
| $c_e$ | $1.374 \cdot 10^{-4}$ $\frac{\text{Nms}}{\text{rad}}$ | $a_I$ | $3.760 \cdot 10^{-2}$ $\frac{\text{Nm rad}}{\text{A}}$ |
| $\varepsilon_c$ | $1713$ [–] | $b$ | $1.260 \cdot 10^{-2}$ rad² |

Analiza bifurkacyjna wskazuje na istnienie okien chaotycznych i prawdopodobnie quasiokresowych. Rozwiązania podejrzewane o bycie quasiokresowymi zaznaczono zielonymi oknami $sq$ na wykresach bifurkacyjnych. W celu potwierdzenia quasiokresowości tego regionu dla wybranej z niego częstotliwości $f = 2.40$ Hz wykonano przekroje Poincarégo oraz obliczono wykładniki Lapunowa. Rys. 3.6a,b przedstawia trajektorie fazowe wraz z przekrojami Poincarégo. Punkty Poincarégo układają się w zamknięte krzywe, co jest charakterystyczne dla quasiokresowości. Wykładniki Lapunowa obliczono na podstawie algorytmu opracowanego przez Benettina i in. [143] oraz Sandriego [144], a czas symulacji przyjęto na 10000 sekund. Na rys. 3.6c można zauważyć, że dla badanego czterowymiarowego układu ($\varphi_1, \dot{\varphi}_1, \varphi_2, \dot{\varphi}_2$), ponieważ wymiar czasu został pominięty, trzy z obliczonych wykładników Lapunowa przyjmuje ujemne wartości, a jeden z nich zbliża się do zera. Średnie wartości tych wykładników są następujące: $\{\lambda_1, \lambda_2, \lambda_3, \lambda_4\} = \{0.000719, 0.004300, 0.034329, 0.055820\}$. Wartości te mogą zostać zakwalifikowane jako spektrum (0, , , ), co według literatury [144, 145] świadczy o istnieniu orbity quasiokresowej. Niestety eksperymentalnie nie udało się potwierdzić quasiokresowości w sposób inny niż poprzez okno $sq$ widoczne na wykresie bifurkacyjnym (rys. 3.4).

Biorąc pod uwagę zaprezentowane wyniki, układ składający się z dwóch stopni swobody charakteryzuje się podobną dynamiką nieliniową jak układ o jednym stopniu swobody. Jednak oprócz zachowań chaotycznych i okresowych pojawiają

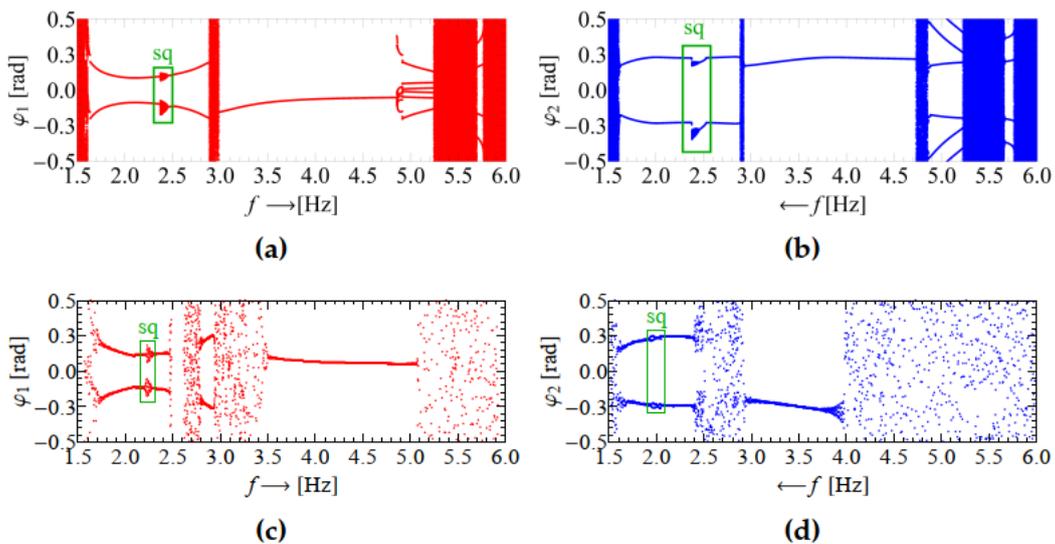

**Rys. 3.4.** Symulacyjne (a,b) i eksperymentalne (c,d) wykresy bifurkacyjne dla układu dwóch sprzężonych wahadeł. Wyjaśnienie zielonych okien $sq$ znajduje się w tekście powyżej.

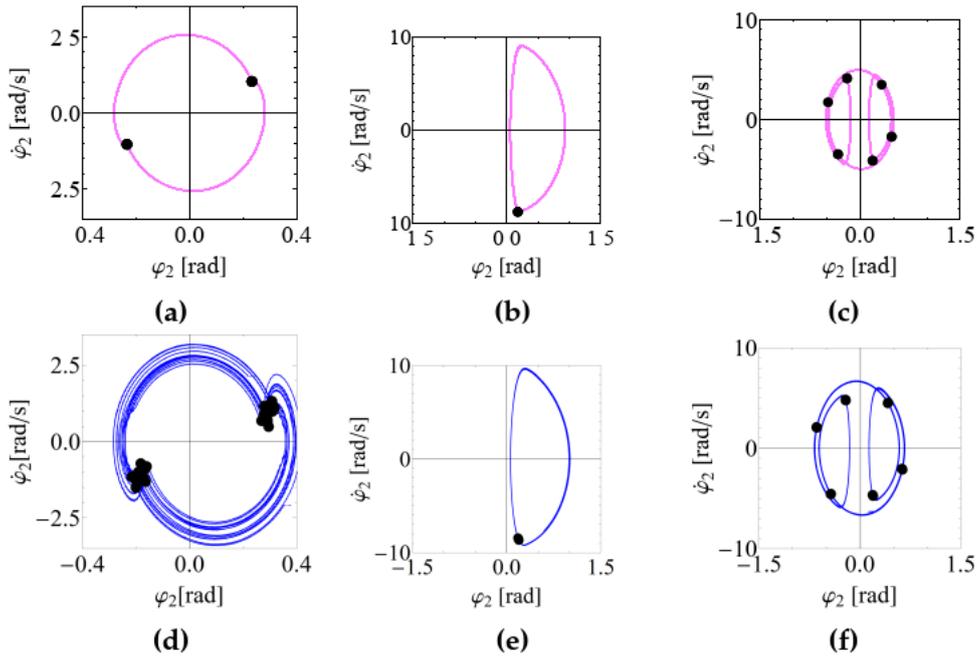

**Rys. 3.5.** Symulacyjne (a,b,c) i eksperymentalne (d,e,f) trajektorie fazowe wraz z przekrojami Poincarégo otrzymane dla następujących rozwiązań okresowych: (a,d) dwuokresowość ($f$ = 2.75 Hz), (b,e) jednookresowość ($f$ = 3 Hz), (c,f) sześciookresowość ($f$ = 5.1 Hz).

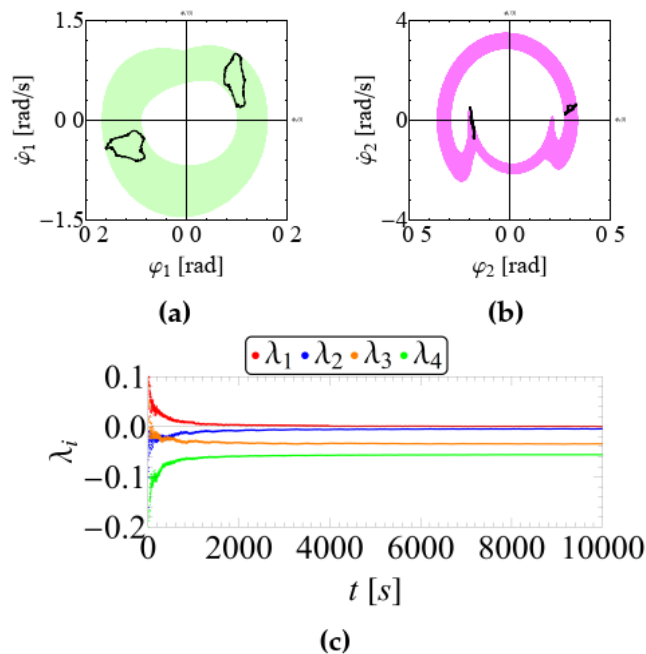

**Rys. 3.6.** Symulacyjne wykresy fazowe wraz z przekrojami Poincarégo (a,b) oraz wykładniki Lapunowa (c) wyznaczone dla rozwiązania quasiokresowego pojawiającego się przy częstotliwości $f$ = 2.40 Hz.

się w nim nowe zjawiska takie jak quasiokresowość. Rozszerzone badania dynamiki tego układu zostały opublikowane w pracach [146–148]. Oprócz badań numerycznych, prace te zawierają spostrzeżenia dotyczące energii potencjalnej układu, ale także uproszczone badania analityczne oparte na metodzie bilansu harmonicznych.

## 3.2 Sterowanie przepływem energii między wahadłami

W tym podrozdziale przedstawiony zostanie sposób sterowania wymianą energii pomiędzy dwoma słabo sprzężonymi wahadłami magnetycznymi spełniającymi warunek rezonansu 1:1. Sterowanie to odbywać się będzie przy użyciu pól magnetycznych generowanych przez cewki znajdujące się pod nimi. Wykazane zostanie, że odpowiednio generowane pola magnetyczne cewek mogą skutecznie zmieniać energie potencjalne poszczególnych wahadeł w taki sposób, że kierunek przepływu energii pomiędzy nimi będzie odbywał się w sposób z góry założony.

### 3.2.1 Stanowisko badawcze

Stanowisko badawcze dwóch słabo sprzężonych wahadeł magnetycznych pokazane zostało na rys. 3.7. Stanowisko to jest prawie takie samo, jak to wykorzystywane do badań dynamicznych z rys. 3.1, ale zawiera pewne modyfikacje. Pierwsza zmiana dotyczy elementu sprzęgającego wały wahadeł, bo w omawianym stanowisku za element sprzęgający posłużyła stalowa sprężyna torsyjna (pozycja 7 na rys. 3.7). Charakteryzuje się ona znacznie mniejszą sztywnością i tłumieniem, niż wykorzystywany wcześniej gumowy element podatny. Sprężyna została wykonana ręcznie z drutu o średnicy 0.6 mm i posiadała 4 zwoje o średnicy zewnętrznej 57 mm. Druga zmiana dotyczy wyposażenia wahadeł, ponieważ oba z nich wyposażone były w magnesy neodymowe. Dodatkowo podczas eksperymentów wykorzystywano dwie różne pod względem rozmiarów pary magnesów neodymowych dużych i małych. Duże magnesy cechowały się 22 mm średnicą, 10 mm wysokością i masą wynoszącą 28.36 g, podczas gdy małe magnesy charakteryzowały się 14 mm średnicą, 10 mm wysokością i 11.42 g masy.

Kolejna modyfikacja dotyczy sposobu zasilania cewek elektrycznych (pozycje 5 i 6) znajdujących się pod wahadłami i generujących pola magnetyczne. W celu zapewnienia kontroli przepływu energii między wahadłami, obie cewki elektryczne muszą mieć zdolność generowania pól magnetycznych zarówno przyciągających, jak i odpychających magnesy z określoną siłą. Z tego względu konieczna była zmiana sposobu zasilania cewek, tak aby sygnały prądowe mogły przyjmować różne przebiegi o różnych znakach. Jako źródło zasilania cewek użyto sterowalnego zasilacza laboratoryjnego Rohde & Schwarz®NGL202. Zasilacz ten pozwala na wygenerowanie sygnału prądowego $i(t)$ o zadanym w programie LabView przebiegu, jednakże sygnał ten może być tylko jednego znaku tzn. tylko dodatni lub tylko ujemny. Dlatego zmiana znaku sygnału prądowego zrealizowana została przy użyciu osobnego, specjalnie wykonanego układu elektronicznego opartego na przekaźnikach. Układ ten działa na zasadzie mostka H, którego zadaniem jest zmienianie kierunku przepływu prądu w obwodzie cewki poprzez fizyczną zmianę polaryzacji jej zasilania. Zmiana znaku sygnału prądowego odbywa

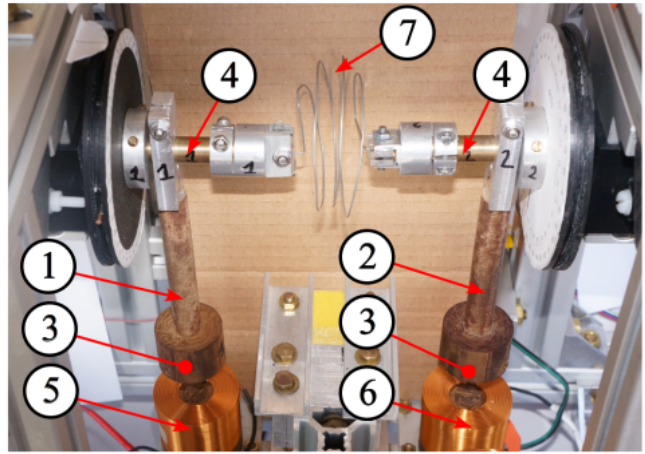

**Rys. 3.7.** Stanowisko eksperymentalne dwóch słabo sprzężonych wahadeł poddanych działaniu pola magnetycznego; gdzie: 1,2 – wahadła, 3 – magnes neodymowy, 4 – wał wahadła, 5,6 – cewka elektryczna i 7 – stalowa sprężyna torsyjna.

się zgodnie z założonym w programie LabView przebiegiem sygnału prądowego. Przypomnijmy, że dodatnia wartość sygnału prądowego cewki generuje pole magnetyczne odpychające magnes, natomiast jego ujemna wartość powoduje przyciąganie magnesu przez cewkę. Wymiana sygnałów sterujących i prądowych zachodząca pomiędzy poszczególnymi elementami stanowiska badawczego, została przedstawiona na poglądowym schemacie na rys. 3.8.

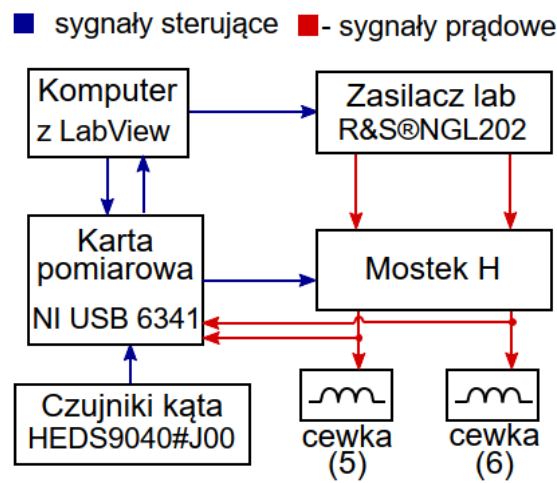

**Rys. 3.8.** Schemat przepływu sygnałów sterujących i prądowych pomiędzy elementami stanowiska dwóch słabo sprzężonych wahadeł.

Należy podkreślić, że w niniejszych badaniach oddziaływania magnetyczne pomiędzy cewką a magnesem nie mają za zadanie stanowić źródła wymuszenia układu. Elementy te mają na celu generowanie „sił sterujących", na skutek których nastąpi zamierzona dystrybucja energii układu pomiędzy oscylatorami. Wzbudzenie drgań układu następować będzie poprzez zapewnianie niezerowych warunków początkowych wahadeł.

### 3.2.2 Model fizyczny i matematyczny układu

Schemat ogólnego modelu fizycznego badanego układu pokazano na rys. 3.9a. Składa się on z dwóch połączonych sprężyną ciał wyposażonych w magnesy. Dodatkowo ciała te znajdują się w polu magnetycznym pochodzącym z cewek elektrycznych, zasilanych sygnałami prądowymi $i_1(t)$, $i_2(t)$. Zakładając, że sprzężenie między oscylatorami i rozpraszanie energii w układzie są małe oraz fakt, że układ znajduje się blisko warunku rezonansu 1:1 spowoduje to, że podczas drgań pewna porcja energii mechanicznej będzie cyklicznie przemieszczać się między oscylatorami, aż do momentu całkowitego rozproszenia się energii układu.

Przekształcając ogólny model fizyczny z rys. 3.9a w odniesieniu do stanowiska badawczego, otrzymamy model fizyczny pokazany na rys. 3.9b, gdzie oscylatory liniowe i sprężyny łączące je z nieruchomymi podporami zostały zastąpione wahadłami. Do dalszej analizy przyjęto, że oba wahadła mają tę samą masę $m$ oraz masowy moment bezwładności $J$. Wahadła połączone są liniową sprężyną skrętną, charakteryzującą się współczynnikiem sztywności $k_{et}$ oraz współczynnikiem tłumienia wiskotycznego $c_{et}$. Badania eksperymentalne przeprowadzone na potrzeby rozważanego problemu pokazały, że przede wszystkim główną przyczyną dyssypacji energii podczas ruchu wahadeł były opory wewnątrz łożysk, które w przybliżeniu zgodne były z oporami Coulomba. Natomiast występujące podczas ruchu wahadeł nieznaczne opory wiskotyczne mogły zostać pominięte. Dlatego opory ruchu zamodelowano równaniem (2.2) o współczynnikach oporów coulombowskich $M_{c1}$ i $M_{c2}$, ale z pominięciem składnika oporów wiskotycznych. Ponadto ze względu na cel prowadzonych w tym podrozdziale badań, wystarczającym będzie rozważanie stosunkowo małych przemieszczeń kątowych $\varphi_i$ wahadeł spełniających warunek $|\varphi_i| \ll \pi/2$ ($i = 1, 2$). Pozwoli to na aproksymację wielomianem trzeciego stopnia charakterystyki momentu grawitacyjnego dzia-

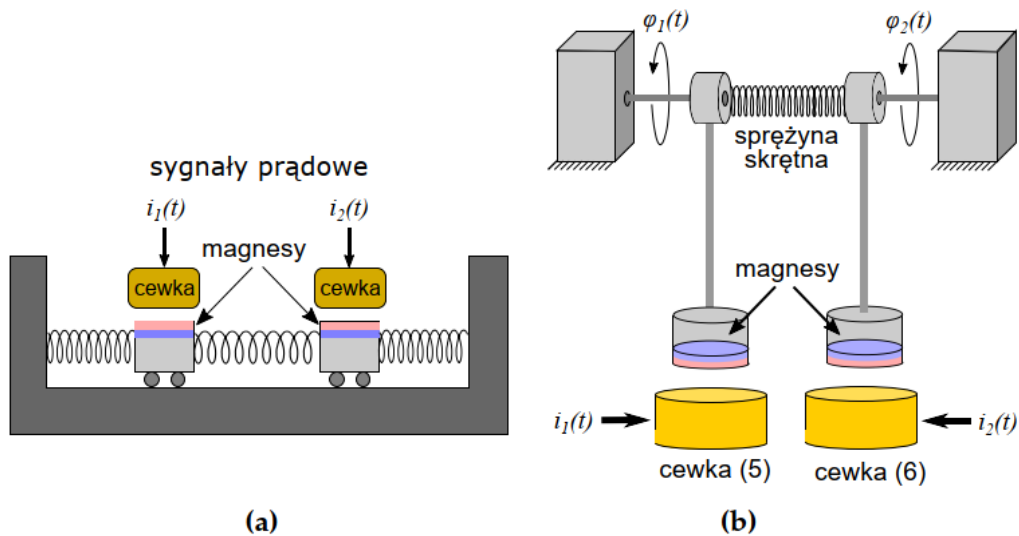

**Rys. 3.9.** Modele fizyczne dwóch słabo sprzężonych oscylatorów liniowych (a) i oscylatorów w postaci wahadeł (b), których ruch kontrolowany jest polami magnetycznymi generowanymi przez cewki elektryczne zasilane sygnałami prądowymi $i_1(t)$ i $i_2(t)$.

łającego na wahadła. Na podstawie przyjętych założeń, dynamiczne równania ruchu badanego układu prezentują się w następującej postaci

$$\begin{aligned}
J\ddot{\varphi}_1 &= M_{c1}\,\mathrm{sgn}(\dot{\varphi}_1) - c_{et}(\dot{\varphi}_1 - \dot{\varphi}_2) - mgs\left(\varphi_1 - \frac{1}{6}\varphi_1^3\right) - k_{et}(\varphi_1 - \varphi_2) + \\
&\quad \widehat{M}_{mag}(\varphi_1, i_1(t)), \\
J\ddot{\varphi}_2 &= M_{c2}\,\mathrm{sgn}(\dot{\varphi}_2) - c_{et}(\dot{\varphi}_2 - \dot{\varphi}_1) - mgs\left(\varphi_2 - \frac{1}{6}\varphi_2^3\right) - k_{et}(\varphi_2 - \varphi_1) + \\
&\quad M_{mag}(\varphi_2, i_2(t)),
\end{aligned} \tag{3.3}$$

gdzie momenty magnetyczne $M_{mag}$ opisane są gaussowskim modelem (2.14). Wartości parametrów występujących w równaniach (3.3) zawarto w tabeli 3.2, uwzględniając rozmiary zastosowanych magnesów.

**Tabela 3.2.** Parametry układu dwóch skrętnie sprzężonych wahadeł z podziałem na rozmiary magnesów.

|  | Duże magnesy |  | Małe magnesy |
|---|---|---|---|
| $a_I$ | $8.036 \cdot 10^{-3}\,\mathrm{Nm \cdot rad \cdot A^{-1}}$ | $a_I$ | $3.635 \cdot 10^{-3}\,\mathrm{Nm \cdot rad \cdot A^{-1}}$ |
| $b$ | $30.810 \cdot 10^{-3}\,\mathrm{rad}^2$ | $b$ | $43.366 \cdot 10^{-3}\,\mathrm{rad}^2$ |
| $M_{c1}$ | $2.500 \cdot 10^{-4}\,\mathrm{Nm}$ | $M_{c1}$ | $3.114 \cdot 10^{-4}\,\mathrm{Nm}$ |
| $M_{c2}$ | $1.600 \cdot 10^{-4}\,\mathrm{Nm}$ | $M_{c2}$ | $2.705 \cdot 10^{-4}\,\mathrm{Nm}$ |
| $c_{et}$ | $9.615 \cdot 10^{-6}\,\mathrm{Nms \cdot rad^{-1}}$ | $c_{et}$ | $9.615 \cdot 10^{-6}\,\mathrm{Nms \cdot rad^{-1}}$ |
| $k_{et}$ | $3.999 \cdot 10^{-3}\,\mathrm{Nm \cdot rad^{-1}}$ | $k_{et}$ | $3.999 \cdot 10^{-3}\,\mathrm{Nm \cdot rad^{-1}}$ |
| $J$ | $6.787 \cdot 10^{-4}\,\mathrm{kgm}^2$ | $J$ | $5.675 \cdot 10^{-4}\,\mathrm{kgm}^2$ |
| $mgs$ | $5.840 \cdot 10^{-2}\,\mathrm{Nm}$ | $mgs$ | $5.018 \cdot 10^{-2}\,\mathrm{Nm}$ |

### 3.2.3 Badania wstępne

W tym paragrafie analizowane będą reguły jakimi powinny rządzić się sygnały prądowe $i_1$ i $i_2$ zasilające cewki i sterujące przepływem energii między wahadłami. Badania wstępne rozpoczęto od wyznaczenia wyrażenia na energię potencjalną zlinearyzowanego wokół punktu równowagi układu zachowawczego wahadeł. W tym celu przyjęto, że w układzie nie występuje tłumienie oraz że wzięte pod uwagę zostaną tylko te człony energii potencjalnych pola grawitacyjnego i magnetycznego, których stopień wielomianu jest nie większy niż dwa. Wyprowadzenie wzorów na poszczególne człony energii potencjalnej zawarto w Załączniku B. Zatem, wyrażenie na energię potencjalną zlinearyzowanego układu zachowawczego odniesione do masowego momentu bezwładności wahadeł, przedstawia

się następująco

$$V_J(\varphi_1, \varphi_2) = \frac{1}{2}\Omega^2 \left[\varphi_1^2 + \beta(\varphi_2 - \varphi_2) + \varphi_2^2\right] - \frac{a_I}{bJ}\left(i_1\varphi_1^2 + i_2\varphi_2^2\right), \quad (3.4)$$

gdzie $\Omega = \sqrt{mgs/J}$ jest częstością drgań własnych zlinearyzowanych wahadeł, a parametr $\beta = k_{et}/mgs$ odpowiada za względną „siłę" sprzężenia. Wartości tych parametrów jaki i innych pozostałych wykorzystywanych podczas badań symulacyjnych zawarto w tabeli 3.3 z podziałem na rozmiary magnesów wahadeł.

**Tabela 3.3.** Zredukowane parametry układu dwóch skrętnie sprzężonych wahadeł z podziałem na rozmiary magnesów.

|          | Duże magnesy                      |          | Małe magnesy                      |
|----------|-----------------------------------|----------|-----------------------------------|
| $\Omega$ | $9.276\,\text{s}^{-1}$            | $\Omega$ | $9.403\,\text{s}^{-1}$            |
| $\beta$  | $6.849 \cdot 10^{-2}\,\text{rad}^{-1}$ | $\beta$ | $7.969 \cdot 10^{-2}\,\text{rad}^{-1}$ |
| $\zeta_1$ | $1.985 \cdot 10^{-2}\,\text{s}^{-1}$ | $\zeta_1$ | $2.918 \cdot 10^{-2}\,\text{s}^{-1}$ |
| $\zeta_2$ | $1.270 \cdot 10^{-2}\,\text{s}^{-1}$ | $\zeta_2$ | $2.534 \cdot 10^{-2}\,\text{s}^{-1}$ |
| $\alpha$ | $7.636 \cdot 10^{-4}\,\text{rad}^{-1}$ | $\alpha$ | $9.009 \cdot 10^{-4}\,\text{rad}^{-1}$ |

Funkcja opisująca energię potencjalną $V_J$ jest zależna od czasu przez występowanie w niej prądów $i_1(t)$ i $i_2(t)$. Wartości sygnałów prądowych wpływają na kształt powierzchni potencjału, a tym samym na lokalne zachowania dynamiczne układu wahadeł, tak jak pokazano to na wykresach z rys. 3.10. O ile obecność pola magnetycznego w układzie nie ma wpływu na położenie punktu stacjonarnego $(\varphi_1, \varphi_2) = (0, 0)$, o tyle wartości sygnałów prądowych mogą wpływać na jego rodzaj. Diagram z rys. 3.10 pokazuje, jak wartości energii potencjalnej $V_J(\varphi_1, \varphi_2) = const.$ układu ewoluują w sąsiedztwie punktu stacjonarnego $(0, 0)$ na skutek zmian wartości sygnałów prądowych. Eliptyczne kształty energii potencjalnej w lewej dolnej części diagramu potwierdzają, że ujemne znaki sygnałów prądowych $i_1$ i $i_2$ prowadzą do minimum energii potencjalnej układu i zapewniają stabilność położenia równowagi. Natomiast zwiększanie wartości jednego lub dwóch sygnałów prądowych może prowadzić do przekształcenia się powierzchni energii potencjalnej układu z wklęsłej na wypukłą, a także siodło. Oznacza to wtedy, że odpychający charakter oddziaływania magnetycznego dominuje w układzie nad momentami przywracającymi układ do położenia równowagi, do których należą momenty sił grawitacji oraz moment sprzęgający wahadła. Granice dla trzech różnych rodzajów punktu stacjonarnego $(0,0)$ z rys. 3.10 wyznaczono na podstawie częstości drgań własnych zlinearyzowanego układu przy założeniu, że sygnały prądowe są stałe (obliczenia znajdują się w Załączniku B):

$$\omega_{1,2}^2 = (1+\beta)\Omega^2 - \frac{a_I}{bJ}(i_1^2 + i_2^2) \mp \sqrt{\beta^2\Omega^4 + \frac{a_I^2}{b^2J^2}(i_1^2 - i_2^2)}. \quad (3.5)$$

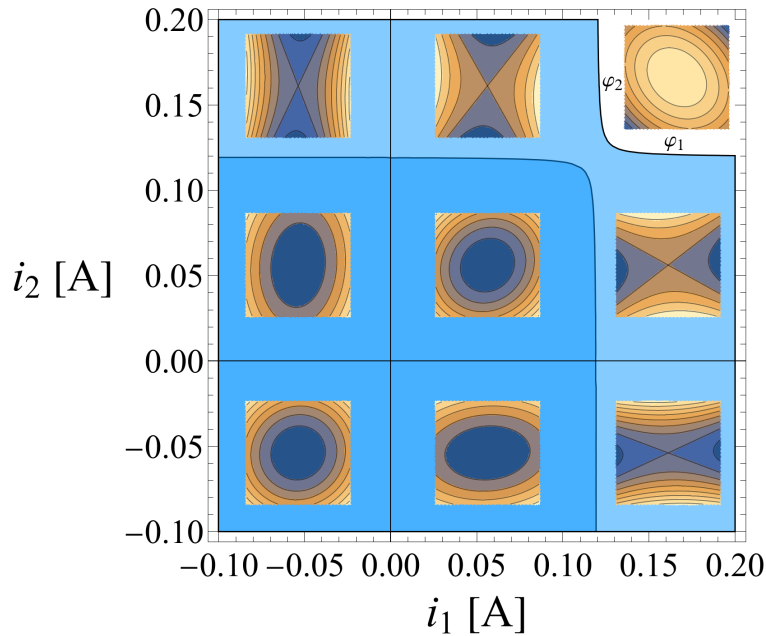

**Rys. 3.10.** Ewolucje energii potencjalnej (3.4) układu w sąsiedztwie punktu równowagi $(0,0)$ obliczone dla różnych stałych wartości prądów $i_1$ i $i_2$ cewek. Ciemnoniebieski region z eliptycznymi kształtami odpowiada minimum energii potencjalnej, podczas gdy jasnoniebieski region odpowiada siodłom. Jasny region znajdujący się w prawym górnym rogu odpowiada maksimum energii potencjalnej.

Obie z obliczonych częstości dla ciemnoniebieskiego regionu znajdującego się w lewym dolnym rogu rys. 3.10 muszą być rzeczywiste, wtedy funkcja (3.4) posiada minimum.

Podsumowując ten paragraf, równania (3.3), (3.4) i (3.5) pokazują, że oba prądy płynące w cewkach mają parametryczny wpływ na zachowanie zlinearyzowanego układu poprzez skuteczną zmianę jego sztywność. Dlatego można przyjąć, że dynamika analizowanego układu kontrolowana jest poprzez zmianę jego parametrów, a nie poprzez dokładanie do niego zewnętrznych sił sterujących. Dodatkowo, sygnałami prądowymi można przesuwać wartości częstości drgań własnych wahadeł w kierunku do lub od warunku rezonansu 1:1 w celu osiągnięcia pożądanego procesu wymiany energii między nimi.

### 3.2.4 Adaptacja modelu matematycznego

W tym paragrafie przedstawione zostanie podejście analityczne dla strategii sterowania przepływem energii między wahadłami. Podejście to opierać się będzie na idei uśredniania układu w odniesieniu do jego tzw. „szybkiej" zmiennej, przy założeniu, że oba wahadła będą zlinearyzowane i charakteryzować się będą tą samą częstość drgań własnych. Założenie to jest konieczne, aby rozważać wymianę energii podczas rezonansu 1:1.

Opracowanie strategii sterowania przepływem energii wymaga odpowiedniego przejścia z układu opisanego podstawowymi zmiennymi stanu na układ

o zmiennych opisowych związanych z energią. W pierwszym kroku należy przedstawić układ (3.3) w postaci czterech równań pierwszego rzędu zawierających zmienne stanu, takie jak przemieszczenia $\varphi_j$ i prędkości kątowe $v_j$ wahadeł:

$$\frac{d\varphi_j}{dt} = v_j, \quad \frac{dv_j}{dt} = \Omega^2 \varphi_j \; f_j; \quad j = 1, 2, \qquad (3.6)$$

gdzie

$$\begin{aligned} f_1 &= 2\Omega \left[ \zeta_1 \, \text{sgn} \, \dot{\varphi}_1 + \alpha \left( \dot{\varphi}_1 \quad \dot{\varphi}_2 \right) \right] + \Omega^2 \left[ \beta \left( \varphi_1 \quad \varphi_2 \right) \quad \frac{1}{6}\varphi_1^3 \right] + \\ &\qquad \frac{i_1(t)}{J} \cdot \frac{2a_I}{b} \exp\left( \frac{\varphi_1^2}{b} \right) \varphi_1, \\ f_2 &= 2\Omega \left[ \zeta_2 \, \text{sgn} \, \dot{\varphi}_2 + \alpha \left( \dot{\varphi}_2 \quad \dot{\varphi}_1 \right) \right] + \Omega^2 \left[ \beta \left( \varphi_2 \quad \varphi_1 \right) \quad \frac{1}{6}\varphi_2^3 \right] + \\ &\qquad \frac{i_2(t)}{J} \cdot \frac{2a_I}{b} \exp\left( \frac{\varphi_2^2}{b} \right) \varphi_2, \end{aligned} \qquad (3.7)$$

oraz parametry, których wartości zawarto w tabeli 3.3:

$$\Omega = \sqrt{\frac{mgs}{J}}, \quad \beta = \frac{k_{et}}{mgs}, \quad \zeta_1 = \frac{M_{c1}}{2J\Omega}, \quad \zeta_2 = \frac{M_{c2}}{2J\Omega}, \quad \alpha = \frac{c_{et}}{2J\Omega}. \qquad (3.8)$$

Należy zauważyć, że znormalizowane współczynniki tłumienia $\zeta_{1,2}$ nie charakteryzują liniowego tłumienia wiskotycznego, ponieważ w rozważaniach przyjęto tylko model z oporami Coulomba. Zastąpienie sgn $\dot{\varphi}_j$ przez $\dot{\varphi}_j$ zapewniłoby tym współczynnikom konwencjonalne znaczenie, gdy przypadek tłumienia wiskotycznego stałby się interesujący.

Jak już wspomniano, podejście analityczne oparte jest na przejściu z układu (3.6)-(3.7) na układ, który w łatwy sposób opisywał będzie proces wymiany energii między wahadłami. Podobna metodologia była już stosowana wcześniej do analizy wymiany energii między różnymi postaciami drgań cieczy przelewającej się w zbiorniku [154] oraz efektów dudnienia drgań wywołanych tarciem w układzie dwóch słabo sprzężonych oscylatorów [155]. Poniżej przedstawiono podstawowe działania prowadzące do transformacji między tymi układami. W szczególności wykazane zostanie, że poziomy wzbudzeń wahadeł i przesunięcie fazowe między nimi zdefiniowane będą przez elementy symetrycznej „macierzy energii" [156]

$$E_{ij} = \frac{1}{2} \left( \frac{d\varphi_i}{dt} \frac{d\varphi_j}{dt} + \Omega^2 \varphi_i \varphi_j \right), \quad i, j = 1, 2, \qquad (3.9)$$

gdzie $\varphi_{i,j}$ ($i, j = 1, 2$) są kątowymi wychyleniami wahadeł, a $\Omega$ jest częstością drgań własnych zlinearyzowanego wahadła. W przypadku, gdy w układzie pominiemy rozpraszanie energii, nieliniowości oraz sprzężenie wahadeł, a także wyłączymy zasilanie cewek, wielkości (3.9) otrzymają konkretne znaczenia fizyczne. Mianowicie wielkości $E_{11}$ i $E_{22}$ reprezentują całkowite energie wahadeł niesprzężonych i są odniesione do wartości masowego momentu bezwładności. Wielkości $E_{12} = E_{21}$ niosą natomiast informację o przesunięciu fazowym. Do ujawnienia ich związku z przesunięciem fazowym załóżmy, że oba wahadła są zlinearyzowane i rozprzęgnięte, a więc ich ruch opisany jest równaniami $\varphi_1 = A_1 \cos \Omega t$ i $\varphi_2 = A_2 \cos(\Omega t + \Delta)$, gdzie $\Delta$ jest przesunięciem fazowym. Podstawiając te równania do (3.9) otrzymamy

$$E_{12} = \frac{1}{2}(\dot{\varphi}_1 \dot{\varphi}_2 + \Omega \varphi_1 \varphi_2) = \frac{1}{2}\Omega^2 A_1 A_2 [\sin(\Omega t)\sin(\Omega t + \Delta) +$$
$$+ \cos(\Omega t)\cos(\Omega t + \Delta)] = \frac{1}{2}\Omega^2 A_1 A_2 \cos\Delta, \quad (3.10)$$

oraz

$$E_{11}E_{22} = \frac{1}{4}\Omega^4 A_1^2 A_2^2. \quad (3.11)$$

Na podstawie równań (3.10) i (3.11) można zauważyć, że przesunięcie fazowe $\Delta$ jest zdefiniowane przez stosunek

$$\frac{E_{12}}{\sqrt{E_{11}E_{22}}} = \cos\Delta. \quad (3.12)$$

Powodem używania wielkości $E_{ij}$ jako zmiennych opisowych układu jest to, że takie zmienne opisują go w jednostkach energii, eliminując mniej informacyjną szybko zmieniającą się skalę czasową $\Omega t$. Z równań (3.13) poniżej widać, że algebraiczne kombinacje różnych elementów $E_{ij}$ dostarczają wystarczających informacji o stanach dynamicznych dwóch identycznych oscylatorów. Niestety wielkości $E_{11}$ i $E_{22}$ tracą swoje znaczenie fizyczne z powodu rozpraszania energii i innych czynników. Niemniej jednak nadal mogą służyć do charakteryzowania poziomów wzbudzenia poszczególnych wahadeł, dlatego termin „energia" pozostanie w użyciu oraz wprowadzone zostaną następujące kombinacje $E_{ij}$:

$$\begin{aligned} E &= E_{11} + E_{22}, \\ P &= \frac{E_{11} - E_{22}}{E}, \quad 1 \le P \le 1, \\ Q &= \frac{E_{12}}{\sqrt{E_{11}E_{22}}} = \cos\Delta, \quad 1 \le Q \le 1, \end{aligned} \quad (3.13)$$

gdzie $E$ jest całkowitą energią wahadeł, $P$ informuje o podziale energii między wahadłami, a $Q$ jest „wskaźnikiem koherencji" charakteryzującym przesunięcie fazowe $\Delta$ między wahadłami. Wartość $P = 0$ oznacza, że energia jest równomiernie rozłożona między wahadłami ($E_{11} = E_{22} = 1/2E$). Wartość $P = 1$ wskazuje, że drga tylko pierwsze wahadło, podczas gdy drugie pozostaje w spoczynku ($E_{11} = E$, $E_{22} = 0$). Sytuacja odwrotna ma miejsce, gdy $P = 1$, wtedy drga tylko wahadło drugie, a pierwsze pozostaje w spoczynku ($E_{22} = E$, $E_{11} = 0$). Wskaźnik koherencji $Q$ określa rodzaj postaci synchronizacji drgań wahadeł, i został zobrazowany na rys. 3.11. Wartość $Q = 1$ ($\Delta = \pi$) odpowiada postaci antyfazy, wartość $Q = 0$ ($\Delta = \pi/2$) odpowiada postaci „eliptycznej", a wartość $Q = 1$ ($\Delta = 0$) odpowiada postaci w fazie. Następnie wykorzystując jawny parametr $\Delta$ przesunięcia fazowego, możemy dokonać transformacji z (3.13) do układu o oryginalnych zmiennych stanu, tj. transformacji $\{E, P, \Delta, \delta\} \rightarrow \{\varphi_1, v_1, \varphi_2, v_2\}$. W tym celu, aby otrzymać bezpośrednie wyrażenia dla takiej transformacji podstawmy

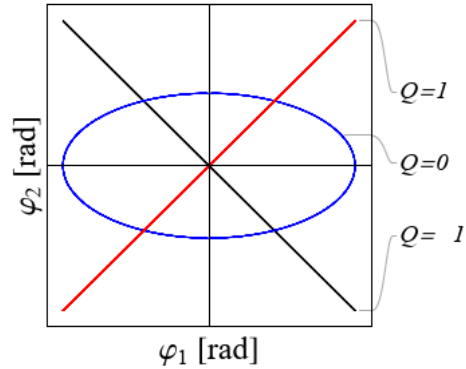

**Rys. 3.11.** Geometryczne znaczenie wskaźnika koherencji $Q$. Postać drgań układu wahadeł: w fazie ($Q = 1$), eliptyczna ($Q = 0$) i w antyfazie ($Q = -1$).

$\varphi_1 = A_1 \cos \Omega t$ i $\varphi_2 = A_2 \cos(\Omega t + \Delta)$ do równań (3.9), otrzymując

$$E = E_{11} + E_{22} = \frac{1}{2}\Omega^2 \left(A_1^2 + A_2^2\right),$$
$$P = \frac{E_{11} - E_{22}}{E} = \frac{A_1^2 - A_2^2}{A_1^2 + A_2^2} = \frac{1}{2E}\Omega^2 \left(A_1^2 - A_2^2\right). \tag{3.14}$$

Rozwiązując równania (3.14), znajdujemy wzory na amplitudy $A_1$ i $A_2$, ostatecznie definiując transformacje $\{E, P, \Delta, \delta\} \to \{\varphi_1, v_1, \varphi_2, v_2\}$:

$$\begin{aligned}
\varphi_1 &= \Omega^{-1}\sqrt{E(1+P)}\cos\delta, \\
v_1 &= -\sqrt{E(1+P)}\sin\delta, \\
\varphi_2 &= \Omega^{-1}\sqrt{E(1-P)}\cos(\delta+\Delta), \\
v_2 &= -\sqrt{E(1-P)}\sin(\delta+\Delta),
\end{aligned} \tag{3.15}$$

gdzie faza $\Omega t$ została zamieniona na tzw. „szybką" fazę $\delta = \delta(t)$.

Podsumowując, transformacja (3.15) została przeprowadzona przy użyciu rozwiązań dwóch niezależnych oscylatorów harmonicznych oraz dla przypadków, gdy wielkości $E$, $P$ i $\Delta$ były stałe w czasie, a $\delta = \Omega t$. W ogólnym przypadku, oscylatory są sprzężone i poddawane różnego rodzaju zaburzeniom. Niemniej jednak, w warunkach, gdy sprzężenie wahadeł jest stosunkowo słabe, a perturbacje małe, transformacja (3.15) nadal może być poddana metodzie wariacji stałych dowolnych [138] przy założeniu, że $E$, $P$ i $\Delta$ nie są już stałymi parametrami, ale wolno zmieniającymi się wielkościami.

### 3.2.5 Procedura uśredniania

Uśrednianie układu (3.6) można przeprowadzić analogicznie jak w metodzie Van der Pola [138]. Prezentowana w tym podrozdziale procedura, polega na podstawieniu równań (3.15) do (3.6), następnie rozwiązaniu tych równań w celu wyznaczenia wyrażeń pochodnych $d\{E, P, \Delta, \delta\}/dt$ i uśrednienia ich prawych stron względem szybko zmieniającej się fazy $\delta$. Operator uśredniania wyrażony jest

w następujący sposób

$$\langle \ldots \rangle = \frac{1}{2\pi} \int_0^{2\pi} \ldots \, d\delta. \tag{3.16}$$

W omawianym przypadku procedura uśredniania uzasadniona jest zarówno fizycznym charakterem badanego zjawiska, jak i założeniami narzuconymi na parametry modelu. Uśrednianie zastosowano dla przypadku drgań o względnie wysokiej częstotliwości, w celu opisania stosunkowo powolnej modulacji amplitudy i przesunięcia fazowego charakteryzujących stopniową wymianę energii między dwoma wahadłami w warunkach bliskich rezonansowi 1:1. Z tego powodu założono, że w układzie (3.6) i (3.7) oddziaływania magnetyczne, efekty rozpraszania, sprzężenie i nieliniowość opisane funkcjami $f_i$ ($i = 1, 2$) są stosunkowo słabe w porównaniu z momentami przywracającymi układ do położenia równowagi. Ponad to pomimo występowania w tych funkcjach nieciągłości w postaci funkcji signum (modelujących opory coulombowskie), procedura uśredniania wciąż może być przeprowadzona. Skomplikowane obliczenia analityczne związane z transformacją układu i uśrednianiem, przeprowadzono w programie *Wolfram Mathematica* (obliczenia zawarto w Załączniku C). Ostateczną postać uśrednionego układu przedstawiają następujące równania różniczkowe pierwszego rzędu

$$\frac{dE}{dt} = 2\alpha\Omega E \left(1 - \sqrt{1-P^2}\cos\Delta\right) - \frac{4}{\pi}\Omega^2 \left(\zeta_1\sqrt{\lambda_{a1}} + \zeta_2\sqrt{\lambda_{a2}}\right),$$

$$\frac{dP}{dt} = \Omega\sqrt{1-P^2}\left(\beta\sin\Delta - 2\alpha P\cos\Delta\right) - \frac{4}{\pi}\frac{\Omega^4}{E^2}\left(\zeta_1\lambda_{a2}\sqrt{\lambda_{a1}} - \zeta_2\lambda_{a1}\sqrt{\lambda_{a2}}\right),$$

$$\frac{d\Delta}{dt} = \frac{1}{8\Omega}EP - \frac{\Omega}{\sqrt{1-P^2}}\left(\beta P\cos\Delta + 2\alpha\sin\Delta\right) +$$

$$+\frac{a_I}{bJ\Omega}\left\{i_1(t)e^{-\frac{\lambda_{a1}}{2b}}\left[I_{B0}\left(\frac{\lambda_{a1}}{2b}\right) - I_{B1}\left(\frac{\lambda_{a1}}{2b}\right)\right] - i_2(t)e^{-\frac{\lambda_{a2}}{2b}}\left[I_{B0}\left(\frac{\lambda_{a2}}{2b}\right) - I_{B1}\left(\frac{\lambda_{a2}}{2b}\right)\right]\right\}$$

$$\tag{3.17}$$

oraz

$$\frac{d\delta}{dt} = \left[1 + \frac{1}{2}\beta - \frac{1}{16}\lambda_{a1} - \frac{1}{2}\sqrt{\frac{\lambda_{a2}}{\lambda_{a1}}}\left(\beta\cos\Delta - 2\alpha\sin\Delta\right)\right]\Omega +$$

$$- \frac{a_I}{bJ\Omega}i_1(t)e^{-\frac{\lambda_{a1}}{2b}}\left[I_{B0}\left(\frac{\lambda_{a1}}{2b}\right) - I_{B1}\left(\frac{\lambda_{a1}}{2b}\right)\right], \tag{3.18}$$

gdzie $\lambda_{a1} = E(1+P)/\Omega^2 = 2E_{11}/\Omega^2$ i $\lambda_{a2} = E(1-P)/\Omega^2 = 2E_{22}/\Omega^2$ są znormalizowanymi energiami poszczególnych wahadeł. Wyrażenia $I_{B0}$ i $I_{B1}$ są zmodyfikowanymi funkcjami Bessela pierwszego rodzaju (rzędu 0 i 1), powstałymi na skutek uśredniania wyrażeń związanych z momentami magnetycznymi (2.14). Poprawność zaprezentowanej procedury uśredniania potwierdzono w paragrafie 3.2.6.

### 3.2.6 Numeryczna weryfikacja uśrednionego modelu

W celu weryfikacji procedury uśredniania prowadzącej do równań (3.17), porównano wyniki otrzymane dla symulacji numerycznych oryginalnych równań (3.6)-(3.7) z wynikami otrzymanymi dla układu uśrednionego (3.17). Ze względu na to, że wielkościami wyjściowymi układu oryginalnego są kąty i prędkości kątowe wahadeł, a modelu uśrednionego wielkości $P$ i $Q = \cos\Delta$, kąty i prędkości zostały przeliczone na $P$ i $Q$ przy użyciu równań (3.9) i (3.13). Symulacje zostały przeprowadzone dla następujących sygnałów prądowych

$$i_1(t) = A + B\sin^2\left(\pi\frac{t}{t_k}\right), \quad i_2(t) = i_1(t), \qquad (3.19)$$

gdzie $A = 0.001$ A, $B = 0.055$ A i $t_k = 40$ s. Parametry sygnałów prądowych (3.19) zostały dobrane empirycznie na podstawie wstępnych badań w taki sposób, aby pożądany efekt wymiany energii między wahadłami stał się widoczny. Z tego względu przyjęto, że wartości sygnałów prądowych zaczynają się od pewnego bardzo niskiego poziomu $A$ (tzw. offsetu), który zawsze może być obecny podczas badań, a następnie osiągają swoje maksimum $A + B$ w chwili $t = t_k/2$ i ostatecznie spadają do poziomu $A$ w chwili $t = t_k$. Wartość $t_k$ jest czasem trwania procesu wymiany energii, zależnym od intensywności jej dyssypacji w układzie oraz początkowego poziomu energii całkowitej układu. Sygnały prądowe opisane równaniami (3.19) z zaimplementowanymi różnymi wartościami parametrów, będą wykorzystywane podczas symulacji numerycznych w dalszych częściach badań nad wymianą energii. Warto przypomnieć, że prądy dodatnie generują odpychające momenty magnetyczne, podczas gdy prądy ujemne wytwarzają momenty przyciągające wahadła.

Porównanie wyników symulacji otrzymanych dla układu oryginalnego i uśrednionego z zaimplementowanymi parametrami odpowiadającymi dużym magnesom pokazano na rys. 3.12. Lewa kolumna rysunku przedstawia wielkości uzyskane na podstawie równań (3.6)-(3.7), natomiast prawa kolumna daje wgląd w dynamikę układu na podstawie wielkości współczynnika rozkładu energii $P$ oraz wskaźnika koherencji $Q = \cos\Delta$. Jednocześnie, prawa kolumna potwierdza dobrą zgodność pomiędzy wielkościami otrzymanymi na podstawie układu oryginalnego i uśrednionego. Jak wynika z wykresów, obecność pól magnetycznych zasadniczo wpływa na dynamikę drgań układu w porównaniu z początkowymi swobodnymi drganiami w antyfazie, gdzie energia początkowa jest prawie równo rozłożona pomiędzy wahadła ($P \approx 0$). W przypadku drgań swobodnych, całkowita energia stopniowo się rozprasza, podczas gdy jej bardzo mała część przemieszcza się z jednego wahadła do drugiego w sposób „rytmiczny". Natomiast stopniowo narastające sygnały prądowe łamią symetrię rozkładu energii w taki sposób, że energia jest prawie całkowicie przenoszona z wahadła będącego pod wpływem odpychającego momentu magnetycznego do wahadła, które drga w przyciągającym polu magnetycznym. Zjawisko to jest najbardziej widoczne na rys. 3.12b, który pokazuje, że początkowe równe rozłożenie energii między wahadłami ($P \approx 0$) nie jest utrzymywane wraz ze wzrostem wartości sygnałów prądowych. Gdy prądy w cewkach osiągną określony poziom, energia zaczyna „płynąć" do wahadła znajdującego się w przyciągającym polu magnetycznym wytwarzanym przez ujemny prąd. Dzieje się tak dla ok. 12 sekundy, kiedy to rozkład energii przyjmuje dolną wartość graniczną $P = (E_{11} - E_{22})/(E_{11} + E_{22}) = -1$, a wahadło oznaczone jako drugie absorbuje całą energię układu (rys. 3.12b,d). Należy zauważyć, że podczas całego procesu wymiany energii wskaźnik koherencji (rys. 3.12f) pozostawał ujemny, $Q = \cos\Delta < 0$. Świadczy to o tym, że taki

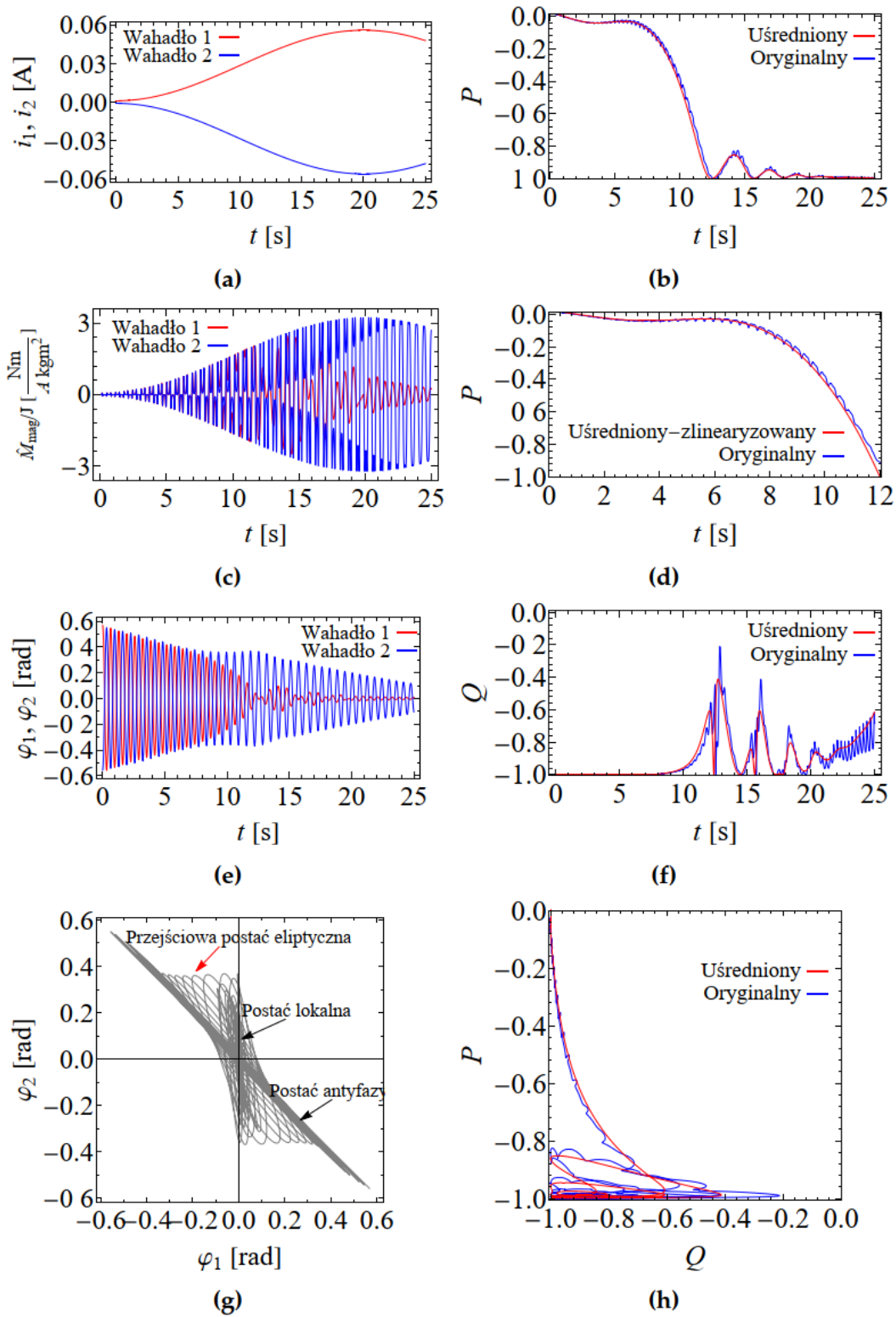

**Rys. 3.12.** Wyniki symulacji numerycznych wykonanych dla układów (3.6)-(3.7) oraz (3.17) z parametrami odpowiadającymi dużym magnesom i sygnałom prądowym (3.19). Po lewo: (a) sygnały prądowe cewek, (c) przebiegi czasowe momentów magnetycznych, (e) przemieszczenia kątowe wahadeł i (g) przejście z drgań w antyfazie do zlokalizowanej postaci drgań (ang. *localized mode*). Po prawo: (b) i (d) wykresy czasowe współczynnika rozkładu energii między wahadłami, (f) wykres czasowy wskaźnika koherencji i (h) zależność rozkładu energii od wskaźnika koherencji. Warunki dla jakich wykonano symulacje: $\varphi_1(0) = 0.568$ rad, $\varphi_2(0) = -0.557$ rad, $v_1(0) = 0$ rad/s, $v_2(0) = -5\,166 \cdot 10^{-3}$ rad/s jednokierunkowy przepływ energii wiąże się z przewagą drgań w przeciwfazie. Rys. 3.12g przedstawia zmiany postaci drgań, które występowały podczas omawianego procesu wymiany energii między wahadłami. Zmiany te przedstawione są na płaszczyźnie $\varphi_1\varphi_2$ i można je rozumieć jako przejście z drgań w antyfazie do drgań o zlokalizowanej nieliniowej postaci, gdzie drgania o postaci eliptycznej $Q \approx 0$ ($\Delta \approx \pi/2$) są drganiami przejściowymi. Uwagi te posłużą dalej jako podstawa do opracowania strategii sterowania przepływem energii. Z praktycznego punktu widzenia ważne jest, aby kontrolować proces do momentu, w którym energia układu po raz pierwszy całkowicie przemieści się do docelowego oscylatora zakładając, że zostanie przez niego wchłonięta lub po prostu szybko rozproszona. Takie podejście pozwala na dalsze uproszczenia układu (3.17) poprzez linearyzację w odniesieniu do zmiennej rozkładu energii $P$, tak jak to opisano w paragrafie 3.2.9 i potwierdzono na rys. 3.12d.

### 3.2.7 Analiza jakościowa trajektorii fazowych na płaszczyźnie $P\Delta$

Wyniki zawarte w paragrafie 3.2.6 ujawniły związek między współczynnikiem rozkładu energii $P$, a postacią drgań związaną z przesunięciem fazowym $\Delta$. Dlatego przeprowadzenie analizy dynamiki układu (3.17) poddanego zmieniającym się polom magnetycznym przy wykorzystaniu płaszczyzny $P\Delta$, może być pomocne do opracowania adekwatnej strategii sterowania przepływem energii. Rys. 3.13 przedstawia trajektorie układu na płaszczyźnie $P\Delta$ oraz punkty stacjonarne wyznaczone dla różnych kombinacji natężeń prądów cewek i przy założeniu, że całkowita energia wahadeł $E$ jest stała. Górny rząd (a, b, c) na rys. 3.13 pokazuje ewolucję trajektorii fazowych w wyniku wzrostu odpychającego pola magnetycznego pod jednym z wahadeł, przy jednoczesnym braku działania pola magnetycznego na drugie wahadło. Punkty stacjonarne $(\Delta, P)=(0,0)$ i $(\Delta, P)=(\pm\pi,0)$ są związane z drganiami układu w fazie i antyfazie (patrz równania (3.15)). Interesującym jest fakt, że czynniki rozpraszające energię mają efekt destabilizujący punkty stacjonarne; dla antyfazy są one zawsze niestabilnym ogniskiem, podczas gdy dla drgań w fazie zmieniają się one ze stabilnego ogniska na niestabilne wraz ze spadkiem całkowitej energii układu (dyskusja o tym znajduje się również w paragrafie 3.2.9). Środkowy rząd (d, e, f) daje wgląd w szczegóły zjawiska anihilacji antyfazy i postaci lokalnych pokazanych na rys. 3.13c. W wyniku zaniku wspomnianych postaci powstaje efekt tzw. „biegnącej fazy" $\Delta$. Dolny rząd (g, h, i) pokazuje sytuację, kiedy oba wahadła poddane są polom magnetycznym generowanym przez prądy o różnych kombinacjach znaków. Z analizy tych wykresów można wnioskować, że najefektywniejsze przejście układu do drgań o postaci lokalnej, tj. gdy $P \approx \pm 1$, możliwe jest wtedy, kiedy prądy cewek mają przeciwne znaki. Oczywistym jest też, że sygnały prądowe cewek w przypadku strategii zamkniętego sterowania opartej na pętli sprzężenia zwrotnego powinny zależeć od przesunięcia fazowego $\Delta$ (paragraf 3.2.10). Stopniowe rozpraszanie całkowitej energii układu nie wpływa jakościowo na trajektorie zaprezentowane na rys. 3.13 pod warunkiem, że wielkości sygnałów prądowych zmniejszają się w odpowiedniej proporcji. Należy jednak pamiętać, że wyniki z rys. 3.13 nie odzwierciedlają dokładnie całej złożoności układu (3.17), który jest jeszcze zależny od zmieniającej

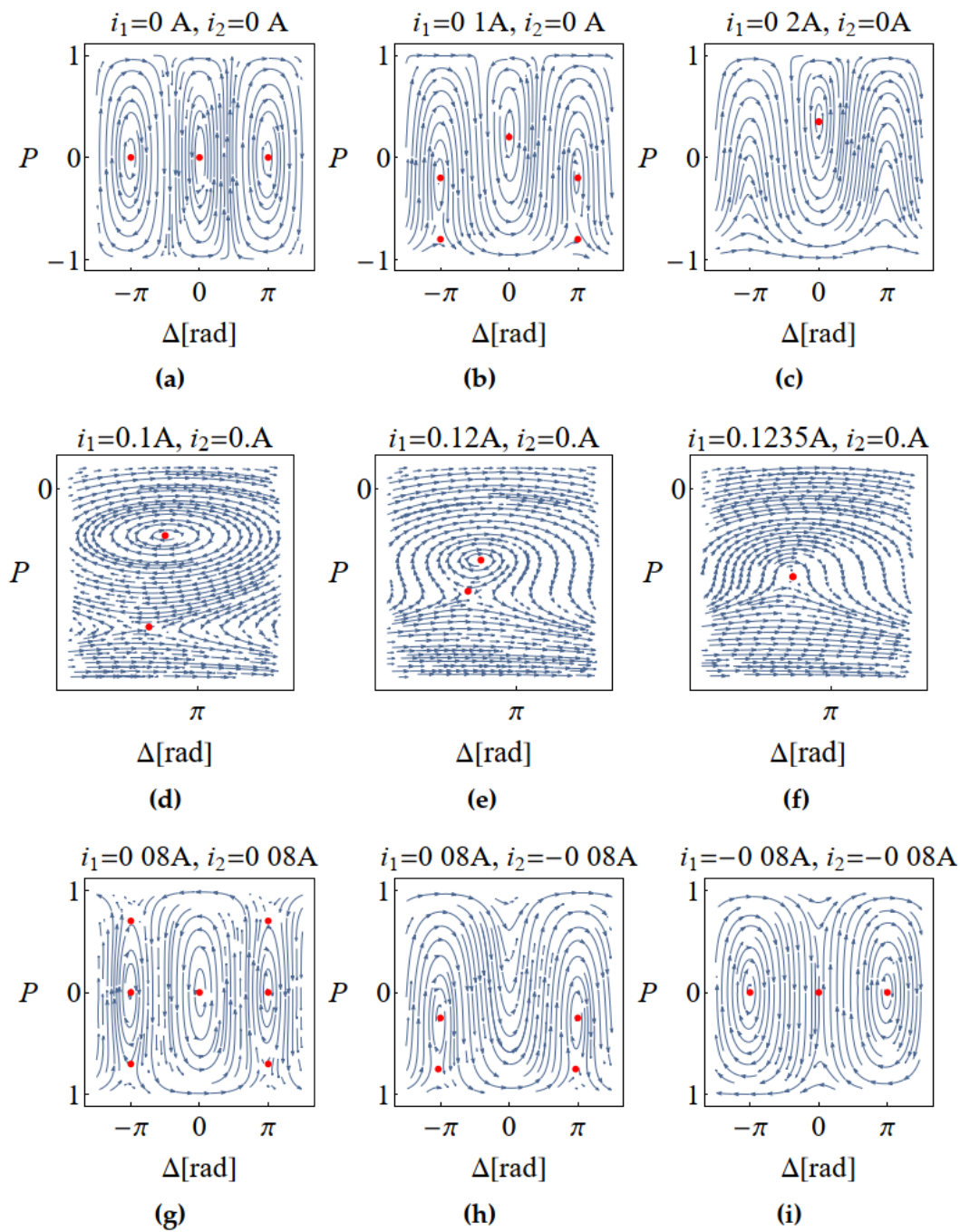

**Rys. 3.13.** Przekroje fazowe układu (3.17) obliczone dla stałej wartości energii całkowitej $E = 20$ J i stałych w czasie wartości prądów w cewkach. Górny rząd przedstawia punkty stacjonarne odpowiadające postaciom drgań w fazie i antyfazie układu swobodnego (a) przy niezasilanych cewkach, ponadto przedstawia (b) powstanie punktów siodłowych odpowiadających postaciom lokalnym drgań oraz (c) anihilację antyfazy i postaci lokalnych, która prowadzi do efektu „biegnącej fazy". Środkowy rząd (d, e, f) przedstawia szczegóły procesu anihilacji (c). Dolny rząd przedstawia (g) efekt odpychania obu wahadeł przez pola magnetyczne cewek, (h) odpychanie jednego z wahadeł i przyciąganie drugiego z nich przez pola magnetyczne oraz (i) przypadek przyciągania obu wahadeł przez pola magnetyczne.

się energii całkowitej $E$ oraz zmiennych w czasie prądów cewek, a przestrzeń fazowa układu (3.17) jest czterowymiarowa: $\{E, P, \Delta, t\}$. Niemniej jednak, wykresy z rys. 3.13 wyjaśniają związek między sygnałami prądowymi cewek a efektem ulokowywania energii układu na jednym z wahadeł. Związek pomiędzy kształtem trajektorii fazowych wykreślonych na płaszczyźnie $P\Delta$ a przebiegami wartości $P$ i $\Delta$ w czasie wyjaśniono na rys. 3.18.

### 3.2.8 Eksperymentalna walidacja modelu o strategii otwartego sterowania

Z rezultatów zawartych w paragrafach 3.2.4 3.2.7 wynika, że informacja o przesunięciu fazowym $\Delta$ uzyskana w czasie rzeczywistym jest niezbędna do opracowania strategii zamkniętego sterowania rozkładem energii $P$. Niemniej jednak, w paragrafie 3.2.6 przeprowadzono numeryczną weryfikację modelu, poddanego strategii otwartego sterowania przepływem energii z wahadła 1 do wahadła 2 (rys. 3.12b), gdzie sygnały prądowe miały z góry założone przebiegi opisane równaniami (3.19). Stało się tak, ponieważ zależność (3.19) zawierała informację o początkowej (antyfazowej) postaci drgań (rys. 3.12g). W przypadku, gdyby początkową postacią drgań były drgania w fazie, energia płynęłaby z wahadła 2 do wahadła 1. Sterowanie otwarte ma duży sens praktyczny ze względu na prostszą implementację w porównaniu ze sterowaniem zamkniętym, ale tylko wtedy, gdy warunki początkowe układu, a konkretnie postaci drgań są z góry znane.

**Strategia otwartego sterowania jednego wahadła**
Rysunki 3.14 i 3.15 przedstawiają symulacyjne i eksperymentalne wykresy czasowe drgań i rozkładów energii, wykonane dla układu wykorzystującego sterowanie otwarte. Wahadła wyposażone są w duże magnesy, ale tylko wahadło 1 poddane jest działaniu odpychającego pola magnetycznego (na skutek przepływu prądu $i_1$), podczas gdy cewka wahadła 2 jest niezasilana. Oba wahadła wprawiane były w ruch z różnych początkowych położeń kątowych i zerowych prędkości, prowadzących układ swobodny do drgań w antyfazie lub bliskich postaci antyfazy. Rys. 3.16 przedstawia wykresy czasowe przesunięcia fazowego $\Delta$ i współczynnika rozkładu energii $P$, odpowiadające przypadkowi drgań z rys. 3.15.

Podsumowując, przedstawione przebiegi czasowe wykazują wystarczającą na potrzeby prowadzonych badań zgodność jakościową i ilościową pomiędzy eksperymentami a wynikami symulacji.

**Strategia otwartego sterowania dwóch wahadeł**
W celu przeprowadzenia badań nad sterowaniem otwartym dla dwóch wahadeł magnetycznych, oba z nich wyposażono w małe magnesy. Następnie do cewek znajdujących się pod wahadłami, doprowadzono sygnały prądowe powodujące odpychanie wahadła 1 i przyciąganie wahadła 2. Podobnie jak wcześniej, oba wahadła wprawiane były w ruch z różnych początkowych położeń kątowych i zerowych prędkości, prowadzących układ swobodny do drgań w antyfazie lub bliskich antyfazie. Rys. 3.17 przedstawia symulacyjne i eksperymentalne przebiegi czasowe drgań i rozkładów energii w układzie, w przypadku zastosowania sterowania otwartego dla obu wahadeł magnetycznych. Rys. 3.18 przedstawia

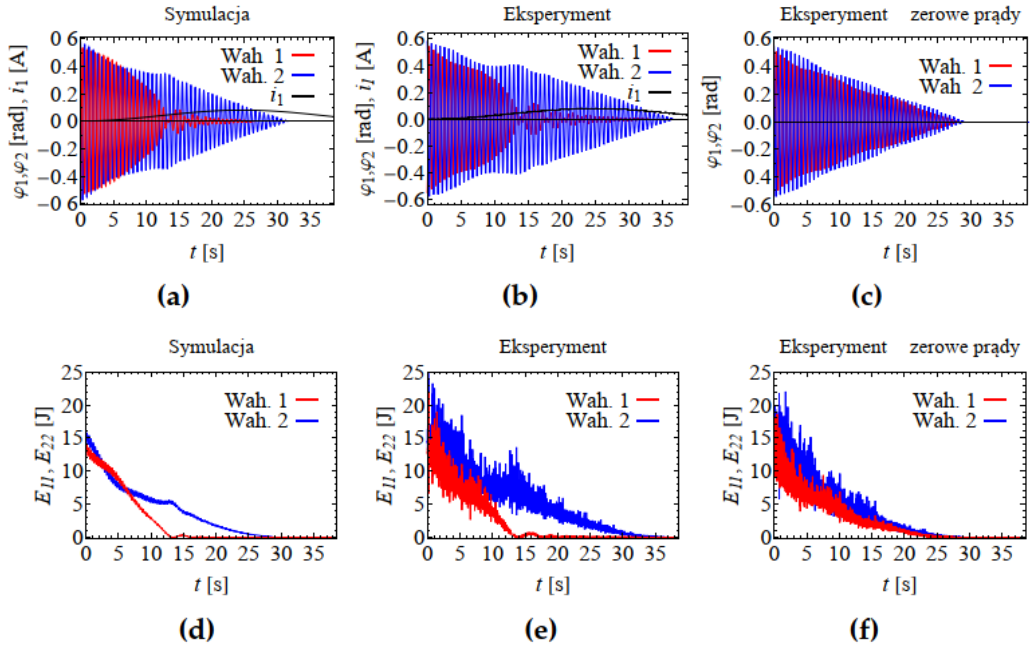

**Rys. 3.14.** Porównanie wyników symulacji numerycznych z eksperymentem. Wykresy czasowe położeń wahadeł (a)-(c) i odpowiadających im rozkładom energii (d)-(f) w czasie. W przypadku (a)-(b) oraz (d)-(e), sygnały prądowe cewek były następujące: $i_1(t) = 0.001 + 0.08\sin^2(\pi t/38.6)$ i $i_2(t) = 0$. Dla porównania przypadki (c) i (f) odpowiadają ruchowi swobodnemu układu, gdyby cewki nie były zasilane, $i_1 = i_2 = 0$. Warunki dla jakich wykonano badania: $\varphi_1(0) = 0.545$ rad, $\varphi_2(0) = 0.580$ rad, $v_1(0) = 0.001$ rad/s, $v_2(0) = 0.001$ rad/s.

zmiany rozkładów energii i przesunięcia fazowego między wahadłami w czasie. Dodatkowo na rys. 3.18c pokazano ewolucję trajektorii fazowych na płaszczyźnie $P\Delta$ w odniesieniu do zmian przesunięcia fazowego $\Delta$ w czasie. Porównanie wyników z rys. 3.18b i 3.18c pozwala na wyciągnięcie wniosków, że przejściu z przypadku równego rozkładu energii między wahadłami ($P \approx 0$) do przypadku jej lokalizacji na wahadle 2 ($P \approx 1$), towarzyszy przejście z drgań mających postać antyfazy do drgań z efektem „biegnącej fazy". Korelacja zaprezentowanych wyników eksperymentalnych i symulacyjnych dla tego rodzaju sterowania jest zadowalająca.

### 3.2.9 Dalsza redukcja układu i uwagi dotyczące zamkniętej strategii sterowania

W tym paragrafie opisane zostaną działania mające na celu dalsza redukcję uśrednionego układu, w wyniku czego ustalone zostaną bezpośrednie zależności pomiędzy sygnałami prądowymi a procesem wymiany energii między wahadłami. Następnie, wstępna analiza jakościowa tych zależności pozwoli na opracowanie strategii zamkniętego sterowania, tzn. ze sprzężeniem zwrotnym, a opisanej dokładniej w paragrafie 3.2.10.

Analizując prawe strony równań (3.17) można zauważyć, że sygnały prądowe wpływają na rozkład energii $P$ w sposób pośredni poprzez przesunięcie fazowe $\Delta$.

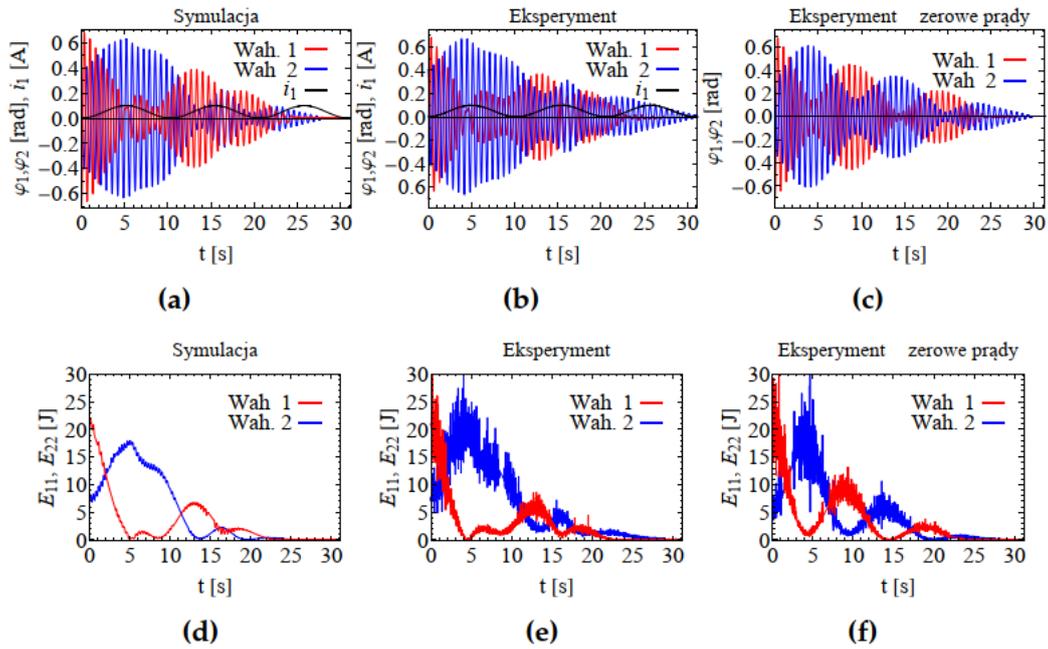

**Rys. 3.15.** Porównanie wyników symulacji numerycznych z eksperymentem. Wykresy czasowe położeń wahadeł (a)-(c) i odpowiadających im rozkładom energii (d)-(f) w czasie. W przypadku (a)-(b) oraz (d)-(e), sygnały prądowe cewek były następujące: $i_1(t) = 0.001 + 0.1\sin^2(\pi t/10.31)$ i $i_2(t) = 0$. Dla porównania przypadki (c) i (f) odpowiadają ruchowi swobodnemu układu gdyby cewki nie były zasilane, $i_1 = i_2 = 0$. Warunki dla jakich wykonano badania: $\varphi_1(0) = -0.697$ rad, $\varphi_2(0) = 0.403$ rad, $v_1(0) = 0.0001$ rad/s, $v_2(0) = 0.0001$ rad/s.

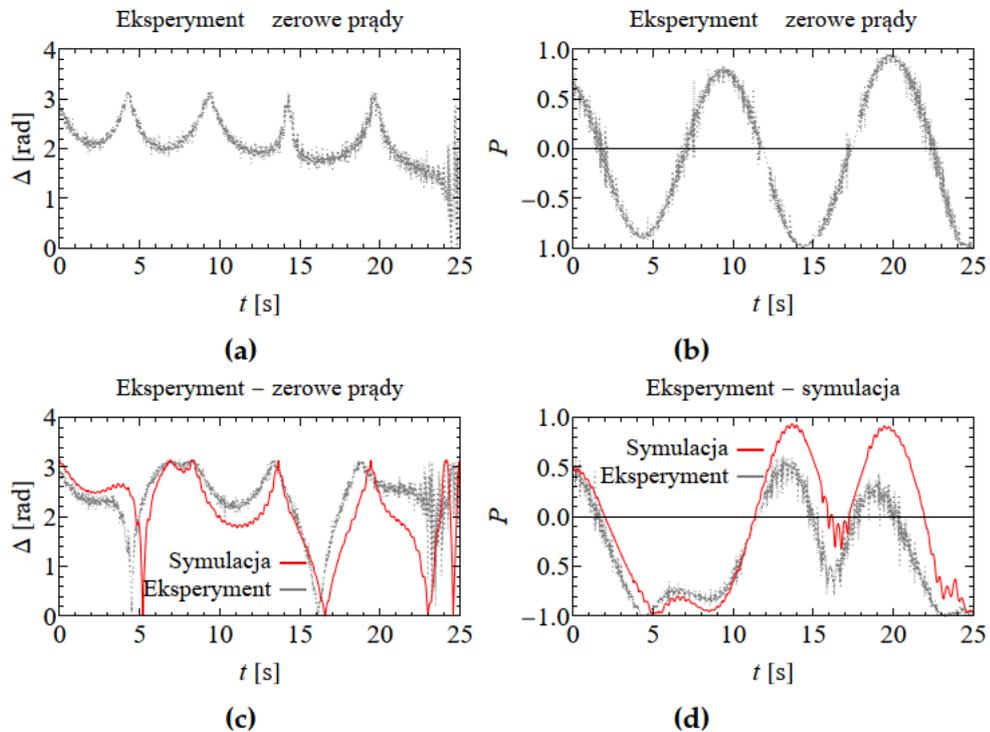

**Rys. 3.16.** Efekty wpływu sygnałów prądowych na rozkład energii $P$ i przesunięcia fazowego $\Delta$ otrzymane dla drgań układu przedstawionych na rys. 3.15. Wykresy czasowe (a)-(b) przedstawiają zachowanie się przesunięcia fazowego oraz współczynnika rozkładu energii, gdy cewki wahadeł są niezasilane ($i_1 = i_2 = 0$). Natomiast wykresy (c)-(d) przedstawiają wpływ pola magnetycznego, wywołanego niezerowym prądem $i_1$ płynącym przez cewkę wahadła 1. Symulacje numeryczne wykonano na podstawie równań oryginalnego układu (3.6)-(3.7) i wyrażeń (3.9)-(3.13).

Zależność tę można ujawnić eliminując przesunięcie fazowe z równań, co niestety jest trudne do wykonania w sposób analityczny ze względu na silną nieliniowość układu. Jednakże, eliminacja przesunięcia fazowego $\Delta$ z układu jest możliwa, ale dopiero po jego linearyzacji. Linearyzację należy przeprowadzić osobno dla układu znajdującego się w pobliżu warunków antyfazy, gdy $\Delta = \pi$ i warunków drgań w fazie, gdy $\Delta = 0$. Ponadto linearyzacja prowadzona jest w pobliżu $|P| \ll 1$.

W wyniku eliminacji $\Delta$ z równań otrzymamy dla każdej z analizowanych postaci drgań po jednym liniowym równaniu różniczkowym drugiego rzędu ze względu na funkcję $P(t)$. Warunek początkowy $dP/dt$ będzie opisany drugim równaniem z układu (3.17), po wcześniejszym obliczeniu wartości $P(0)$. Analiza otrzymanych równań doprowadzi do opracowania zależności analitycznych potrzebnych do procesu sterowania ze sprzężeniem zwrotnym.

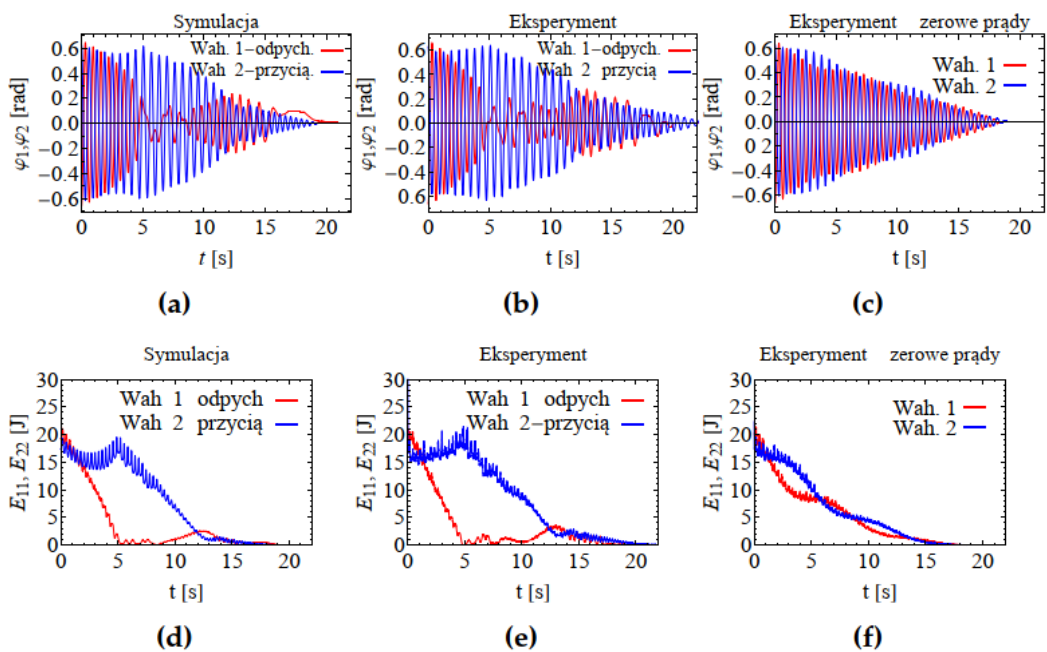

**Rys. 3.17.** Porównanie wyników symulacji numerycznych z eksperymentem. Wykresy czasowe położeń wahadeł (a)-(c) i odpowiadających im rozkładom energii (d)-(f) w czasie. W przypadku (a)-(b) oraz (d)-(e), sygnały prądowe cewek były następujące: $i_1(t) = 0.001 + 0.4\sin^2(\pi t/11.236)$ i $i_2(t) = i_1(t)$. Dla porównania przypadki (c) i (f) odpowiadają ruchowi swobodnemu układu gdyby cewki nie były zasilane, $i_1 = i_2 = 0$. Dane eksperymentalne z (e)-(f) zostały wygładzone przez zastosowanie filtra w postaci średniej ruchomej o liczbie uśrednianych punktów równej 60. Warunki dla jakich wykonano badania: $\varphi_1(0) = 0.660$ rad, $\varphi_2(0) = 0.627$ rad, $v_1(0) = 0.001$ rad/s, $v_2(0) = 0.001$ rad/s.

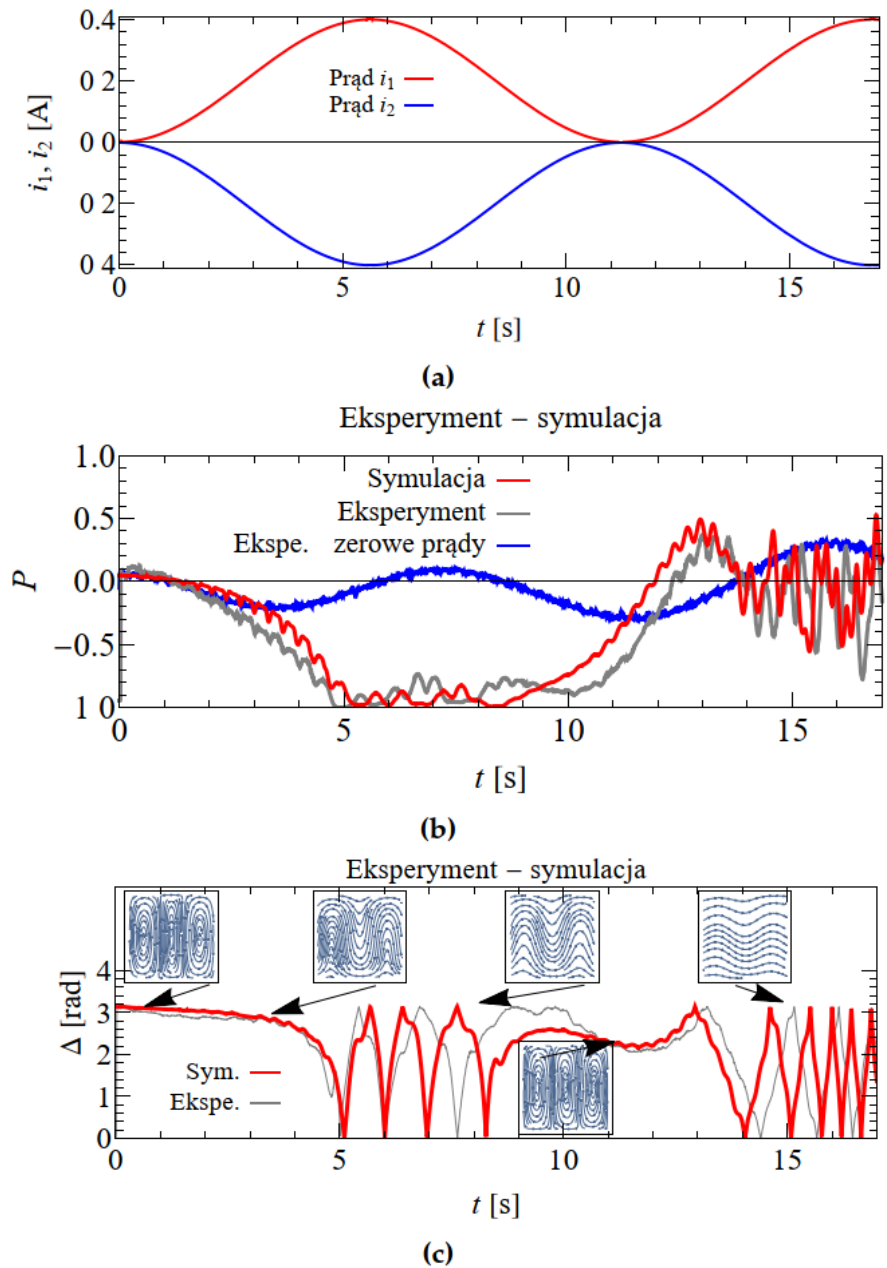

**Rys. 3.18.** Efekty wpływu sygnałów prądowych na współczynnik rozkładu energii $P$ i przesunięcia fazowego $\Delta$ otrzymane dla drgań układu z rys. 3.17. Wykresy czasowe sygnałów prądowych (a), współczynnika rozkładu energii (b) i przesunięcia fazowego (c) z wstawionymi wykresami trajektorii fazowych wykreślonych na płaszczyźnie $P\Delta$. Znaczenie trajektorii fazowych wyjaśnione jest na rys. 3.13. Na powyższym wykresie zmiany portretów fazowych mają pomóc w wyjaśnieniu przejścia między jakościowo różnymi przedziałami czasowymi przesunięcia fazowego $\Delta$ i rozkładu energii $P$. Symulacje numeryczne wykonano na podstawie równań oryginalnego układu (3.6)-(3.7) i wyrażeń (3.9)-(3.13).

**Linearyzacja dla przypadku warunków początkowych bliskich postaci drgań w antyfazie**

Zakładając, że zmniejszanie się całkowitej energii $E(t)$ w czasie procesu wymiany energii między wahadłami jest powolne, dokonamy linearyzacji drugiego i trzeciego równania układu (3.17) w odniesieniu do przesunięcia fazowego bliskiego $\Delta = \pi$ i gdy $P = 0$. Zatem zakładając, że $E$ jest stałe i eliminując przesunięcie fazowe, otrzymamy następujące równanie

$$\frac{d^2P}{dt^2} - 2\Omega\left(2\alpha + \frac{\zeta_1 + \zeta_2}{\sqrt{E}\pi}\right)\frac{dP}{dt} + \left\{\frac{1}{8}E\beta + \left[\beta^2 + 4\alpha\left(\alpha + \frac{\zeta_1 + \zeta_2}{\sqrt{E}\pi}\right)\right]\Omega^2 \right.$$
$$\left. + \frac{\alpha\beta}{Jb^2\Omega^2}e^{-\frac{E}{2b\Omega^2}}\left[\left(b\Omega^2 + E\right)I_{B1} - EI_{B0}\right](i_1 + i_2)\right\}P = 8\alpha\Omega^2\frac{\zeta_1 - \zeta_2}{\sqrt{E}\pi} \quad (3.20)$$
$$- \frac{\alpha\beta}{bJ}e^{-\frac{E}{2b\Omega^2}}(I_{B0} - I_{B1})(i_1 - i_2),$$
$$I_{Bn} = I_{Bn}\left(\frac{E}{2b\Omega^2}\right), \quad n = 1, 2.$$

Przekształcenia matematyczne prowadzące do otrzymania równania 3.20 zawarto w Załączniku D.

Ujemny współczynnik tłumienia w równaniu (3.20) potwierdza uwagę z paragrafu 3.2.7, dotyczącą lokalnej niestabilności niektórych punktów stacjonarnych widocznych na płaszczyznach $P\Delta$ z rys. 3.13. Fakt występowania różnicy sygnałów prądowych po prawej stronie równania (3.20) pokazuje, że prądy o przeciwnych znakach mają silniejszy wpływ na proces wymiany energii między wahadłami.

**Linearyzacja dla przypadku warunków początkowych bliskich postaci drgań w fazie**

Postępując podobnie jak w przypadku postaci antyfazy, dokonamy teraz linearyzacji względem przesunięcia fazowego bliskiego $\Delta = 0$, w wyniku czego otrzymamy następujące równanie

$$\frac{d^2P}{dt^2} + 2\Omega\left(2\alpha + \frac{\zeta_1 + \zeta_2}{\sqrt{E}\pi}\right)\frac{dP}{dt} + \left\{-\frac{1}{8}E\beta + \left[\beta^2 + 4\alpha\left(\alpha - \frac{\zeta_1 + \zeta_2}{\sqrt{E}\pi}\right)\right]\Omega^2 \right.$$
$$\left. - \frac{\alpha\beta}{Jb^2\Omega^2}e^{-\frac{E}{2b\Omega^2}}\left[\left(b\Omega^2 + E\right)I_{B1} - EI_{B0}\right](i_1 + i_2)\right\}P = \quad (3.21)$$
$$8\alpha\Omega^2\frac{\zeta_1 - \zeta_2}{\sqrt{E}\pi} + \frac{\alpha\beta}{bJ}e^{-\frac{E}{2b\Omega^2}}(I_{B0} - I_{B1})(i_1 - i_2),$$
$$I_{Bn} = I_{Bn}\left(\frac{E}{2b\Omega^2}\right), \quad n = 1, 2.$$

Można zauważyć, że w tym przypadku współczynnik tłumienia może być chwilowo dodatni, dopóki całkowita energia $E$ będzie wystarczająco duża, tj. będzie spełniała warunek $\sqrt{E} > (\zeta_1 + \zeta_2)/(2\alpha\pi)$.

Rys. 3.19 przedstawia porównanie wyników symulacji zlinearyzowanych równań (3.20)-(3.21), otrzymanych przy założeniu $|P| \ll 1$, z numerycznymi rozwiązaniami układu oryginalnego (3.6) (3.7). Na ich podstawie można wnioskować, że

zlinearyzowane równania z powodzeniem mogą być stosowane nawet dla szer szego przedziału wartości współczynnika rozkładu energii $P$.

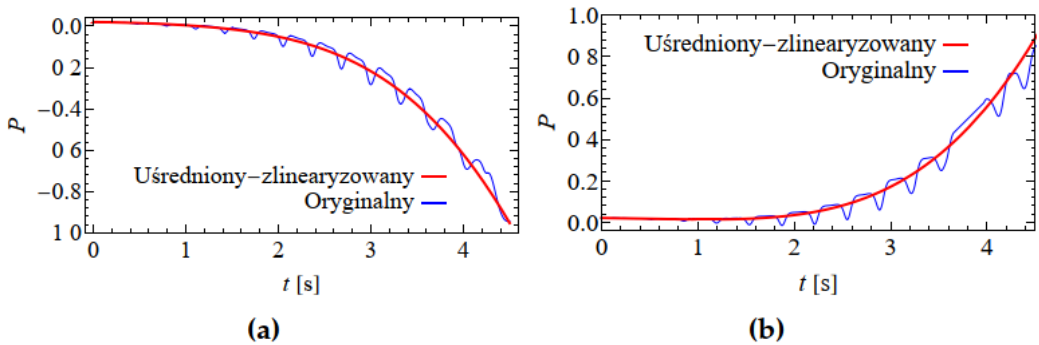

**Rys. 3.19.** Porównanie wartości współczynnika $P$ rozkładu energii obliczonego dla układu oryginalnego (3.6) (3.7) i zlinearyzowanego (3.20)-(3.21), przy warunkach początkowych odpowiadających drganiom w antyfazie (a) i w fazie (b). Symulacje przeprowadzono dla układu poddanego sygnałom prądowym o przebiegach jak na rys. 3.18a. Warunki dla jakich wykonano badania: $E(0) = 30$ J, $P(0) = 0.02$; antyfaza $\Delta(0) = \pi$ 0.001 rad, w fazie $\Delta(0) = $ 0.001 rad.

### 3.2.10 Strategia zamkniętego sterowania

Analiza prawych stron równań układu uśrednionego (3.17) oraz jego zlinearyzowanych wersji (3.20)-(3.21) w odniesieniu do współczynnika rozkładu energii prowadzi do wniosku, że efektywna kontrola jednokierunkowego przepływu energii pomiędzy wahadłami może być zapewniona, kiedy cewki zasilane są sygnałami prądowymi o przeciwnych znakach i będącymi funkcją przesunięcia fazowego. Dla przypadku kierunkowej kontroli przepływu energii z wahadła 1 do wahadła 2, sygnały prądowe przyjmują następująca postać

$$i_1(t) = \ i_A Q(t) = \ i_A \cos\Delta(t), \qquad i_2(t) = i_A Q(t) = i_A \cos\Delta(t), \qquad (3.22)$$

gdzie $i_A = const$.

Trzecie równanie układu (3.13) pokazuje, że $\cos\Delta$ może zostać wyznaczony przez elementy macierzy energii (3.9), których wartości obliczane są na podstawie sygnałów ($\varphi_{i,j}$, $\dot\varphi_{i,j}$) pochodzących ze sprzężenia zwrotnego

$$\cos\Delta(t) = Q(t) = \frac{E_{12}(t)}{\sqrt{E_{11}(t)E_{22}(t)}}. \qquad (3.23)$$

Rysunek 3.20 przedstawia wyniki symulacji układu (3.6) (3.7) z zaimplementowanymi sygnałami prądowymi opisanymi równaniami (3.22)-(3.23). Przyjęto, że wahadła wyposażone są w małe magnesy i wprawiane są w ruch z różnych położeń początkowych, odpowiadających postaciom drgań: w antyfazie, w fazie, eliptyczna. Położenia początkowe są tak dobrane, żeby początkowa całkowita energia układu $E$ wynosiła 30 J. Dodatkowo w symulacjach przyjęto amplitudę sygnałów prądowych $i_A = 0.35$ A oprócz przypadku z rys. 3.20d, gdzie przyjęto $i_A = 0.5$ A. Analizując lewą kolumnę rys. 3.20 widać, że przy różnych początkowych współczynnikach rozkładu energii $P$ i przesunięciach fazowych $\Delta$ między wahadłami, energia przenoszona jest z pierwszego wahadła do drugiego, $P \approx 1$. Natomiast prawa kolumna rys. 3.20 potwierdza, że całkowita energia układu pozostaje w drugim wahadle, ilekroć jest w nim początkowo skumulowana. Występowanie pewnych wahań wartości $P$ w przypadku drgań o postaci eliptycznej (d), może być zredukowane przez zwiększenie amplitudy $i_A$ sygnałów prądowych. Wszystkie analizowane na rys.3.20 przypadki drgań, charakteryzuje dobra zgodność wyników symulacji opartych na oryginalnym układzie (3.6)-(3.7) z wynikami uzyskanymi dla jego uśrednionej wersji (3.17).

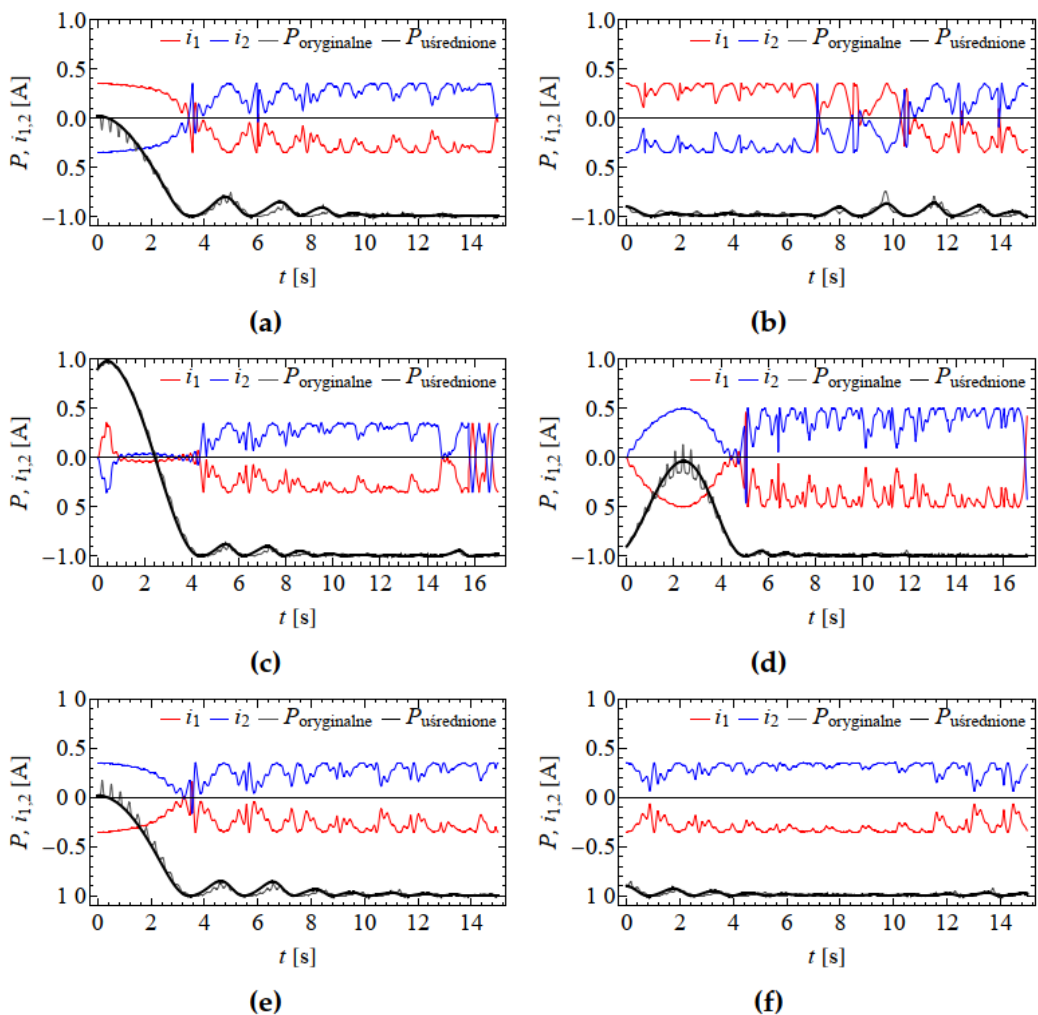

**Rys. 3.20.** Wyniki symulacji numerycznych wykonanych dla układu (3.6)-(3.7) ze sprzężeniem zwrotnym, tj. z zaimplementowanymi sygnałami prądowymi opisanymi równaniami (3.22) (3.23). Wyniki opracowano dla warunków początkowych odpowiadających drganiom o postaci: antyfazy (a b), eliptycznej (c-d), w fazie (e-f). Dokładne warunki dla jakich wykonano badania: $E(0) = 30$ J, $P(0) = 0.02$ lub $P(0) = 0.9$; antyfaza $\Delta(0) = \pi$ 0.001 rad, postać eliptyczna $\Delta(0) = \frac{\pi}{2}$ 0.001 rad, w fazie $\Delta(0) = -0.001$ rad.

Przypomnijmy, że współczynnik koherencji (3.23) charakteryzuje przesunięcie fazowe $\Delta$ między dwoma wahadłami jako $Q = \cos \Delta$, a zatem jest niezależny od całkowitej energii układu $E$. W rezultacie „siła" sprzężenia zwrotnego zależna od prądów (3.22) utrzymywana jest pomimo stopniowego rozpraszania się energii układu. Jak wynika z symulacji numerycznych, fakt ten nie ma niepożądanych skutków dla końcowej fazy dynamiki układu, co potwierdzają wyniki przedstawione na rys. 3.20. W rzeczywistości jednak, różne niedoskonałości strukturalne i nieuwzględnione czynniki fizyczne sprawiają, że układ wahadeł jest coraz bardziej wrażliwy na zewnętrzne zakłócenia, gdy całkowita energia spada do poziomu odpowiadającego równowadze wahadeł. Dlatego na potrzeby eksperymentu, w celu uniknięcia związanego z tym faktem zjawiska wzbudzania się wahadeł, do równań (3.22) zostanie wprowadzony energetyczny składnik tłumiący

$$
\begin{aligned}
i_1(t) &= i_A \left\{1 - \exp\left[-\eta E(t)\right]\right\} \cos \Delta(t), \\
i_2(t) &= i_A \left\{1 - \exp\left[-\eta E(t)\right]\right\} \cos \Delta(t),
\end{aligned}
\qquad (3.24)
$$

gdzie $E(t)$ jest całkowitą energią wahadeł zdefiniowaną przez pierwsze równanie (3.13), a $\eta$ jest numerycznym parametrem określanym w sposób empiryczny. Wyniki eksperymentów przeprowadzonych dla układu z zaimplementowanymi sygnałami prądowymi (3.24) otrzymano dla parametrów zbliżonych do tych przyjętych podczas symulacji numerycznych (rys. 3.20) i przedstawiono je na rys. 3.21. Potwierdzają one efekty przewidziane przez symulacje numeryczne dla trzech różnych przypadków warunków początkowych. W szczególności, wyniki z rys. 3.21a,c,e przedstawiają wyraźną i nieodwracalną tendencję dążenia współczynnika rozkładu energii do wartości $P \approx -1$, co świadczy o jednokierunkowym przepływie energii z pierwszego wahadła do drugiego. Należy wziąć pod uwagę, że pojawiający się w końcowych fazach drgań układu wzrost wartości $P$, spowodowany jest zanikaniem prądów sterujących. Trajektorie fazowe przedstawione na rys. 3.21b,d,f potwierdzają dążenie układu do osiągnięcia lokalnej postaci drgań $\varphi_1 \approx 0$ rad.

Podsumowując, zaprezentowane w tym paragrafie strategie sterowania oparte są na następujących obserwacjach:

- W przypadku drgań wahadeł w antyfazie, energia przemieszcza się z wahadła znajdującego się w odpychającym polu magnetycznym do wahadła poddanego działaniu przyciągającego pola magnetycznego.

- W przypadku drgań wahadeł w fazie, energia przemieszcza się z wahadła znajdującego się w przyciągającym polu magnetycznym do wahadła poddanego działaniu odpychającego pola magnetycznego.

Strategia otwartego sterowania wymaga znajomości warunków początkowych (postaci początkowej drgań) oraz wartość czasu ruchu wahadeł jak upłynie, aż do chwili jego ustania na skutek dyssypacji energii. Strategia sterowania zamkniętego natomiast wymaga tylko informacji o przesunięciu fazowym między wahadłami, która to jest pozyskiwana z pomiarów położeń kątowych wahadeł. Z tego względu metoda ta dzięki pętli sprzężenia zwrotnego, może być stosowana w przypadkach, gdy ruch układu wynika nie tylko z niezerowych warunków początkowych, ale również z zakłóceń zewnętrznych takich jak np. siły czy momenty sił.

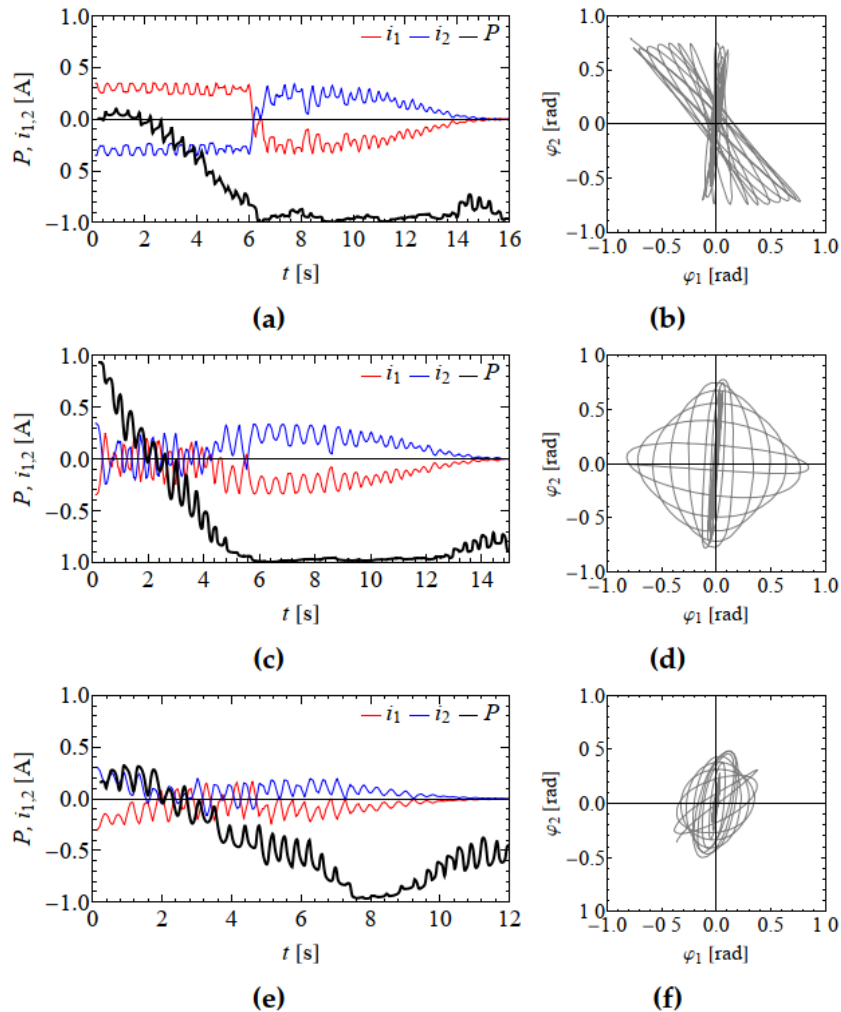

**Rys. 3.21.** Wyniki eksperymentalne wykonane dla układu ze sprzężeniem zwrotnym, tj. z zaimplementowanymi sygnałami prądowymi opisanymi równaniami (3.23)-(3.24). Wyniki opracowano dla warunków początkowych odpowiadających drganiom o postaci: antyfazy (a-b), eliptycznej (c-d), w fazie (e-f). Dokładne warunki dla jakich wykonano badania: Antyfaza $\varphi_1(0) = \frac{\pi}{4}$ rad, $\varphi_2(0) = \frac{\pi}{4}$ rad, $\eta = 0.3$; postać eliptyczna $\varphi_1(0) = \frac{5\pi}{18}$ rad, $\varphi_2(0) = 0$ rad, $\eta = 0.2$; w fazie $\varphi_1(0) = \varphi_2(0) = \frac{\pi}{9}$ rad, $\eta = 0.2$. We wszystkich przypadkach $v_1(0) = 0.001$ rad/s, $v_2(0) = 0.001$ rad/s.

Zaletą proponowanych strategii sterowania jest to, że szybkość zmian oddziaływań magnetycznych jest dyktowana szybkością dudnień, które względem samych drgań wahadeł są znacznie wolniejsze. Pomimo tego, że zaprezentowane badania dotyczyły określonego modelu fizycznego sprzężonych wahadeł, opracowana metodologia transferu energii nie jest ograniczona przez specyfikę tego układu. Wahadła mogą mieć różne parametry geometryczne, a jedyne wymogi jakie muszą spełniać to zachowanie częstości drgań własnych w proporcji bliskiej 1:1 oraz mały wpływ efektów rozpraszania energii, sprzężenia i pól magnetycznych względem efektu sił przywracających układ do położenia równowagi. Zaprezen-towane w podrozdziale wyniki badań zostały opublikowane w pracy [157].

# Rozdział 4

# Podsumowanie i wnioski

W pracy przedstawiono modelowanie matematyczne oraz badania symulacyjne i eksperymentalne dynamiki dwóch układów mechatronicznych, których elementem bazowym było wahadło magnetyczne, tj. wahadło z zamocowanym na końcu magnesem. Pierwszy układ składał się z pojedynczego wahadła magnetycznego umieszczonego w niestacjonarnym polu magnetycznym, generowanym przez cewkę elektryczną znajdują się pod nim. Drugi układ był rozszerzeniem pierwszego i składał się z dwóch takich samych wahadeł magnetycznych, których wały sprzężone były ze sobą elementem podatnym. Ze względu na dwa układy o różnych stopniach swobody, pracę podzielono na dwie części i dla każdej z nich postawiono inną tezę badawczą.

    Proces modelowania matematycznego dotyczył głównie opracowania modeli oporów ruchu i oddziaływania magnetycznego pomiędzy magnesem a cewką. Zarówno w układzie o jednym jak i o dwóch stopniach swobody stosowano te same modele oporów ruchu oraz oddziaływania magnetycznego. W oparciu o eksperyment wykazano, że uzasadnionym będzie uwzględnienie w równaniach ruchu występowania oporów statycznych i/lub coulombowskich. Model oddziaływania magnetycznego pomiędzy cewką a magnesem, sprowadzony został do momentu sił magnetycznych działającego na wahadło. Ponieważ literaturowe modele tego oddziaływania dawały wyniki dalekie od przeprowadzonego eksperymentu, opracowano dwa rodzaje nowych empirycznych modeli momentu magnetycznego. Pierwszy z nich bazował na przekształconej funkcji Gaussa, natomiast drugi oparty były na funkcjach wielomianowych. Ze względu na małe różnice między przebiegami tych modeli, oba rodzaje wykorzystywane były podczas badań symulacyjnych.

    Za ważne cechy charakterystyczne momentu magnetycznego, które miały wpływ na dynamikę badanych układów, uznano jego silną i nieliniową zależność od położenia magnesu wahadła względem cewki oraz zależność od sygnału prądowego. Im dalej magnes znajdował się od cewki, tym wpływ jej pola magnetycznego na wahadło był mniejszy. Dzięki tej cesze przyjęto, że istnieje pewna strefa wychyleń kątowych, wewnątrz której wahadło ulega działaniu pola magnetycznego cewki, a po opuszczeniu której zmiany tego pola będą miały znikomy wpływ na jego dynamikę. Strefę tę nazwano strefą aktywną, a jej granice zależały od wartości sygnału prądowego płynącego w cewce. Kierunek sygnału prądowe-

go decydował zaś o przyciąganiu bądź odpychaniu magnesu przez cewkę. Dzięki tym dwóm cechom, moment magnetyczny w równaniach ruchu wchodził w skład czynnika odpowiadającego za sztywność układu, która okazała się być nieliniową, a ze względu na przebieg sygnału prądowego mogła się zmieniać w czasie. Wnioskować więc można, że układy wahadeł magnetycznych należą do specyficznej rodziny oscylatorów parametrycznych.

Badania dynamiki pojedynczego wahadła magnetycznego prowadzono dla przypadku, gdy cewka zasilana była prostokątnym pulsującym sygnałem prądowym o stałej amplitudzie i regulowanej częstotliwości oraz wypełnieniu. Polaryzacja prądu cewki powodowała odpychanie od niej magnesu, przez co potencjał układu miał charakter dwudołkowy. Analizę dynamiczną przeprowadzono przy zastosowaniu metod numerycznych i opracowanej metody analityczno-numerycznej opartej na metodzie uśredniania. Metoda analityczno-numeryczna wykorzystywała funkcje piłokształtną do aproksymacji nieliniowych przebiegów zmian amplitudy i przesunięcia fazowego w czasie dla przypadku drgań okresowych wahadła. W odniesieniu do tradycyjnej metody uśrednienia, wprowadzenie funkcji piłokształnych pozwoliło na jeszcze dokładniejsze opisanie dynamiki wspomnianej amplitudy i przesunięcia fazowego dla przypadku układów z pulsującym wymuszeniem, kiedy to w rozwiązaniu okresowym pojawiają się „podskoki" amplitudy (rys. 2.23). W ramach analizy dynamiki układu wykonano szeregi wykresów czasowych, amplitudowo-czestotliwościowych, bifurkacyjnych, fazowych oraz przekrojów Poincarégo. Pojedyncze wahadło magnetyczne wykazywało liczne drgań okresowe o różnych okresowościach, ale także typowe dla dynamiki nieliniowej zjawiska takie jak multistabilność, zachowania chaotyczne czy gwałtowne skoki amplitud.

Dokładnym badaniom poddany został specyficzny rodzaj jednookresowych drgań wahadła magnetycznego odbywających się w jednym dołku potencjału. Okresowość drgań wahadła rozumiana jest jako liczba okresów wymuszenia (sygnału prądowego) przypadająca na jeden okres drgań wahadła. Założenie istnienia wspomnianej strefy aktywnej, pozwoliło zastąpić ciągły nieautonomiczny model matematyczny układu poprzez model dyskretny w postaci układu dwóch przełączających się między sobą autonomicznych równań różniczkowych. Przedstawiono trzy różne scenariusze analizowanego ruchu jednookresowego oraz określono warunki ich istnienia w postaci warunków początkowych (położenie i prędkość kątowa wahadła) dla równań różniczkowych. Następnie przy pomocy tych warunków, dla pierwszego ze scenariuszy wyznaczono parametry sygnału prądowego takie jak wypełnienie i częstotliwość, dla których ruch ten występuje, a także zbadano jego dwu i trój okresowe modyfikacje. Analiza numeryczna oraz teoretyczna wyjaśniła, dlaczego znaczna zmiana parametru wypełnienia i częstotliwości sygnału prądowego nie musi wpływać na zmianę przebiegu trajektorii fazowej badanego ruchu. Na jej podstawie wywnioskowano, że ze względu na szczególny charakter oddziaływania magnetycznego układu i istnienie tzw. strefy aktywnej, w przypadku drgań okresowych wahadła magnetycznego odbywających się w jednym dołku potencjału, możliwe jest „sztuczne" zmienianie okresowości tych drgań bez wyraźnego naruszenia przebiegu ich trajektorii fazowej. Wniosek ten udowadnia pierwszą z postawionych tez badawczych.

Badania dynamiki układu o dwóch stopniach swobody prowadzone dla przy-padku, gdy tylko jedno z wahadeł poddane było niestacjonarnemu polu magnetycznemu cewki. Układ ten charakteryzował się bogatszą dynamiką nieliniową niż układ o jednym stopniu swobody. Przede wszystkim oprócz zachowań chaotycznych i okresowych pojawiły się w nim nowe zjawiska, takie jak quasiokresowość, która została potwierdzona zarówno na wykresach fazowych jak i przy pomocy wykładników Lapunowa.

Dalsze badania nad układem dwóch sprzężonych wahadeł magnetycznych skupiły się na możliwości sterowania przepływem energii między nimi, na skutek odpowiedniego generowania pól magnetycznych ich cewek. Oprócz typowego modelu matematycznego układu zawierającego klasyczne zmienne stanu, opracowano model zawierający zmienne odnoszące się do energii i charakteryzujące proces jej transferu. Zaproponowano dwie metody sterowania przepływem energii: otwartą bez sprzężenia zwrotnego i zamkniętą ze sprzężeniem zwrotnym. Obie z nich oparto na obserwacjach, że w przypadku drgań w antyfazie energia przepływa z wahadła poddanego odpychającemu polu magnetycznemu do wahadła znajdującego się w przyciągającym polu. Natomiast podczas drgań w fazie przepływ energii jest odwrotny. Na podstawie obserwacji oraz analizy numerycznej dynamiki układu wykazano, że podczas sterowania bez sprzężenia zwrotnego konieczne jest znanie a priori typu drgań układu (w fazie, w antyfazie) oraz czasu po jakim energia układu zostanie całkowicie rozproszona. Natomiast w przypadku sterowania ze sprzężeniem zwrotnym konieczne jest tylko określenie przesunięcia fazowego między wahadłami na podstawie pomiaru ich położeń. Opierając się na przeprowadzonych badaniach numerycznych oraz eksperymentalnych można wnioskować, że przy odpowiednim sterowaniu pól magnetycznych cewek, a z mechanicznego punktu wiedzenia nieliniową sztywnością poszczególnych wahadeł, możliwe jest zapewnienie kierunkowego transferu energii między sprzężonymi oscylatorami. Wniosek ten udowadnia drugą z postawionych w pracy tez badawczych.

Według autora, innowacyjnymi elementami pracy, dzięki którym jej cel naukowy został osiągnięty są:

1. Opracowanie modeli dynamicznych układów o jednym i dwóch stopniach swobody składających się z wahadeł magnetycznych oraz empirycznych modeli matematycznych oddziaływania magnetycznego pary cewka-magnes.

2. Opracowanie analityczno-numerycznej metody opartej na metodzie uśredniania, wykorzystującej funkcje piłokształtne do aproksymacji nieliniowych przebiegów amplitudy i przesunięcia fazowego rozwiązania okresowego dla układu wahadła magnetycznego.

3. Wykazanie, że podczas drgań wahadła magnetycznego w jednym dołku potencjału, którego cewka zasilana jest prostokątnym sygnałem prądowym, możliwe jest zmienianie okresowości drgań bez zmieniania przebiegu ich trajektorii fazowej.

4. Opracowanie modelu matematycznego dla układu dwóch słabo sprzężonych wahadeł magnetycznych, wykorzystującego zmienne związane z energią wahadeł oraz jej podziale między nimi.

5. Opracowanie dwóch metod sterowania przepływem energii między dwoma słabo sprzężonymi wahadłami magnetycznymi, wykorzystujących odpowiednią zmianę pól magnetycznych ich cewek. Pierwsza metoda oparta jest na otwartej pętli sterowania bez sprzężenia zwrotnego, natomiast druga na zamkniętej pętli ze sprzężeniem zwrotnym.

W perspektywie dalszych badań planowane jest rozszerzenie analizy dynamiki pojedynczego wahadła magnetycznego drgającego w jednym dołku potencjału o przypadek drgań pomiędzy dwoma dołkami potencjału. Zaprezentowane w rozprawie badania dotyczące pojedynczego wahadła magnetycznego, mogą stanowić podstawy do nowego sposobu modelowania silników krokowych.

Mechanizmy propagacji, pułapkowania (ang. *trapping*) i rozpraszania energii stanowią jedne z podstawowych problemów fizycznych zarówno na poziomie mikro jak i makro świata. Dlatego opracowane w rozprawie metody sposobu sterowania przepływem energii w układach połączonych wahadeł magnetycznych, mogą stanowić bazę do przyszłych badań, w zakresie problemów występujących podczas projektowania struktur molekularnych o pożądanych właściwościach ukierunkowanego transferu energii (ang. *targeted energy transfer — TET*) [149, 150], urządzeń odzyskujących energię z drgań mechanicznych [151, 152] lub mechanicznych pochłaniaczy energii wpływających na dynamikę konstrukcji [158].

# Załączniki

## A  Metoda wariacji stałych dowolnych

Analizując drgania układu opisane równaniem różniczkowym rzędu drugiego o postaci

$$\varphi'' + \varphi = G(\varphi, \varphi'), \ (*)' = \frac{d(*)}{d\tau_t} \tag{A.1}$$

wiadomym jest, że przy dostatecznie małej wartości nieliniowej funkcji $G(\varphi, \varphi')$ opisuje ono drgania quasi-liniowe. Przy $G(\varphi, \varphi') = 0$ rozwiązanie takiego równania jest następujące

$$\varphi = k\sin(\theta), \ \theta = \Omega_t \tau_t + u, \tag{A.2}$$

a jego pierwsza pochodna pochodna wynosi

$$\varphi' = k\Omega_t \cos(\theta). \tag{A.3}$$

Przyjmując, że przy małym $G(\varphi, \varphi')$ drgania układu opisane są równaniem (A.2), w którym można się spodziewać powolnych zmian amplitudy $k$ oraz przesunięcia fazowego $u$ w czasie. Biorąc to pod uwagę, drgania takiego układu opisać można poprzez chwilowe wartości amplitudy $k(\tau_t)$ i fazy $u(\tau_t)$.

Przyjmując zmienne $k(\tau_t)$ i $u(\tau_t)$ oraz biorąc pod uwagę zasady różniczkowania funkcji złożonej, pochodna (A.2) wyrażona jest wzorem

$$\varphi' = k'\sin(u + \tau_t\Omega_t) + k(u' + \Omega_t)\cos(u + \tau_t\Omega_t) \tag{A.4}$$

Przyrównując do siebie równania (A.3) i (A.4) otrzymamy zależność

$$k'\sin(u + t\Omega_t) + k(u' + \Omega_t)\cos(u + t\Omega_t) - k\Omega_t \cos(u + t\Omega_t) = 0. \tag{A.5}$$

Ponownie różniczkując równanie (A.3) przy takich samych założeniach, tzn. $k$ i $u$ są funkcjami czasu otrzymamy

$$\varphi'' = \Omega_t k'\cos(u + t\Omega_t) - \Omega_t k(u' + \Omega_t)\sin(u + t\Omega_t). \tag{A.6}$$

Wstawiając równanie (A.6) i (A.2) do równania (A.1) otrzymamy

$$\Omega_t k'\cos(u + t\Omega_t) - \Omega_t k(u' + \Omega_t)\sin(u + t\Omega_t) + k\sin(u + t\Omega_t) = G(\varphi, \varphi') \tag{A.7}$$

Rozwiązując układ równań liniowych (A.5) i (A.7) dla niewiadomych $k'$ i $u'$, otrzymamy

$$\begin{aligned} k' &= \frac{\cos(u + t\Omega_t)\left(G + \left(\Omega_t^2 - 1\right)k\sin(u + t\Omega_t)\right)}{\Omega_t}, \\ u' &= -\frac{\sin(u + t\Omega_t)\left(G + \left(\Omega_t^2 - 1\right)k\sin(u + t\Omega_t)\right)}{\Omega_t k}, \end{aligned} \tag{A.8}$$

który po przekształceniach da nam układ równań (2.32).

# B  Energia potencjalna i częstości drgań własnych zlinearyzowanego układu zachowawczego dwóch sprzężonych wahadeł

Równanie (3.4) opisuje energię potencjalną odniesioną do momentu bezwładności wahadeł i obliczoną dla zlinearyzowanego, i zachowawczego układu. Całkowita energia potencjalna układu dwóch słabo sprzężonych wahadeł jest sumą energii potencjalnych pochodzących od pola grawitacyjnego, pola magnetycznego i sprężyny. Wzięte pod uwagę zostaną tylko te człony energii potencjalnych pola grawitacyjnego i magnetycznego, których stopień wielomianu jest nie większy niż dwa. Poszczególne energie potencjalne można więc wyrazić następującymi formułami:

- Energia potencjalna pochodząca od pola grawitacyjnego działającego na wahadło wynosi

$$V_{graw} = mgs \int \left( \varphi - \frac{\varphi^3}{6} \right) d\varphi = mgs \left( \frac{\varphi^2}{2} - \frac{\varphi^4}{24} \right) \approx \frac{1}{2} mgs \varphi^2 \qquad \text{(B.1)}$$

- Energia potencjalna gromadzona w sprężynie zależy od względnego kąta skręcenia, który jest różnicą kątów wychyleń wahadeł. Wyraża się ją wzorem

$$V_{spr} = \frac{1}{2} k_{et} (\varphi_1 - \varphi_2)^2. \qquad \text{(B.2)}$$

- Energia potencjalna pola magnetycznego opisana jest wzorem

$$V_{pM} = a_I \left[ 1 - e^{-\frac{\varphi^2}{b}} \right] i, \qquad \text{(B.3)}$$

gdzie *i* to prąd cewki. Rozwijając w szereg Taylora wyrażenie eksponencjalne zawarte we wzorze (B.3) otrzymamy

$$e^{-\frac{\varphi^2}{b}} \approx 1 - \frac{\varphi^2}{b} + \frac{\varphi^4}{2b^2}, \qquad \text{(B.4)}$$

a biorąc pod uwagę tylko składniki o stopniu nie większym niż dwa, przybliżoną wartość energii potencjalnej pola magnetycznego możemy zapisać jako

$$V_{pM} \approx a_I \left[ 1 - \left( 1 - \frac{\varphi^2}{b} \right) \right] i = \frac{a_I}{b} i \varphi^2. \qquad \text{(B.5)}$$

Odnosząc przedstawione powyżej energie potencjalne do wartości masowego momentu bezwładności wahadeł, otrzymamy wyrażenie na całkowitą energię potencjalną zlinearyzowanego układu zachowawczego wahadeł (3.4).

Obliczenia kwadratów częstości $\omega_{1,2}$ drgań własnych układu, oparto na następującym układzie równań

$$\ddot{\varphi}_1 = -\Omega^2\varphi_1 - \Omega^2\beta(\varphi_1 - \varphi_2) + \frac{2a_I}{bJ}i_1\varphi_1,$$
$$\ddot{\varphi}_2 = -\Omega^2\varphi_2 - \Omega^2\beta(\varphi_2 - \varphi_1) + \frac{2a_I}{bJ}i_2\varphi_2. \tag{B.6}$$

Rozwiązań tego układu równań poszukujemy w postaci [153]

$$\varphi_1 = A_1\cos(\omega t - \phi),$$
$$\varphi_2 = A_2\cos(\omega t - \phi), \tag{B.7}$$

a po podstawieniu ich do równania (B.6) otrzymujemy

$$A_1\frac{(2a_I i_1 + bJ(\omega^2 - (\beta+1)\Omega^2))}{bJ} + A_2\beta\Omega^2 = 0,$$
$$A_1\beta\Omega^2 + A_2\frac{(2a_I i_2 + bJ(\omega^2 - (\beta+1)\Omega^2))}{bJ} = 0. \tag{B.8}$$

Układ ten zawsze ma rozwiązania zerowe, gdy $A_1 = 0$ i $A_2 = 0$, które odpowiadają położeniu równowagi. Natomiast rozwiązanie niezerowe jest możliwe, gdy wyznacznik charakterystyczny układu (B.8) jest zerem

$$\begin{vmatrix} \frac{(2a_I i_1 + bJ(\omega^2-(\beta+1)\Omega^2))}{bJ} & \beta\Omega^2 \\ \beta\Omega^2 & \frac{(2a_I i_2 + bJ(\omega^2-(\beta+1)\Omega^2))}{bJ} \end{vmatrix} = 0. \tag{B.9}$$

W efekcie otrzymujemy następujące równanie częstości

$$\frac{4a_I^2 i_1 i_2}{b^2 J^2} + \frac{2a_I(i_1+i_2)(\omega^2 - (\beta+1)\Omega^2)}{bJ} + (\omega^2 - \Omega^2)(\omega^2 - (2\beta+1)\Omega^2) = 0, \tag{B.10}$$

które po rozwiązaniu i przekształceniach matematycznych da dwa różne kwadraty $\omega_{1,2}^2$ częstości drgań swobodnych układu, odpowiadające równaniu (3.5).

# C Metoda Van der Pola dla dwóch wahadeł magnetycznych

W celu otrzymania równań (3.17 3.18) zastosowano metodę analogiczną do metody Van der Pola. Stosując metodę wariacji stałych dowolnych zażądajmy, aby równania prędkości $v_1$ i $v_2$ z (3.15) były równe prędkościom obliczonym z różniczkowania równań na $\varphi_1$ i $\varphi_2$, gdy $E$, $P$, $\Delta$ i $\delta$ są zależne od czasu. W ten sposób otrzymamy następujące dwa równania

$$\sqrt{E(P+1)}\sin(\delta) - \left[\frac{(E'P + E' + P'E)\cos(\delta) - 2\delta' E(P+1)\sin(\delta)}{2\Omega\sqrt{E(P+1)}}\right] = 0,$$
$$\sqrt{E(1-P)}\sin(\delta+\Delta) + \tag{C.1}$$
$$\left[\frac{2E(P-1)(\delta'+\Delta')\sin(\delta+\Delta) + (-PE' + E' - P'E)\cos(\delta+\Delta)}{2\Omega\sqrt{E-EP}}\right] = 0,$$

gdzie $(*') = \frac{d}{dt}(*)$. Przy tym samym założeniu, że $E$, $P$, $\Delta$ i $\delta$ zależą od czasu, obliczamy przyspieszenia różniczkując równania na prędkości $v_1$ i $v_2$ z (3.15), a obliczone formuły wstawiamy do równań (3.6-3.7) otrzymując

$$\left[\frac{2\delta' E(P+1)\cos(\delta) - (E'P + E' + P'E)\sin(\delta)}{2\sqrt{E(P+1)}}\right] + \Omega^2 \frac{\left(\sqrt{E(P+1)}\cos(\delta)\right)}{\Omega} +$$
$$+ f_1\left(\frac{\sqrt{E(P+1)}\cos(\delta)}{\Omega}, \sqrt{E(P+1)}\sin(\delta)\right) = 0,$$

$$\left[\frac{2E(P-1)(\delta' + \Delta')\cos(\delta + \Delta) + (E'(P-1) + P'E)\sin(\delta + \Delta)}{2\sqrt{E-EP}}\right] +$$
$$+ \Omega^2 \frac{\sqrt{E(1-P)}\cos(\delta + \Delta)}{\Omega} + f_2\left(\frac{\sqrt{E(1-P)}\cos(\delta + \Delta)}{\Omega}, \sqrt{E(1-P)}\sin(\delta + \Delta)\right) = 0,$$
(C.2)

gdzie $f_1$ i $f_2$ są funkcjami wychyleń i prędkości kątowych z (3.7).

Rozwiązując układ równań (C.1-C.2) po zmiennych $d\{E, P, \delta, \Delta\}/dt$ otrzymujemy następujący układ

$$E'_u = E(P+1)\Omega \sin(\delta) \left[\cos(\delta)\left(\beta - \frac{2a_I i_1 e^{-\frac{E(P+1)\cos^2(\delta)}{b\Omega^2}}}{bJ\Omega^2}\right) - \frac{E(P+1)\cos^3(\delta)}{6\Omega^2}\right.$$
$$- \frac{2\zeta_1\sqrt{\text{sgn}(P+1)}\text{sgn}(\sin(\delta))}{\sqrt{E(P+1)}}$$
$$\left. + \frac{2\alpha\sqrt{1-P^2}\sin(\delta + \Delta) + \beta\left(-\sqrt{1-P^2}\right)\cos(\delta + \Delta) - 2\alpha(P+1)\sin(\delta)}{P+1}\right]$$
$$- \frac{E(P-1)\sin(\delta + \Delta)}{6bJ\Omega\sqrt{E-EP}}\left[-12a_I i_2\sqrt{E-EP}\cos(\delta + \Delta)e^{-\frac{E(P-1)\cos^2(\delta + \Delta)}{b\Omega^2}}\right.$$
$$- 12\alpha bJ\Omega^2\sqrt{E-EP}\sin(\delta + \Delta) + 12\alpha bJ\Omega^2 \sin(\delta)\sqrt{E(P+1)}$$
$$+ 6b\beta J\Omega^2 \sqrt{E-EP}\cos(\delta + \Delta) - 6b\beta J\Omega^2 \cos(\delta)\sqrt{E(P+1)}$$
$$- bEJ\sqrt{E-EP}\cos^3(\delta + \Delta) + bEJP\sqrt{E-EP}\cos^3(\delta + \Delta)$$
$$\left. - 12b\zeta_2 J\Omega^2\sqrt{\text{sgn}(1-P)}\text{sgn}(\sin(\delta + \Delta))\right],$$

$$P'_u = \frac{1}{48\Omega} \left[ \frac{1}{bJ} \left( 48 a_I i_1 P^2 \sin(2\delta) e^{-\frac{E(P+1)\cos^2(\delta)}{b\Omega^2}} - 48 a_I i_1 \sin(2\delta) e^{-\frac{E(P+1)\cos^2(\delta)}{b\Omega^2}} \right. \right.$$
$$\left. - 48 a_I i_2 P^2 \sin(2(\delta+\Delta)) e^{-\frac{E(P-1)\cos^2(\delta+\Delta)}{b\Omega^2}} + 48 a_I i_2 \sin(2(\delta+\Delta)) e^{-\frac{E(P-1)\cos^2(\delta+\Delta)}{b\Omega^2}} \right)$$
$$- 48\alpha\Omega^2 \cos(2(\delta+\Delta)) + 48\alpha\Omega^2 \cos(2\delta) - 24\beta\Omega^2 \sin(2(\delta+\Delta)) + 24\beta\Omega^2 \sin(2\delta)$$
$$+ 2E\sin(2(\delta+\Delta)) + E\sin(4(\delta+\Delta)) - 2E\sin(2\delta) - E\sin(4\delta) + 2EP^3 \sin(2(\delta+\Delta))$$
$$+ EP^3 \sin(4(\delta+\Delta)) - 2EP^3 \sin(2\delta) - EP^3 \sin(4\delta) - 2EP^2 \sin(2(\delta+\Delta))$$
$$- EP^2 \sin(4(\delta+\Delta)) + 2EP^2 \sin(2\delta) + EP^2 \sin(4\delta)$$
$$+ \frac{96\zeta_2 \sqrt{1-P^2}\,\Omega^2 \sqrt{E(P+1)}\sqrt{\operatorname{sgn}(1-P)} \sin(\delta+\Delta)\operatorname{sgn}(\sin(\delta+\Delta))}{E}$$
$$- 2EP\sin(2(\delta+\Delta)) - EP\sin(4(\delta+\Delta)) - 2EP\sin(2\delta) - EP\sin(4\delta)$$
$$+ \frac{96\zeta_1(P-1)\Omega^2 \sin(\delta)\sqrt{E(P+1)}\sqrt{\operatorname{sgn}(P+1)}\operatorname{sgn}(\sin(\delta))}{E} + 48\alpha P^2 \Omega^2 \cos(2(\delta+\Delta))$$
$$+ \frac{96\alpha \sqrt{1-P^2}\,P^2 \Omega^2 \cos(2\delta+\Delta)}{P+1} + \frac{96\alpha\sqrt{1-P^2}\,P\Omega^2 \cos(2\delta+\Delta)}{P+1} - 48\alpha P^2 \Omega^2 \cos(2\delta)$$
$$- \frac{96\alpha\sqrt{1-P^2}\,P^2\Omega^2 \cos(\Delta)}{P+1} - \frac{96\alpha\sqrt{1-P^2}\,P\Omega^2 \cos(\Delta)}{P+1} + 24\beta P^2 \Omega^2 \sin(2(\delta+\Delta))$$
$$+ \frac{48\beta\sqrt{1-P^2}\,P^2\Omega^2 \sin(2\delta+\Delta)}{P+1} + \frac{48\beta\sqrt{1-P^2}\,P\Omega^2 \sin(2\delta+\Delta)}{P+1} - 24\beta P^2 \Omega^2 \sin(2\delta)$$
$$\left. + \frac{48\beta\sqrt{1-P^2}\,P\Omega^2 \sin(\Delta)}{P+1} + \frac{48\beta\sqrt{1-P^2}\,\Omega^2 \sin(\Delta)}{P+1} \right],$$

$$\Delta'_u = \frac{1}{48(P+1)\Omega} \left\{ \frac{1}{bJ(P-1)} \left[ -4\left(P^2-1\right)\cos(2\delta) e^{-\frac{E(P+1)\cos^2(\delta)}{b\Omega^2}} \left(12 a_I i_1 \right. \right. \right.$$
$$\left. + bJ\left(-6\beta\Omega^2 + EP + E\right) e^{\frac{E(P+1)\cos^2(\delta)}{b\Omega^2}} \right) - 48 a_I i_1 P^2 e^{-\frac{E(P+1)\cos^2(\delta)}{b\Omega^2}} + 48 a_I i_1 e^{-\frac{E(P+1)\cos^2(\delta)}{b\Omega^2}}$$
$$+ 48 a_I i_2 P^2 e^{-\frac{E(P-1)\cos^2(\delta+\Delta)}{b\Omega^2}} + 48 a_I i_2 P^2 \cos(2(\delta+\Delta)) e^{-\frac{E(P-1)\cos^2(\delta+\Delta)}{b\Omega^2}}$$
$$- 48 a_I i_2 e^{-\frac{E(P-1)\cos^2(\delta+\Delta)}{b\Omega^2}} - 48 a_I i_2 \cos(2(\delta+\Delta)) e^{-\frac{E(P-1)\cos^2(\delta+\Delta)}{b\Omega^2}} - 4bEJ\cos(2(\delta+\Delta))$$
$$- bEJ\cos(4(\delta+\Delta)) - 4bEJP^3 \cos(2(\delta+\Delta)) - bEJP^3 \cos(4(\delta+\Delta)) - 6bEJP^3$$
$$+ 4bEJP^2 \cos(2(\delta+\Delta)) + bEJP^2 \cos(4(\delta+\Delta)) + 4bEJP\cos(2(\delta+\Delta))$$
$$+ bEJP\cos(4(\delta+\Delta)) - bEJ(P-1)(P+1)^2 \cos(4\delta) + 6bEJP$$
$$- 48\alpha bJ\Omega^2 \sin(2(\delta+\Delta)) + 48\alpha bJ\Omega^2 \sin(2\delta) + 24b\beta J\Omega^2 \cos(2(\delta+\Delta))$$
$$+ 48\alpha bJP^2\Omega^2 \sin(2(\delta+\Delta)) + 96\alpha bJ\sqrt{1-P^2}\,P\Omega^2 \sin(2\delta+\Delta)$$
$$- 48\alpha bJP^2\Omega^2 \sin(2\delta) - 96\alpha bJ\sqrt{1-P^2}\,\Omega^2 \sin(\Delta) - 24b\beta JP^2\Omega^2 \cos(2(\delta+\Delta))$$
$$\left. - 48b\beta J\sqrt{1-P^2}\,P\Omega^2 \cos(2\delta+\Delta) - 48b\beta J\sqrt{1-P^2}\,P\Omega^2 \cos(\Delta) \right]$$
$$+ \frac{96\zeta_2\sqrt{1-P^2}\,\Omega^2 \sqrt{E(P+1)}\sqrt{\operatorname{sgn}(1-P)} \cos(\delta+\Delta)\operatorname{sgn}(\sin(\delta+\Delta))}{E(P-1)}$$
$$\left. + 96\zeta_1 \Omega^2 \cos(\delta) \sqrt{\frac{P+1}{E}} \sqrt{\operatorname{sgn}(P+1)}\operatorname{sgn}(\sin(\delta)) \right\},$$

$$\begin{aligned}
\delta'_u &= \cos^2(\delta)\left(\beta\Omega - \frac{2a_I i_1 e^{\frac{E(P+1)\cos^2(\delta)}{b\Omega^2}}}{bJ\Omega}\right) - \frac{E(P+1)\cos^4(\delta)}{6\Omega} \\
&+ \Omega - \frac{2\zeta_1\Omega\cos(\delta)\sqrt{\operatorname{sgn}(P+1)\operatorname{sgn}(\sin(\delta))}}{\sqrt{E(P+1)}} \\
&- \frac{\Omega\cos(\delta)\left(-2\alpha\sqrt{1-P^2}\sin(\delta+\Delta)+\beta\sqrt{1-P^2}\cos(\delta+\Delta)+2\alpha(P+1)\sin(\delta)\right)}{P+1}.
\end{aligned}$$
(C.3)

Teraz należy uśrednić prawe strony równań (C.3) stosując operator (3.16), w efekcie otrzymamy

$$\begin{aligned}
\frac{dE}{dt} &= \frac{1}{2\pi}\int_0^{2\pi} E'_u\, d\delta = 2\alpha E\sqrt{1-P^2}\Omega\cos(\Delta) \\
&- \frac{2\Omega}{\pi}\left(\pi\alpha E + 2\zeta_1\sqrt{E(P+1)} + 2\zeta_2\sqrt{E-EP}\right), \\
\frac{dP}{dt} &= \frac{1}{2\pi}\int_0^{2\pi} P'_u\, d\delta = \frac{\Omega}{\pi E}\left(\pi E\sqrt{1-P^2}(\beta\sin(\Delta)-2\alpha P\cos(\Delta))\right. \\
&\left. + 4\zeta_1(P-1)\sqrt{E(P+1)} + 4\zeta_2(P+1)\sqrt{E-EP}\right),
\end{aligned}$$

$$\begin{aligned}
\frac{d\Delta}{dt} &= \frac{1}{2\pi}\int_0^{2\pi}\Delta'_u\, d\delta = \frac{1}{8\Omega bJ}\left\{e^{-\frac{E(P+1)}{2b\Omega^2}}\left[8a_I i_1\left(I_{B0}\left(\frac{E(P+1)}{2b\Omega^2}\right)-I_{B1}\left(\frac{E(P+1)}{2b\Omega^2}\right)\right)\right.\right. \\
&\left.\left. -8a_I i_2 e^{\frac{EP}{b\Omega^2}}\left(I_{B0}\left(\frac{E(P-1)}{2b\Omega^2}\right)-I_{B1}\left(\frac{E-EP}{2b\Omega^2}\right)\right)\right]\right\} + \frac{EP}{8\Omega} - \frac{2\alpha\Omega\sin(\Delta)}{\sqrt{1-P^2}} - \frac{\beta P\Omega\cos(\Delta)}{\sqrt{1-P^2}}, \\
\frac{d\delta}{dt} &= \frac{1}{2\pi}\int_0^{2\pi}\delta'_u\, d\delta = \frac{a_I i_1 e^{-\frac{E(P+1)}{2b\Omega^2}}\left(I_{B1}\left(\frac{E(P+1)}{2b\Omega^2}\right)-I_{B0}\left(\frac{E(P+1)}{2b\Omega^2}\right)\right)}{bJ\Omega} \\
&+ \frac{1}{2}(\beta+2)\Omega - \frac{E(P+1)}{16\Omega} - \frac{1}{2}\sqrt{\frac{2}{P+1}}-1\Omega(\beta\cos(\Delta)-2\alpha\sin(\Delta)),
\end{aligned}$$
(C.4)

Podstawiając wartości $\lambda_{a1} = E(1+P)/\Omega^2 = 2E_{11}/\Omega^2$ i $\lambda_{a2} = E(1-P)/\Omega^2 = 2E_{22}/\Omega^2$ otrzymamy układ równań (3.17).

# D  Procedura linearyzacji dla przypadku drgań o warunkach początkowych bliskich antyfazie

Poniżej przedstawiony zostanie tok postępowania prowadzący do otrzymania zlinearyzowanego wzoru na $P$ dla przypadku drgań o warunków początkowych bliskich antyfazie (równanie (3.20)). Przyjęto, że zmiana energii całkowitej w czasie $dE/dt$ jest mała z czego wynika, że $E$ jest stałe. Linearyzacja została wykonana przy założeniu, że przesunięcie fazowe $\Delta$ jest zaburzone wokół swojego położenia stacjonarnego $\Delta_0$, a współczynnik rozkładu energii $P$ jest zaburzony wokół przyjętego położenia stacjonarnego $P_0$, co można zapisać jako

$$\begin{aligned}\Delta &= \Delta_0 + \epsilon\Delta, \\ P &= P_0 + \epsilon P,\end{aligned} \quad (D.1)$$

gdzie $\epsilon$ jest małym parametrem zaburzenia. Następnie zależności (D.1) zostały wstawione do drugiego i trzeciego równania uśrednionego układu (3.17) w wyniku czego przyjęły następującą postać

$$\begin{aligned}P' =\Omega &\left(\left(\beta\sin(\Delta\epsilon+\Delta_0) - 2\alpha\cos(\Delta\epsilon+\Delta_0)(P\epsilon+P_0)\right)\sqrt{1-(P\epsilon+P_0)^2}\right.\\ &\left.+\frac{4\Omega\left(E\zeta_2\sqrt{\frac{E(-1+P\epsilon+P_0)}{\Omega^2}}(1+P\epsilon+P_0)+E\zeta_1(-1+P\epsilon+P_0)\sqrt{\frac{E(1+P\epsilon+P_0)}{\Omega^2}}\right)}{E^2\pi}\right),\\ \Delta' =\frac{1}{bJ\Omega}&e^{-\frac{E(1+P\epsilon+P_0)}{2b\Omega^2}}\left(-ae^{\frac{E(P\epsilon+P_0)}{b\Omega^2}}i_2\left(I_{B0}\left(\frac{E(-1+P\epsilon+P_0)}{2b\Omega^2}\right)+I_{B1}\left(\frac{E(-1+P\epsilon+P_0)}{2b\Omega^2}\right)\right)\right.\\ &\left.+ai_1\left(I_{B0}\left(\frac{E(1+P\epsilon+P_0)}{2b\Omega^2}\right)-I_{B1}\left(\frac{E(1+P\epsilon+P_0)}{2b\Omega^2}\right)\right)\right)\\ &+\frac{E(P\epsilon+P_0)}{8\Omega}-\frac{\Omega\left(2\alpha\sin(\Delta\epsilon+\Delta_0)+\beta\cos(\Delta\epsilon+\Delta_0)(P\epsilon+P_0)\right)}{\sqrt{1-(P\epsilon+P_0)^2}}.\end{aligned} \quad (D.2)$$

W kolejnym kroku, równania (D.2) zostały rozwinięte w szeregi Taylora wokół $\epsilon = 0$. Podstawiając wartości $\Delta_0 = \pi$ (warunek drgań w antyfazie) i $P_0 = 0$ oraz biorąc pod uwagę tylko dwa pierwsze wyrazy szeregu (odpowiadające $\epsilon^0$ i $\epsilon^1$) otrzymamy

$$\begin{aligned}P' =&\frac{4(-\zeta_1+\zeta_2)}{\pi\sqrt{\frac{E}{\Omega^2}}}+\frac{\epsilon\left(2P\zeta_1+2P\zeta_2+\pi(2P\alpha-\beta\Delta)\sqrt{\frac{E}{\Omega^2}}\Omega\right)}{\pi\sqrt{\frac{E}{\Omega^2}}},\\ \Delta' =&\frac{ae^{-\frac{E}{2b\Omega^2}}(i_1-i_2)\left(I_{B0}\left(\frac{E}{2b\Omega^2}\right)-I_{B1}\left(\frac{E}{2b\Omega^2}\right)\right)}{bJ\Omega}+\epsilon\left(\frac{EP}{8\Omega}+P\beta\Omega+2\alpha\Delta\Omega\right.\\ &\left.+\frac{ae^{-\frac{E}{2b\Omega^2}}(i_1+i_2)P\left(-EI_{B0}\left(\frac{E}{2b\Omega^2}\right)+(E+b\Omega^2)I_{B1}\left(\frac{E}{2b\Omega^2}\right)\right)}{b^2J\Omega^3}\right).\end{aligned} \quad (D.3)$$

Oczywiście dla przypadku drgań w fazie, tok postępowania jest analogiczny, należy tylko przyjąć $\Delta_0 = 0$. Następnie przyjmując $\epsilon = 1$ i upraszczając równania

(D.3) otrzymamy następujące wyrażenia

$$P' = \beta\Delta\Omega - \frac{4(\zeta_1 - \zeta_2)\Omega}{\sqrt{E}\pi} + P\left(2\alpha\Omega + \frac{2(\zeta_1 + \zeta_2)\Omega}{\sqrt{E}\pi}\right),$$

$$\Delta' = 2\alpha\Delta\Omega + \frac{ae^{-\frac{E}{2b\Omega^2}}(i_1 - i_2)\left(I_{B0}\left(\frac{E}{2b\Omega^2}\right) - I_{B1}\left(\frac{E}{2b\Omega^2}\right)\right)}{bJ\Omega}$$

$$+ P\left(\frac{E}{8\Omega} + \beta\Omega + \frac{ae^{-\frac{E}{2b\Omega^2}}(i_1 + i_2)\left(-EI_{B0}\left(\frac{E}{2b\Omega^2}\right) + (E + b\Omega^2)I_{B1}\left(\frac{E}{2b\Omega^2}\right)\right)}{b^2J\Omega^3}\right). \quad (D.4)$$

Do otrzymania wyrażenia na $P$ z wyrugowanym przesunięciem fazowym $\Delta$ należy obliczyć drugą pochodną $P$ korzystając z pierwszego równania (D.4)

$$P'' = \left(2\alpha\Omega + \frac{2(\zeta_1 + \zeta_2)\Omega}{\sqrt{E}\pi}\right)P' - \beta\Omega\Delta' \quad (D.5)$$

i z tego samego równania wyznaczyć wzór na $\Delta$

$$\Delta = \frac{4\zeta_1\Omega + 4\zeta_2\Omega + 2\sqrt{E}\pi\alpha\Omega P + 2\zeta_1\Omega P + 2\zeta_2\Omega P - \sqrt{E}\pi P'}{\sqrt{E}\pi\beta\Omega}. \quad (D.6)$$

Podstawiając do równania (D.5) wyrażenie na $\Delta'$ z (D.4) otrzymamy

$$P'' = -\beta\Omega\left[\frac{ae^{-\frac{E}{2b\Omega^2}}(i_1 - i_2)\left(I_{B0}\left(\frac{E}{2b\Omega^2}\right) - I_{B1}\left(\frac{E}{2b\Omega^2}\right)\right)}{bJ\Omega}\right.$$

$$+ \left(\frac{E}{8\Omega} + \beta\Omega + \frac{ae^{-\frac{E}{2b\Omega^2}}(i_1 + i_2)\left(-EI_{B0}\left(\frac{E}{2b\Omega^2}\right) + (E + b\Omega^2)I_{B1}\left(\frac{E}{2b\Omega^2}\right)\right)}{b^2J\Omega^3}\right)P$$

$$\left. + 2\alpha\Omega\Delta\right] + 2\left(\alpha + \frac{\zeta_1 + \zeta_2}{\sqrt{E}\pi}\right)\Omega P', \quad (D.7)$$

gdzie ostatecznie możemy wyeliminować $\Delta$ wstawiając równanie (D.6), otrzymując

$$P'' = \frac{8\alpha(\zeta_1 - \zeta_2)\Omega^2}{\sqrt{E}\pi} + \frac{ae^{-\frac{E}{2b\Omega^2}}(i_1 - i_2)\beta\left(-I_{B0}\left(\frac{E}{2b\Omega^2}\right) + I_{B1}\left(\frac{E}{2b\Omega^2}\right)\right)}{bJ}$$

$$+ \left(-\frac{1}{8}(E\beta) - (4\alpha^2 + \beta^2)\Omega^2 - \frac{4\alpha(\zeta_1 + \zeta_2)\Omega^2}{\sqrt{E}\pi}\right.$$

$$\left. + \frac{ae^{-\frac{E}{2b\Omega^2}}(i_1 + i_2)\beta\left(EI_{B0}\left(\frac{E}{2b\Omega^2}\right) - (E + b\Omega^2)I_{B1}\left(\frac{E}{2b\Omega^2}\right)\right)}{b^2J\Omega^2}\right)P$$

$$+ 2\left(2\alpha + \frac{\zeta_1 + \zeta_2}{\sqrt{E}\pi}\right)\Omega P'. \quad (D.8)$$

Po przekształceniach matematycznych (D.8) otrzymamy równanie (3.20).

# Literatura

LITERATURA